%% file: mfiwidesurvey.tex
    \documentclass[a4paper,fleqn,usenatbib]{mnras}
    
    \usepackage[T1]{fontenc}
    \usepackage{ae,aecompl}

    \usepackage{graphicx}	
    \usepackage{amsmath}	
    \usepackage{amssymb}	
    \usepackage{multirow}
    
    \usepackage{newtxtext,newtxmath}

    \newcommand\xpol{{\sc Xpol}}
    \newcommand{\picasso}{{\sc PICASSO}}
    \newcommand{\healpix}{{\sc HEALPix}}
    \newcommand{\namaster}{{\sc NaMaster}}
    \newcommand{\nside}{{N_{\rm side}}}
    \newcommand{\rcond}{{r_{\rm cond}}}
    \newcommand{\nhit}{{N_{\rm hit}}}
    \newcommand{\fsky}{{f_{\rm sky}}}
    \newcommand{\omegapix}{\Omega_{\rm pix}}
    \newcommand{\omegadeg}{\Omega_{\text{1-deg}}}
    \newcommand{\sigmadeg}{\sigma_{\text{1-deg}}}

    \title[QUIJOTE MFI wide survey]{QUIJOTE scientific results -- IV. A northern sky survey in intensity and polarization at 10--20\,GHz with the Multi-Frequency Instrument}
    
    \input{authors_and_institutes_mfiwidesurvey.tex}

    \date{Accepted 2022 November 11. Received 2022 November 1; in original form 2022 July 29}
    
    \pubyear{2022}
    
\begin{document}
    \label{firstpage}
    \pagerange{\pageref{firstpage}--\pageref{lastpage}}
    \maketitle

    \begin{abstract}
    We present QUIJOTE intensity and polarization maps in four frequency bands centred around 11, 13, 17 and 19\,GHz, and covering approximately $29\,000$\,deg$^2$, including most of the Northern sky region. These maps result from $9\,000$\,h of observations taken between May 2013 and June 2018 with the first QUIJOTE instrument (MFI), and have angular resolutions of around $1^\circ$, and sensitivities in polarization within the range 35--40\,$\mu$K per 1-degree beam, being a factor $\sim 2$--$4$ worse in intensity. We discuss the data processing pipeline employed, and the basic characteristics of the maps in terms of real space statistics and angular power spectra. A number of validation tests have been applied to characterise the accuracy of the calibration and the residual level of systematic effects, finding a conservative overall calibration uncertainty of 5\,\%. We also discuss flux densities for four bright celestial sources (Tau A, Cas A, Cyg A and 3C274) which are often used as calibrators at microwave frequencies. The polarization signal in our maps is dominated by synchrotron emission. The distribution of spectral index values between the 11\,GHz and WMAP 23\,GHz map peaks at $\beta=-3.09$ with a standard deviation of $0.14$. The measured BB/EE ratio at scales of $\ell=80$ is $0.26\pm 0.07$ for a Galactic cut $|b|>10^\circ$. We find a positive TE correlation for 11\,GHz at large angular scales ($\ell \lesssim 50$), while the EB and TB signals are consistent with zero in the multipole range $30 \lesssim \ell \lesssim 150$. The maps discussed in this paper are publicly available. 
    \end{abstract}
    
    
    
    
    \section{Introduction}
    
    Measurements of the Cosmic Microwave Background (CMB) anisotropies provide one of the most powerful tools in modern cosmology, playing a fundamental role in our current understanding of the physics of the early Universe and structure formation \citep{Bennett2013, Planck2018-i}. 
    Moreover, CMB polarization observations open a window to probe the amplitude of primordial gravitational waves generated during the inflationary epoch \citep{Kamion1997, ZS1997}. Following this scientific motivation, observations of B-modes at large angular scales have progressed substantially over the last few years. Current best upper limits on the tensor-to-scalar ratio come from the BICEP/Keck 2018 CMB polarization data \citep{BK2021}, 
    and give  $r < 0.036$ at 95\% confidence level, which improves to $r < 0.032$ when adding the latest Planck PR4 data \citep{Tristam2022}. Upcoming ground-based experiments like Simons Observatory \citep{so_19} or CMB-S4 \citep{s4_22}, and space missions like LiteBIRD \citep{litebird2022} will improve these constraints in the coming years.
    
    Due to the low amplitude of this primordial B-mode signal, the control and removal of diffuse Galactic foreground contamination in polarization is becoming a key challenge for current and future CMB experiments. 
    Basically there are two main Galactic foregrounds that are known to emit linearly polarized radiation: the synchrotron emission resulting from cosmic ray electrons accelerated around the Galactic magnetic field lines, and the thermal radiation from interstellar dust grains also aligned with the magnetic field \citep{Bennett2013, Planck2015-x, Planck2018-iv}. Anomalous microwave emission (AME) has been also detected in intensity, but no polarization has been measured up to date \citep{AME2012, AME2018}. Although there are theoretical motivations to expect negligible polarization levels if AME is produced by spinning dust grains \citep{DraineHensley2016}, improved low frequency observations will be needed to consolidate our understanding of this physical process. 
    
    The Planck satellite \citep{Planck2018-i} produced seven full sky polarization maps covering the frequency range between 30 and 353\,GHz. The Wilkinson Microwave Anisotropy Probe (WMAP) satellite \citep{Bennett2013} scanned the full sky in polarization in five bands between 23 and 94\,GHz. The analysis of these data shows that, for a B-mode signal with amplitude $r =10^{-3}$ (which is the target of the LiteBIRD space mission), there is no frequency domain or sky region where the sum of the synchrotron and thermal dust foregrounds is subdominant with respect to the expected CMB B-mode signal \citep{PlanckInt2016-xxx, Krachmalnicoff2016}. 
    Moreover, further analyses of these and other datasets show increasing evidence of complexity in the spectral and spatial behaviour of the Galactic dust and synchrotron emissions \citep{ChoiPage2015,PlanckInt2017-L,Krachmalnicoff2018,Fuskeland2021,Weiland2022,deBelsunce2022}. 
    
    The situation is particularly complex for the polarized synchrotron emission. The sensitivity of the low frequency channels from Planck and WMAP does not allow the detection of polarized synchrotron signal at intermediate and high Galactic latitudes, and therefore we are lacking a detailed spectral modelling of this emission precisely in the regions of cosmological interest. In this context, there is a need for complementing the existing satellite observations with measurements at lower frequencies in order to improve our description of the foregrounds at the required level for B-mode studies. 
    There are only a limited number of radio surveys that preserve the large-scale structure of Galactic emission, and most of them provide only intensity measurements \citep{Haslam1982,dwingeloo,reich,hartrao}, but this situation is now changing. The S-band Polarization All-Sky Survey \citep[S-PASS;][]{Carretti2019} recently provided the first map of the polarized radio emission over the southern sky at declinations below $-1^\circ$ taken with the Parkes radio telescope at 2.3\,GHz. The C-Band All Sky Survey \citep[C-BASS;][]{CBASS} will cover the full sky at 5\,GHz, and the maps of the northern sky will be soon available.

    With the aim of providing spectral coverage complementary to WMAP and Planck at intermediate frequencies, the Q-U-I JOint Tenerife Experiment \citep[QUIJOTE,][]{Rubino10} is a scientific collaboration between the Instituto de Astrofisica de Canarias (IAC), the Instituto de Fisica de Cantabria (IFCA), the Universities of Cantabria, Manchester and Cambridge, and the IDOM company. It has the goal of characterising the polarization of the CMB and other Galactic and extragalactic physical processes in the frequency range $10$--$40$\,GHz and at large angular scales ($\ga 1^\circ$). QUIJOTE has been designed to have the required sensitivity to detect a primordial gravitational-wave component if the tensor-to-scalar ratio is larger than $r=0.05$.
    The experiment is located at the Teide Observatory (altitude of 2,400\,m a.s.l) in Tenerife (Canary Islands), and consists of two telescopes equipped with three instruments: the Multi-Frequency Instrument (hereafter, MFI), operating at 10--20GHz, the Thirty-GHz Instrument (TGI) and the Forty-GHz Instrument (FGI).
    The two QUIJOTE telescopes, QT-1 \citep{QT1} and QT-2 \citep{QT2,QT2b}, are based on an offset crossed-Dragone design with
    projected apertures of $2.25$ and $1.89$\,m for the primary and secondary mirrors respectively, and provide optimal polarization properties 
    (polarization leakage $\le -25$dB), low sidelobes ($\le -40$\,dB) and highly symmetric beams (ellipticity $\le 2$\,\%).
    
    MFI is a multi-channel instrument that has been operating between November 2012 and October 2018 mounted on the first QUIJOTE telescope, QT-1. 
    MFI consists of four polarimeters (also called here "horns"). Horns 1 and 3 operate in the band 10--14\,GHz, while horns 2 and 4 operate at 16--20\,GHz.
    Using frequency filters in the back-end module (hereafter BEM) of the instrument, each horn provides outputs in two frequency sub-bands, each one with an approximate bandwidth of $\Delta \nu = 2$\,GHz. There are a total of 8 outputs for each polarimeter, and these are then fed into the Data Acquisition Electronics (DAE). In total, the MFI provides four frequency bands centred around 11, 13, 17 and 19\,GHz, with each band covered by two independent horns.
    The approximate angular resolution, given in terms of the full width at
    half-maximum, is 52\,arcmin for the low-frequency bands (11 and 13\,GHz), and 38\,arcmin for the 17 and 19\,GHz channels.
    During the lifetime of the instrument, we had basically two instrumental configurations for the MFI. The main difference of the second configuration with respect to the first one is the integration of $90^\circ$ hybrid couplers in each polarimeter, giving correlated outputs in all four detectors. 
    A more detailed description of the instrument can be found in
    \cite{MFIstatus12, status2016SPIE}, and will be included in a future paper (Hoyland et al., in prep).
    A complete description of the MFI instrument characteristics, as well as the MFI data processing pipeline, is included in an accompanying paper 
    \citep[][]{mfipipeline}.
    
    As described in \cite{Rubino10}, most of the QUIJOTE-MFI observing time was dedicated to two main surveys: a shallow Galactic survey (hereafter the "wide survey") covering all the visible sky from Tenerife at elevations larger than $30^\circ$, and a deep cosmological survey covering approximately $3\,000$\,deg$^2$ in three separated sky patches in the northern sky. 
    In addition to those two main surveys, a fraction of the MFI observing time was dedicated to raster scan observations in some selected Galactic regions.
    Data from some of those MFI raster scan observations were already presented in three QUIJOTE collaboration
    publications \citep{Perseus,W44,Taurus}, where we characterised the presence of AME towards several Galactic molecular complexes, as
    the Perseus region, W43, W47 or Taurus, and towards a supernova remnant, W44. In particular, the study of W43 provides the strongest upper limits to date on the polarization fraction of the AME \citep{W44}. Additional raster scan observations were carried out in W51, IC443, rho-Ophiucus, and M31, among others.
    
    A preliminary version of the MFI wide survey maps, in combination with C-BASS North data, were used in the study of the $\lambda$-Orionis region \citep{LambdaOrionis}. 
    %
    This paper presents the final maps of the QUIJOTE-MFI wide survey. Section~\ref{sec:data} describes the observations and the data processing pipeline.
    The final maps are presented in Section~\ref{sec:maps}. The validation and characterisation of these maps is presented in 
    Section~\ref{sec:validation}. An assessment of the overall calibration uncertainty of the maps is discussed in Section~\ref{sec:cal}, while
    Section~\ref{sec:sims} describes the generation of specific noise simulations for the QUIJOTE MFI wide survey. 
    Sections~\ref{sec:spectra}, \ref{sec:properties} and \ref{sec:sources} discuss some of the basic properties of the maps both in real and harmonic space, including 
    the photometry results of some bright radio sources. 
    Finally, Section~\ref{sec:release} describes the data products and associated scientific papers accompanying this paper. All of them are devoted to the understanding of the low frequency Galactic foregrounds in intensity and polarization, either in the full QUIJOTE MFI footprint or in localised regions, and using various analysis techniques. The conclusions of this work are presented in Section~\ref{sec:conclusions}.

    \section{The QUIJOTE-MFI wide survey data}
    \label{sec:data}
    
\begin{figure*}
    \centering
    \includegraphics[trim={0 10 5 20}, clip,width=13cm]{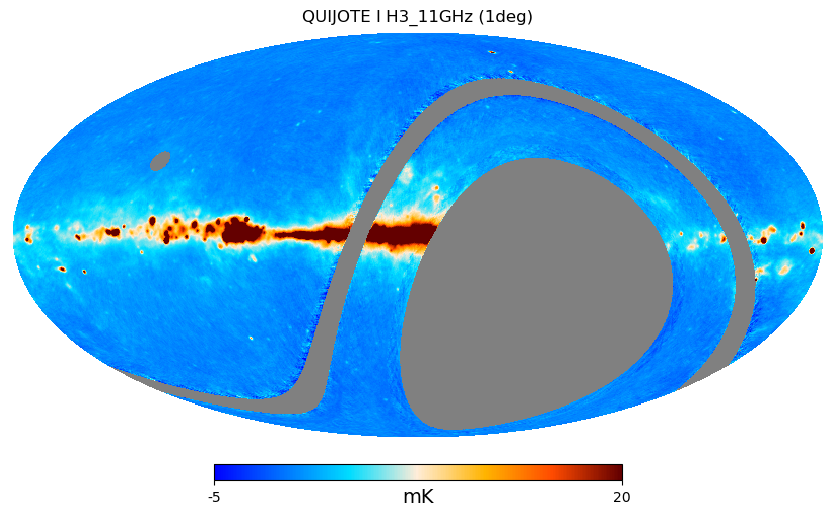}
    \includegraphics[trim={0 10 5 20}, clip,width=13cm]{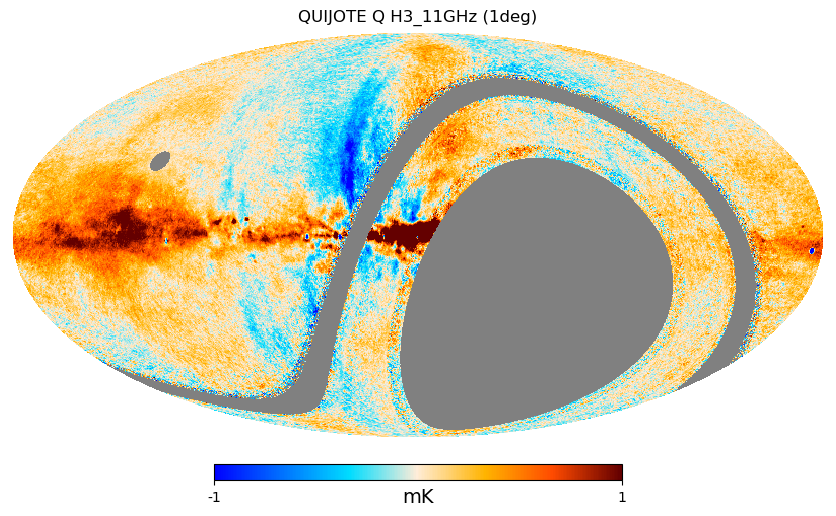}
    \includegraphics[trim={0 10 5 20}, clip,width=13cm]{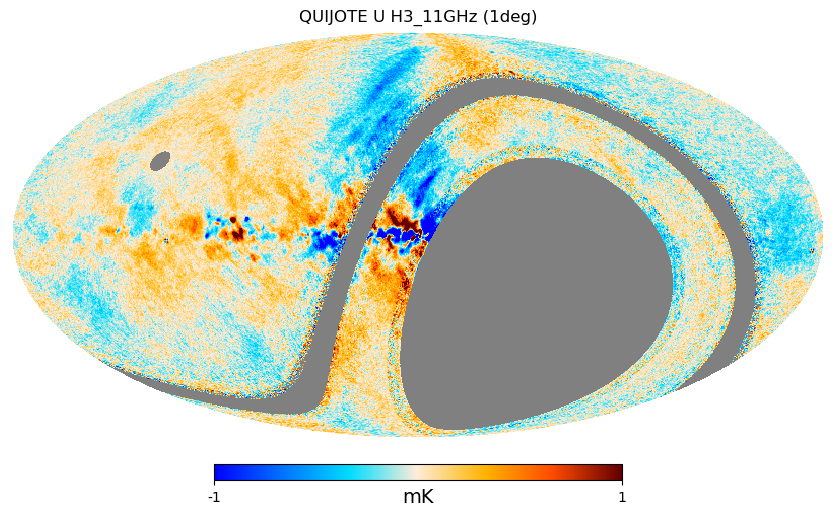}
    \caption{QUIJOTE MFI maps at 11\,GHz in Galactic coordinates, smoothed to 1 degree resolution and using $\nside = 512$. Top: intensity $I$. Middle: polarization $Q$ component. Bottom: polarization $U$ component.  }
    \label{fig:iqumaps_11ghz_1deg}
    \end{figure*}

\begin{figure*}
    \centering
    \includegraphics[trim={0 10 5 20}, clip,width=13cm]{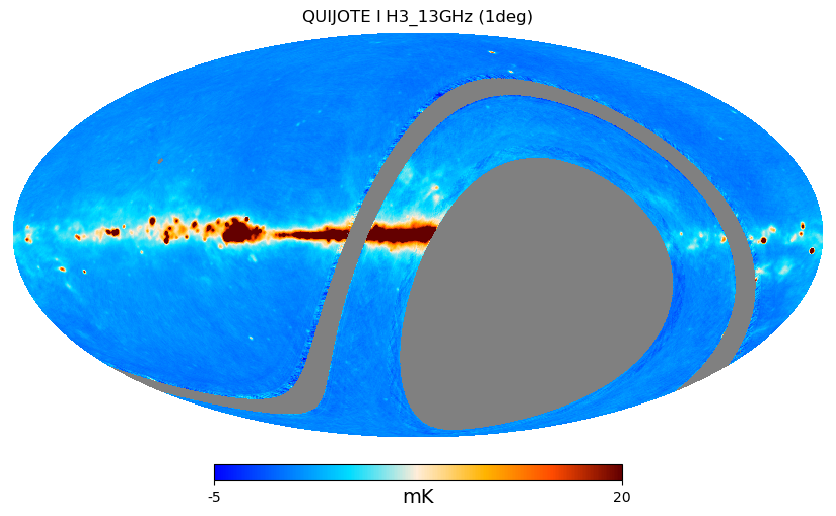}
    \includegraphics[trim={0 10 5 20}, clip,width=13cm]{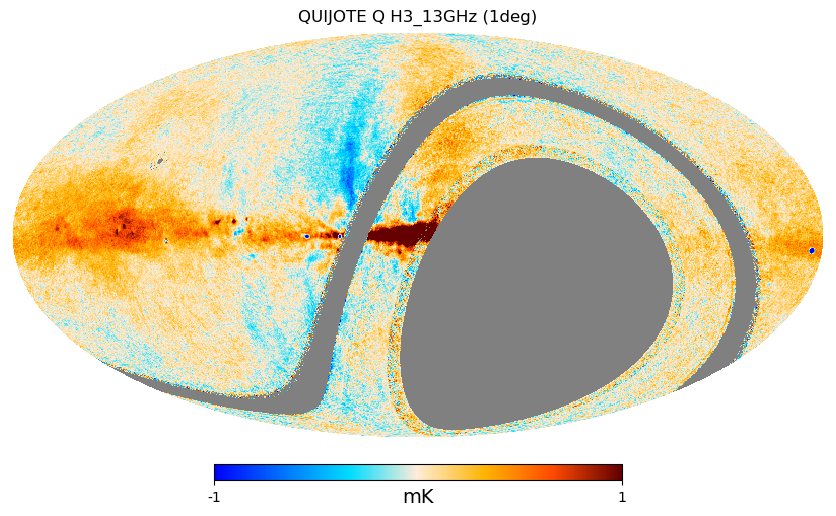}
    \includegraphics[trim={0 10 5 20}, clip,width=13cm]{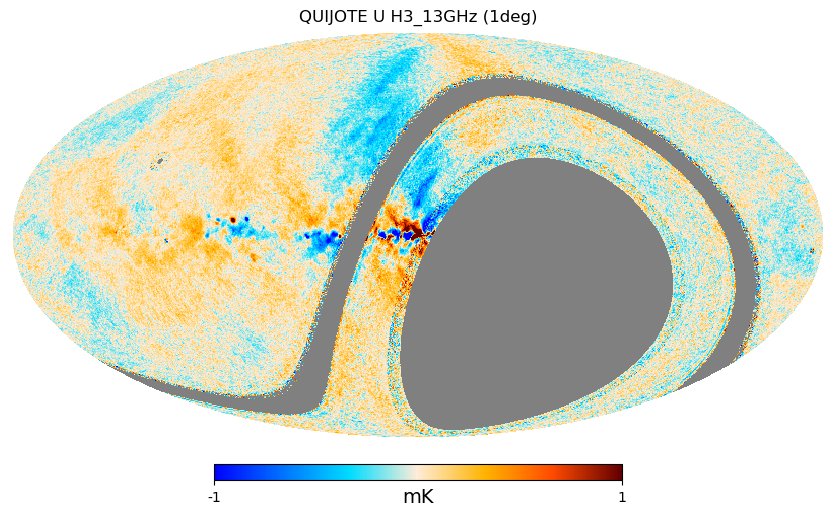}
    \caption{QUIJOTE MFI maps at 13\,GHz smoothed to 1 degree resolution. Top:
      intensity $I$. Middle: polarization $Q$ component. Bottom: polarization $U$ component.  }
    \label{fig:iqumaps_13ghz_1deg}
    \end{figure*}

\begin{figure*}
    \centering
    \includegraphics[trim={0 10 5 20}, clip,width=13cm]{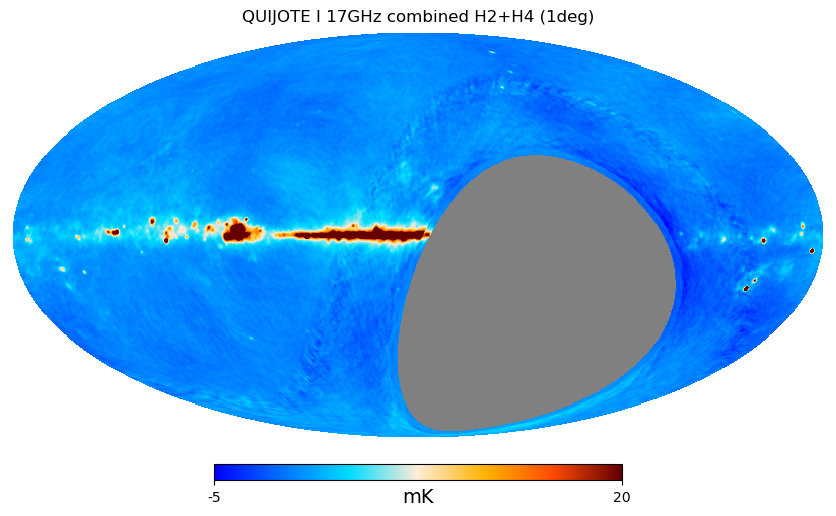}
    \includegraphics[trim={0 10 5 20}, clip,width=13cm]{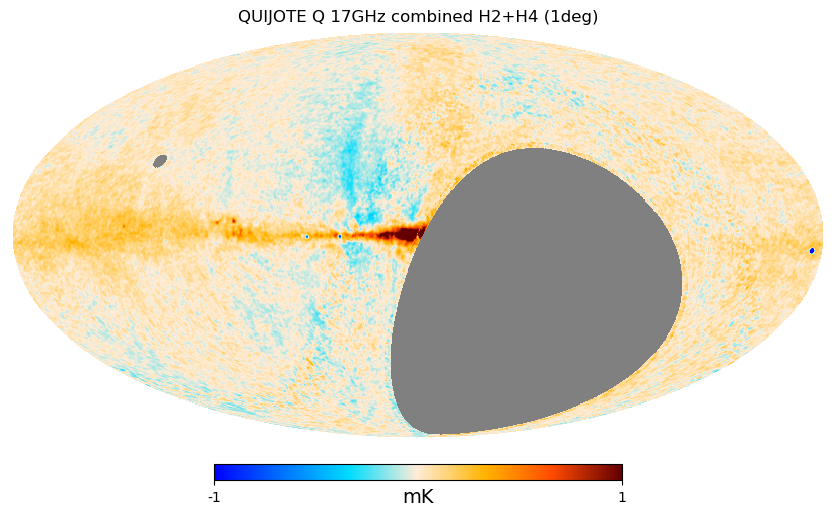}
    \includegraphics[trim={0 10 5 20}, clip,width=13cm]{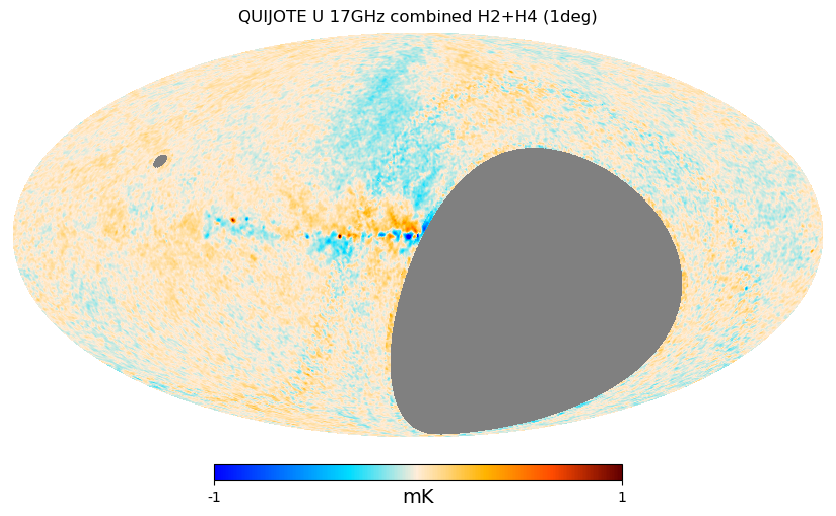}
    \caption{QUIJOTE MFI maps at 17\,GHz smoothed to 1 degree resolution. Top:
      intensity $I$. Middle: polarization $Q$ component. Bottom: polarization $U$ component.  }
    \label{fig:iqumaps_17ghz_1deg}
    \end{figure*}

\begin{figure*}
    \centering
    \includegraphics[trim={0 10 5 20}, clip,width=13cm]{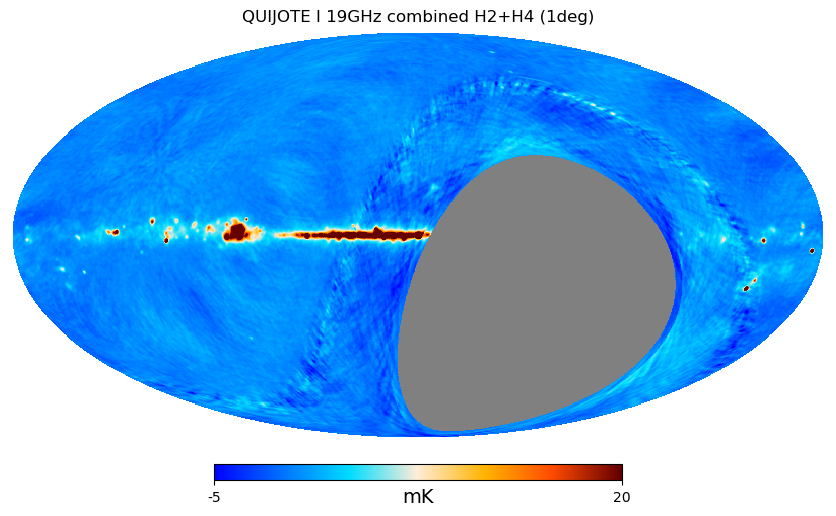}
    \includegraphics[trim={0 10 5 20}, clip,width=13cm]{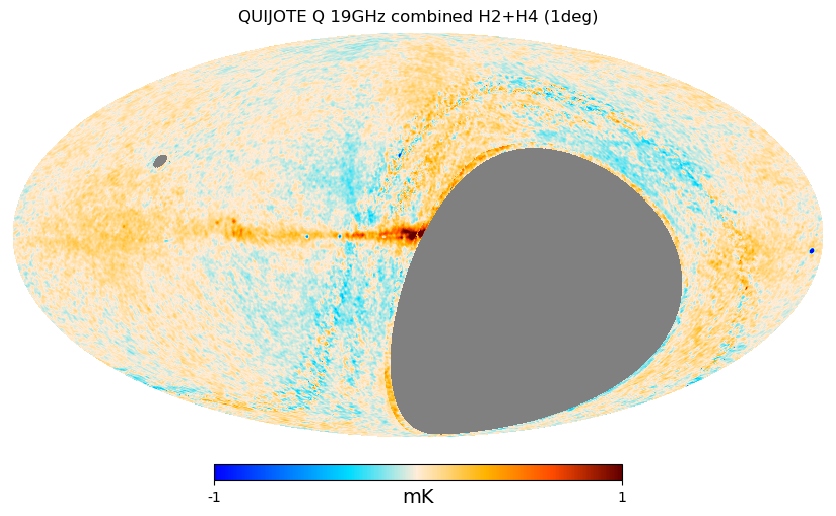}
    \includegraphics[trim={0 10 5 20}, clip,width=13cm]{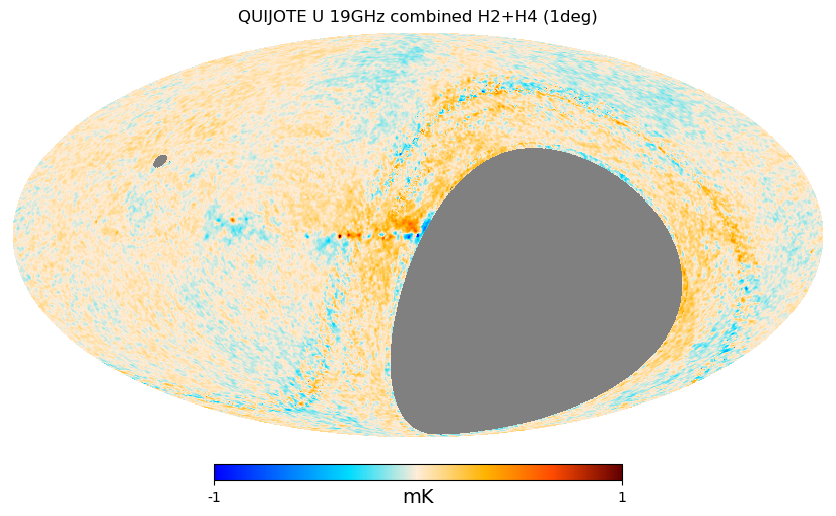}
    \caption{QUIJOTE MFI maps at 19\,GHz smoothed to 1 degree resolution. Top:
      intensity $I$. Middle: polarization $Q$ component. Bottom: polarization $U$ component.  }
    \label{fig:iqumaps_19ghz_1deg}
    \end{figure*}

\begin{figure*}
    \centering
    \includegraphics[trim={0 10 5 20}, clip,width=13cm]{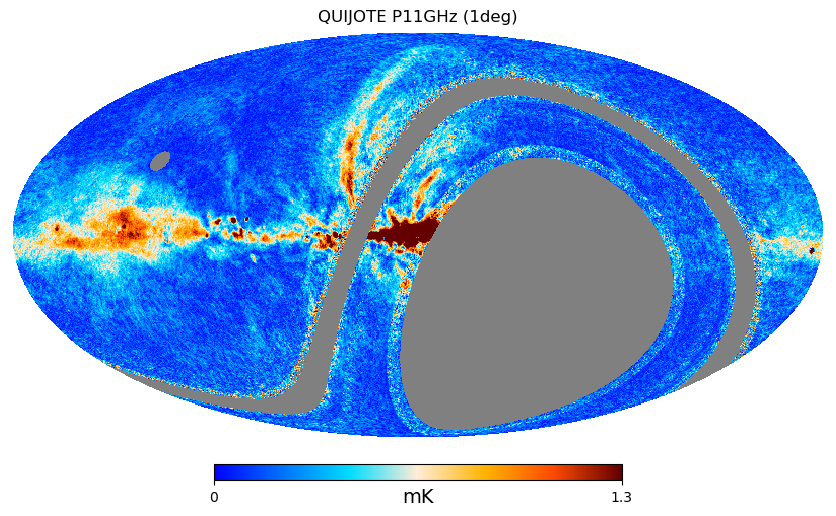}
    \includegraphics[trim={0 5 5 20}, clip,width=13cm]{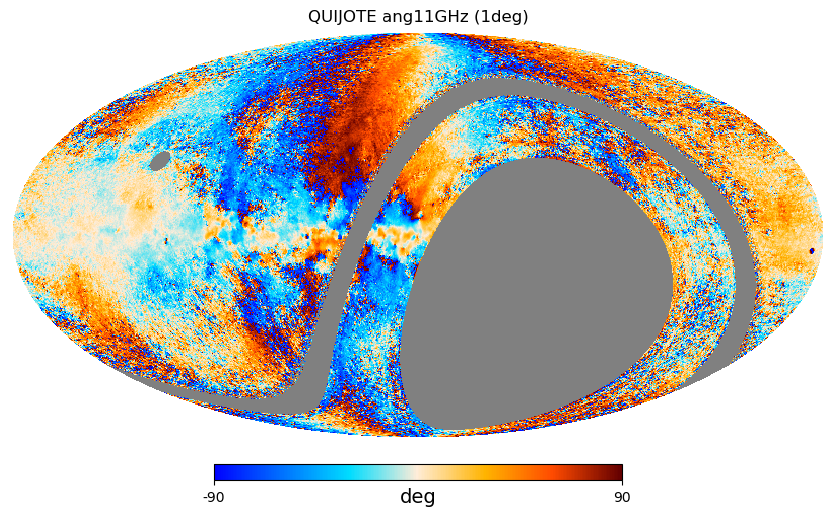}
    \includegraphics[trim={0 10 5 20}, clip,width=13cm]{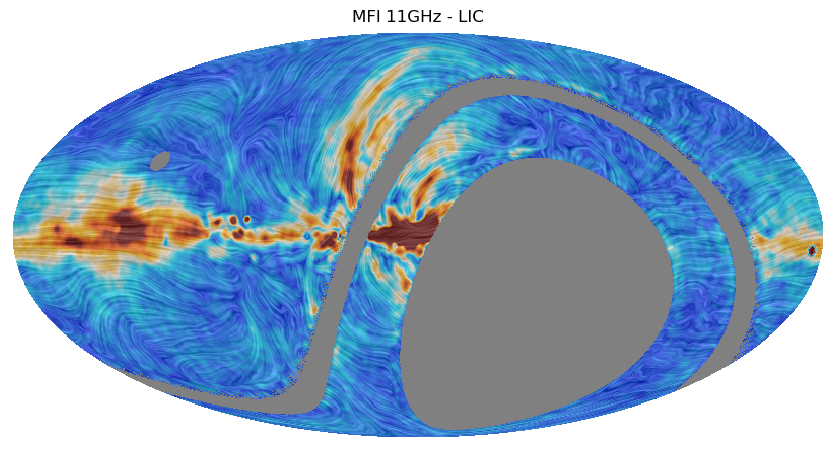}
    \caption{QUIJOTE MFI maps at 11\,GHz smoothed to 1 degree resolution. Top: polarized 
    intensity $P=\sqrt{Q^2+U^2}$. Middle: polarization angle. Bottom: Polarization angle at 11\,GHz, 
    rotated by $90^\circ$ to indicate the direction of the Galactic magnetic field projected on the plane of the sky. 
    The colours represent the polarized intensity signal. The "drapery" pattern was obtained with 
    the healpy routine {\it line\_integral\_convolution}, and it is smoothed to $2^\circ$ for display purposes. }
    \label{fig:p11_ang11_1deg}
    \end{figure*}

    The QUIJOTE wide survey is a shallow survey which covers all the visible sky from the Teide Observatory (latitude $+28.3^\circ$) with elevations greater than $30^\circ$ (more than $29\,000$\,deg$^2$). This was one of the main scientific objectives of QUIJOTE \citep{RubinoSPIE12}, and in particular, of the MFI instrument. 
    This paper presents the QUIJOTE MFI wide survey maps, which were obtained with approximately $9\,000$\,h of observing time. The four final maps at nominal frequencies 11, 13, 17 and 19\,GHz, smoothed to 1 degree resolution, are shown in Figs.~\ref{fig:iqumaps_11ghz_1deg}, \ref{fig:iqumaps_13ghz_1deg}, \ref{fig:iqumaps_17ghz_1deg} and \ref{fig:iqumaps_19ghz_1deg}, respectively.  All maps were generated using the \healpix\footnote{\url{https://healpix.sourceforge.io}} pixelization scheme \citep{healpix} with $\nside = 512$. In \healpix\ the sphere is divided into $12\nside^2$ pixels of equal area. In particular, $\nside = 512$ corresponds to a pixel size of approximately $6.9$ arcmin on the sky.
    Figure~\ref{fig:p11_ang11_1deg} also shows the polarized intensity ($P=\sqrt{Q^2+ U^2}$), the polarization angle direction\footnote{QUIJOTE polarization maps use the COSMO convention from \healpix, so we use a minus sign in the definition of $\gamma$ to recover the IAU convention for the angle.} ($\gamma = 0.5 \arctan(-U/Q)$), and the direction of magnetic field lines for the 11\,GHz map. 
    In the following subsections we describe the observations, the data processing pipeline, the map-making and the specific post-processing and recalibration applied to these maps. 
    
    \subsection{Observations}
    
    The maps described in this paper are based on MFI observations carried out between May 2013 and June 2018 using the so-called "nominal mode", which consists of continuous ($360^\circ$) azimuth scans at a constant telescope elevation. 
    The default azimuth scan speed was $v_{\rm AZ} = 6$\,deg\,s$^{-1}$ from the beginning of the survey until January 9th 2014,
    but this was increased to $v_{\rm AZ} = 12$\,deg\,s$^{-1}$ after this date, in order to reduce the $1/f$ noise contribution in the intensity maps. In this observing mode, every day each MFI horn covers a continuous band of $360^\circ$ in right ascension, and a certain declination range specified by the elevation of the telescope. As in all QUIJOTE-MFI observations, and in order to minimize systematic effects in the polarization parameters, observations are carried out in four discrete positions of the polar modulators $\theta_{\rm pm} =$($0^\circ$, $22.5^\circ$, $45^\circ$ and $67.5^\circ$). In the wide survey, each observation at a given elevation and modulator angle position has a typical duration of 24\,h. 
    
    The combination of multiple elevations allows us to obtain a more homogeneous sampling of the sky. Table~\ref{tab:elevations} contains the final set of telescope elevations considered here to produce the maps, together with the total number of hours observed and used in each case. In total, there are approximately $9\,200$\,h of observations, equivalent to 383 observing days. Almost all of this observing time was suitable for use in the preparation of the intensity maps. However, the final polarization maps only use of the order of $5\,700$\,h, as explained below.
    
    Observations are also separated in periods of several months. The definition of each period is usually associated with changes either in the MFI instrument configuration, telescope configuration, or simply to new observing cycles after instrument maintenance. A complete description of those periods, as well as the associated instrument changes, can be found in \citet{mfipipeline}.  We note that for the MFI wide survey, we conducted observations only during periods 1, 2, 5 and 6. The global dates and effective epoch (year) for each of those periods are listed in Table~\ref{tab:periods}.

    As noted in this table, an extended shielding  was installed in the first QUIJOTE telescope (QT-1) at the beginning of period 2. The main reason for this was to minimize the impact of far sidelobes due to the emission of geo-stationary satellites, which were particularly important for horn 1 \citep{mfipipeline}. In addition, during the operations horn 1 was either not operative (periods 5 and 6) or had problems with the positioning of the polar modulator (period 2). Because of these reasons, although wide-survey maps of horn 1 have been produced for internal consistency tests, they have not been used for this paper because they are significantly affected by systematic effects.

\input{table1_elevations.tex}

\input{table_periods.tex}

    \subsection{Data processing pipeline}
    A complete description of the MFI data processing pipeline can be found in the MFI pipeline paper \citep{mfipipeline}. Here, we summarize the basic characteristics of the MFI data, and we discuss those aspects which are specific of the MFI wide survey. 

    Each MFI polarimeter is divided into a lower and upper band of approximately 2\,GHz bandwidth which is defined by the bandpass filters. Each sub-band has four outputs, which are labelled as $(V_{\rm x+y}, V_{\rm x-y}, V_{\rm x}, V_{\rm y})$. The first two outputs are called "correlated" channels because in the first (original) configuration of the instrument they passed through a $180^\circ$-hybrid, and therefore they have correlated (common) $1/f$ noise properties. The second pair is called "uncorrelated" channels, and in the original configuration provided two outputs with independent noise. 
    The first instrument configuration \citep{MFIstatus12} was used during periods 1 and 2 (see Table~\ref{tab:periods}), but a new configuration was later implemented using $90^\circ$-hybrids \citep{status2016SPIE}. In this second configuration, all MFI channels are formally correlated, but for historical reasons we maintain the notation of correlated and uncorrelated channels. 
    
    The sum of pairs of channels provides two independent measurements of the intensity. For example, for the first MFI configuration, we have
    \begin{align}
    \label{eq:mfi_response_i_u}
    V_{\rm x} + r_{\rm u} V_{\rm y} &= s_{\rm x} g^2 I\\
    V_{\rm x+y} + r_{\rm c} V_{\rm x-y} &= s_{\rm x+y} g^2 I,
    \end{align}
    while the difference of the pairs of channels provides two measurements of the linear polarization
    \begin{align}
    \label{eq:mfi_response_u} 
    V_{\rm x} - r_{\rm u} V_{\rm y} &=  s_{\rm x} g^2 \Big( Q \cos(4\theta_{\rm pm}+2\gamma_{\rm p}) - U \sin(4\theta_{\rm pm}+2\gamma_{\rm p}) \Big) \\
    \label{eq:mfi_response_c} 
    V_{\rm x+y} - r_{\rm c} V_{\rm x-y} &= s_{\rm x+y} g^2 \Big( Q\sin(4\theta_{\rm pm}+2\gamma_{\rm p})+U\cos(4\theta_{\rm pm}+2\gamma_{\rm p}) \Big), 
    \end{align}
    where $V_i$ represents the output voltage for channels $i \in \{ {\rm x}, {\rm y}, {\rm x+y}, {\rm x-y} \}$,  $s_{\rm x}$ and $s_{\rm x+y}$ are the responsivities of those branches in the MFI instrument, $g$ represents the voltage gain of the two MFI Low Noise Amplifiers (here taken to be the same in the two LNAs for simplicity), $r_{\rm c}$ and $r_{\rm u}$ are the so-called r-factors which measure the possible gain and responsivity imbalance in the pair of channels, $\theta_{\rm pm}$ is the position angle of the polar modulator, and $\gamma_{\rm p}$ is the parallactic angle \citep[see details in][]{mfipipeline}. When the two channels in the pair have correlated noise, then the difference cancels significantly the $1/f$ component. 
    In the MFI pipeline, maps for correlated and uncorrelated channels are produced separately, and combined afterwards. Due to their noise properties, in polarization we use only those pair of channels with common $1/f$ properties, i.e. the "correlated" channels during periods 1 and 2, and both of them ("correlated" and "uncorrelated" channels) for periods 5 and 6.
    
    \input{table_general_mfi_parameters.tex}

\input{table_cc_coeff.tex}

    The MFI data sampling rate is 1\,ms. For the wide survey, all time streams (hereafter Time-Ordered Data or TODs) are binned in 40\,ms samples. Note that this is different from the binning scheme of 60\,ms used for raster scan observations in the past \citep[e.g.][]{W44}, due to the higher azimuth scan speed. The binning process allows us to assign a variance $\sigma_i^2$ to each binned sample $i$, which we used to define the associated weights ($w_i = 1/\sigma_i^2$). 
    When propagated through the entire pipeline, the resulting weight maps are used for the combination of maps from correlated and uncorrelated channels, and will be used also in the noise characterization. 
    
    Table~\ref{tab:mfi_parameters} contains the summary of basic parameters (central frequencies, beams, solid angles) for all MFI horns, extracted from \cite{mfipipeline}. We also include the calibration uncertainties discussed in Sect.~\ref{sec:cal}, and representative noise characteristics (knee frequencies and $1/f$ slopes) that we have obtained from this data. Table~\ref{tab:mfi_cc} also presents the colour corrections for these maps, derived from the associated bandpasses as explained in \cite{mfipipeline}. Colour corrections are presented here in terms of second order polynomials as a function of the spectral index $\alpha$. For a sky emission having a flux density law $S_\nu \propto \nu^{\alpha}$, the coefficients $C(\alpha, \nu_0)$ provide the multiplicative correction factor to the measured flux density for the MFI frequency map at nominal frequency $\nu_0$. These corrections are identical for intensity and polarization. 
    
    Throughout the paper, we use the following notation to refer to specific MFI maps per horn and frequency. We will use three numbers, the first one refering to the horn number (i.e. 2, 3 or 4), and the other two indicating the nominal frequency (i.e. 11, 13, 17 or 19). For example, the 19\,GHz map for horn 4 will be cited either as $m_{4,19}$, $m_{419}$, or directly, 419 map. We recall that each map will be made, in principle, from the contribution of both the correlated and uncorrelated channels. In some case, we use the same notation to refer to channels. For example, the correlated channels of 419 are obtained from the $V_{\rm x+y}$ and $V_{\rm x-y}$ outputs of horn 4 at 19\,GHz.

    In the following, we discuss specific additions to the MFI pipeline in the case of the wide survey. 
    In particular, we discuss the gain model for wide survey data and the specific data flagging applied in ''nominal mode". 
    After this, we present our approach to correct for Radio Frequency Interference (RFI) signals and atmospheric contamination in the MFI wide survey data. 
    For these corrections, the general philosophy adopted in our pipeline follows a two step approach. We first implement specific methods to 
    detect and mitigate the effect of RFI and atmospheric signals both at the TOD (see Sect.~\ref{sec:rficorr} and \ref{sec:atmos}) and at the 
    map-level in the post-processing stage (Sect.~\ref{sec:postprocessing}). Then, a detailed assessment is made later of residual signals in the maps by a variety of techniques (Sect.~\ref{sec:validation}). In practice, the values of uncertainties in calibration and other error budgets are increased appropriately if there is clear evidence of residual effects still being present in the maps (Sect.~\ref{sec:cal}).

    \subsubsection{Gain model}
    Gain calibration and the associated relative gain factors ($r_{\rm c}$ and $r_{\rm u}$) between pairs of channels are based on Cas A and Tau A observations taken during each period. Relative gain variations with respect to the mean gain value $G_0$ during the full period are traced using the signal of a thermally stabilized calibration diode, located at the centre of the secondary mirror. Every 30\,s, the diode injects a signal during 1\,s, which is used to measure the relative gain of each channel, $\delta G(t) \equiv G(t) - G_0$ \citep[see][for details]{mfipipeline}. 
    Nominal mode observations used for the wide survey usually have a duration of one day for each polarimeter position. Specifically for this nominal mode data, a smooth (interpolated) gain model is obtained by applying a top-hat smoothing kernel on the individual gain measurements. The width of this kernel is 30 minutes for low frequency channels, and 120 minutes for high frequency ones, due to the different signal-to-noise ratio of the diode signal in the different channels. We have checked that the typical MFI gain variations occur on timescales longer than those. These interpolated models are used to correct the instrument gain as
    \begin{equation}
        G(t) = G_0 \Bigg(1 + \frac{\delta G(t)}{G_0} \Bigg).
    \end{equation}
    \noindent
    Once these interpolated gain models are generated for the entire survey, they are inspected in order to find residual features (peaks or jumps) in the models. These features are introduced in flagging tables which are later applied during the generation of the calibrated TOD.

    \subsubsection{Data flagging}
    \cite{mfipipeline} describes the basic data flagging that is applied by default to all MFI observations, including flags due to voltage ranges, house-keeping parameters, emission of the Sun and Moon (using a $10^\circ$ exclusion radius), and also the emission of geo-stationary satellites. In particular, this last flagging produces the empty strip around declination zero degrees that is seen in the 11 and 13\,GHz maps (Figures~\ref{fig:iqumaps_11ghz_1deg} and \ref{fig:iqumaps_13ghz_1deg}), and also the noise increase in the same region in the 17 and 19\,GHz maps (Figures~\ref{fig:iqumaps_17ghz_1deg} and \ref{fig:iqumaps_19ghz_1deg}), due to the lower number of independent crossings in the area. 
    
    For the wide survey, a specific flagging based on the root mean square (rms) of the data in each scan has been implemented as follows. A first version of the wide survey maps is produced with the default pipeline.  
    From here, and separately for each period, we compute the rms of the data minus the reprojected version of that map onto the TOD, in scales of 30\,s. This time value corresponds to the length of one azimuth scan at the default scanning speed, and to the length of half azimuth scan for the scanning speed used in part of period 1. Histograms with the distribution of these rms values are built for each channel and period, and are used to flag those scans with extreme rms values (either above 1.7 times the median rms value in the entire period, or below 0.5 times that median rms). The fraction of excluded data using this procedure depends on the channel, but typically is of the order of 10--20 per cent. Once this flagging is applied, no obvious residual spikes or rings are visible in the reconstructed maps. 
    Finally, for the final wide survey maps we also exclude Jupiter, Venus and Mars, using a $2^\circ$ exclusion radius directly in the TOD. 
    Appendix~\ref{app:dataflagged} contains detailed tables with the percentage of used (and flagged) data for each MFI channel in every observing period. Those fractions of used data apply to the total number of used hours in each case, which were listed in Table~\ref{tab:elevations}. 
    On average we are using 61\,\% of the data after applying all the different flags. Out of the flagged 39\,\%, most of it (approximately three quarters) is excluded in the specific post-processing stage described in this subsection. 
    The percentage of used data was slightly lower in period 2 (52\,\%), and higher in period 6 (68\,\%).

    \subsubsection{RFI correction}
    \label{sec:rficorr}
    
    \begin{figure*}
    \centering
    \includegraphics[width=0.65\columnwidth]{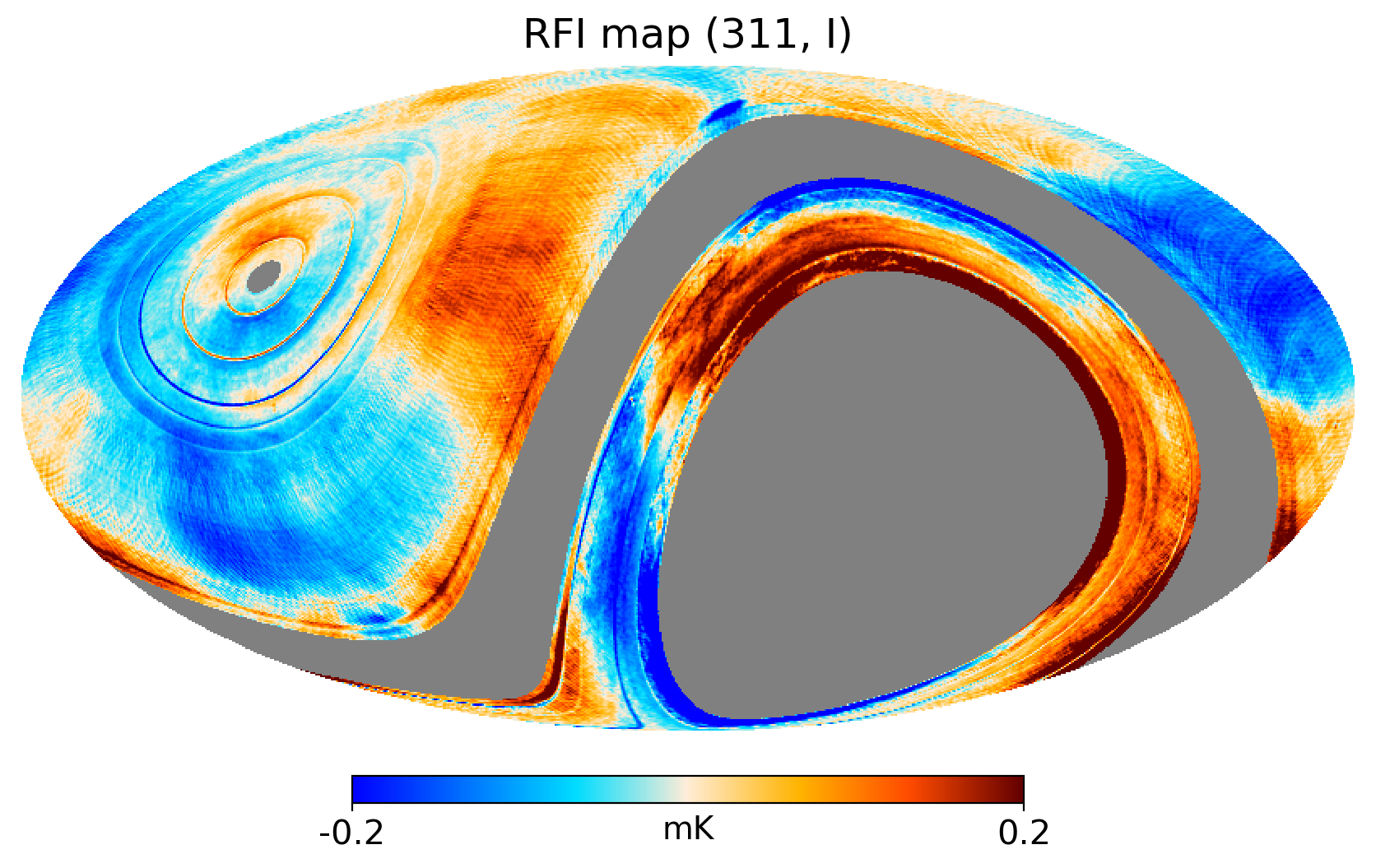}%
    \includegraphics[width=0.65\columnwidth]{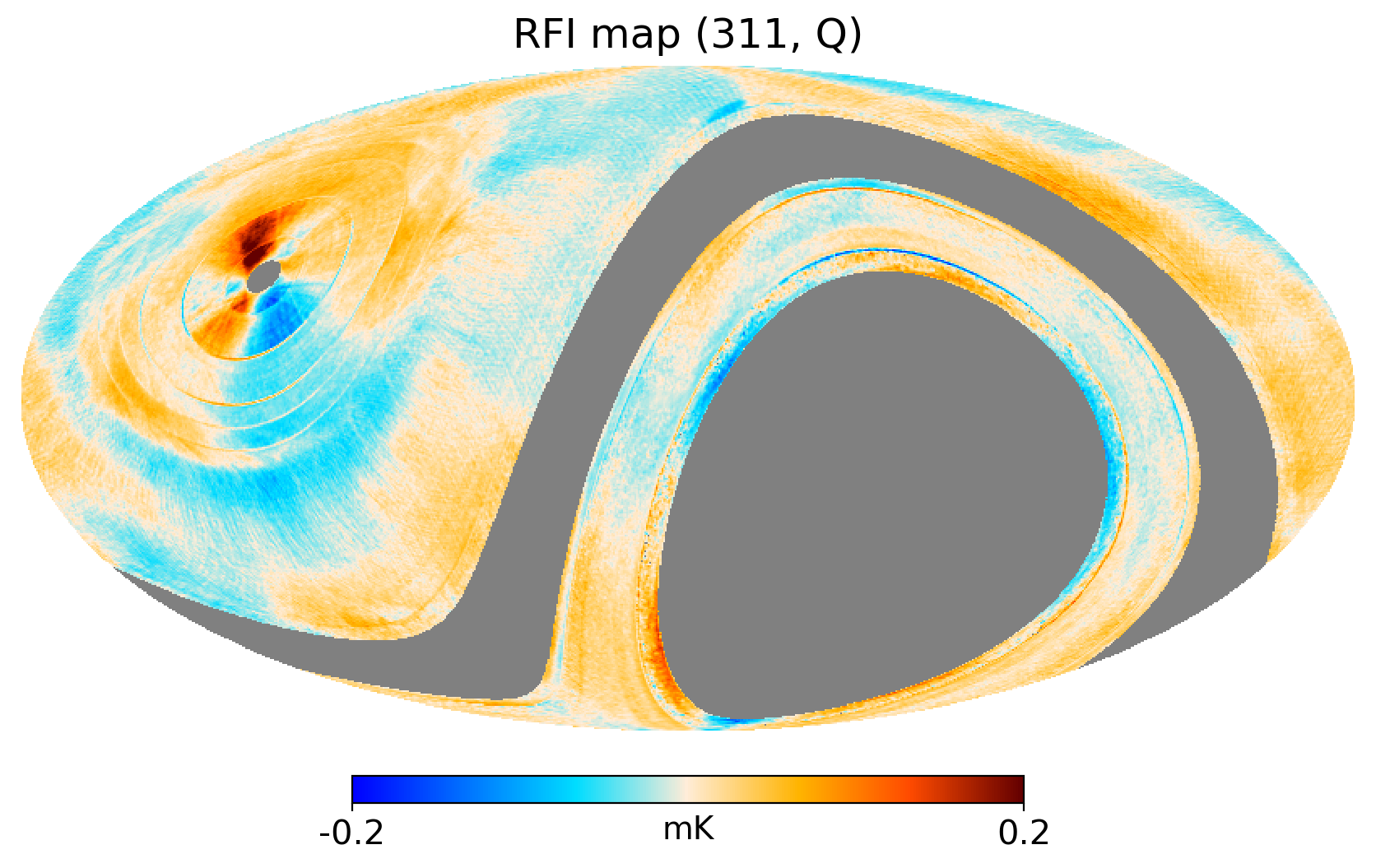}%
    \includegraphics[width=0.65\columnwidth]{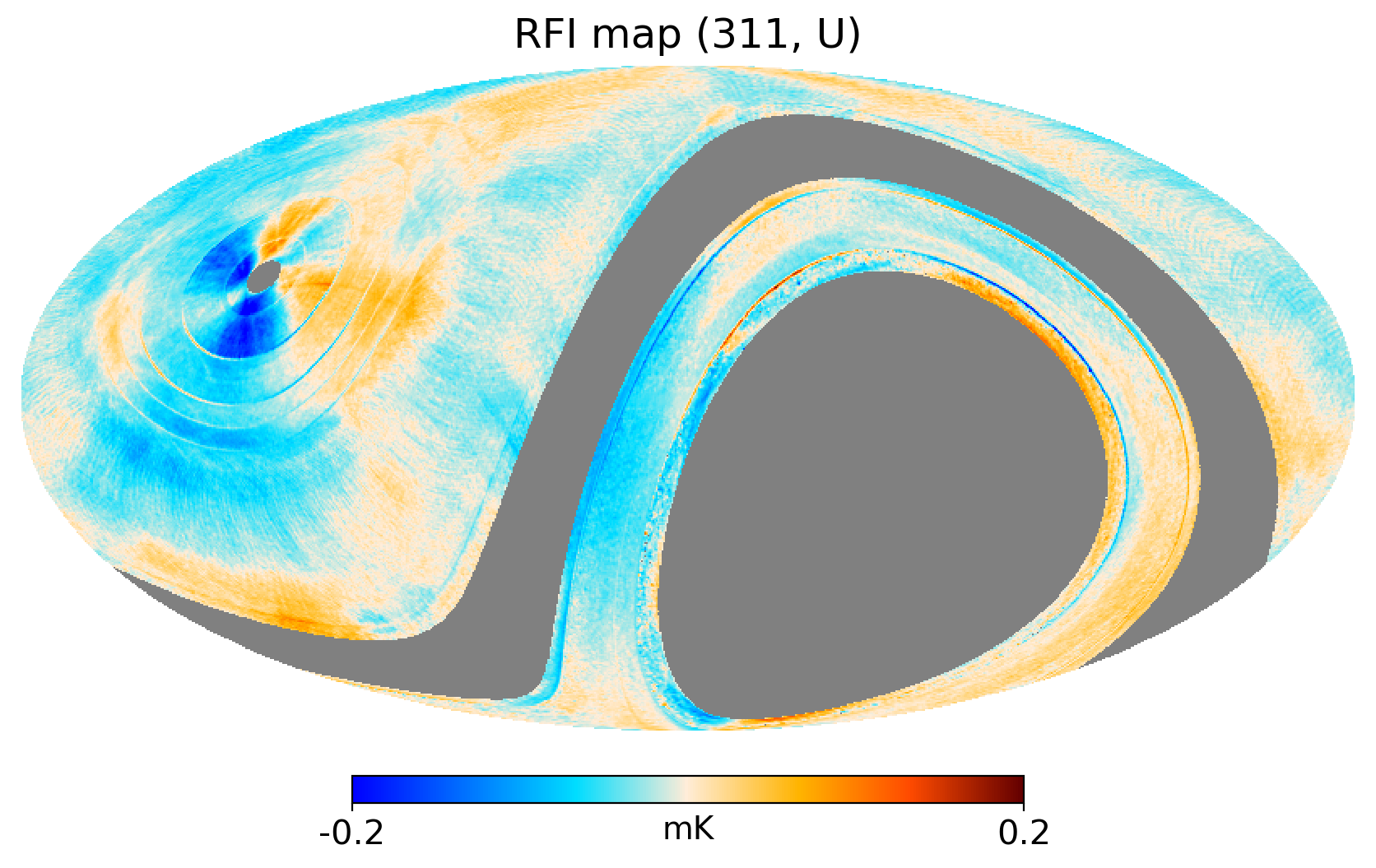}
    \includegraphics[width=0.65\columnwidth]{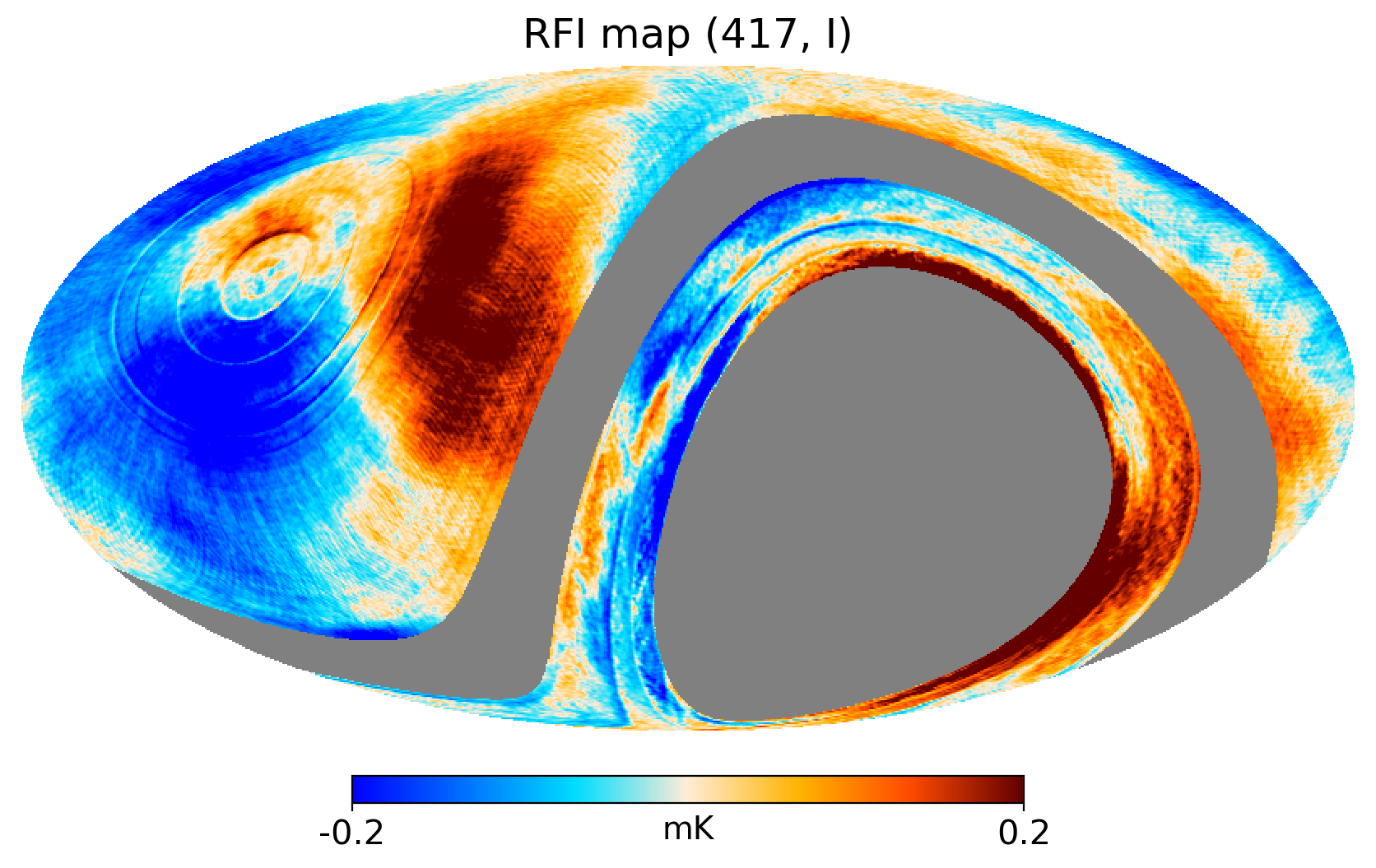}%
    \includegraphics[width=0.65\columnwidth]{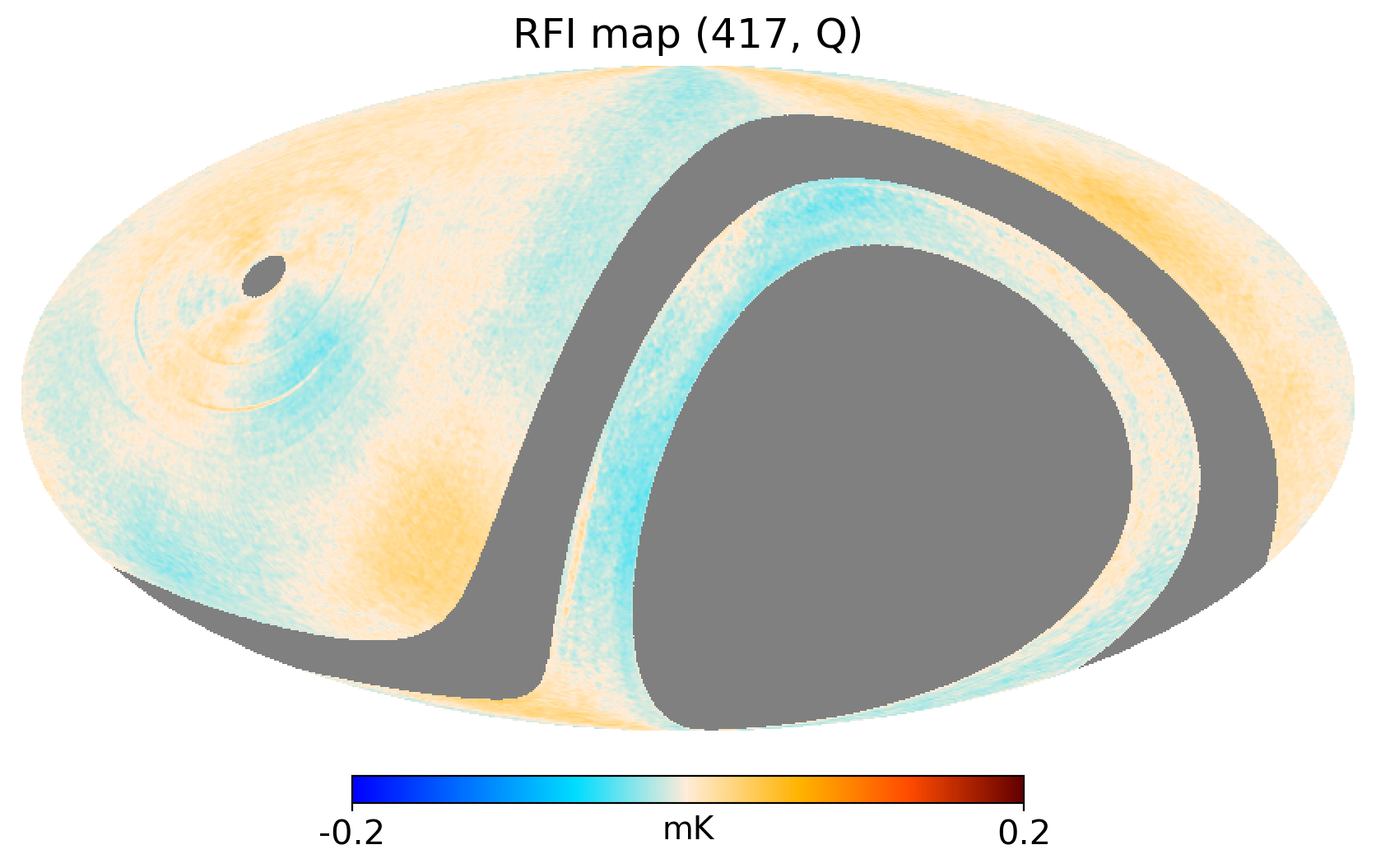}%
    \includegraphics[width=0.65\columnwidth]{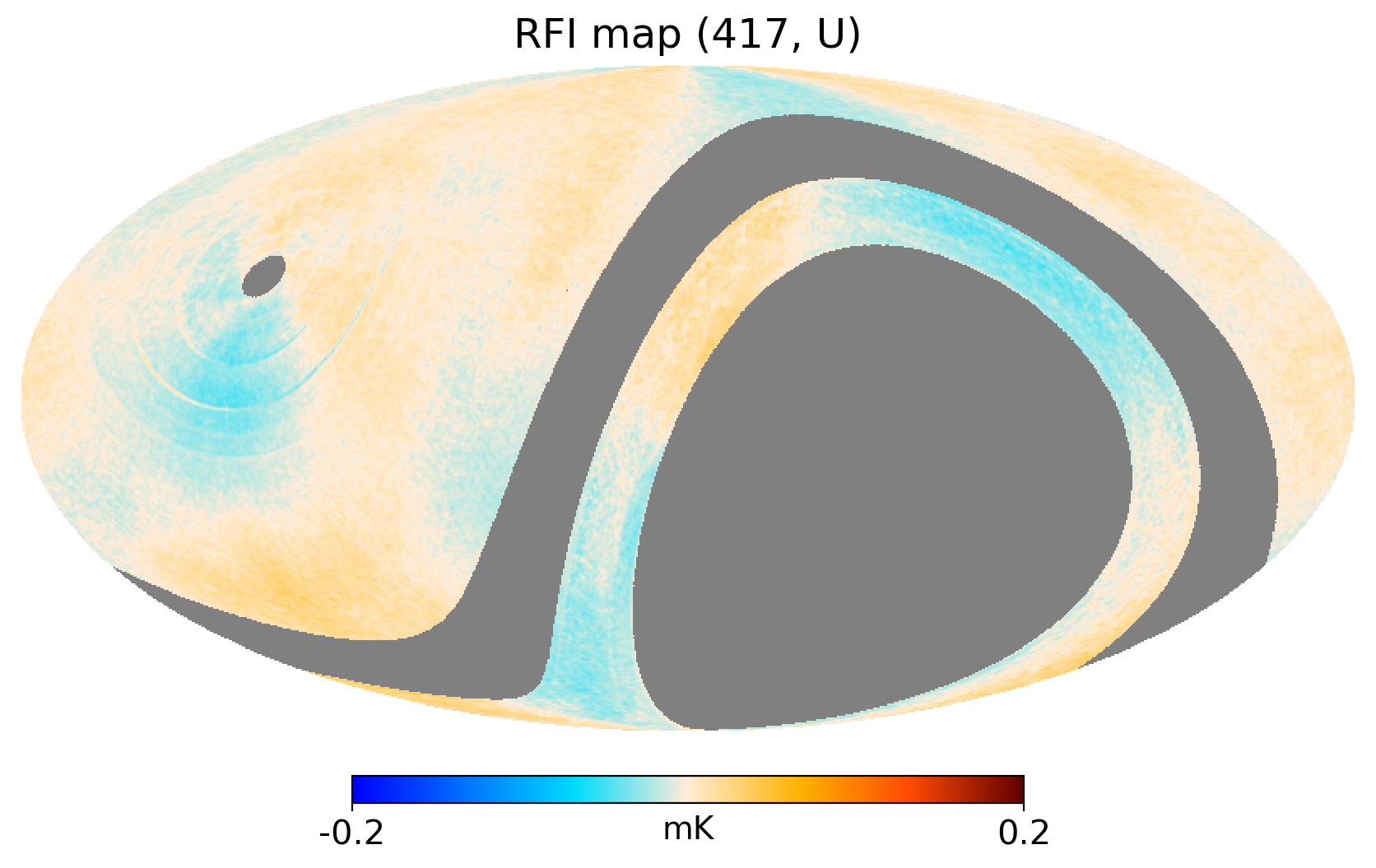}
    \caption{RFI patterns removed from the maps of the QUIJOTE MFI wide survey. Top row corresponds to the RFI emission at 11\,GHz (horn 3, labelled as "311"), while the bottom row corresponds to 17\,GHz (horn 4, labelled as "417"). From left to right, we show the residuals for intensity, Stokes Q and Stokes U parameters. The colour scale is the same in all six panels, corresponding to the temperature range $\pm 0.2$\,mK. For visualization purposes, all maps are smoothed to one degree resolution.  }
    \label{fig:rfidiff}
    \end{figure*}
    
    Specifically for the wide survey data, residual random spikes as well as possible RFI signals from satellites not identified in our standard pipeline are flagged using a dedicated matched-filter code that is applied to the one-dimensional TOD. The only assumption is that the object to be detected is unresolved, and thus should match the beam profile. The code\footnote{\url{https://gitlab.com/HerranzD/quijote-satdet}} excludes the location of the known bright radio-sources, which are also easily detected in the TOD. 
    
    Residual RFI signals appear at fixed azimuth (AZ) locations. In the case of QUIJOTE MFI, most of these signals are due to the radio emission of geo-stationary satellites entering through the beam far sidelobes. These signals were particularly visible in period 1 and at low frequencies (horns 1 and 3), until the installation of the extended shielding of the first QUIJOTE telescope was completed. All other periods are much less affected, due to the significant suppression of the far sidelobes. Because of this reason, period 1 was used for the intensity maps only, and not for polarization. 
    In order to remove these RFI signals, we generate spatial templates in the azimuth direction, by obtaining stacks of the TOD signal as a function 
    of AZ, $f({\rm AZ})$. These templates are computed for each period and each elevation separately, and thus rely on the assumption that the RFI signal is stable in time during the whole period. The templates are generated both for the sum and difference of MFI channels, and thus, they are applied to the intensity and the polarization TOD. Finally, a smoothed version of these templates (in scales of $10^\circ$) is subtracted from the TOD.  
    Figure~\ref{fig:rfidiff} shows two examples of the global RFI patterns removed using this procedure. These figures are obtained as the difference between the end-to-end MFI maps with and without applying the RFI correction at the TOD level.   
    We also note that once the final maps are produced, any residual RFI signals are effectively corrected in the post-postprocessing stage, using a function of the declination as described in Sect.~\ref{sec:postprocessing}. 
    
    Some remaining RFI features and glitches are removed after a careful inspection of the final maps. For this purpose, separate maps for each elevation and
    period are produced. Once a particular RFI feature is identified in these maps, the corresponding location is introduced in specific flagging tables for each period and elevation, which are later applied to the calibrated TOD.

    \subsubsection{Atmospheric correction}
    \label{sec:atmos}
    Although the observations are done at (nominal) constant elevation, there are still some residual variations due to changes in the atmospheric contribution along different directions. These variations are seen in the data as correlated patterns repeating in azimuth on very large angular scales, and with the amplitude increasing strongly with frequency, as expected for MFI frequencies due to the proximity of the 22\,GHz atmospheric water line \citep[see e.g.][]{AM}.
    It also evolves and changes on the scale of several hours, which is expected due to varying integrated
    water vapour content along lines of sight as weather systems blow over the site. It is possible to try to remove these effects especially at the more troublesome higher frequencies by a Principal Component Analysis (PCA) decomposition to look for these correlated signals. 
    
    To model this atmospheric component in the MFI intensity data, only broad scale features are removed by using baselines up to only 5 harmonics over the azimuth scans. A mask is used to avoid bright emission from the Galactic plane and strong point sources. The baseline atmospheric patterns are generated over an hour, as a compromise between good signal to noise and the time evolution of the atmosphere. The PCA decomposition method used is implemented in {\sc Python}, using the {\sc sklearn} module \citep{scikit-learn} on all the channels. 
    
    The first most significant component found is one that increases strongly with frequency, with the spectrum expected for water vapour.
    A histogram of the ratios between 17 and 19\,GHz, the two most strongly affected frequencies, shows a clear broad
    peak at $0.42$ near the values expected from atmospheric models for the Teide Observatory of $0.49$ \citep[see e.g.][ and typical PWV conditions of $3$--$4$\,mm]{AM}, although this sits on a smaller but much broader distribution. Points outside the range 0.3 to 0.6 appear to be for dryer conditions, with the implication that the water vapour signal is too weak to be reliably recovered. It was decided to use this range ratio of 17 to 19\,GHz signal as an indicator of a usable atmospheric signal that can be removed. The removal is done by subtracting the PCA template with the coefficient found for each frequency channel at the TOD level. 
    
\begin{figure}
    \centering
    \includegraphics[width=0.9\columnwidth]{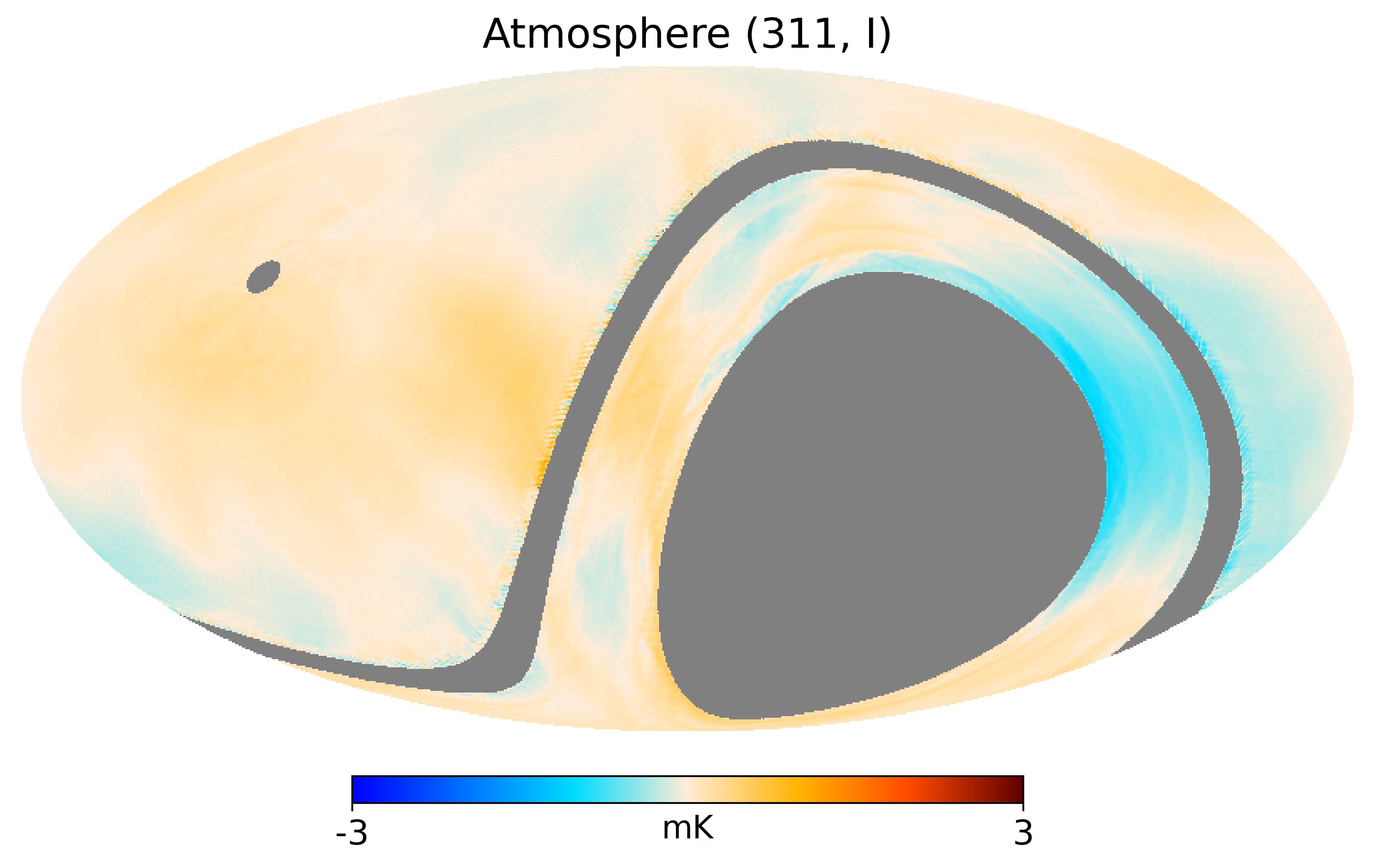}
    \includegraphics[width=0.9\columnwidth]{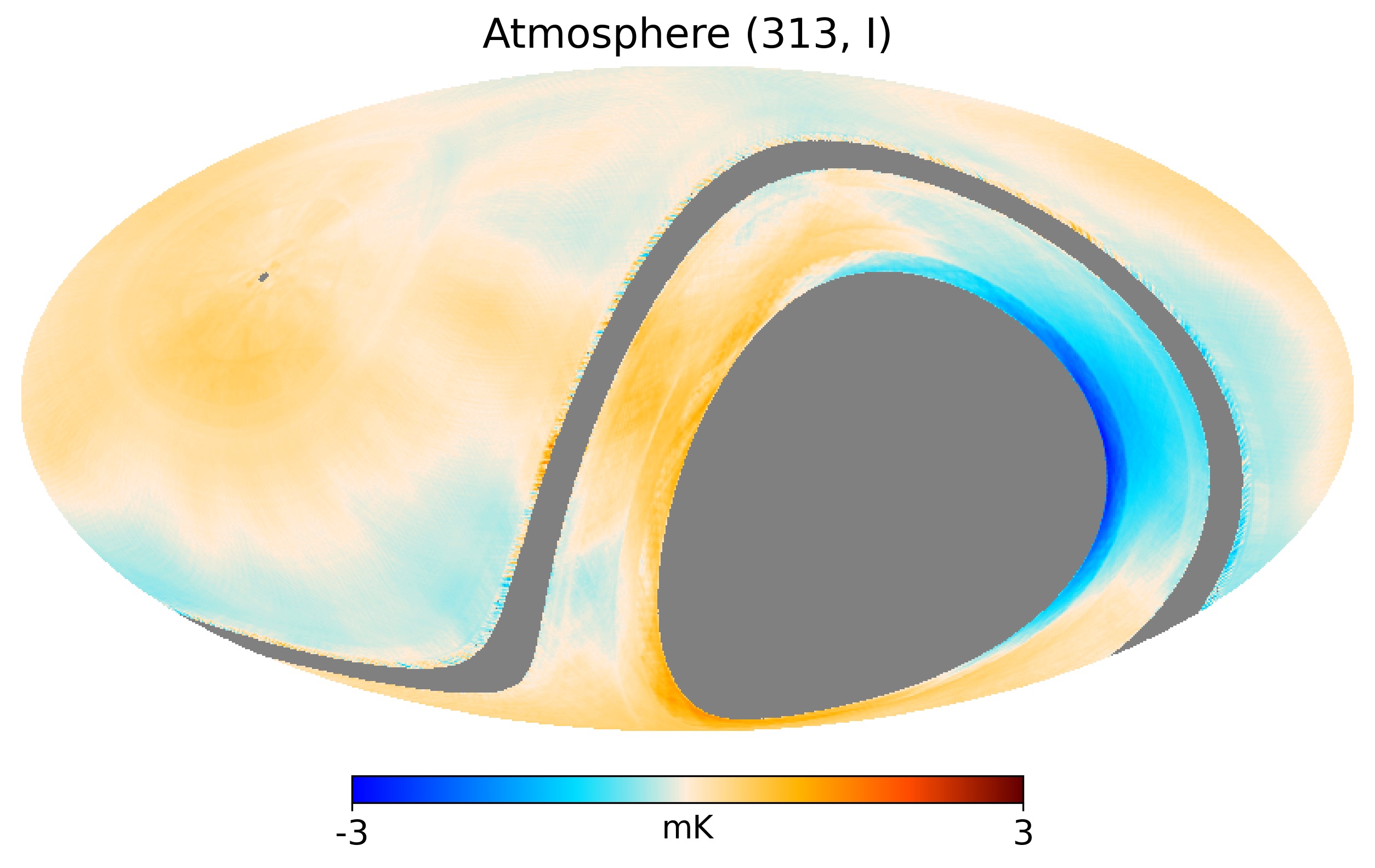}
    \includegraphics[width=0.9\columnwidth]{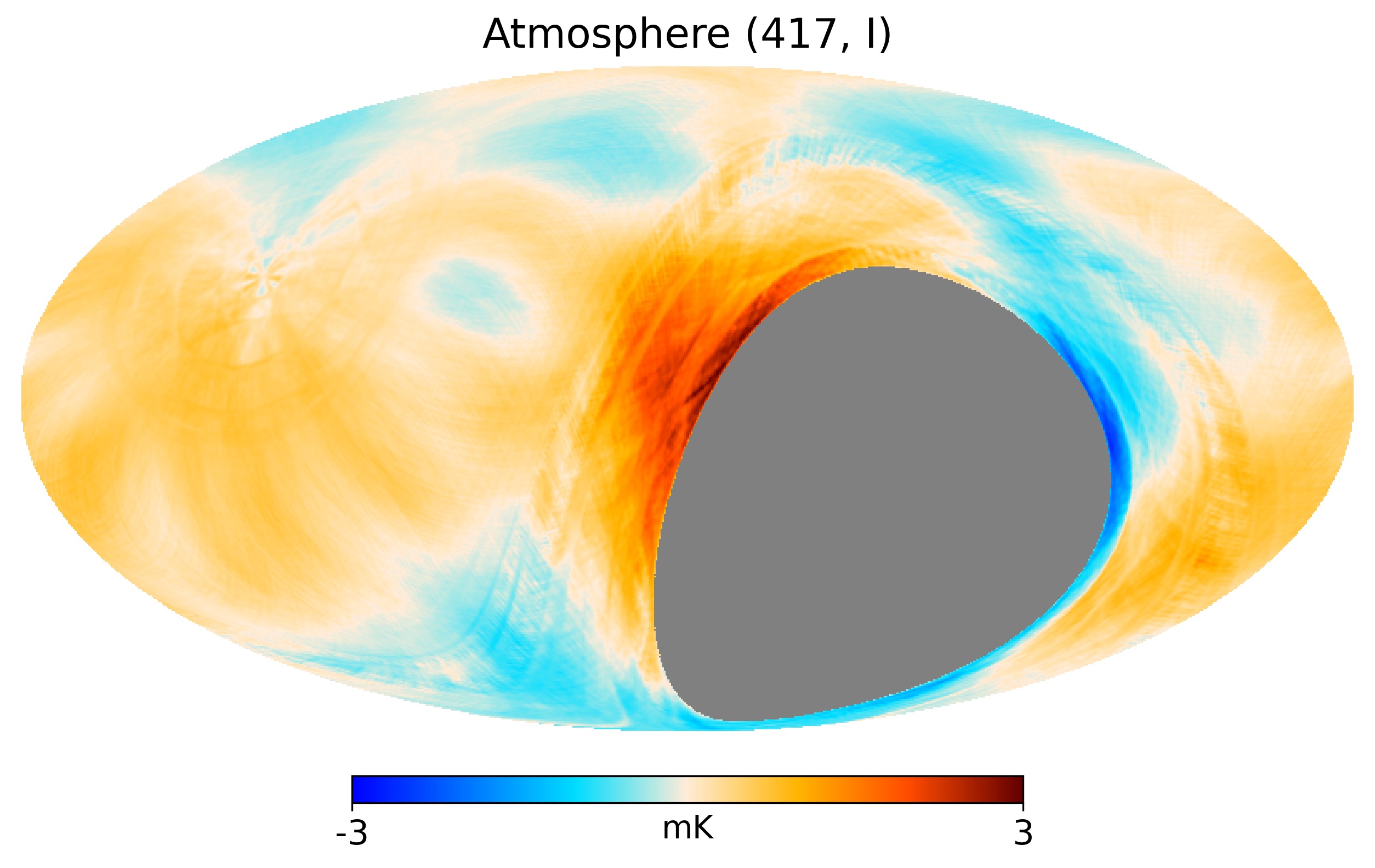}
    \includegraphics[width=0.9\columnwidth]{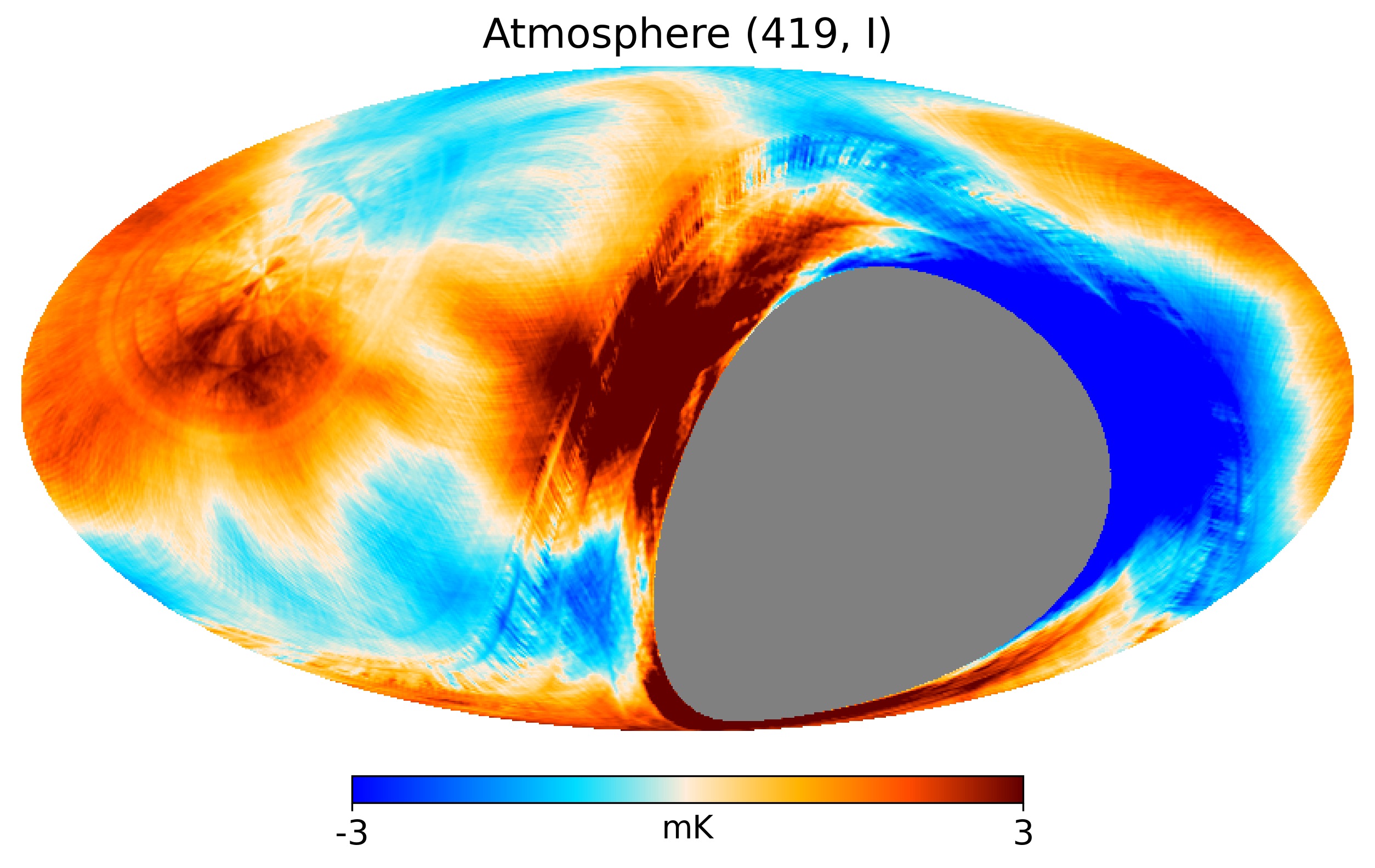}
    \caption{Atmospheric pattern removed from the intensity maps of the QUIJOTE MFI wide survey. From top to bottom, we have the atmospheric emission at 11\,GHz (horn 3), 13\,GHz (horn 3), 17\,GHz (horn 4) and 19\,GHz (horn 4). The colour scale is the same in the four panels ($\pm 3$\,mK), in order to visualise the increasing contribution of the atmospheric emission at higher MFI frequencies. For visualization purposes, all maps are smoothed to one degree resolution.  }
    \label{fig:atmos}
    \end{figure}

    Maps of this atmospheric emission can be produced running the full pipeline with and without this atmospheric correction, and then taking the difference of the two resulting maps. The atmospheric emission maps for horns 3 and 4 are shown in Figure~\ref{fig:atmos}. The map for horn 2 is similar to the one for horn 4, so it is omitted for clarity. As expected, this atmospheric contribution is more relevant at higher MFI frequencies, and affects large angular scales. 
    As shown below (see Sect.~\ref{sec:transfer}), when doing a spherical harmonic expansion of the maps, this correction is only relevant in the intensity maps at multipoles $\ell \lesssim 15$ for 11\,GHz, and $\ell \lesssim 25$ for 19\,GHz.
    No atmospheric correction is needed in polarization for the MFI wide survey maps. When a similar procedure is applied to the polarization data, the results are consistent with essentially unpolarized atmospheric emission.

    \subsection{Map-making}
    The QUIJOTE MFI wide survey maps are produced using the \picasso\ code \citep{destriper}, a destriping algorithm based on the {\sc MADAM} approach \citep{madam, madam2} but specifically implemented and optimised for QUIJOTE MFI. The destriping technique corrects for a correlated noise component by modelling the $1/f$ drifts in the TOD with a set of consecutive offsets with a given time length $t_{\rm b}$, the so-called baselines. The \picasso\ code has been tested extensively using realistic simulations matching the actual observations of the MFI wide survey and with realistic noise properties \citep{destriper}. In these conditions, the reconstructed maps preserve all angular scales with high fidelity, and in particular, we expect a signal error better than 0.001 per cent at $20 < \ell < 200$.
    
    Those realistic simulations were also used to set the reference parameters adopted for the production of the final MFI wide survey maps. In particular, we use a baseline length of $t_{\rm b}=2.5$\,s for the entire survey. Maps are generated using the \healpix\ pixelization scheme with $\nside = 512$. The specific priors for the $1/f$ noise properties (knee frequency $f_k$, slope $\gamma$, and cutoff frequency $f_{\rm cut}$) are shown in Table~\ref{tab:picasso}, both for the intensity and polarization maps. In the later case, the parameters are different depending on the noise levels of the corresponding pair of channels (i.e. if they are correlated or uncorrelated channels). As discussed in \cite{destriper}, those priors are assumed to be stationary parameters for the whole survey. 
        
    \begin{table}  
    \caption{Map-making parameters and related information. We consider three different cases of use with the \picasso\ code: intensity maps, polarization maps with correlated channels, and polarization maps with uncorrelated channels. For each case, we quote the prior values for the knee frequency $f_{\rm k}$, the slope of the $1/f$ noise component $\gamma$, and the low cut-off frequency $f_{\rm cut}$,  as well as the $\nside$ value of the \healpix\ map and the baseline length $t_{\rm b}$ (in seconds). See text for details.   }
    \label{tab:picasso}
    \centering
    \begin{tabular}{|c|c|c|c|c|c|}
    \hline 
    Case &  $f_{\rm k}$ & $\gamma$ &  $f_{\rm cut}$ & $\nside$ & $t_{\rm b}$ \\ 
    &  [Hz] &  & [Hz] & & [s] \\ 
    \hline
      I &  40.0 & 1.5  & 0.033 & 512 & 2.5 \\ 
    Q,U corr &   0.3 & 1.8  & 0.033 & 512 & 2.5 \\ 
    Q,U uncorr &   40.0 & 1.5  & 0.033 & 512 & 2.5 \\ 
    \hline
    \end{tabular}
    \end{table}

    \subsection{Post-processing of MFI wide survey maps}
    \label{sec:postprocessing}
    
    \subsubsection{Combination of maps}
    \label{subsec:combine_cu}
    For each horn and frequency sub-band, maps for the correlated and uncorrelated pairs are produced running the \picasso\ code separately for
    each one of them. These maps are combined at this post-processing stage, using the weight maps which are also produced by the map-making code as the propagation of the individual weights for each sample in the binned TOD. The combination of correlated ($x_{\rm c}$) and uncorrelated ($x_{\rm u}$) maps is done with the usual formula for the weighted arithmetic mean:
    \begin{equation}
    \label{eq:weimean}
    m = \frac{w_{\rm c} x_{\rm c} + w_{\rm u} x_{\rm u} }{w_{\rm c} + w_{\rm u}}. 
    \end{equation}
    Given that both correlated and uncorrelated channels share the same amplifier, we expect a high level of correlation between the two intensity measurements. As shown below in Sect.~\ref{subsec:crossnoisechan}, this correlation is indeed of the order of $90$--$95$ per cent for the intensity channels, and consistent with zero for the polarization ones. Although in principle it is possible to construct a minimum variance estimator accounting for these correlations in the intensity pairs, we still use equation~\ref{eq:weimean} for the combination of the intensity (correlated and uncorrelated) maps, in order to have a more robust estimate of the combination \citep[see e.g. ][]{Schmelling95}. 
    
    From equation~\ref{eq:weimean}, we can derive the expression for the weight map of the linear combination as
    \begin{equation}
    w = \frac{ ( w_{\rm c} + w_{\rm u} )^2}{ w_{\rm c} + w_{\rm u} + 2\rho \sqrt{w_{\rm c} w_{\rm u} } }  ,
    \end{equation}
    where $\rho$ stands for the correlation fraction between correlated and uncorrelated channels. 
    
    The map-making code also produces an estimate of the covariance matrix in polarization, $cov(Q,U)$, as well as the condition number ($\rcond$) map 
    \citep[see equations 44 and 45 in ][]{destriper}. Before forming the combination of the polarization maps in the wide survey, those pixels with $\rcond > 3$ are excluded. In practice, this only affects a small number of pixels close to the boundary of the satellite strip, as well as to the north celestial pole. In particular, for the 419 map (i.e. horn 4 at 19\,GHz) the number of affected pixels is slightly larger in those areas. 
    Appendix~\ref{app:maps} contains the $\rcond$ maps for all the MFI wide survey maps, together with other relevant maps, as discussed in Sect.~\ref{sec:maps}.
    Once the combination of the correlated and uncorrelated maps is carried out in polarization, the corresponding weight maps ($w_Q$, $w_U$) and covariance matrices $cov(Q,U)$ are also derived. Appendix~\ref{app:maps} also presents images of the $cov(Q,U)$ maps for all horns and frequencies. These maps show that, as expected, the normalized covariance $cov(Q,U)/(\sigma_Q \sigma_U)$ is very small (well below 0.01\,\%), so effectively each pair of $Q$ and $U$ maps are almost independent.

    \subsubsection{Residual interference: the FDEC filtering}
    After the map-making step, the resulting maps still present some residual RFI and large-scale patterns, which are corrected during this post-processing stage. 
    As described in Sect.~\ref{sec:rficorr}, residual RFI signals appear at fixed azimuth locations, so during the map-making process these features are projected onto the maps in stripes of constant declination. 
    This residual RFI is removed using a function of the declination, $f(\delta)$ (hereafter FDEC\footnote{\url{https://github.com/jarubinomartin/sancho.git}}), which is extracted directly from the maps as the median of all pixels with the same declination. This template function is built using a $|b|<10^\circ$ mask to exclude the Galactic emission, and specific masks in intensity and polarization for each frequency channel excluding the 10 per cent of the brightest pixels. 
    The procedure is applied both in intensity and polarization. In polarization, the maps are first rotated to local (equatorial) coordinates in order to extract the correction function. In this way, the RFI contamination from static sources in local coordinates appears as a constant signal in a given declination band.
    
\begin{figure}
    \centering
    \includegraphics[width=0.95\columnwidth]{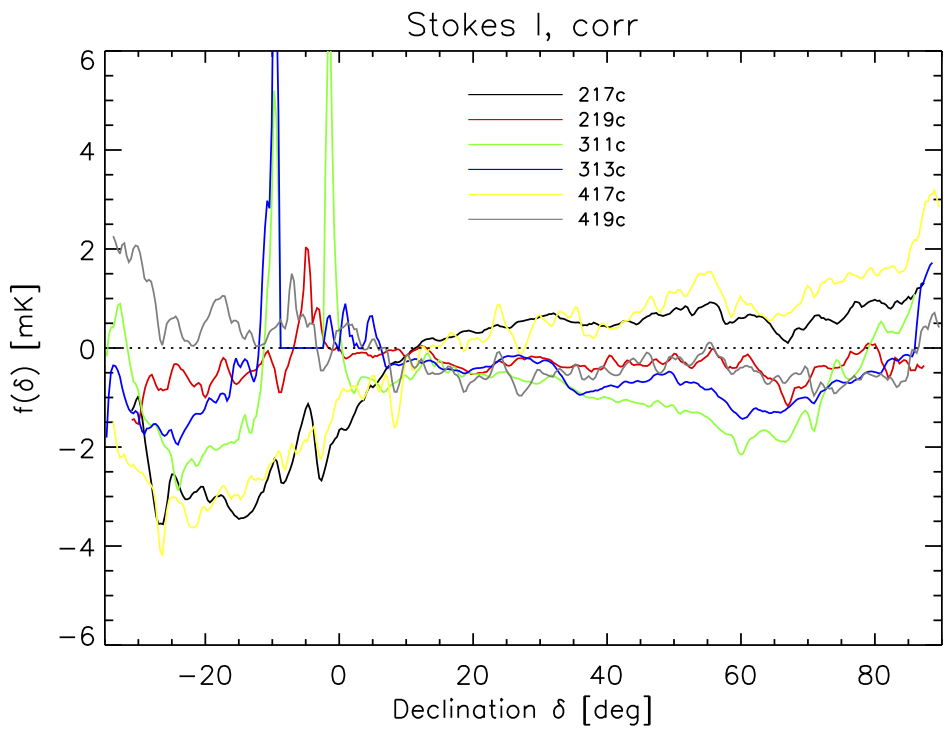}
    \includegraphics[width=0.95\columnwidth]{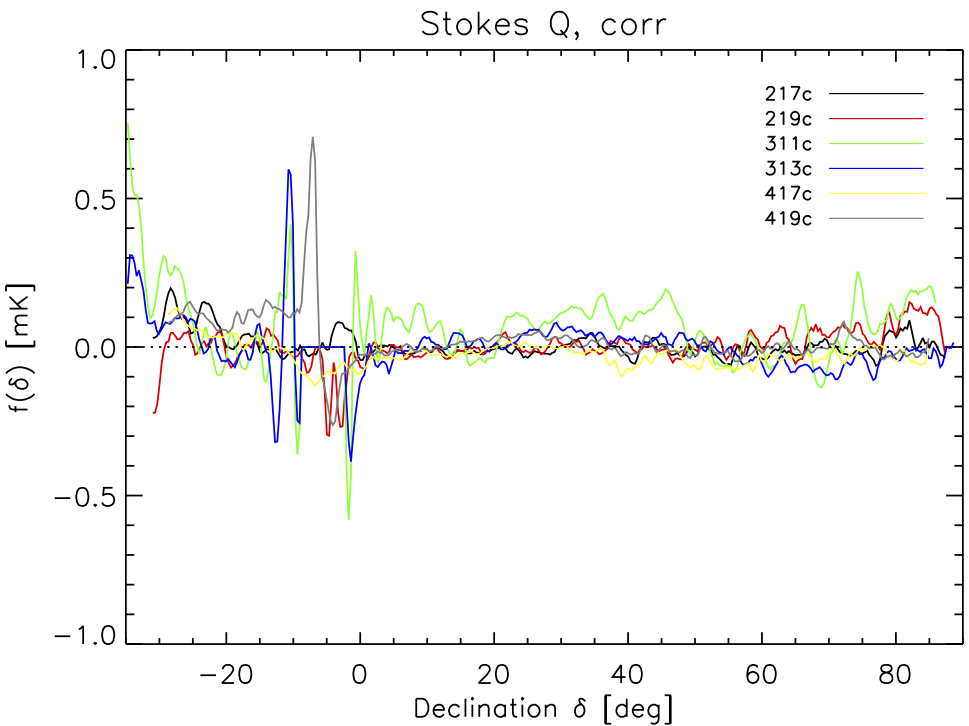}
    \caption{Examples of $f(\delta)$ correction functions (FDEC) to remove residual RFI in the MFI maps. Top: Stokes I FDEC for correlated channels. Bottom: Stokes Q parameter in equatorial (RADEC) coordinates for correlated channels. }
    \label{fig:fdec}
    \end{figure}

    Figure~\ref{fig:fdec} shows the correction functions for intensity and polarization for all MFI maps based on correlated channels. Similar curves are obtained for uncorrelated channels.  Note that in this figure, the panel for Stokes Q parameter corresponds to equatorial coordinates. 
    As expected for RFI signals, these correction functions are larger in the vicinity of the geo-stationary strip (around declination zero) and at low declinations (corresponding to low elevation values of the telescope, where the RFI is expected to be larger). We also note that they are also larger in intensity than in polarization. 
    
    \begin{figure}
    \centering
    \includegraphics[width=0.95\columnwidth]{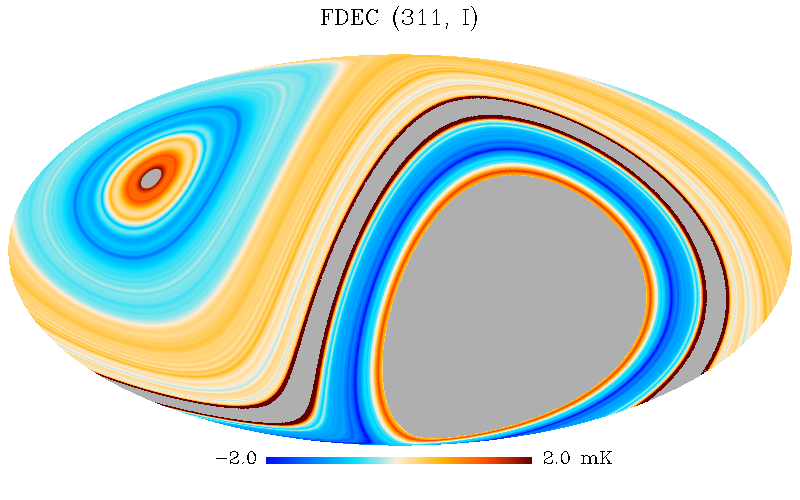}
    \includegraphics[width=0.95\columnwidth]{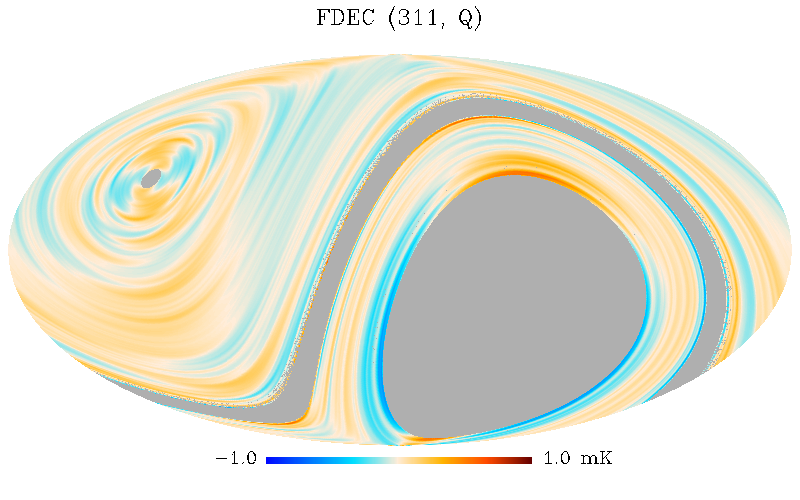}
    \includegraphics[width=0.95\columnwidth]{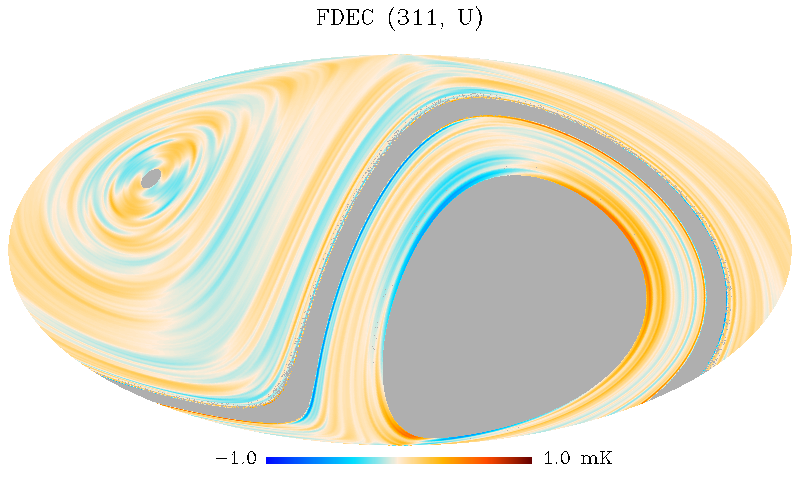}
    \caption{Example of the effective correction map based on a function declination (FDEC) for the 311 map (horn 3 at 11\,GHz). Top: Stokes I, with a colour scale in the range $\pm 2$\,mK. Middle and bottom: Stokes Q and U parameters, with a colour scale in the range $\pm 1$\,mK. }
    \label{fig:fdec2}
    \end{figure}
    
    Once these correction functions $f(\delta)$ are derived, they are reprojected onto a map in order to produce a RFI template. These templates 
    are subtracted from the data before carrying out the combination of correlated and uncorrelated maps. Figure~\ref{fig:fdec2} illustrates the final FDEC correction applied to the maps of horn 3 at 11\,GHz, after combining the correlated and uncorrelated maps in intensity. 
    
    \subsubsection{Monopole and dipole removal}
    Finally, a monopole and a dipole component are subtracted from the correlated and uncorrelated maps before their combination, using 
    the {\tt remove\_dipole} routine of \healpix\ with a Galactic mask excluding the region $|b|<10^\circ$. The removed dipole is consistent with the expected CMB dipole, as discussed in Sect.~\ref{sec:dipole}.

    \subsection{Effective transfer function}
    \label{sec:transfer}
    
\begin{figure}
    \centering
    \includegraphics[width=\columnwidth]{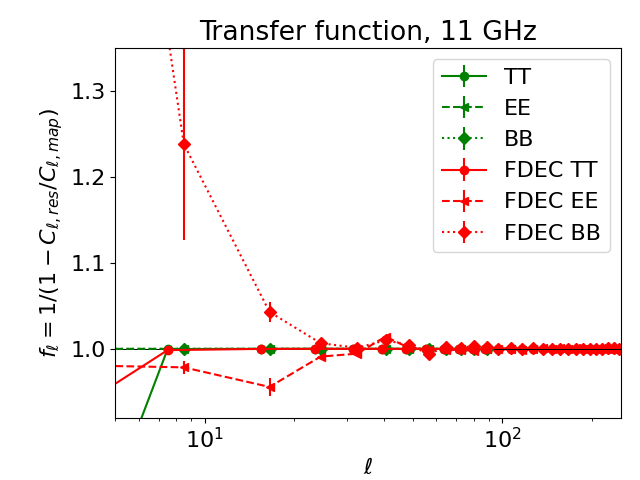}
    \caption{Transfer function (TF) of the QUIJOTE MFI wide survey map at 11\,GHz, after accounting for the post-processing stage of a subtraction of a function of the declination (FDEC). The TF for TT is marked with circles connected by red solid lines; the EE case with triangles and red dashed lines, and the BB with diamonds and red dotted lines. As a reference, in green we also include the TF of the \picasso\ map-making code \citep{destriper}.  }
    \label{fig:tf11}
    \end{figure}

    The \picasso\ map-making code essentially preserves all angular scales in the MFI wide survey maps. The expected signal error is better than 0.001 per cent in the multipole range $20 < \ell < 200$ both for intensity and polarization maps, and stays well within one per cent 
    down to $\ell=10$ \citep[][and see also Fig.~\ref{fig:tf11}]{destriper}.  
    However, some of the specific procedures applied in the MFI pipeline to correct for RFI signals and atmospheric contributions might have an impact on the effective transfer function of the wide survey. In particular, we should consider the impact of the RFI (Sect.~\ref{sec:rficorr}) and atmospheric  (Sect.~\ref{sec:atmos}) corrections at the TOD level, and the RFI correction at the post-processing stage using a function of the declination FDEC (see previous subsection). In terms of their amplitudes at the map level, the largest correction corresponds to the third case (subtracting a function of declination), so we discuss the transfer function of this case in detail. 
    
    It is important to note that, by construction, after applying this FDEC correction, the zero mode at constant declination will be missing from the maps. To characterize its impact on the effective transfer function of the wide survey, we follow the methodology described in Sect.~6.3 of \cite{destriper}. Here, we use simulations in the ideal case including CMB and foregrounds, but without a noise component. 
    The transfer function is then computed in terms of the power spectra of the map with residuals $C_\ell^{\rm res}$ (i.e. reconstructed map minus input sky) and that of the reconstructed map $C_\ell^{\rm map}$, both computed within the same mask, using this expression:
    \begin{equation}
    f_\ell = \frac{1}{1-C_\ell^{\rm res}/C_\ell^{\rm map} }.
    \end{equation}
    Figure~\ref{fig:tf11} presents the result obtained for the 311 case.  As expected, we find that the FDEC correction is 
    affecting low multipoles ($\ell \lesssim 15$). 
    The reconstruction of the sky signal is better than one per cent down to $\ell \approx 10$ in intensity. In polarization, the correction stays within one per cent down to $\ell \approx 30$, being at $\ell = 10$ of the order of 20\,\% for BB, and 5\,\% for EE.
    Because of this reason, and although we are able to reconstruct the sky signal to lower multipoles, as a conservative approach the power spectra analyses in this paper will be restricted to $\ell \ge 30$, so no transfer function correction will be needed. Appendix~\ref{app:fdec} contains a more detailed discussion on how a given map is affected by the FDEC filtering. 
    The impact of the RFI and atmospheric corrections at the TOD level is discussed in detail in Sect.~\ref{sec:validares}, although we anticipate that their impact is lower than the $f(\delta)$ discussed here (except maybe for 19\,GHz, where the atmospheric contribution becomes comparable to the FDEC correction).

    \subsection{Recalibration of the wide survey maps using Tau A}
    \label{sec:recal}
    
    Once the MFI wide survey maps are produced using the pipeline described above, we re-evaluate three aspects of the calibration using Tau A: i) the global calibration scale in intensity, ii) the polarization angle calibration, and iii) the polarization efficiency.  We discuss them in detail here.
    
    \subsubsection{Global recalibration in intensity}
    Tau A and Cas A are the two main primary calibrators of QUIJOTE MFI \citep{mfipipeline}. Daily observations of these sources in raster scan mode are used to obtain the overall gain scale in intensity for each MFI channel in every observing period. However, as daily calibrator observations might suffer from $1/f$ noise and other uncertainties, we recalibrate the MFI wide survey maps in the post-processing stage. For this recalibration, we use Tau A as the reference source, because it is located on a cleaner background than Cas A.
    
    For this, we first generate wide survey maps for each individual period (four maps in total for each horn and frequency). These four maps per period are degraded to one degree angular resolution, and then we apply beam fitting photometry (hereafter BF1d) on Tau A. 
    The derived flux densities are compared, accounting for colour corrections, with a spectral emission model that we have specifically obtained for Tau A, using WMAP and Planck data together with some ancillary measurements, and applying the same BF1d methodology. The new model will be presented and discussed in detail in a separate paper (G\'enova-Santos \& Rubi\~no-Mart\'{\i}n, in preparation), and builds on that presented in \citet{Weiland2011}, but including several improvements: i) improved treatment of the colour-corrections and beam effects on WMAP data, ii) inclusion of Planck data, iii) improved variability model. The adopted Tau A model for the recalibration of MFI maps has the shape 
    \begin{equation}
    S_\nu({\rm Tau\, A}) = 358.3 \Big( \frac{\nu}{22.8\, {\rm GHz}} \Big)^{-0.297} \, {\rm Jy}.
    \end{equation} 
    This model is evaluated at epoch $2016.3$, which corresponds to the effective central epoch of the wide survey, and we use a secular decrease of $-0.218$\,\%\,yr$^{-1}$ \citep{Weiland2011}. 
    From this comparison, we derive global recalibration factors for each MFI frequency map and for each individual period, accounting for the secular decrease of Tau A and the effective epochs in each period (see values in Column~4 of Table~\ref{tab:periods}). 
    The mean value of these recalibration factors results in an overall 4 per cent recalibration of the wide survey maps. 
    The accuracy of the MFI wide survey intensity calibration is discussed in Sect.~\ref{sec:internalcal}.

    \subsubsection{Polar angle recalibration}
    The reference angle for each MFI polarimeter (i.e. the reference for $\theta_{\rm pm}$ in equations~\ref{eq:mfi_response_u} and \ref{eq:mfi_response_c}) changes across the spectral band, and thus from band to band. For this reason, the reference angle for each frequency map is calibrated separately, despite of the fact that the two frequency bands of the same horn share the same polar modulator. This procedure is based on daily Tau A observations, and it is described in \cite{mfipipeline}. In particular, the adopted model for the Tau A angle in Galactic coordinates is given by
    \begin{equation}
    \gamma_{\rm Tau\, A} = \gamma_0 + RM \lambda^2 ,
    \end{equation}
    where $RM=-1406 \pm 12$\,deg\,m$^{-2}$ and $\gamma_0=-88.31^\circ \pm 0.25^\circ$. 
    Our daily calibration provides a reference polar angle for Tau A with a statistical error of approximately $1^\circ$ within a period. But similarly to the intensity calibration, daily observations of Tau A might suffer from $1/f$ noise or other effects, so the polar angles of the final wide survey maps are recalibrated in each period with Tau A again. As for the global recalibration in intensity, we also use BF1d in Tau A to extract the fluxes in Stokes Q and U parameters in the maps per period. 
    From there, recalibration offsets in the reference angles are computed for each channel and each period, and applied in order to generate the final maps. 
    The accuracy of the angle calibration in the MFI wide survey is discussed in Sect.~\ref{sec:polangle}.

    \subsubsection{Polar efficiency}
    Detailed measurements of the polar efficiency of the MFI polarimeters in horns 2, 3 and 4 were obtained in period 6, once the MFI observations concluded. 
    The description of the instrumental setup and the final measurements are presented in \cite{mfipipeline}, and summarized in Table~\ref{tab:poleff}.
    
    \input{table_poleff.tex}

    In order to transfer this polar efficiency information to the other observing periods where we do not have laboratory measurements, we use again BF1d photometry on Tau A, using the MFI wide survey maps per period. The polar efficiency in each period $p$ is transferred from period 6 according to the relative value of the Tau A polarized intensity $P_{\rm Tau A}(p)$ in that period and in period 6, i.e. using the ratio $P_{\rm Tau A}(p) / P_{\rm Tau A}(6) $. On average, this photometry method introduces errors of approximately 1\,\% for horn 3, and 2\,\% for horns 2 and 4. 
    
    \begin{table}
    \caption{Change in the polarization efficiency for horns 2, 3 and 4 in period 6 due to errors in the $r$-factor. See text for details. }
    \label{tab:poleff2}
    \centering
    \begin{tabular}{@{}lrrr}
    \hline
    Channel & \multicolumn{1}{c}{Horn 2} & \multicolumn{1}{c}{Horn 3} & \multicolumn{1}{c}{Horn 4}\\
    \hline
    Low freq, corr  &       $-0.075$ &    0.021 &  $-0.006$ \\
    High freq, corr &       $-0.113$  &  0.029 &    $0.016$ \\
    \hline
    Low freq, uncorr  &  $0.028$   & $0.004$ &  $0.005$ \\
    High freq, uncorr  &  $-0.020$  & $-0.002$ &  $0.011$ \\
    \hline
    \end{tabular}
    \end{table}

    Finally, we also account for possible errors in the determination of the $r$ factors in equations~\ref{eq:mfi_response_u} and \ref{eq:mfi_response_c}, using wide survey data as follows. As shown in Appendix~\ref{app:rfactors}, an error $\epsilon$ in the determination of the $r$ factors translates into a modification of the polar efficiency, and the appearance of a small leakage term in the TOD polarization timeline which is proportional to the intensity map.  
    We use the \picasso\ map-making code to fit for an intensity-to-polarization leakage global component in period 6 data, in a two step process. First, we solve for the intensity map $I$ for each case (i.e. horn, frequency and channel), and then we use it to fit for an additional term $\alpha I $ when solving for the polarization map in equations~\ref{eq:mfi_response_u} and \ref{eq:mfi_response_c}. These values are used to correct for the polar efficiency of each channel in period 6, using the equations derived in Appendix~\ref{app:rfactors}. Table~\ref{tab:poleff2} shows the effective correction terms $\alpha \equiv \epsilon/(2r)$. We can see that in the case of horn 3, this correction introduces a change of 2--3 per cent in correlated channels, and below 1 per cent for uncorrelated channels. Horn 4 is almost  unaffected, while the largest correction factor appears for the correlated channels in horn 2. The accuracy of the polar efficiency calibration in the MFI wide survey is discussed again in Sect.~\ref{subsec:knownsys}.

\section{MFI wide survey maps: Intensity and polarization}
\label{sec:maps}
    
    Following the methodology described in the previous section, we produced intensity and polarization maps for each MFI horn and frequency. Images of these individual maps (per horn and frequency) are shown in Appendix~\ref{app:maps}, at their original resolution (i.e. the angular resolution listed in Table~\ref{tab:mfi_parameters}). 
    The resulting maps cover a sky fraction of $\fsky=0.75$, $0.71$ and $0.73$ (equivalent to sky areas of $30\,900$, $29\,300$ and $30\,100$\,deg$^2$) for horns 2, 3 and 4, respectively. 
    All MFI maps are produced in CMB thermodynamic units (mK$_{\rm CMB}$). For simplicity, throughout this paper we drop the subindex CMB and use the notation mK. Nevertheless, we recall that the correction to Rayleigh-Jeans units is very small at MFI frequencies (at most 1 per cent at 19\,GHz). 
    Smoothed maps at $1^\circ$ resolution are generated by convolving those original maps with the corresponding transfer function $T_\ell \equiv W^{1\,\rm deg}_{\ell}/W^{\rm MFI}_\ell$, which converts the spherical harmonic window function for each horn ($W^{\rm MFI}_\ell$) into a gaussian beam with FWHM$=1^\circ$ ($W^{1\,\rm deg}_{\ell}$). 
    All maps are displayed in Galactic coordinates. We recall that QUIJOTE-MFI Stokes Q and U parameter maps and data follow the COSMO convention for polarization angles from \healpix. Grey regions correspond to the sky areas not observed by QUIJOTE MFI: the southern sky (approximately below $\delta=-34^\circ$); a small area around the North Celestial Pole (NCP) for some of the horns (depending on their location in the MFI focal plane); and the band of geostationary satellites close to declination zero degrees, which mainly emit at 11 and 13\,GHz.

    Appendix~\ref{app:maps} also contains the associated number of hits ($\nhit$) and weight maps. Both set of maps are outputs of the \picasso\ map-making code. The hit maps ($\nhit$) correspond to the total number of 40\,ms samples in each \healpix\  pixel of $\nside=512$ resolution. The weight maps correspond to the propagation through the map-making process of the errors (weights) associated with each individual 40\,ms sample. Both sets of maps clearly show the imprint of the scanning strategy of the QUIJOTE MFI wide survey. The ring structures around the North Celestial Pole correspond to the boundaries of the different elevations considered in the survey. Due to projection effects, the number of hits is significantly larger in those borders (and thus, the noise levels are smaller). 
    In the low declination band of the maps (below the masked area due to geostationary satellites), the number of hits is significantly lower due to the combined effect of a lower number of observations at these low elevations (mainly $30^\circ$, $35^\circ$ and $40^\circ$), and projection effects. 
    We recall that the number of hits in the intensity maps is larger than in polarization due to the fact that some intensity data are not used in polarization (period 1 data are not used for any polarization maps; data from period 2 are not used in polarization for horn 4; and data from period 5 are not used in polarization for horn 2; see summary information in Table~\ref{tab:mfi_notation}).
    
\input{table_notation}

    The final QUIJOTE MFI wide survey maps at 11 and 13\,GHz presented in Fig.~\ref{fig:iqumaps_11ghz_1deg} and \ref{fig:iqumaps_13ghz_1deg} are directly the maps from horn 3, smoothed to  $1^\circ$ resolution. The final maps at 17 and 19\,GHz in Fig.~\ref{fig:iqumaps_17ghz_1deg} and \ref{fig:iqumaps_19ghz_1deg} have been produced as a linear combination of those for horns 2 and 4. For simplicity in the computation of effective beams, frequencies and colour corrections, we adopted constant weights for this combination. We have checked that the resulting maps have comparable noise levels to the maps obtained using spatially-varying weights based on the actual weight maps for each individual map in the combination. Thus, the combined maps 
    at 17\,GHz can be obtained as 
    \begin{equation}
    m_{17} = w_{2,17}m_{2,17} + w_{4,17}m_{4,17}
    \end{equation}
    for $m={I,Q,U}$, and similarly for 19\,GHz, we have
    \begin{equation}
    m_{19} = w_{2,19}m_{2,19} + w_{4,19}m_{4,19}. 
    \end{equation}
    Table ~\ref{tab:weih2h4} contains the final weights used for this linear combination. These values have been derived from the white noise level of the individual frequency maps for each horn, using optimal (inverse variance) weights. We note that horn 2 dominates the linear combination in intensity, while horn 4 contributes with a higher weight to the polarization maps. The actual values of noise levels for these maps are discussed in Sect.~\ref{sec:noiselevels}. 
    
    The final maps in polarization (Figs.~\ref{fig:iqumaps_11ghz_1deg}--\ref{fig:iqumaps_19ghz_1deg}) are dominated by the Galactic synchrotron emission (the spectral index of the observed signal is discussed below in Sect.~\ref{sec:spectra} and \ref{sec:properties}). Large scale features such as the Fan region or the North Polar Spur are clearly seen in the four frequency maps. The MFI instrument is not optimized to measure the intensity signal, and thus the intensity maps present worse noise properties. In particular, the two highest frequency channels show clear large scale $1/f$ residuals, particularly at negative declinations, due to the fact that they are observed only with the lower elevations (higher air masses).

    \begin{table}
    \caption{Constant weight factors used to produce the combined 17 and 19\,GHz MFI wide survey maps. We include only the weight factors for horn 4, as those for horn 2 can be obtained as  $w_{2,17}=1-w_{4,17}$ and $w_{2,19}=1-w_{4,19}$. }
    \label{tab:weih2h4}
    \centering
    \begin{tabular}{@{}cccc}
    \hline
     & I & Q & U  \\
    \hline
    $w_{4,17}$  &  0.362  &   0.732  &  0.732\\
    $w_{4,19}$  &  0.419  &  0.788   &  0.788\\
    \hline
    \end{tabular}
    \end{table}

    %
    \subsection{Analysis masks}
    \label{sec:masks}
\begin{figure}
    \centering
    \includegraphics[width=\columnwidth]{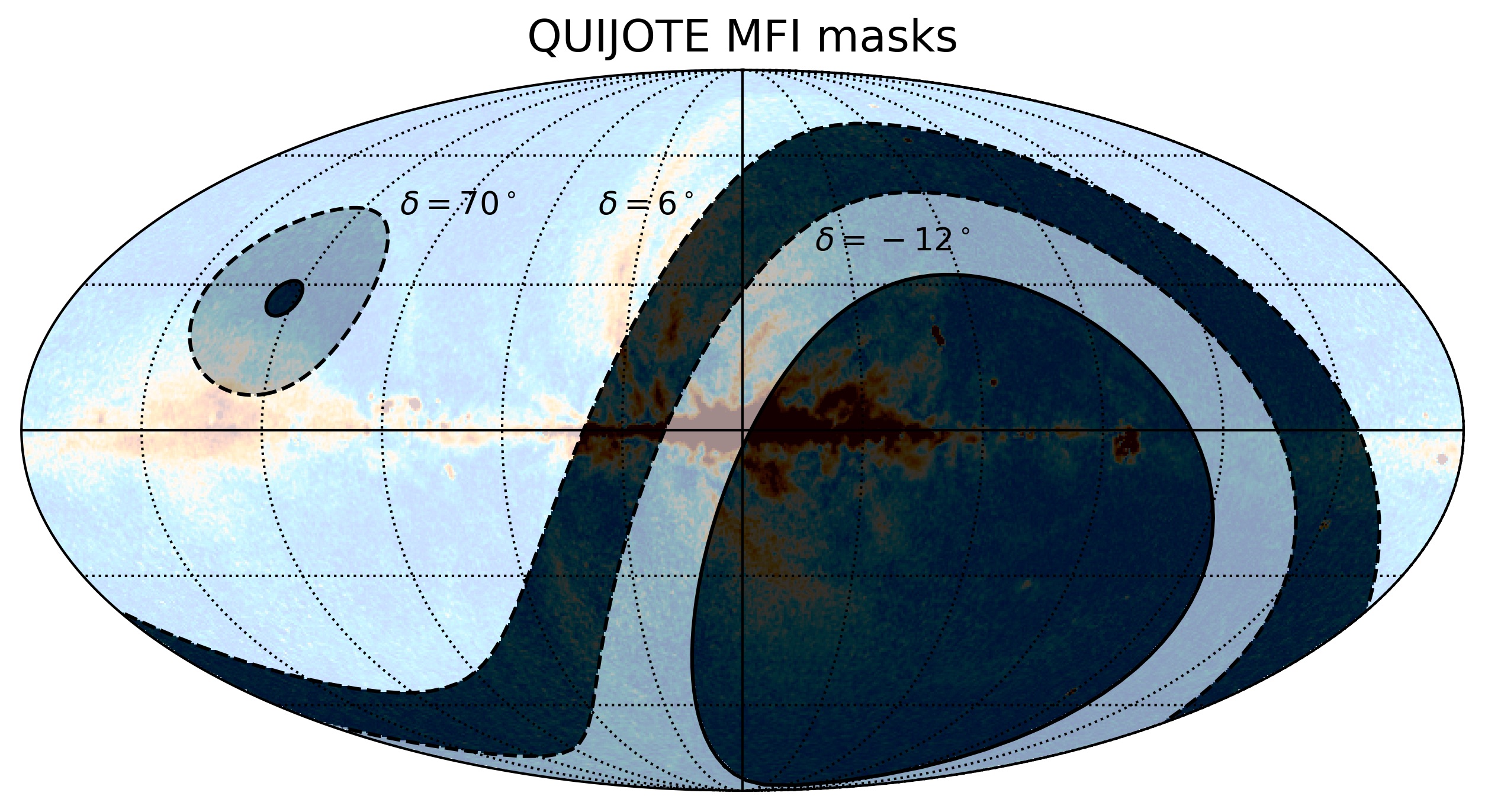}
    \caption{Footprint of the wide survey in Galactic coordinates, and proposed analysis masks. The background image corresponds to the 9-yr WMAP-K band polarized intensity map \citep{Bennett2013}. Light colours indicate the observed MFI wide survey regions. The band excluded due to satellite contamination corresponds to $-12^\circ < \delta < 6^\circ$. The default mask adopted for the analyses in this paper preserves the band $6^\circ < \delta < 70^\circ$, which is marked as the brightest region in the image. This mask is labelled as sat+NCP+lowdec (see text for details).  }
    \label{fig:masks}
    \end{figure}

    Figure~\ref{fig:masks} shows the footprint of the different analysis masks which are specific for the QUIJOTE wide survey. There are three distinct regions that are considered when building these masks: 
    \begin{itemize}
    \item Satellite band ("sat"). The masked region around declination zero is used to block the RFI contamination of geostationary satellites mainly affecting 11 and 13\,GHz maps. In the MFI pipeline, the emission from each geostationary satellite is flagged at the TOD level using a mask of $5^\circ$ radius around each satellite. Other satellites or RFI signals are flagged as described in Section~\ref{sec:data}. After this process, the resulting masked area (with zero number of hits) is located approximately between declinations $-10^\circ$ to $-2^\circ$ (note that geostationary satellites are seen at slightly negative declinations from the Teide Observatory). The proposed mask to remove the satellite band ($-12^\circ < \delta < 6^\circ$) is a conservative choice based on a close inspection of the final maps, extending the unobserved area by two degrees in the negative declination direction, and by eight degrees in the positive direction. This choice accounts for low-level RFI residuals in the intensity maps (some of the residual RFI signals corrected during the post-processing stage are located in that area), while keeping a relatively high number of hits per pixel. 
    \item North Celestial Pole ("NCP") region. Given the latitude of the Teide Observatory ($28.30^\circ$N) and the minimum elevation observed with QUIJOTE MFI (EL$=30^\circ$), some of the maps present a small area of unobserved pixels around the NCP, depending on the location of the MFI horns in the focal plane. The maximum observed declination is approximately $86^\circ$ for horn 3, and $87.5^\circ$ for horn 2. Horn 4 covers up to $90^\circ$ in declination. 
    In any case, the pixels surrounding this NCP area are only accesible with the lowest elevation bands, which usually present the largest levels of atmospheric contamination in the intensity maps, particularly at 19\,GHz. For this reason, for some of the analysis we mask the region above $\delta=70^\circ$, in order to keep a sky area that is observed practically by all the elevations considered in the survey. 
    \item Low (negative) declinations ("lowdec"). Similarly to the NCP area, this region is only observed when using low elevations (below $40^\circ$), and thus the corresponding intensity maps, specially at the two highest frequencies, are more affected by $1/f$ residuals from atmospheric emission (see Fig.~\ref{fig:iqumaps_17ghz_1deg} and \ref{fig:iqumaps_19ghz_1deg}, and also the individual maps for horns 2 and 4 in Appendix~\ref{app:maps}). The proposed mask to exclude this area covers all declinations below $\delta=-12^\circ$. 
    \end{itemize}
    All different combinations of those three masked regions produce the reference set of specific masks for the MFI wide survey used in this and all accompanying papers. In particular, unless otherwise stated, the default analysis mask used in most of the scientific analyses in this paper, and in particular, in all power spectrum computations, corresponds to the superposition of the three regions (sat+NCP+lowdec). This mask preserves a sky fraction of $\fsky = 0.418$, equivalent to approximately $17\,200$\,deg$^2$. 
    
    \section{Data validation}
    \label{sec:validation}
    
    In order to characterize the properties of the wide survey maps, we carry out a number of tests and studies in this section. 
    Most of them rely on different types of null tests, which can be used to detect possible remaining systematic effects in the data, including 
    residual RFI signals, calibration issues, changes in the operational or instrumental conditions, or even unknown effects. 
    
    \subsection{Null tests}
    \label{sec:nulltests}
    A ``null test'' is defined as the difference between the maps produced from two independent sub-sets of files from the full data base, which are expected to give the same signal under the assumption of a perfect calibration and no systematic effects. Null tests have been shown to be a powerful mean to assess the contribution of residual systematic effects in CMB analyses \cite[e.g.][]{Planck2013-iii, Planck2015-iii}. For the characterization of the QUIJOTE MFI wide survey data, we produced the following set of null tests:
    \begin{enumerate}
    \item Half mission. The full database is divided in two halves. The separation is done according to the calendar date inside each period and each elevation, producing maps labelled as ``half1'' and ``half2''. In this way, both null test maps contain data from all periods, and have a similar sky coverage. This is the reference null test used to characterize the overall noise properties. 
    \item Rings. The MFI wide survey maps are produced using the so-called nominal observing mode, in which the QUIJOTE telescope scans the sky using a circular  scanning strategy with a continuous movement in azimuth direction while maintaining a constant elevation. Each azimuth scan is called a "ring".
    For this null test, the full database is divided in odd (``rings1'') and even (``rings2'') rings. With the nominal azimuth scan speed of 12\,deg\,s$^{-1}$, each ring is completed in 30\,s, so this null test can be used to test for instrumental variations in these short time scales. As the instrument gain is stable in time scales much longer than one minute, this null test is not expected to reflect gain variations, and will essentially contain white noise plus a $1/f$-noise component in scales of 30\,s. 
    \item Daynight. In order to evaluate possible residual systematic effects due to day-night variations of the system gain or calibration factors, this null test
      is produced by dividing the full database into day observations (``daynight1'') and night observations (``daynight2''). For simplicity, we define here ``day''
      as all observations from 8 AM to 8 PM (UT). 
    \item PWV. Using the information from GPS measurements at the Teide Observatory of the precipitable water vapour (PWV) content of the atmosphere during each individual observation\footnote{The GNSS antenna that provides these PWV measurements is located at the Iza\~na Atmospheric Observatory (IZO) just $1.4$\,km away from QUIJOTE, and virtually at the same altitude ($\approx 10$\,m below).}, we divide the full data base in two sets of low (``pwv1'') and high (``pwv2'') pwv values. As in the case of the half mission null test, the separation is done inside each period and elevation, to guarantee that both splits contain a similar sky coverage. As a reference, the resulting median pwv in these two data splits is 2\,mm and $5.2$\,mm, for "pwv1" and "pwv2", respectively. 
    \item Halfrings. This null test separates the data by dividing each ring in two halves. Data taken with telescope azimuth values $0^\circ \le AZ \le 180^\circ$ correspond to "halfring1", while data with $AZ > 180^\circ$ are part of "halfring2". Although these maps are expected to be noisier than the other null tests due to $1/f$ contributions (note that in this case we are basically decreasing by a factor of two the number of independent crossings in each pixel when solving the conjugate gradient inside the map-making algorithm), they are still extremely useful to detect residual RFI signals arising from local structures, which usually appear at fixed AZ values. Moreover, these maps can be also used to test residual pointing errors.
    \item $T_{\rm BEM}$. As explained in \cite{mfipipeline}, the overall gain of the instrument is strongly correlated with the physical temperature in the electronic boxes containing the Back-End Module (BEM) of the MFI. As a further test to explore possible residual variations after our gain model correction, we use the values of one of the temperature sensors $T_{\rm BEM}$, which is monitored every second as part of the house-keeping data, to separate the data in two halves, according to low  ("tbem1") and high ("tbem2") values of the BEM temperature. As a reference, the median temperature for these two data splits is $8.1^\circ$C and  $16.1^\circ$C, respectively. As for the half mission and PWV null tests, we do the division in two halves for each period and elevation configuration separately, and then we combine the sub-lists. For simplicity, we refer to this case as "tbem null test" in the text. 
    \end{enumerate}

    Two separated lists of calibrated TOD files are produced for each one of those six null tests cases, and the corresponding maps $h_1$ and $h_2$ are produced with fully independent runs of the map-making code. The post-processing of each null test is identical to the procedure applied to the full maps. 
    From this point, a "null-test difference map" can be produced for each case, as 
    \begin{equation}
    \label{eq:nwei}
      n = \frac{h_1 - h_2}{w},
    \end{equation}
    where the normalizing weight is computed as
    \begin{equation}
      w = \sqrt{ (w_1 + w_2) (w_1^{-1} + w_2^{-1}) }.
    \end{equation}
    Here $w_1$ and $w_2$ are the individual weight maps of the null tests $h_1$ and $h_2$, respectively. They are computed as $w_i = 1/\sigma^2_i$, with $i=1,2$. 
    Defined in this way, equation~\ref{eq:nwei} provides a map with similar noise levels as the residual noise for the weighted-sum of the two halves \citep[see e.g. ][]{Planck2013-ii, Planck2015-vi}.

    \subsubsection{Null tests with a common baseline solution}
    
    \begin{figure*}
        \centering
        \includegraphics[width=6cm]{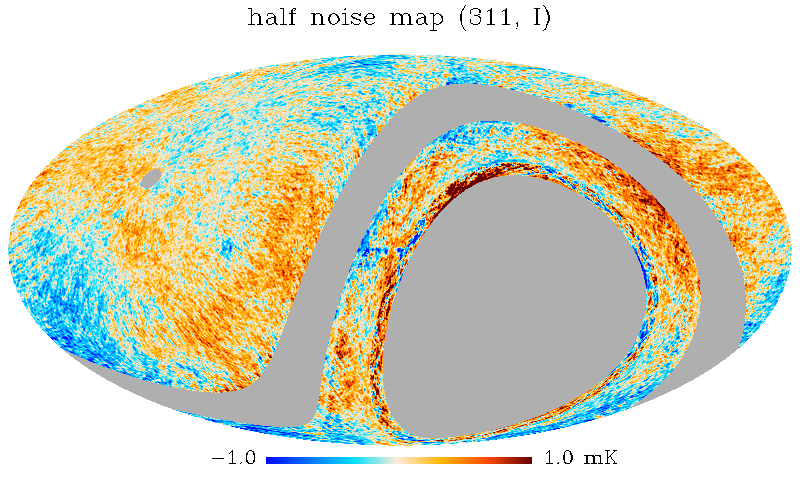}%
        \includegraphics[width=6cm]{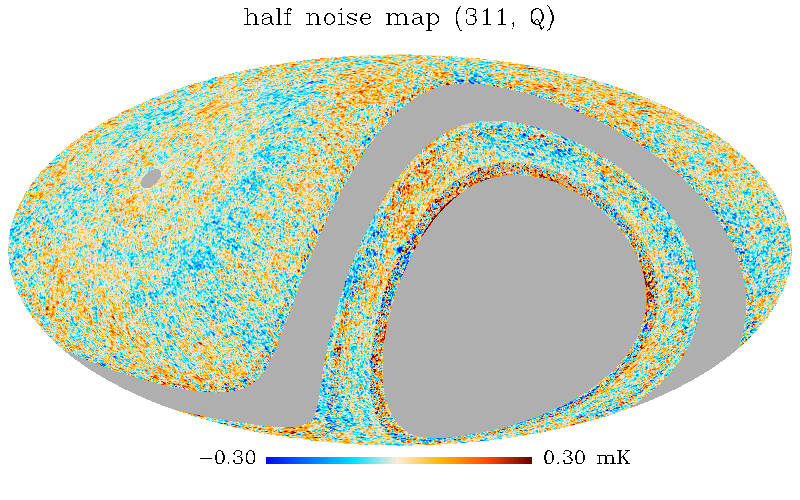}%
        \includegraphics[width=6cm]{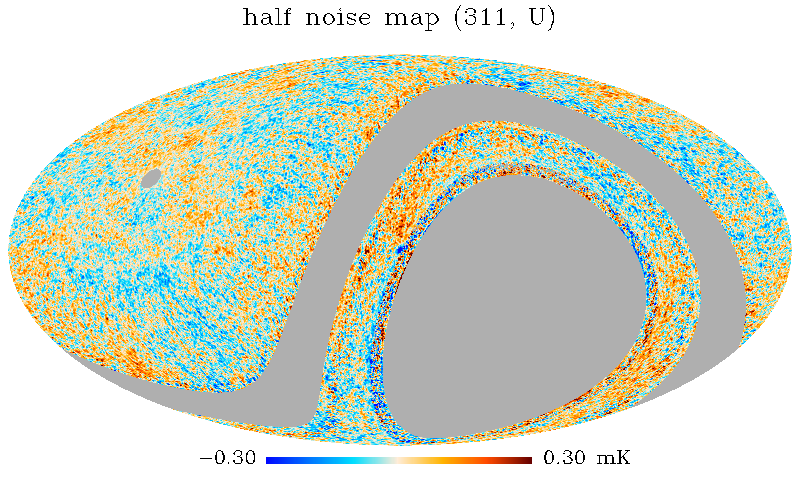}
        \includegraphics[width=6cm]{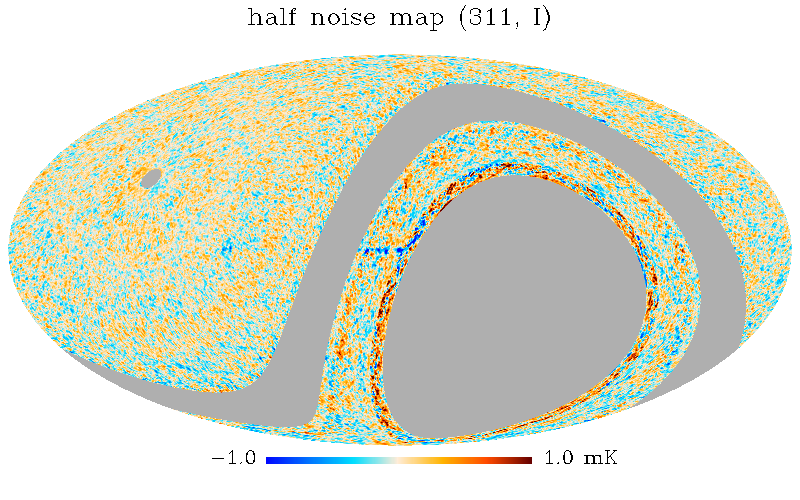}%
        \includegraphics[width=6cm]{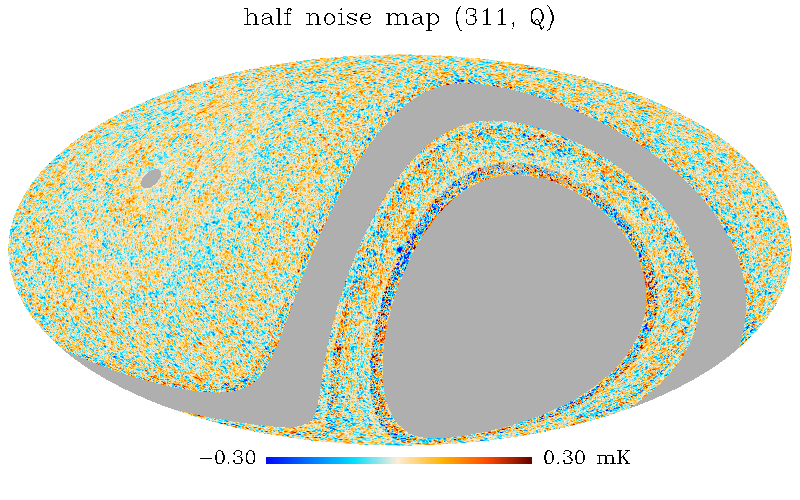}%
        \includegraphics[width=6cm]{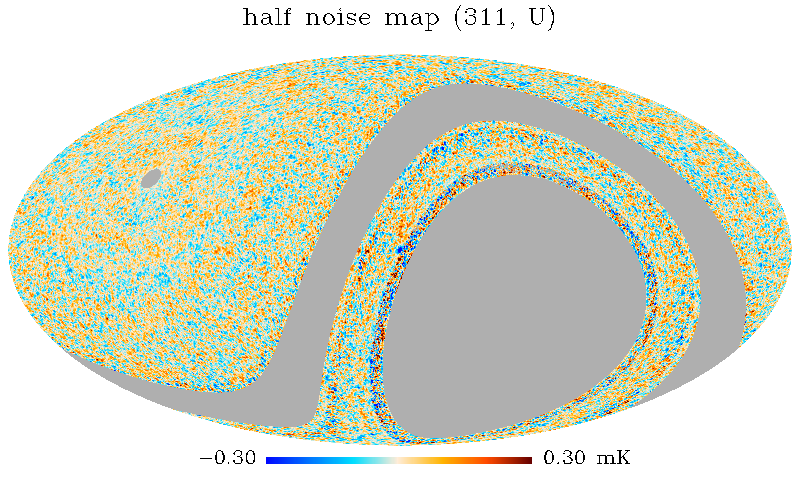}
        \caption{Half-mission null test difference maps for horn 3 11\,GHz. Top row shows the Stokes I (left), Q (centre) and U (right) difference maps for the case of "independent baselines". Bottom row corresponds to the case of "common baselines" (see text for details). For display purposes, all maps are smoothed to 1 degree resolution. The colour scale corresponds to $\pm 1$\,mK for the intensity maps, and $\pm 0.3$\,mK for polarization.  }
        \label{fig:noise311}
    \end{figure*}
    
    For those six cases listed above we have also produced a different set of null test maps, named as "null test with common baselines", as follows. 
    First, we run the map-making code for the complete database, and record the baseline solutions. Then, each pair of null test maps is generated using that recorded solution, instead of solving for the baselines with half of the data only, as it was the case before. 
    By construction, this procedure cancels out an important part of the $1/f$ noise contribution associated with long time-scale variations, partly due to the fact that the baseline solution is better constrained when using the full database. 
    Differences between the two halves $h_1$ and $h_2$ now will be entirely due to the fact that each half uses different input data, and not to the possible uncertainties in the determination of the baseline solution.
    For this reason, these null test maps are found to be particularly useful to study those variations in the data which can be (mainly) ascribed to calibration uncertainties, instrument changes or to variability of the sky signal. Thus, these maps will be used specifically in Section~\ref{sec:internalcal} to assess the internal calibration of the wide survey. For all the remaining analyses,  and in particular, for assessing the noise levels in the wide survey maps, we will always use the default set of null tests maps ("with independent baselines"). 
    
\begin{figure*}
        \centering
        \includegraphics[width=6cm]{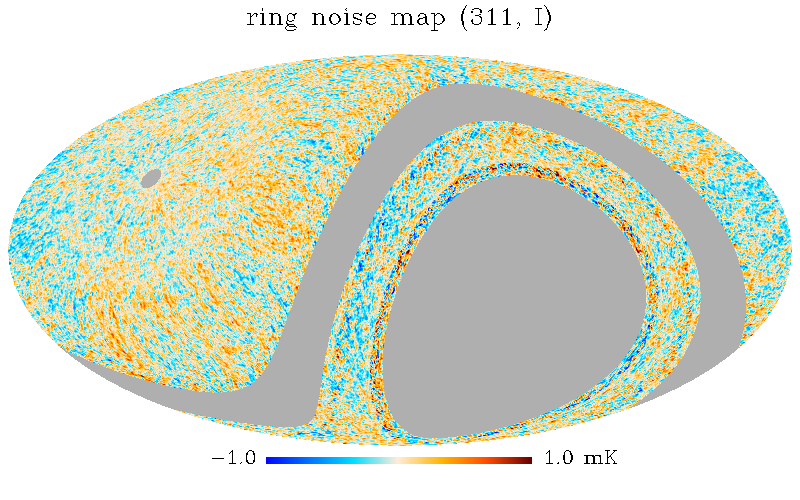}%
        \includegraphics[width=6cm]{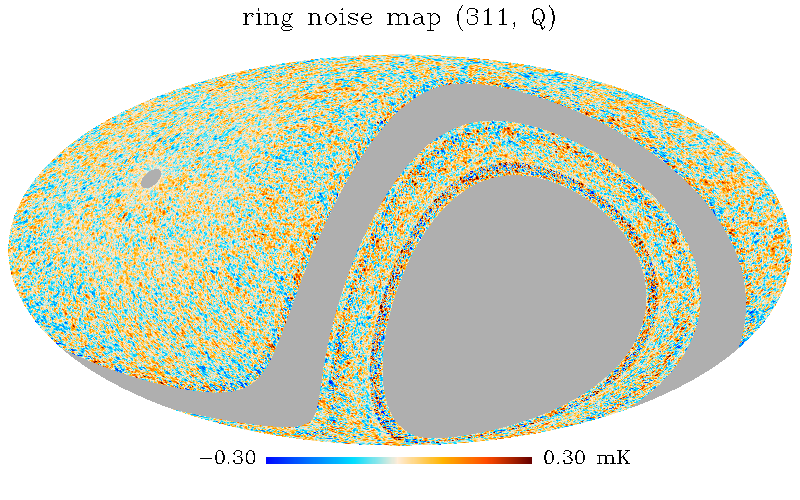}%
        \includegraphics[width=6cm]{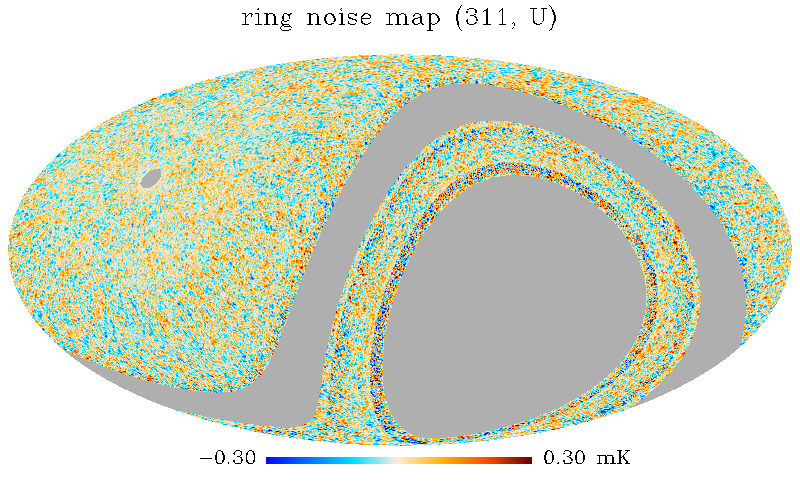}
        \includegraphics[width=6cm]{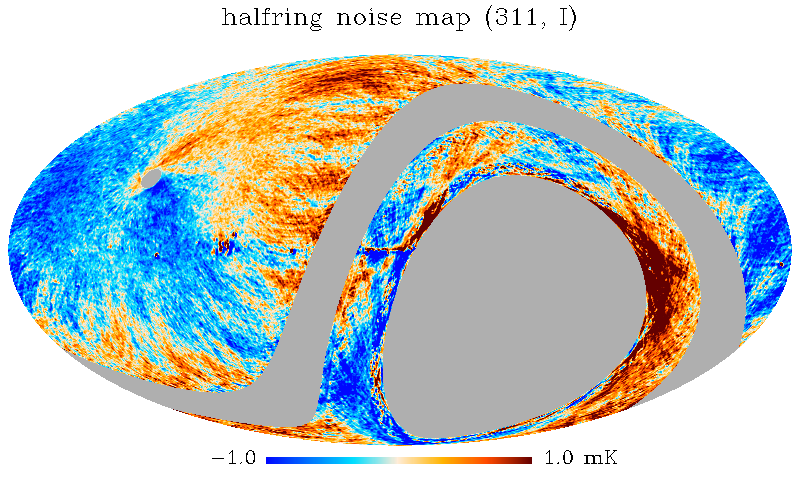}%
        \includegraphics[width=6cm]{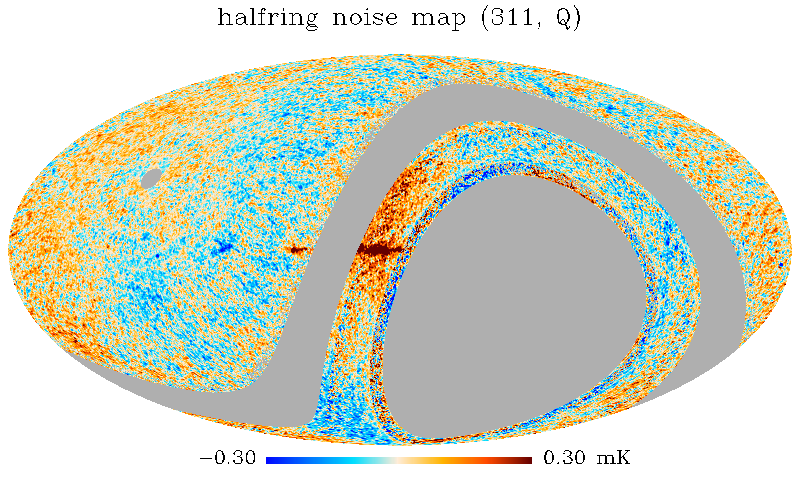}%
        \includegraphics[width=6cm]{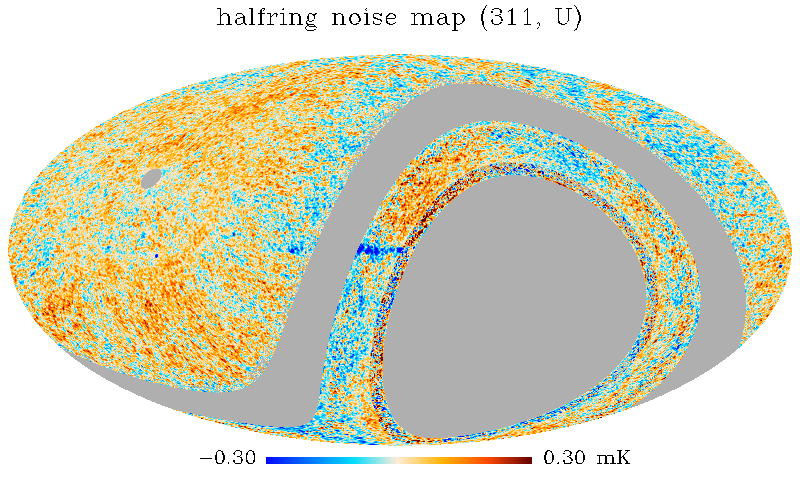} 
        \includegraphics[width=6cm]{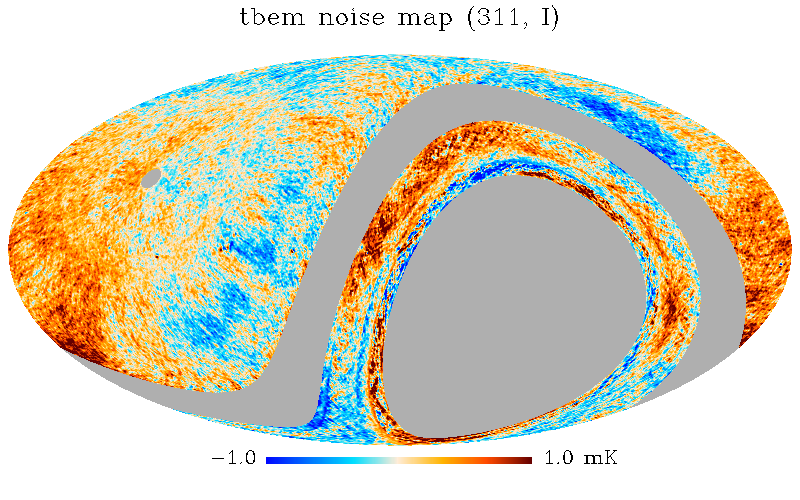}%
        \includegraphics[width=6cm]{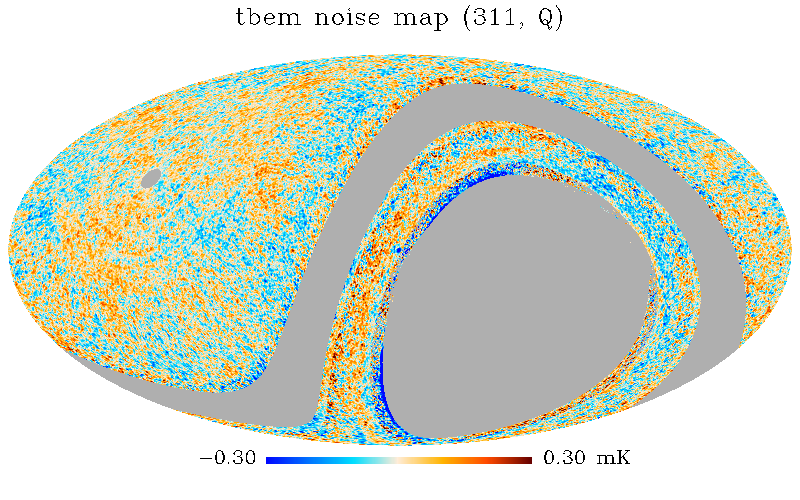}%
        \includegraphics[width=6cm]{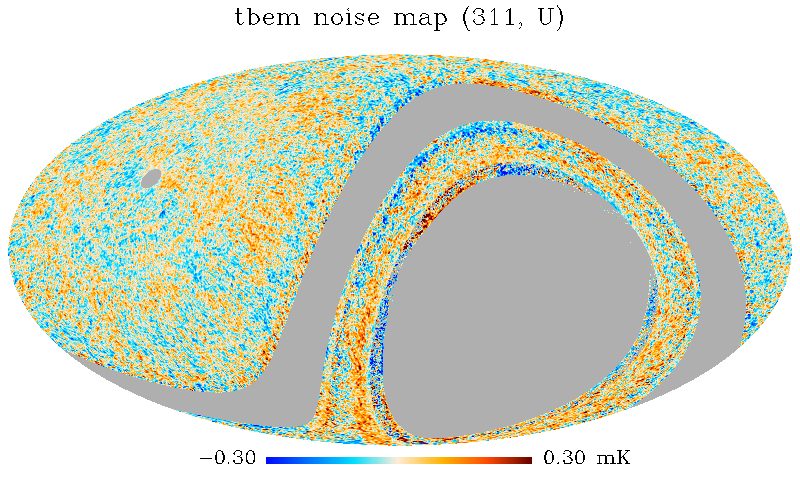}
        \caption{Three examples of null test difference maps for horn 3 11\,GHz, for the case of "independent baselines": ring (top), halfring (centre) and 
        tbem (bottom). From left to right, each row shows the Stokes I (left), Q (centre) and U (right) difference maps. For display purposes, all maps are smoothed to 1 degree resolution. The colour scale corresponds to $\pm 1$\,mK for the intensity maps, and $\pm 0.3$\,mK for polarization.  }
        \label{fig:noise311c}
    \end{figure*}

\begin{figure*}
        \centering
        \includegraphics[width=6cm]{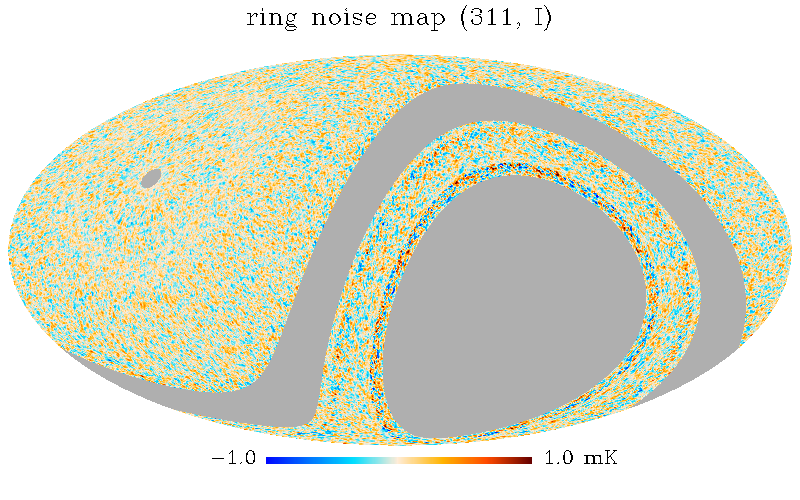}%
        \includegraphics[width=6cm]{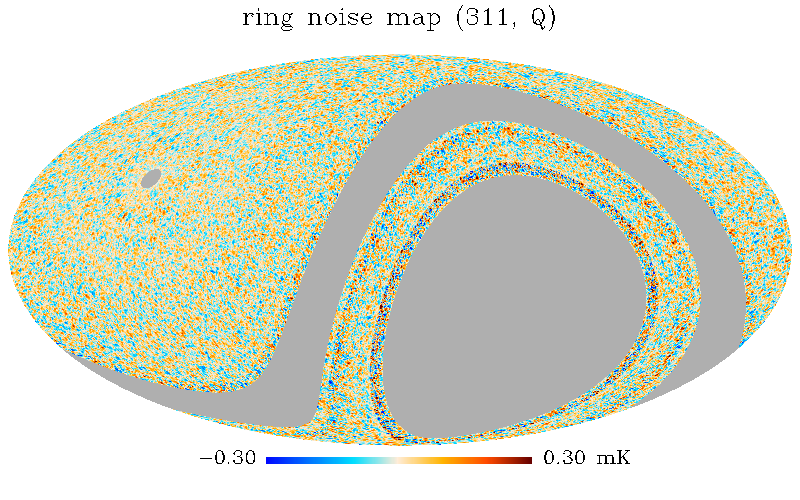}%
        \includegraphics[width=6cm]{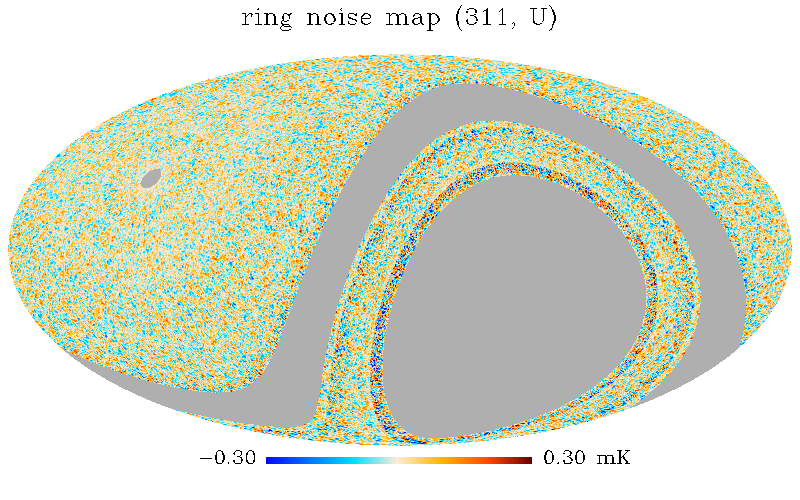}
        \includegraphics[width=6cm]{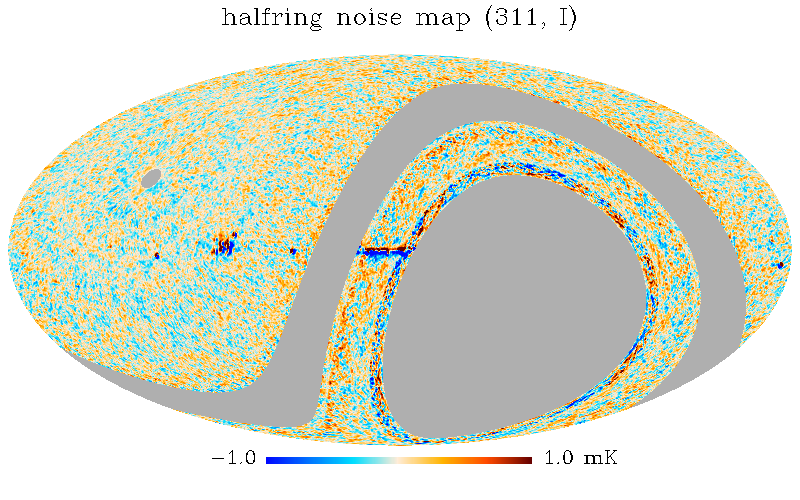}%
        \includegraphics[width=6cm]{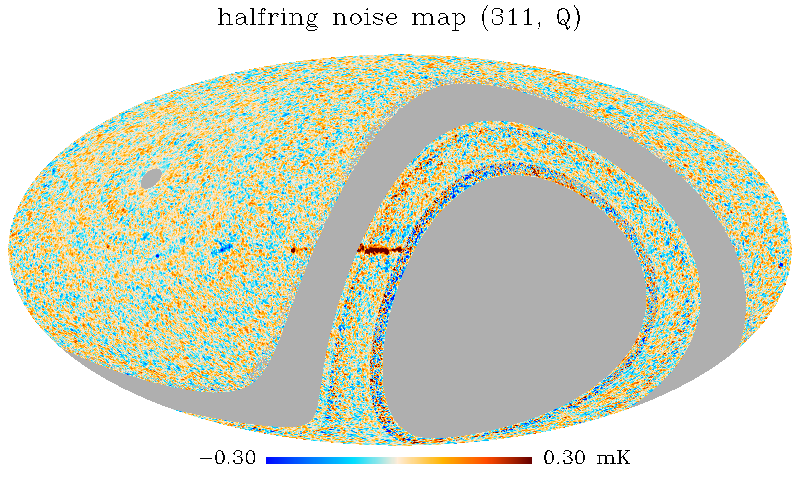}%
        \includegraphics[width=6cm]{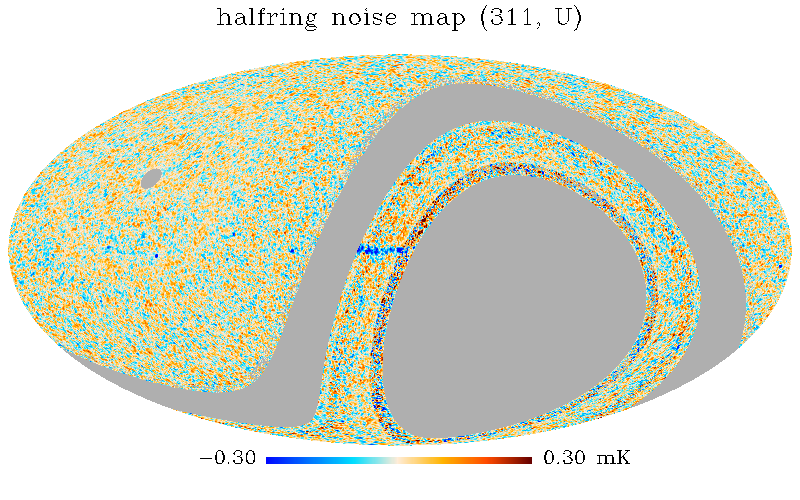} 
        \includegraphics[width=6cm]{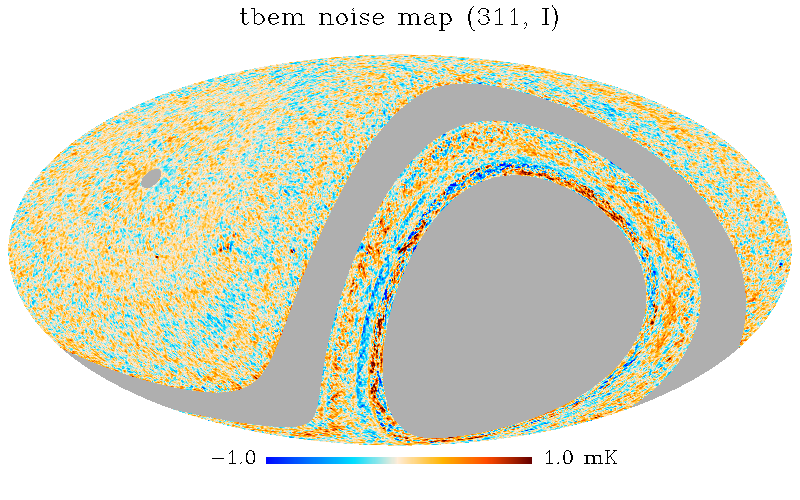}%
        \includegraphics[width=6cm]{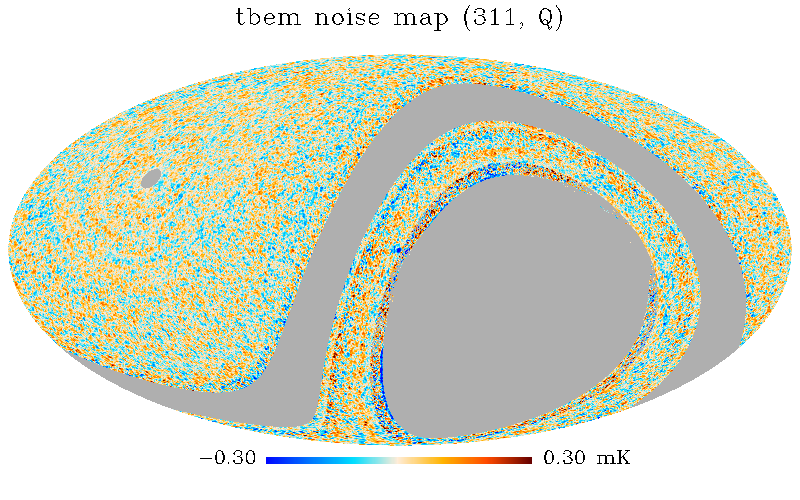}%
        \includegraphics[width=6cm]{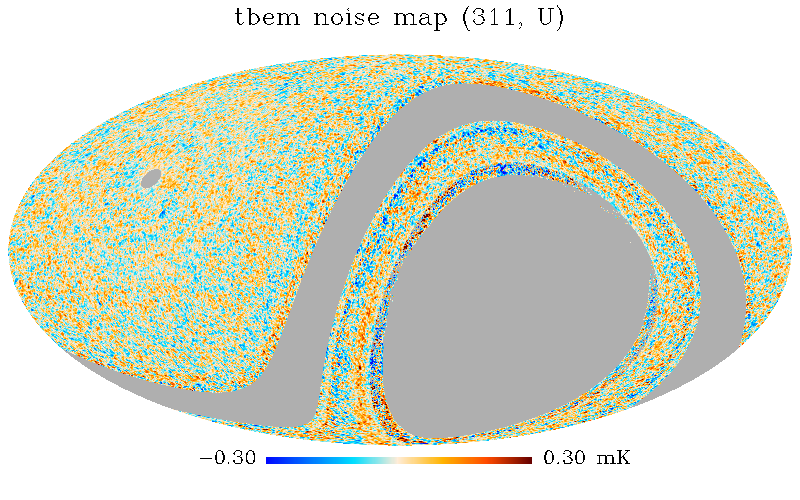}
        \caption{Same as Fig.~\ref{fig:noise311c}, but for the case of "common baselines" difference maps. The colour scale corresponds also to $\pm 1$\,mK for the intensity maps, and $\pm 0.3$\,mK for polarization. }
        \label{fig:noise311b}
    \end{figure*}

    As illustration, Figs.~\ref{fig:noise311}, \ref{fig:noise311c} and \ref{fig:noise311b} present few examples of null test difference maps for horn 3 11\,GHz, after smoothing to one degree resolution. Fig.~\ref{fig:noise311} shows the half mission difference map both for the "independent baselines" and the "common baselines" cases. 
    Fig.~\ref{fig:noise311c} contains the ring, halfring and tbem null tests for the case of independent baselines, while  Fig.~\ref{fig:noise311b} shows the same three cases for the "common baselines" solution. 
    
    \subsubsection{Other data splits}
    \label{subsec:perperiods}
    In addition to the null tests described above, other data splits have been considered and generated for the MFI wide survey. 
    In particular, we generated the four "maps per period", in correspondence to periods 1, 2, 5 and 6, both for the case of "independent baselines", and also with "common baselines". Although these four maps per period do not have exactly the same sky coverage (e.g. elevation 30 is only used in period 5) or the same format (e.g. polarization maps are not generated in period 1), they are still very useful for validation purposes (RFI residuals, gain model, calibration), as shown in the following sections. Moreover, these maps are also used for the study of transients and in particular, to characterise the potential variability of some bright point sources \citep[see e.g. ][]{sourceswidesurvey}.

    \subsection{Assessing systematic effects with null tests in power spectra and maps}
    
    \subsubsection{Power spectra}
    
    \begin{figure*}
    \centering
    \includegraphics[width=8cm]{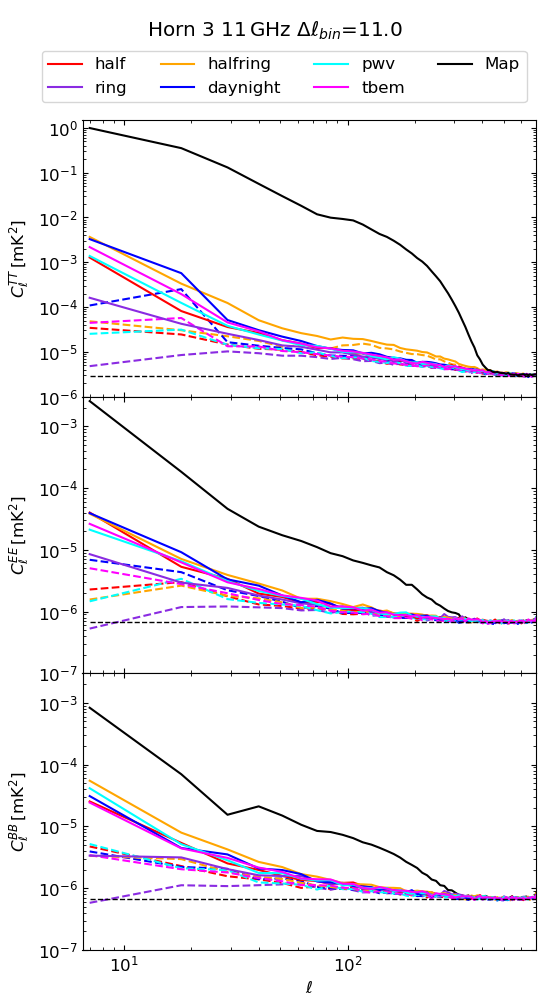}%
    \includegraphics[width=8cm]{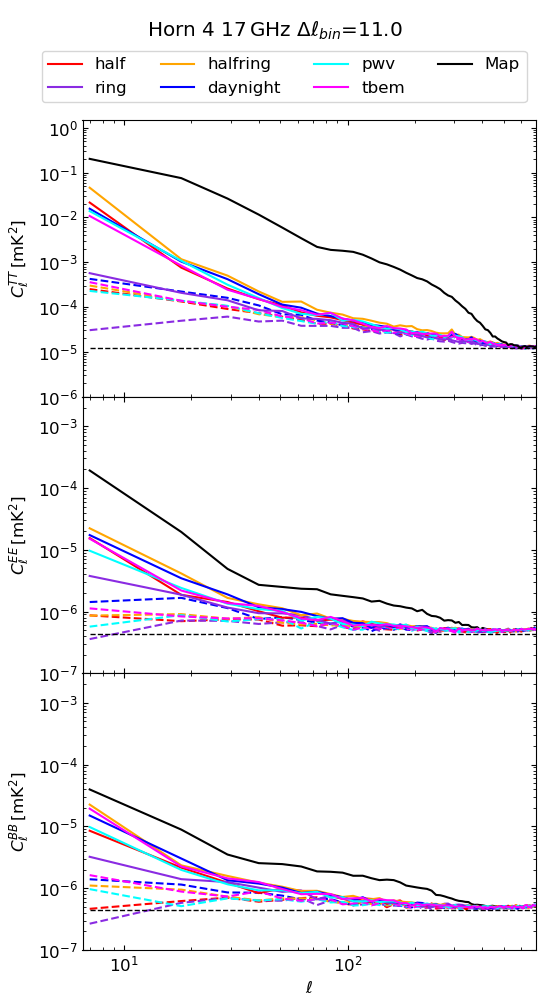}
    \caption{Binned raw power spectra ($\Delta \ell=11$) of the six null test difference maps discussed in the text, for horn 3 at 11\,GHz (left) and horn 4 at 17\,GHz (right). For comparison, we also include as dashed lines the spectra of the null test difference maps for the case of "common baselines". Black solid lines depict the spectra of the signal maps, while the horizontal dashed lines indicate the ideal white noise level for each map (see text for details). }
    \label{fig:nulltests}
    \end{figure*}
    
    Fig.~\ref{fig:nulltests} presents the binned raw power spectra (i.e. uncorrected for the beam and pixel window functions) of the six null-test difference maps described in the previous section and computed using eq.~\ref{eq:nwei}, compared to the raw power spectra of the final maps for each horn and frequency. For simplicity, we show only two cases, for horn 3 (11\,GHz) and horn 4 (17\,GHz). The equivalent figures for other horns and frequencies provide qualitatively similar results. 
    In this section, the $C_{\ell}$'s are computed with the publicly available code \xpol\footnote{\href{https://gitlab.in2p3.fr/tristram/Xpol}{https://gitlab.in2p3.fr/tristram/Xpol}}, which is based on a pseudo-$C_{\ell}$ estimator, and accounts for incomplete sky coverage (\citealt{xpol}). The mask adopted for this computation is the default one described in Section~\ref{sec:masks} (NCP+sat+lowdec), using a $5^\circ$ apodization with a cosine function, as implemented in the \namaster\ library \citep{namaster}. 
    In all panels, we show as a reference the angular power spectrum of the final map in black, and the spectra of the different "null test difference maps" (eq.~\ref{eq:nwei}) in various colours. For completeness, these figures also include the power spectra (as dotted lines) of the null-test difference maps for the case of "common baselines". We also include the ideal white noise level for each map, computed from the normalized weights (see Sect.~\ref{sec:noisereal} for details).
    
    All six null test difference maps present a similar behaviour, being asymptotically flat at high multipoles when reaching the white noise level, and increasing at low multipoles (large angular scales) as expected for residual $1/f$ noise. A comparison of these six null test power spectra provides a useful tool to identify and isolate different sources of systematic effects or calibration errors. In polarization, all null test spectra are basically consistent among them, except the ring case, which presents a slightly lower level of $1/f$ residuals at low multipoles. This behaviour is expected because the ring null test maps probe noise variations in scales of one minute, while the others cases (half, daynight, tbem, pwv) probe longer time scales. We also note that the halfring null test tends to be slightly above the other noise estimates, but again this is expected as this null test uses basically half of the possible crossings for each pixel, and thus the baseline solution is less constrained. However, this is not the case of halfring null test with common baselines, as in this case the baseline solution was obtained with the complete dataset. 
    For the intensity maps, the qualitative behaviour is similar to polarization, although the scatter among the null tests in the $1/f$ residuals at low multipoles is larger, particularly at 11\,GHz where the RFI contamination due to geostationary satellites was higher. In this case, the largest $1/f$ residuals at low multipoles correspond to the tbem, daynight and halfring cases, as expected. By construction, the halfring case amplifies the presence of residual RFI signals. In the case of tbem and daynight, this might indicate some low-level RFI residual which becomes visible when splitting the data according to the daily gain variations. We have confirmed that this is indeed the case, by constructing a new set of maps excluding period 1 in intensity, which was the period most affected by RFI due to the absence of the extended shielding in the telescope. 
    When generating the halfring null test for the case of no period 1, that small excess disappears. 
    Finally, we note that the power spectra for the null test difference maps with "common baselines" present a significantly lower level of $1/f$ residuals, as anticipated.

    \subsubsection{Maps}
    \label{sec:sysmaps}
    Visual inspection of the null test difference maps provides complementary information to the one obtained from the power spectra analysis, in terms of identifying localised features due to systematic effects. For example, the halfring null test maps (see the example for horn 3 at 11\,GHz in Fig.~\ref{fig:noise311c} and \ref{fig:noise311b}) can be used to assess the residual systematic effects due to uncertainties in the pointing model.
    As described in \citet{mfipipeline}, the pointing model solution for each MFI horn provides a reconstruction of the pointing with an overall 1 arcmin accuracy. Any residual pointing error will produce a characteristic feature in the halfring null-test map, as each one of the two sub-maps (halfring1 and halfring2) uses totally different ranges of local coordinates of the telescope. Indeed, the morphology and amplitude of the features appearing in the intensity map along the Galactic plane, both around the Galactic centre and the Cygnus area, match the expected residual signals for a shift of 1 arcmin between the halfring1 and halfring2 sub-maps.

\input{table_system_corr.tex}

    Null test difference maps can also be used for assessing the level of residuals in real space. For example, a cross correlation analysis of each null test difference map ($n$) with the corresponding signal map ($m$) can be used to trace the presence of both errors in the overall gain model or time-dependent RFI residuals. As usual, a cross-correlation coefficient $\alpha$ can be obtained as the minimum variance estimator that minimizes $n - \alpha m$ \citep[see e.g.][]{HMRM2004}. 
    Table~\ref{tab:systemcorr} presents the correlation coefficients $\alpha$, in percent units, for the case of the half mission null tests both for common and independent baselines. The analysis is carried out using the standard mask NCP+sat+lowdec defined in Sect.~\ref{sec:masks}.  These numbers are consistent with the power spectra analyses described in the previous subsection, and lie below the calibration uncertainty of the wide survey (see details  in Sect.~\ref{sec:cal} and Table~\ref{tab:summarycal}). In particular, for horn 3, these values are within one per cent, both in intensity ($I$) and polarization ($Q$, $U$). Moreover, in polarization all values are below $1.4$ per cent.

    \subsection{Noise characterization: $1/f$ noise and correlations}
    \label{sec:noiselevels}
    
    Noise parameters for the MFI instrument have been described in \citep{mfipipeline}, and are summarized in Table~\ref{tab:mfi_parameters}. Those values determine some of the noise properties of the final wide survey maps. Here, we use the half-mission difference maps (hereafter HMDM), constructed as in equation~\ref{eq:nwei} and for the case of "independent baselines", to assess the overall noise properties of the MFI wide survey, including white noise levels, $1/f$-type components and correlation properties. 
    The analyses are done both in harmonic (Sect.~\ref{sec:noiseharmonic}) and real (Sect.~\ref{sec:noisereal}) space, using the standard mask defined as NCP+sat+lowdec in Sect.~\ref{sec:masks}, which contains the region in the declination range $6^\circ <\delta < 70^\circ$. 
    In addition, and due to the MFI receiver design, there are well-known noise correlations at the TOD level (also called "common mode $1/f$ noise") between channels of the same horn, which are inherited by the final maps. We use the cross-spectra of different HMDM to characterize these noise correlations at the map level, both between the two frequencies of the same horn (Sect.~\ref{sec:noisecorr_freqs}) and between the correlated and uncorrelated channels contributing to a given map (Sect.~\ref{subsec:crossnoisechan}). 
    
    \subsubsection{Noise properties in harmonic space}
    \label{sec:noiseharmonic}
\begin{figure}
    \centering
    \includegraphics[width=\columnwidth]{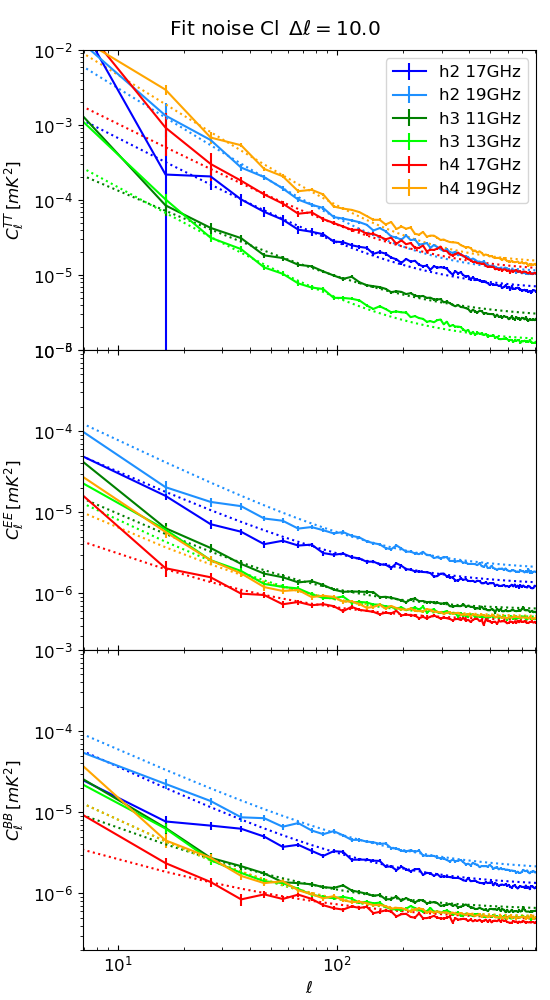}
    \caption{Best-fit solutions to the power spectra of the half-mission difference maps (HMDM). Using eq.~\ref{eq:fit_cl}, we obtain the best-fit models depicted here as dotted lines. The corresponding coefficients are listed in Table~\ref{tab:noiseps}.  }
    \label{fig:noisepsfit}
    \end{figure}

    Our analysis of the noise properties in harmonic space is shown in Fig.~\ref{fig:noisepsfit} and Table~\ref{tab:noiseps}. 
    The power spectra for the HMDM are computed using \namaster\, and then fitted with the following empirical model:
    \begin{equation}
        C_{\ell}= C_{\rm w}\left(1+\left(\frac{\ell_{\rm k}}{\ell}\right)^{\alpha}\right),
        \label{eq:fit_cl}
    \end{equation}
    which accounts for a $1/f$ noise component projected on sky. We fit for the three parameters in this equation in two steps. First, we obtain the white noise level $C_{\rm w}$ as the average level of the angular power spectrum at high multipoles ($\ell \in [700,800]$ for TT, and $\ell \in [600,800]$ for EE and BB). Then, the knee-multipole $\ell_{\rm k}$ and the slope $\alpha$ are obtained analytically after fitting for a linear relation in $\log_{10}(C_{\ell}-C_{\rm w})\,vs\, \log_{10}(\ell)$, in the multipole range $\ell \in [20,100]$ for both intensity and polarization. To have a better fit in the high multipole range for the EE and BB case of horn 2, we use here the range $\ell \in [80,300]$.
    The parameter $C_{\rm w}$, which represents the white noise level of the full maps, can be translated into the commonly used quantity $\sigmadeg$, the equivalent noise level (rms) of the map for a 1-degree beam, with the relation $\sigmadeg =\sqrt{C_{\rm w}/ \omegadeg}$, where $\omegadeg$ is the solid angle of a Gaussian beam with a FWHM of 1-degree, which corresponds to $0.345$\,msr$=1.133$\,deg$^2$. These numbers (third column in Table~\ref{tab:noiseps}) can be directly compared to those obtained with real space statistics in the next subsection\footnote{Note that if we want to quote the map sensitivity in the usual units of $\mu$K.arcmin (or $\mu$K.deg), we can not use directly $\sigmadeg$, as we have to account for the $\sqrt{\omegadeg}$ factor. For instance, the  white noise level of the MFI 311 map in polarization is $42.2\,$$\mu$K per 1-degree beam, or equivalently, $44.9$\,$\mu$K.deg $=2695.1$\,$\mu$K.arcmin, consistently with the reported $C_{\rm w}$ value. }.
    
    In summary, for the intensity spectra, horn 3 presents the lowest noise levels both for the $1/f$ and the white noise components, while horn 4 is the most noisy one. However, in polarization, horn 4 has a much better performance, yielding the lowest noise levels, while horn 2 is the noisiest in this case. Although the noise levels for horn 3 in polarization are slightly higher than those for horn 4, given that the sky signal is significantly brighter at lower frequencies (see Fig.~\ref{fig:nulltests}), the wide survey polarization maps of horn 3 (11 and 13\,GHz) have the better signal-to-noise ratios. 
    Regarding the correlated noise component, we find that the noise spectra in intensity are dominated by the $1/\ell$ component down to scales of 1 degree, as a consequence of the large $1/f$ noise in the intensity TODs. In polarization, we find typical knee-multipoles of $\ell_{\rm k} = 54$--$86$ for horns 3 and 4, as expected for the significantly lower correlated noise component. 
    
    \input{table6_fit2noisespectrum_nochi2.tex}


    \subsubsection{Noise properties in real space}
    \label{sec:noisereal}
    First, we normalize the HMDM by dividing each individual pixel by the square root of its covariance as computed from the map weights (i.e., $\sigma_i = w_i^{-1/2}$). We recall that those weights are propagated through the pipeline and the map-making code, and were computed from the variance of each individual 40\,ms sample in the TOD. For this normalized map, we fit for the standard deviation within the reference mask. The results are shown in Table~\ref{tab:noise}. As expected, these values are reasonably close to unity for the case of the polarization maps, while in intensity these factors are greater than 3 in all cases. These deviations from unity are generally consistent with the level of $1/f$ noise in each case (see e.g. Table~\ref{tab:noiseps}). This set of values could be used to renormalize the weight maps, so they would be representative of the actual noise levels, while preserving the underlying spatial distribution of the hit maps. Indeed, these factors are used to estimate the ideal white noise of each map at the power spectrum level. For example, the dashed lines in Fig.~\ref{fig:nulltests} are computed with these rescaled weight maps. Moreover, these rescaled weight maps can be used to produce signal-to-noise maps for each frequency (see Appendix~\ref{app:snr}).

\input{table3_noise.tex}

    As a second analysis, we repeat the same procedure but now we normalize each difference map according to the square root of the number of hits. Taking into account that hits correspond to 40\,ms samples, we can obtain from here representative normalization values to describe the noise standard deviation as 
    \begin{equation}
    \sigma = \frac{\sigma_0}{\sqrt{\nhit}} .
    \end{equation}
    Our results are shown in Table~\ref{tab:noise2}. The values obtained for the MFI wide survey in polarization are comparable to those obtained for raster scan observations with the MFI in smaller regions \citep[see e.g. last column in Table~1 from][]{W44}, and represent the actual sensitivity of the instrument.

\input{table4_sigma0.tex}

    Finally, we can also estimate the noise variance directly from the HMDM, using apertures of 1-degree radius across the same mask. The average values obtained from this analysis are given in Table~\ref{tab:noise3}. To facilitate the comparison with the numbers in the previous subsection, these values are re-scaled by the factor $\sqrt{\omegapix / \omegadeg}$, so they represent $\sigmadeg$. In summary, the final combined maps of the MFI wide survey in polarization present sensitivities within the range 35--40\,$\mu$K per 1-degree beam for the four frequencies.

\input{table5_rms.tex}

    \subsubsection{Noise correlations between frequencies of the same horn}
    \label{sec:noisecorr_freqs}
    
    Two MFI frequency channels from the same horn have a correlated ("common mode") $1/f$ noise component, due to the fact that they share the same LNA. This fact is particularly relevant for the intensity maps, which are strongly dominated by correlated noise. Because of this reason, our final wide 
    survey maps at 11 and 13\,GHz have correlated noise between them, as is the case for the maps at 17 and 19\,GHz.

    In order to characterize the actual degree of correlation between two wide-survey maps obtained from the same horn, we use the normalized cross-spectra between the corresponding null-test difference maps. As in the previous section, we use as a reference the HMDM for the case of independent baselines. 
    Following the notation in Sect.~\ref{sec:nulltests}, here $n_{h,f}$ represents the half-mission difference map for horn $h$ and frequency $f$ (see eq.~\ref{eq:nwei}). Then, for a given horn $h(=2,3,4)$, the normalized correlation between the lowest frequency band $f_1$ and the highest frequency band $f_2$, is given by
    \begin{equation}
    \label{eq:rhoell}
    \rho_\ell \equiv \frac{C^{n_{h,f_1} \times n_{h,f_2}}_\ell}{ \sqrt{ C^{n_{h,f_1}}_\ell C^{n_{h,f_2}}_\ell } },
    \end{equation}
    where $C^{n_{h,f_1} \times n_{h,f_2}}_\ell$ is the cross-spectrum between the two difference maps, and $C^{n_{h,f_i}}_\ell$ for $i=1,2$ represents the auto-spectra.  

\begin{figure*}
    \centering
    \includegraphics[width=1.5\columnwidth]{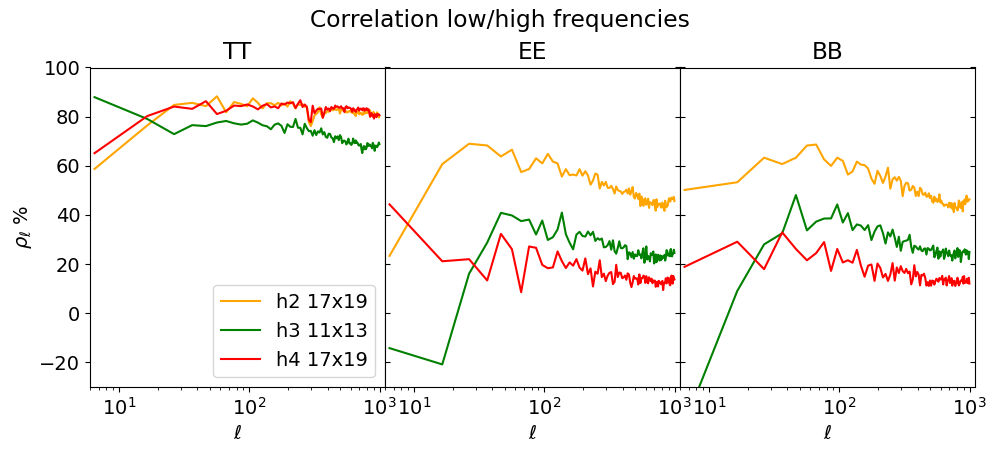}
    \caption{Cross-correlation spectra of the half-mission difference maps between the two frequencies of the same horn, for TT (left), EE (centre) and BB (right). }
    \label{fig:crossnoise}
    \end{figure*}

    Figure~\ref{fig:crossnoise} shows this normalized cross-spectrum $\rho_\ell$ in the final MFI wide survey maps for horns 2, 3 and 4,  both in intensity and polarization.
    In intensity, the resulting noise correlation is of the order of $75$--$85$ per cent for the three horns, being relatively flat in the multipole range $20 \lesssim \ell \lesssim 300$. 
    In polarization, the correlation is found to be $\sim 20$--$60$ per cent depending on the horn, with a moderate dependence on the multipole, being slightly lower at higher multipoles (smaller scales). In order to obtain a representative value for this correlation, we compute the average (and standard deviation) of $\rho_\ell$ in the multipole range $[20, 200]$. 
    For TT, we obtain $85.0 \pm 0.3$\,\%, $76.5\pm0.4$\,\% and $84.1\pm0.3$\,\% for horns 2, 3 and 4, respectively. 
    In polarization, for EE we obtain $60.7 \pm 1.0$\,\%, $32.8\pm1.4$\,\% and $20.9\pm1.2$\,\%, and 
    for BB we have $60.8 \pm 0.9$\,\%, $36.2\pm1.1$\,\% and $21.7\pm1.2$\,\%, again for horns 2, 3 and 4. 
    This high degree of correlation has to be taken into account when doing combined analyses of the two frequency maps of the same horn. 
    
    As a consistency check, and in order to test that these inter-frequency correlations are entirely due to instrumental (common mode) $1/f$ noise, and not to external correlated signals produced either by the atmosphere or by RFI, we performed the same analysis but now comparing two frequencies coming from two different horns. In particular, we evaluated the cross-correlation of horn 2 at 17\,GHz with horn 4 at 19\,GHz, obtaining $-0.64\pm2.47$\,\%,  $0.47\pm0.96$\,\%  and  $-0.49\pm0.96$\,\% for TT, EE, and BB, respectively. In addition, the cross-correlation of horn 4 at 17\,GHz with horn 2 at 19\,GHz gives  $-0.32\pm 2.13$\,\%, $0.99\pm0.83$\,\% and  $0.72\pm0.76$\,\%, again for TT, EE and BB. In both cases, the results are consistent with zero within the error bar. 
    
    \subsubsection{Noise correlations between channels}
    \label{subsec:crossnoisechan}
    
    As described above, for any given horn and frequency sub-band of MFI, we produce two versions of the intensity and polarization maps, the so-called correlated ($x_{\rm c}$) and uncorrelated ($x_{\rm u}$) maps. Due to the MFI design, we expect a high degree of correlation between the noise affecting those two versions of the intensity maps, due to the fact that they all share the same LNAs and there is no cancellation of the $1/f$ noise in any of the sums of channels contributing to $x_{\rm c}$ and $x_{\rm u}$. 
    We can use the same methodology applied in the previous sub-section to characterize this correlation level of the noise between correlated and uncorrelated channels maps for a given horn and frequency. We also use the half-mission null test maps as a reference for this analysis. But now, in the post-processing stage, we generate two independent versions for each individual map, using either the correlated or the uncorrelated information only. With these maps, and using again eq.~\ref{eq:nwei}, for a given horn and frequency we can produce $n_{\rm c}$ and $n_{\rm u}$, the half-mission difference maps of the correlated and uncorrelated channels, respectively. 
    In analogy to equation~\ref{eq:rhoell}, we now compute
    \begin{equation}
    \rho_\ell \equiv \frac{ C^{n_{\rm c} \times n_{\rm u}}_\ell }{ \sqrt{ C^{n_{\rm c}}_\ell C^{n_{\rm u}}_\ell } } ,
    \end{equation}
    where $C^{n_{\rm c} \times n_{\rm u}}_\ell$ is the cross-spectrum between the two difference maps, and $C^{n_{\rm c}}_\ell$ and $C^{n_{\rm u}}_\ell$ are the auto-spectra.  
    
    \begin{table}
    \caption{Average inter-channel correlations $<\rho_\ell>$ of the half-mission difference maps between the correlated and uncorrelated channels for a given horn and frequency. The values correspond to the mean and standard deviation of the $\rho_\ell$ displayed in Figure~\ref{fig:crossnoisechan}, computed in the multipole range 20--200. }
    \label{tab:cross_internchan}
    \centering
    \begin{tabular}{@{}cccc}
    \hline
    Channel & TT  (\%) & EE  (\%) & BB  (\%)  \\
    \hline
    217 & $97.14\pm0.08$ & $-1.36\pm1.10$ & $-0.91\pm1.37$  \\
    219 & $87.91\pm0.34$ & $ 2.40\pm1.19$ & $-0.60\pm1.04$  \\
    311 & $85.82\pm0.26$ & $ 3.81\pm0.75$ & $ 2.38\pm1.18$ \\
    313 & $79.65\pm0.42$ & $-0.32\pm0.89$ & $-2.01\pm 0.91$ \\
    417 & $97.95\pm0.04$ & $-3.26\pm1.07$ & $-1.22\pm 1.12$ \\
    419 & $91.56\pm0.22$ & $-2.81\pm0.80$ & $-0.61\pm 1.15$ \\
    \hline
    \end{tabular}
    \end{table}
    
    \begin{figure*}
    \centering
    \includegraphics[width=1.5\columnwidth]{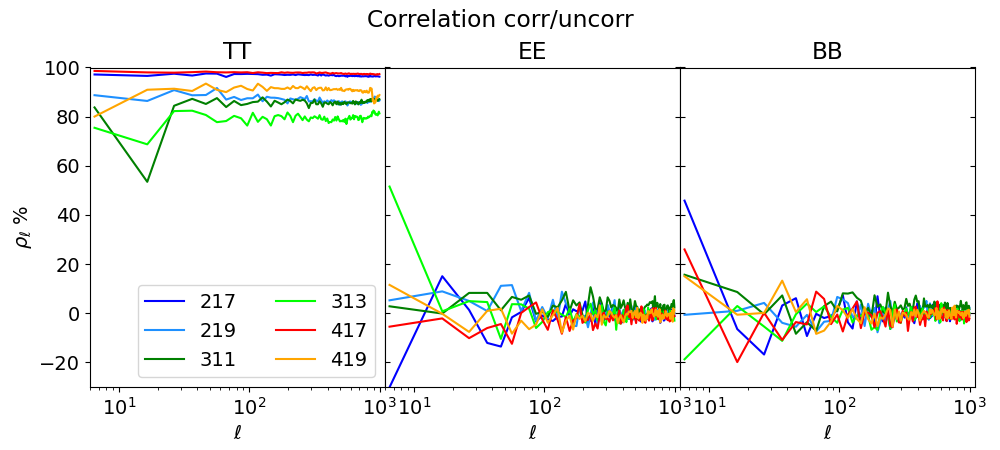}
    \caption{Cross-correlation spectra of the half-mission difference maps between the correlated and uncorrelated channels from the same horn and frequency, for TT (left), EE (centre) and BB (right). }
    \label{fig:crossnoisechan}
    \end{figure*}
    
    Fig.~\ref{fig:crossnoisechan} shows the resulting correlation level between correlated and uncorrelated channels. As expected, we find a very high degree of correlation (of the order of 90 per cent) in intensity, and a signal consistent with zero in polarization (both for EE and BB spectra). Again, as a representative value for this correlation, we compute the average and standard deviation of $\rho_\ell$ in the multipole range $[20, 200]$. The results are shown in Table~\ref{tab:cross_internchan}.
    These average correlation values in intensity are used in the pipeline in order to produce the final combinations of correlated and uncorrelated channels, as described in Section~\ref{subsec:combine_cu}.

    \subsection{Impact of residuals on the power spectra: atmospheric and RFI corrections}
    \label{sec:validares}
    
    As described in Sect.~\ref{sec:data}, the MFI wide-survey pipeline incorporates several steps tailored to correct for the contribution of atmospheric and RFI signals in the final maps. 
    Atmospheric corrections are applied at the TOD level (see Sect.~\ref{sec:atmos}), and for intensity maps only. When projected on maps, they appear as large scale patterns with an increasing amplitude in frequency (see Fig.~\ref{fig:atmos}). 
    RFI signals are corrected both in the intensity and polarization maps, in two stages. First, RFI signals at the TOD level are corrected using spatial templates as described in Sect.~\ref{sec:rficorr}. When projected on sky, they also appear as large scale patterns with a moderate amplitude ($\lesssim 0.5$\,mK) and presenting a higher amplitude in intensity (see Fig.~\ref{fig:rfidiff}). 
    Later, in the post-processing stage (Sect.~\ref{sec:postprocessing}), any residual RFI signals emerging after co-adding all data in the map-making process are corrected using a function of the declination. In terms of relative amplitude, this is by far the largest correction applied to the MFI wide-survey polarization data, with its amplitude being higher in the 11 and 13\,GHz channels due to the emission of geo-stationary satellites entering through the far sidelobes. Indeed, the effective transfer function of the MFI wide survey in polarization is mainly determined by this effect (see Sect.~\ref{sec:transfer}).
    
    \begin{figure*}
    \centering
    \includegraphics[width=0.66\columnwidth]{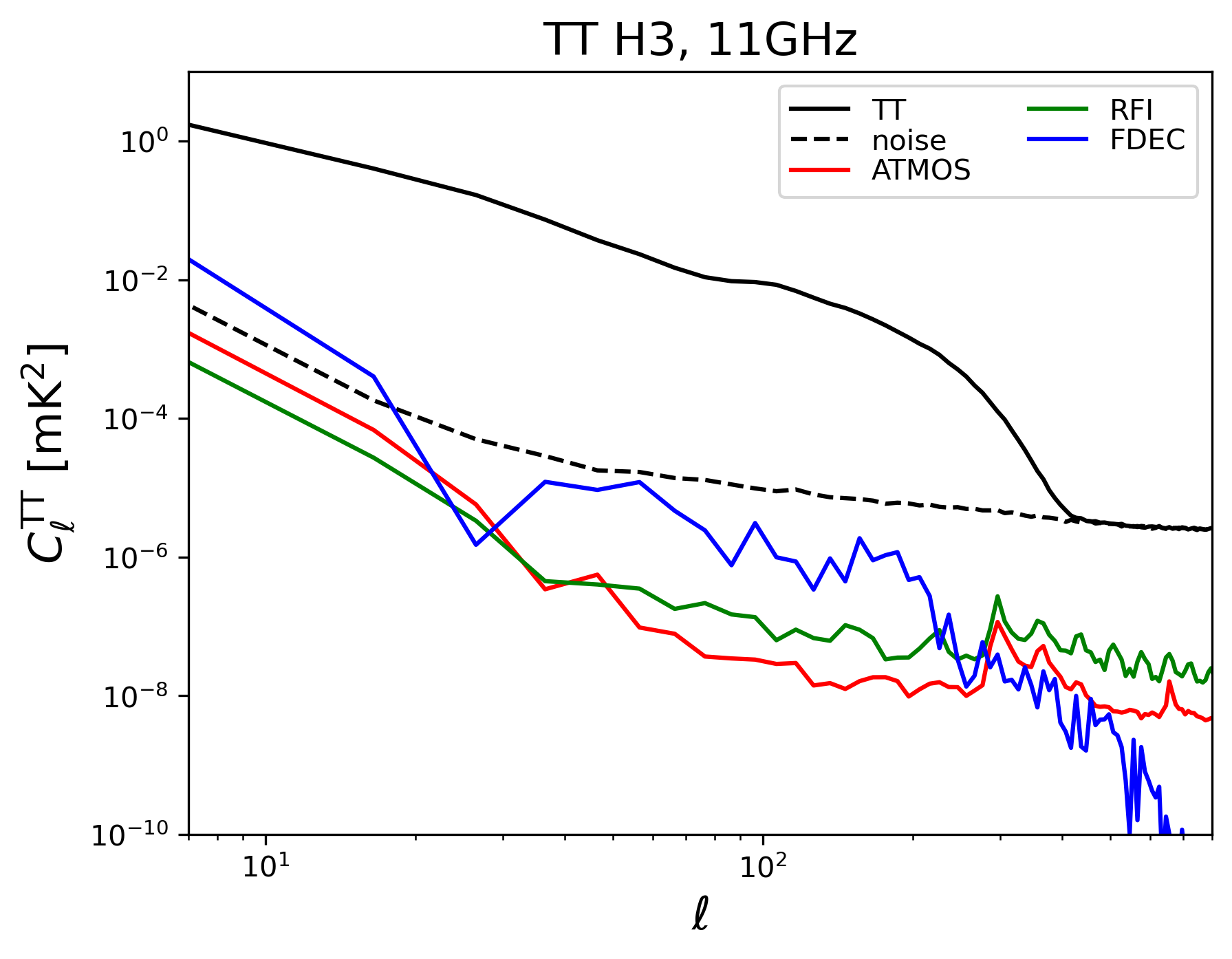}%
    \includegraphics[width=0.66\columnwidth]{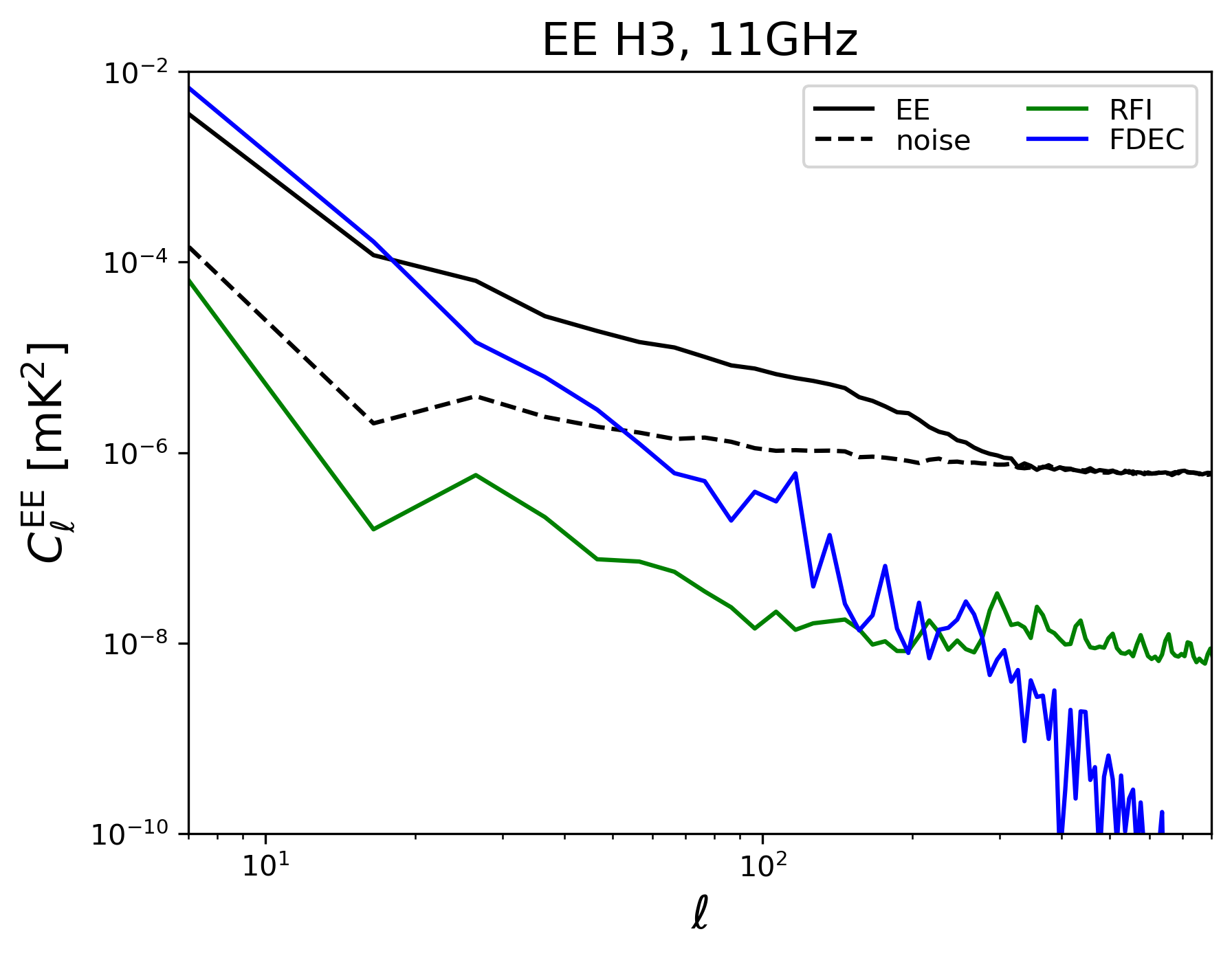}%
    \includegraphics[width=0.66\columnwidth]{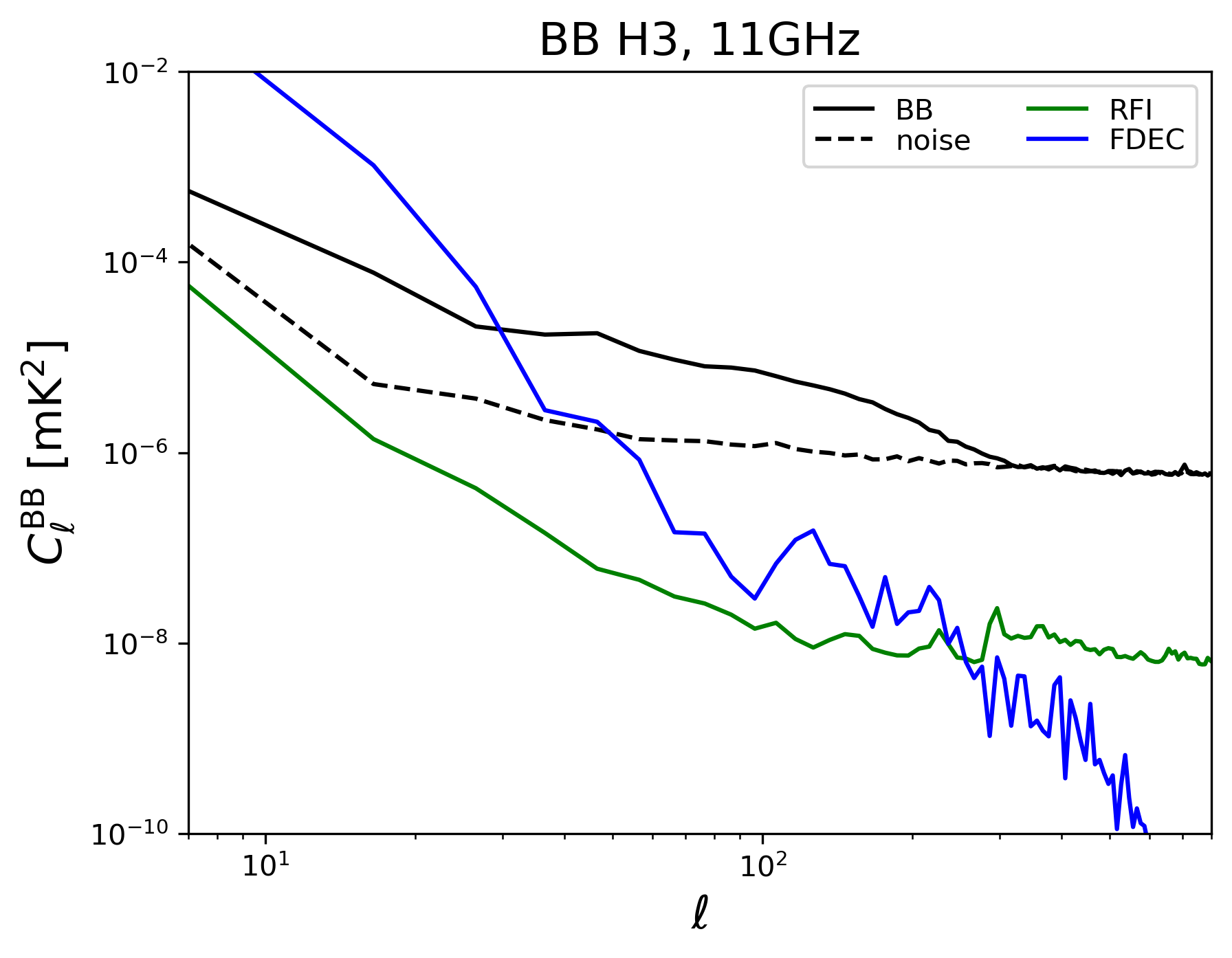}
    \includegraphics[width=0.66\columnwidth]{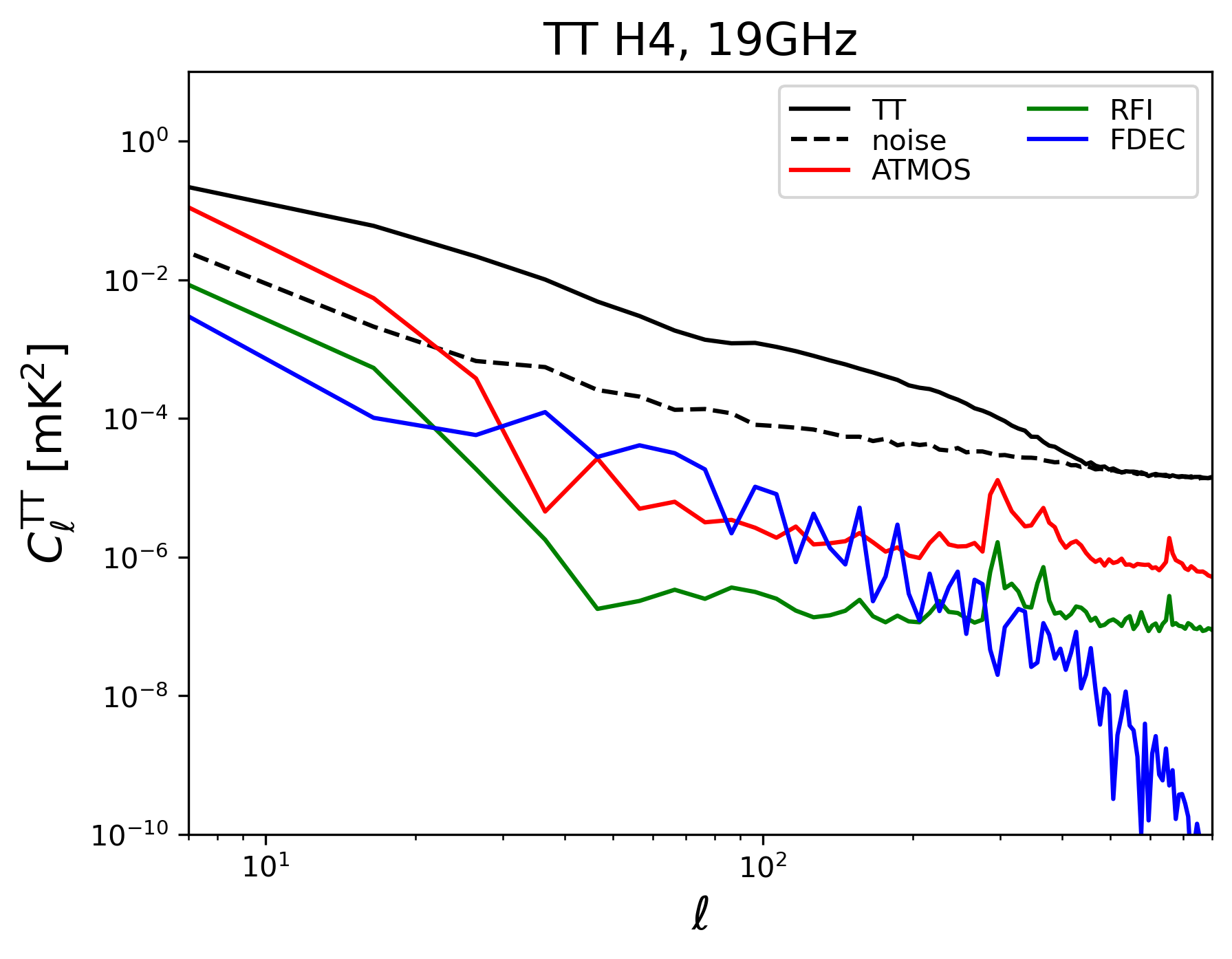}%
    \includegraphics[width=0.66\columnwidth]{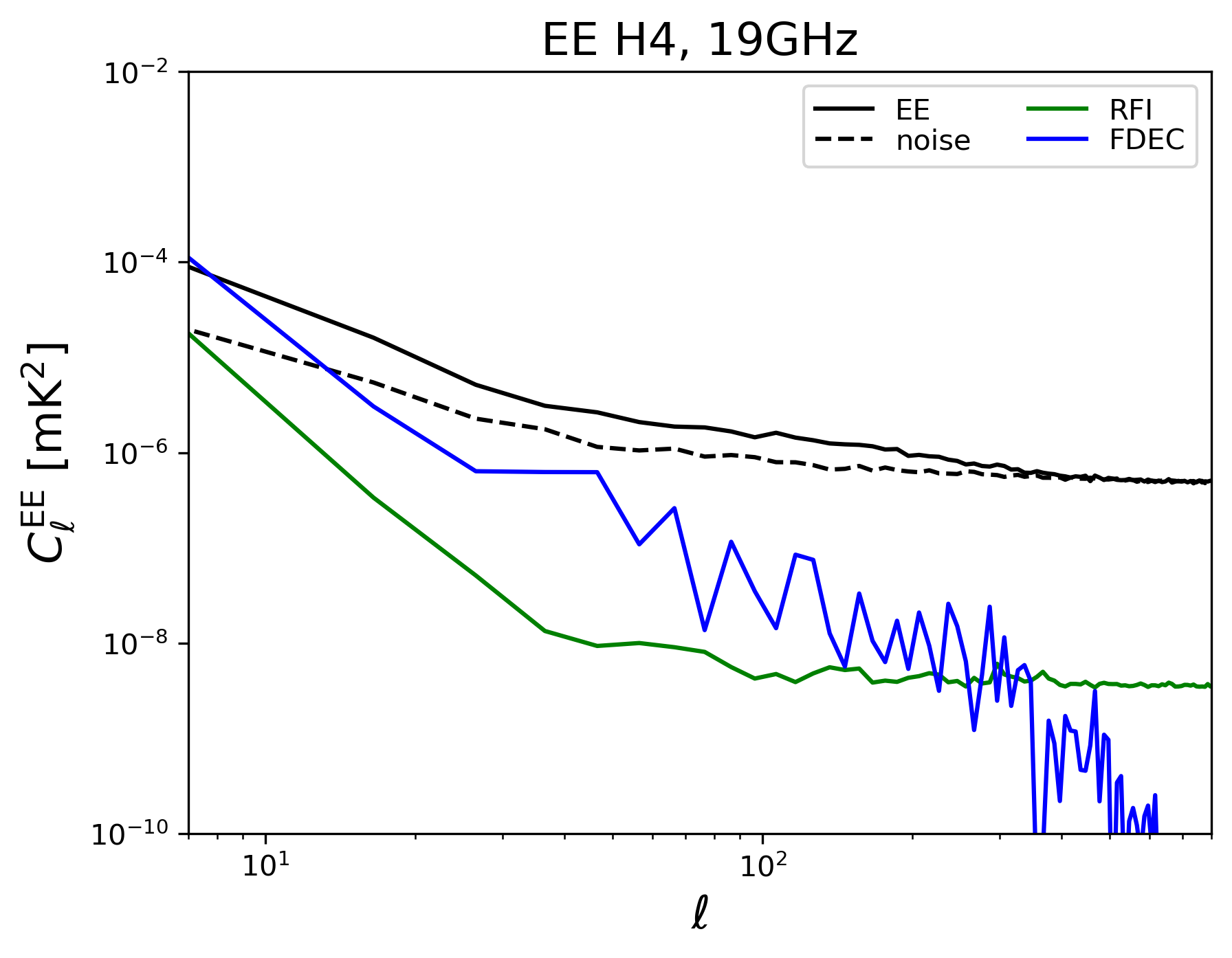}%
    \includegraphics[width=0.66\columnwidth]{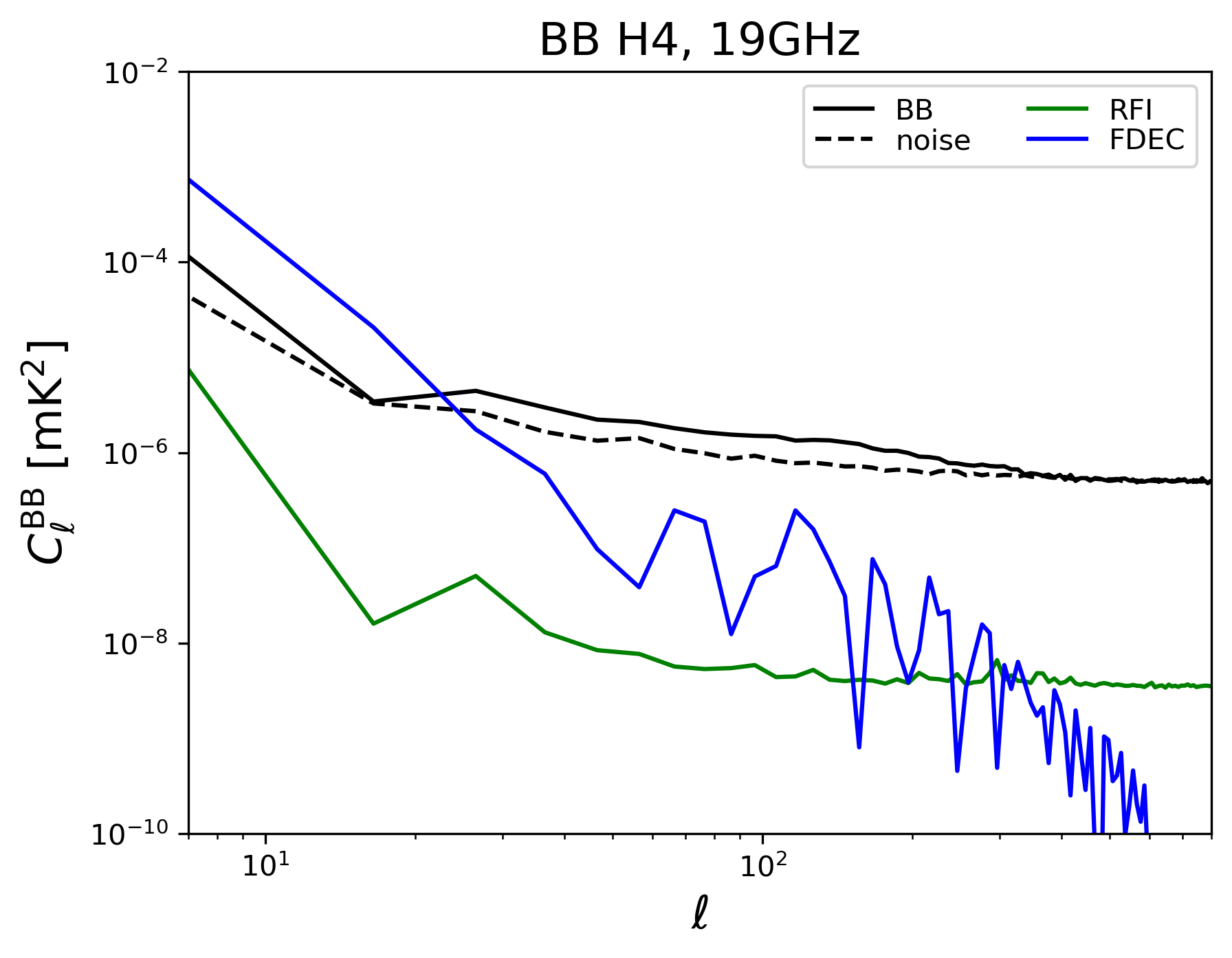}
    \caption{Raw angular power spectra of the ATMOS (red), RFI (green) and FDEC (blue) patterns removed from the MFI wide survey maps, for horn 3 at 11\,GHz (top row) and horn 4 at 19\,GHz (bottom row). For each case, we represent TT (left), EE (centre) and BB (right) spectra. Solid black lines correspond to the angular power spectra of the corresponding wide survey maps, while dashed lines correspond to the half-mission difference maps. All spectra are computed using the default analysis mask (NCP+sat+lowdec). }
    \label{fig:norfi_noatmos}
    \end{figure*}
    
    In order to quantify the relative importance of these three corrections, and to evaluate the possible impact of any residual systematic effects due to uncorrected contamination in the final wide survey maps, we have computed the angular power spectra of those patterns that are removed from the maps, 
    and we have compared them with the spectra of the final maps and the half-mission noise levels.
    Figure~\ref{fig:norfi_noatmos} shows the resulting power spectra for the two extreme frequency values (11 and 19\,GHz) taken here as representative cases, with 11\,GHz being the one with highest RFI contamination, and 19\,GHz the one with the highest atmospheric contamination. In this plot, we use the notation of ATMOS, RFI and FDEC for "atmospheric", "RFI at TOD level using a function of azimuth", and "RFI at the map level using function of declination" corrections, respectively. 

    Regarding the atmospheric contribution (ATMOS) to the intensity power spectra, the removed pattern is subdominant at all angular scales in the 11\,GHz case when compared to the noise level. At 19\,GHz, we have a similar behaviour at small angular scales ($\ell \ga 20$). However, the atmospheric residuals become comparable to the noise levels for multipoles $\ell \lesssim 20$, as can be anticipated from the visual inspection of Fig.~\ref{fig:atmos}. 
    
    For the RFI contribution at the TOD level, the removed patterns both in intensity and polarization are always below the noise levels at all frequencies, although they become comparable to the noise at large angular scales $\ell \lesssim 20$. Thus, in this RFI case, as well as for ATMOS, any residual systematic effect with an amplitude being a fraction of the applied correction will have negligible impact at the power spectrum level. 
    
    Finally, for the removed FDEC patterns, the largest amplitude is found at 11\,GHz, as anticipated from Fig.~\ref{fig:fdec}. At this frequency, the removed pattern in intensity is above the correlated noise level for multipoles $\ell \lesssim 20$. Moreover, in polarization, the applied correction is found to be critical, in the sense that its amplitude is above the sky signal for multipoles $\ell \lesssim 30$. When looking at the 19\,GHz FDEC patterns, in intensity the corrected amplitude is always below the noise levels for all multipoles, while in polarization again becomes comparable to the sky signal for $\ell \lesssim 20$. 
    In this case, although the underlying assumption for modelling residual RFI signals using a function of the declination is very robust and well tested, it is important to keep in mind that residual contributions might have an impact on the polarization maps of the MFI wide survey on large angular scales. In addition, as explained in Sect.~\ref{sec:transfer}, the FDEC procedure also affects the same multipole range by introducing a signal error in the reconstructed sky. For these reasons, in the following sections involving scientific analyses based on power spectra of the polarization signals in the wide survey, we adopt the conservative choice of restricting the study to multipoles $\ell \ge 30$. 
    
    \subsection{Inter-frequency comparison of the MFI maps}
    
    \begin{figure*}
    \centering
       \includegraphics[width=6cm]{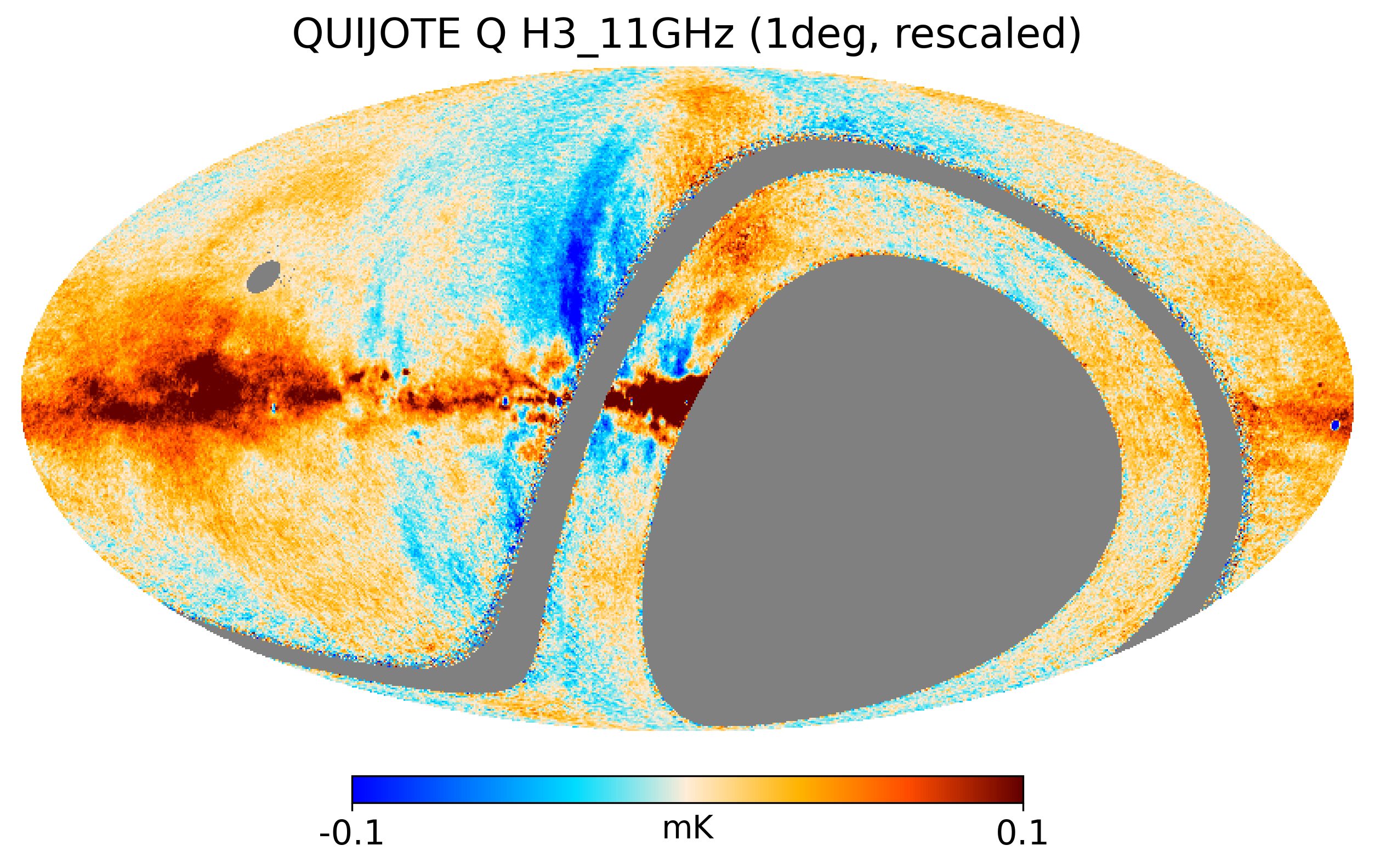}%
       \includegraphics[width=6cm]{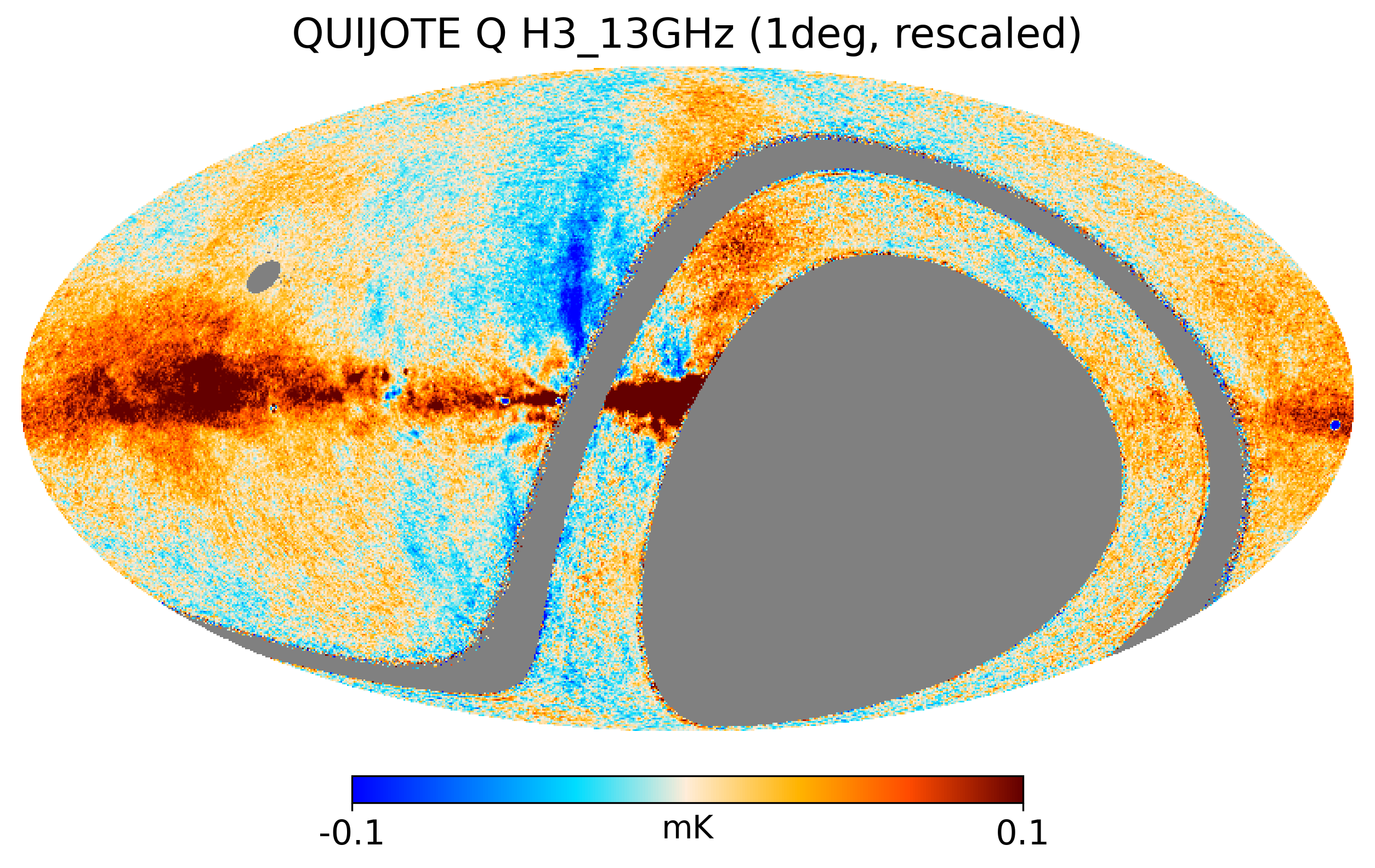}%
       \includegraphics[width=6cm]{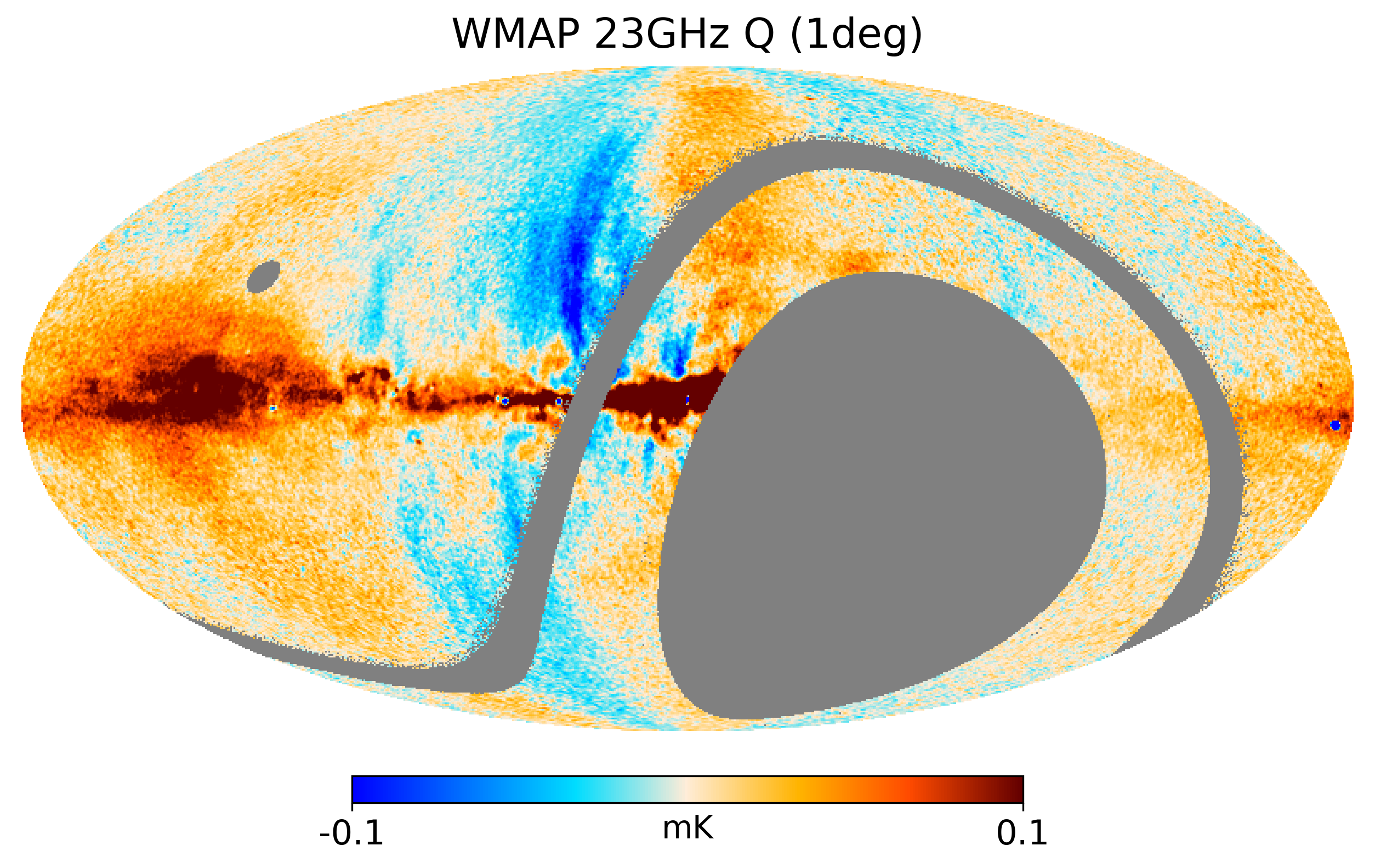}
       \includegraphics[width=6cm]{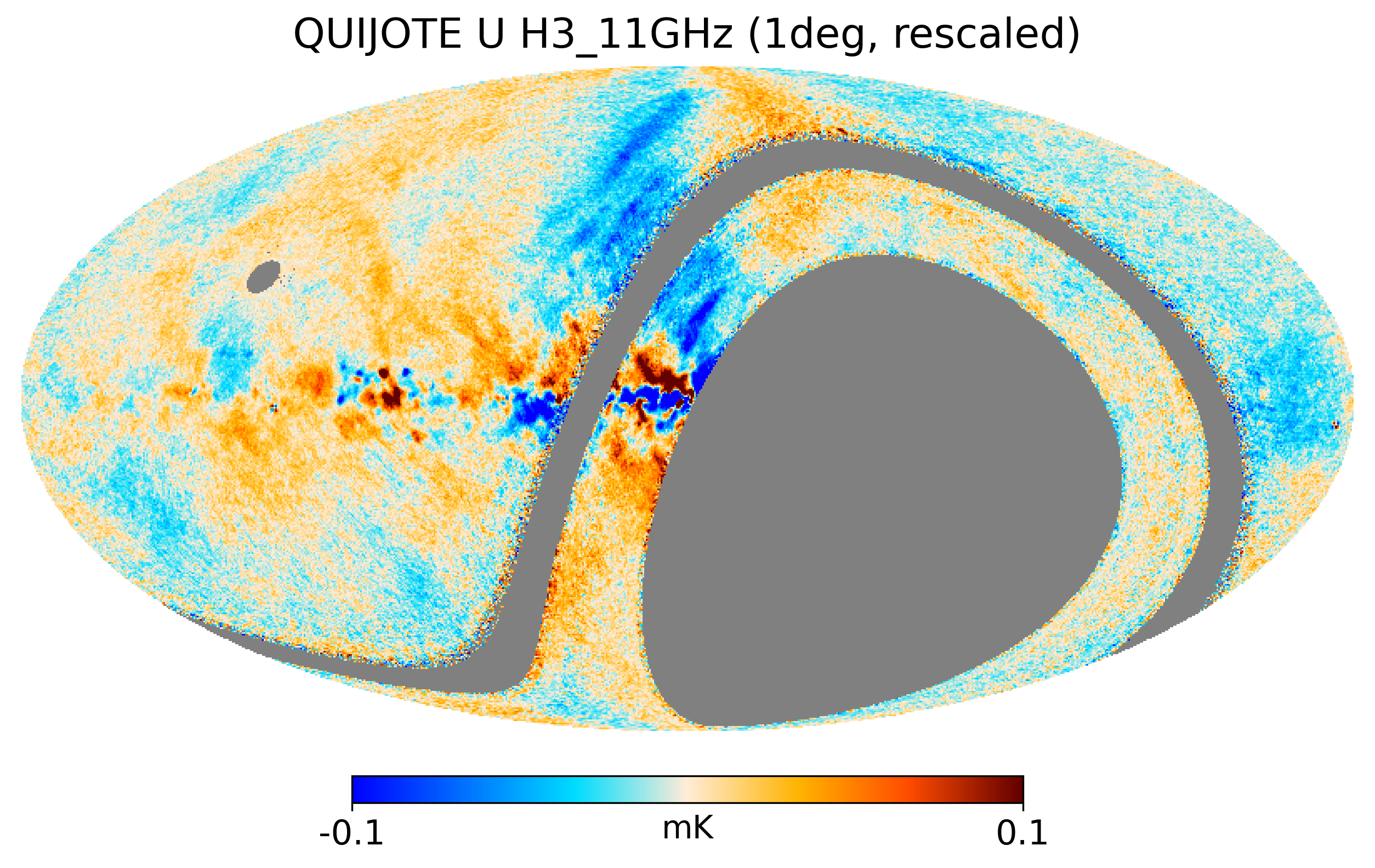}%
       \includegraphics[width=6cm]{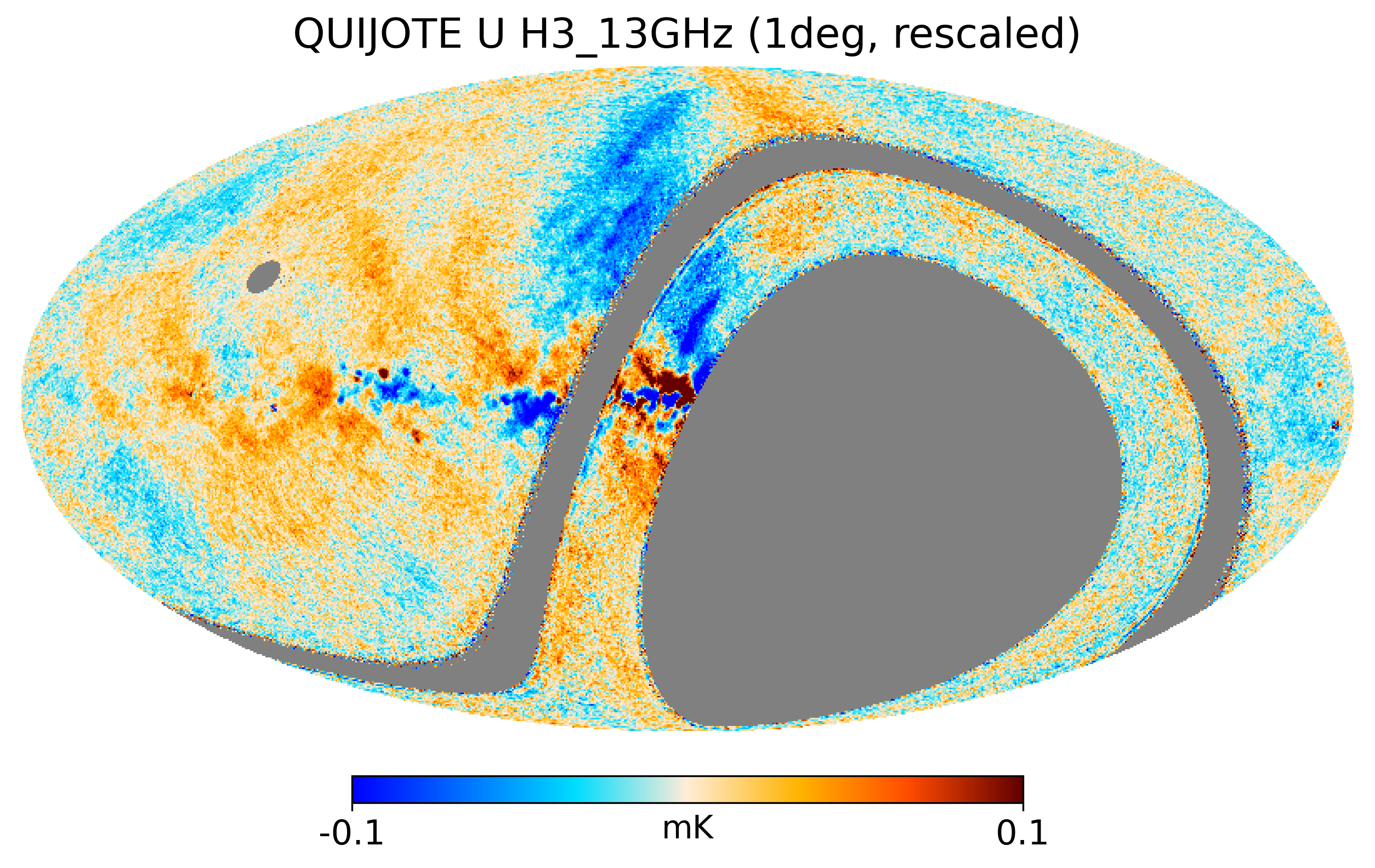}%
       \includegraphics[width=6cm]{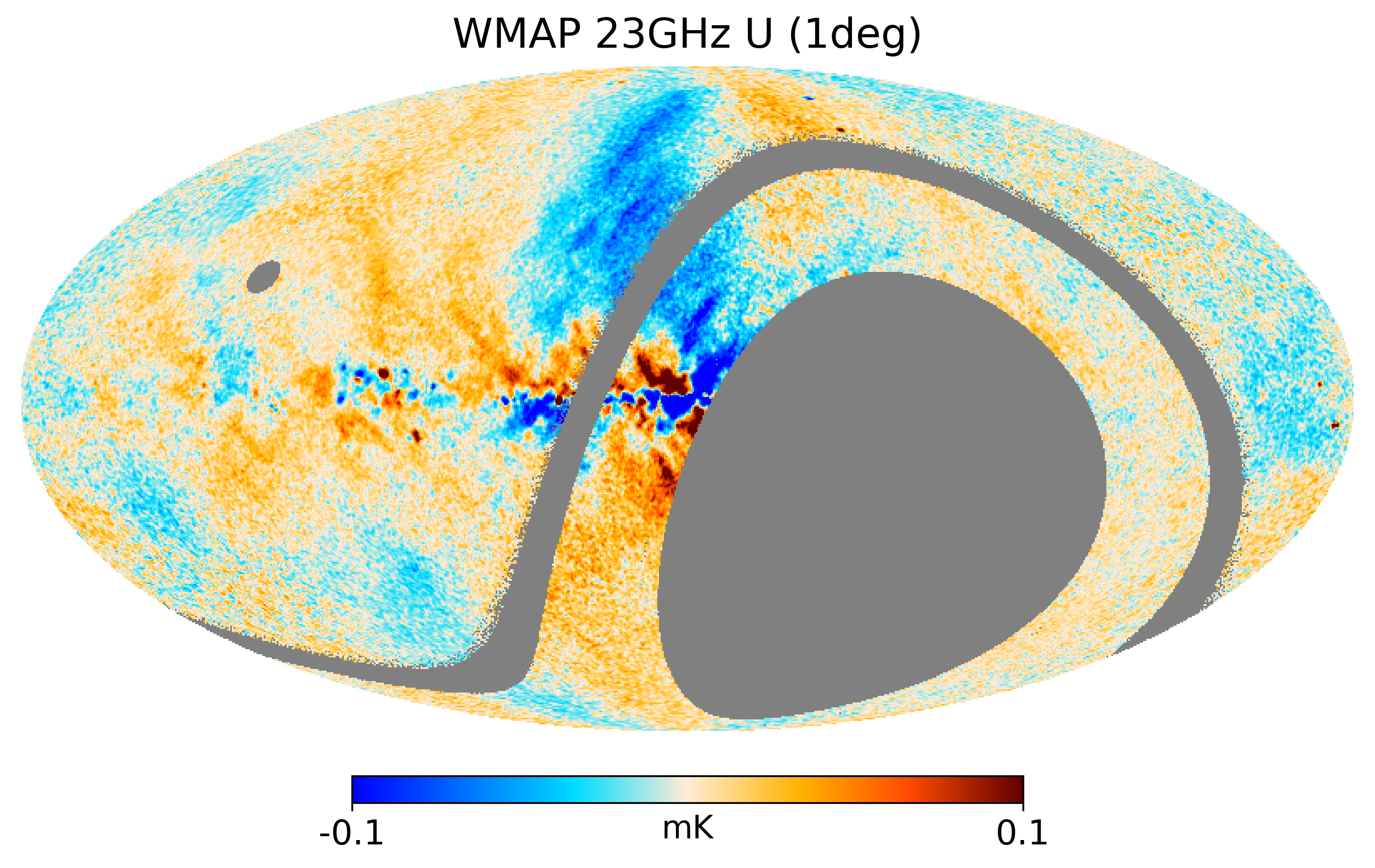}
    \caption{Comparison of the rescaled polarization MFI maps at 11 and 13\,GHz with the 9-yr WMAP-K band map \citep{Bennett2013}. MFI maps are rescaled to 23\,GHz using an average spectral of $\beta=-3.1$, and accounting for colour corrections. All maps use the same colour scale, saturated at $\pm 0.1$\,mK. From left to right, we show MFI 11\,GHz (rescaled), MFI 13\,GHz (rescaled) and WMAP-K. Top row: Stokes Q maps. Bottom row: Stokes U maps. For display purposes, to facilitate the comparison of the different structures near the mask edges, we applied here the QUIJOTE MFI sky mask to the WMAP map. }
    \label{fig:interfreqMFI}
    \end{figure*}
    
    \begin{figure*}
    \centering
       \includegraphics[width=6cm]{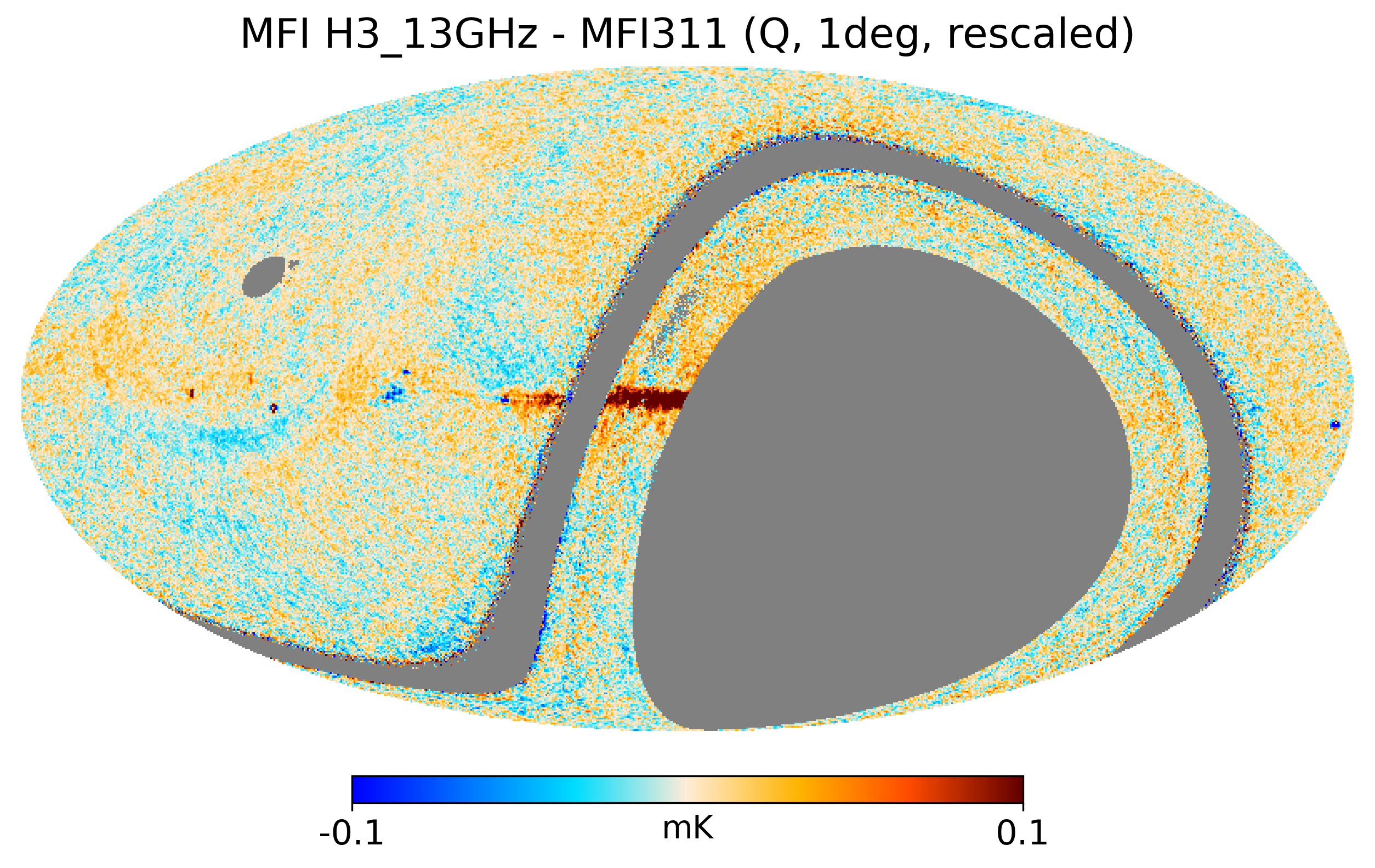}%
       \includegraphics[width=6cm]{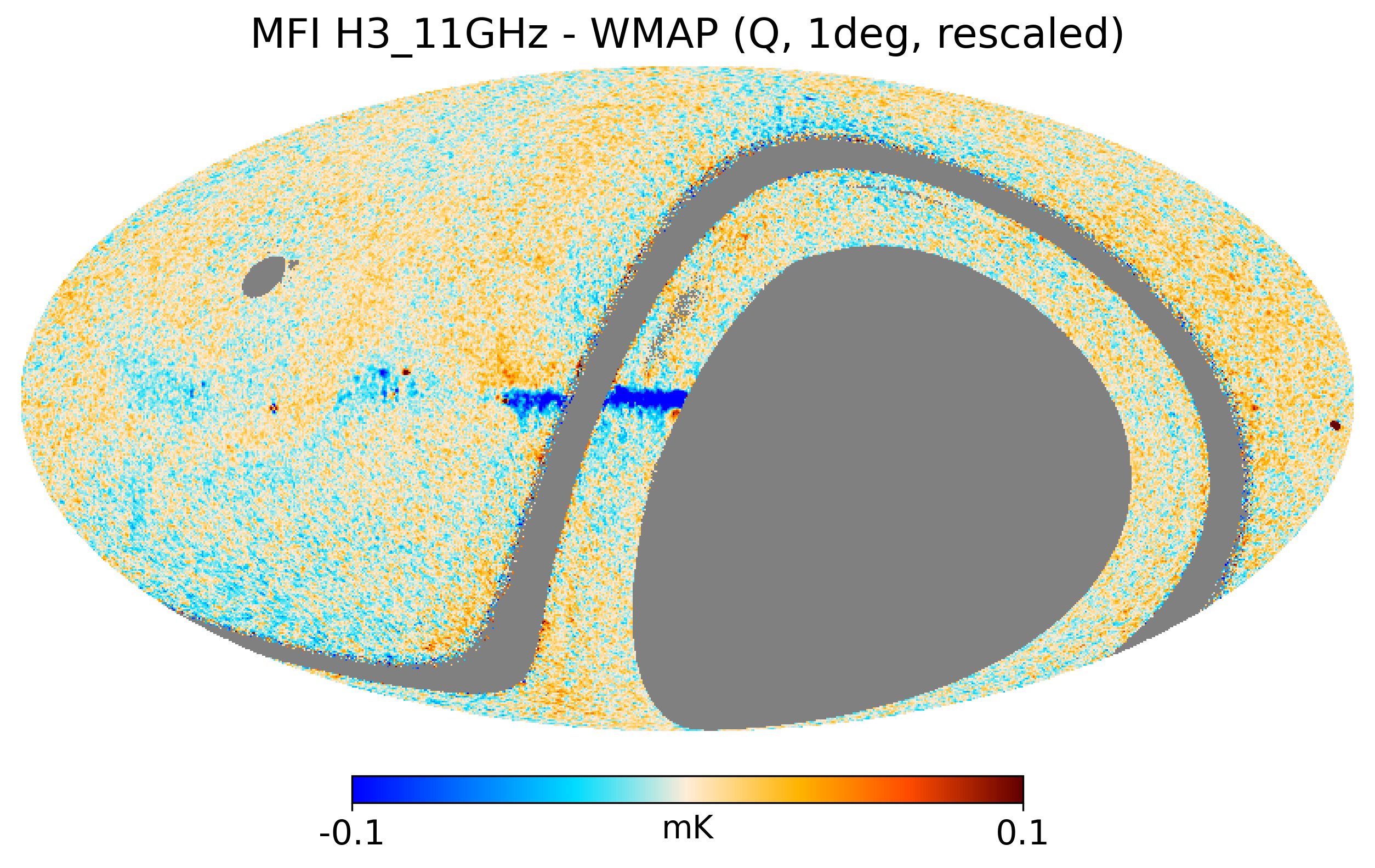}%
       \includegraphics[width=6cm]{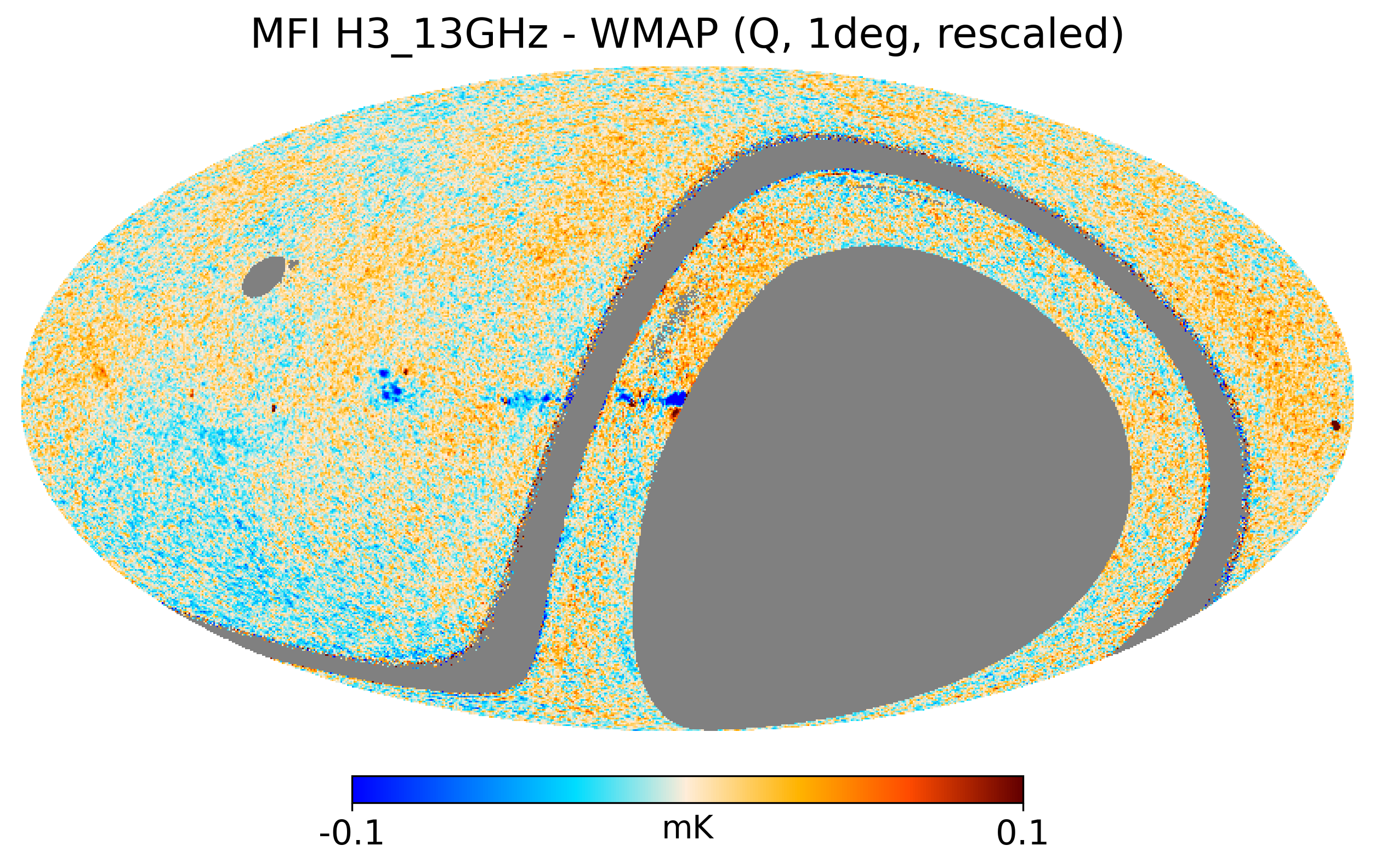}
       \includegraphics[width=6cm]{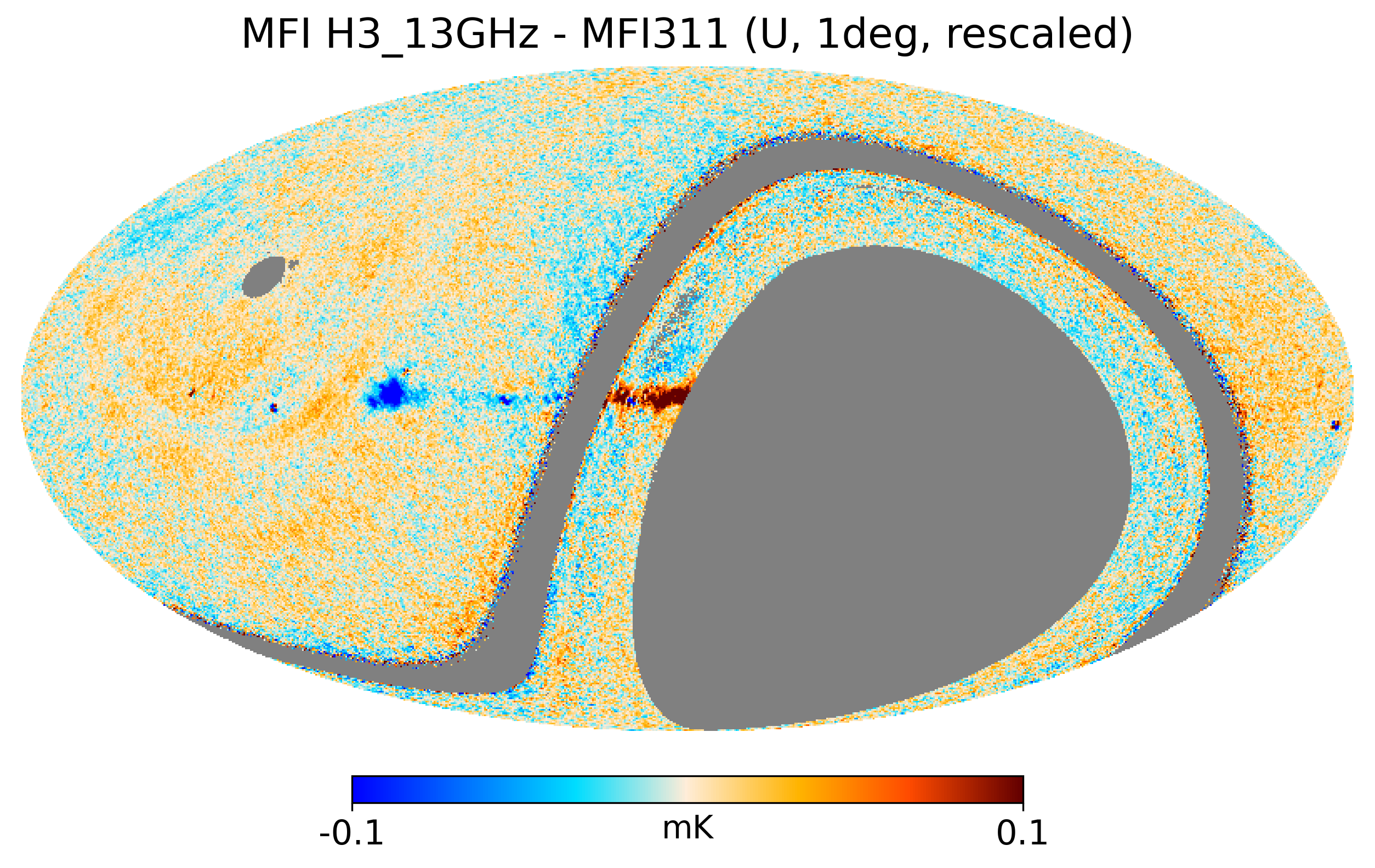}%
       \includegraphics[width=6cm]{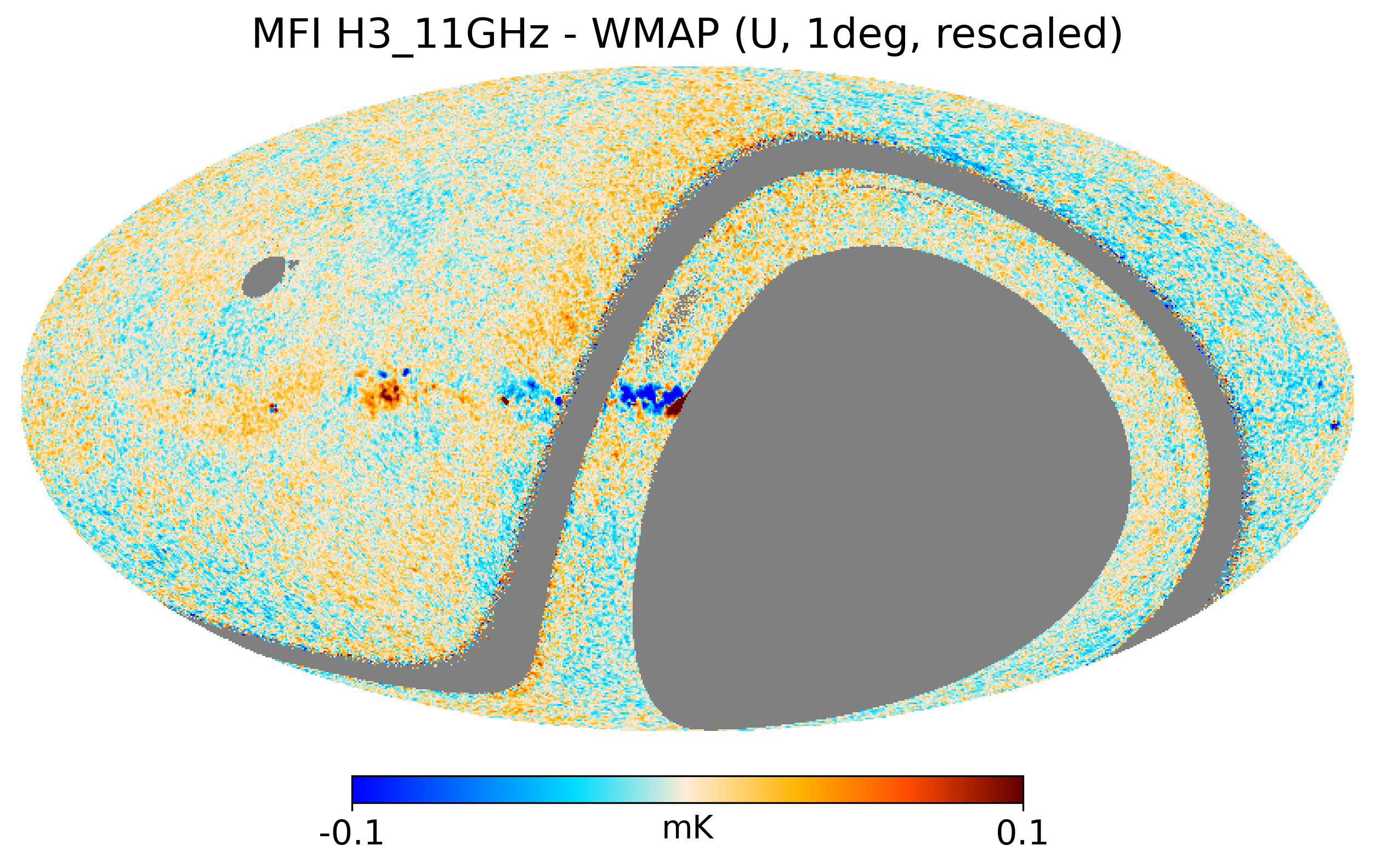}%
       \includegraphics[width=6cm]{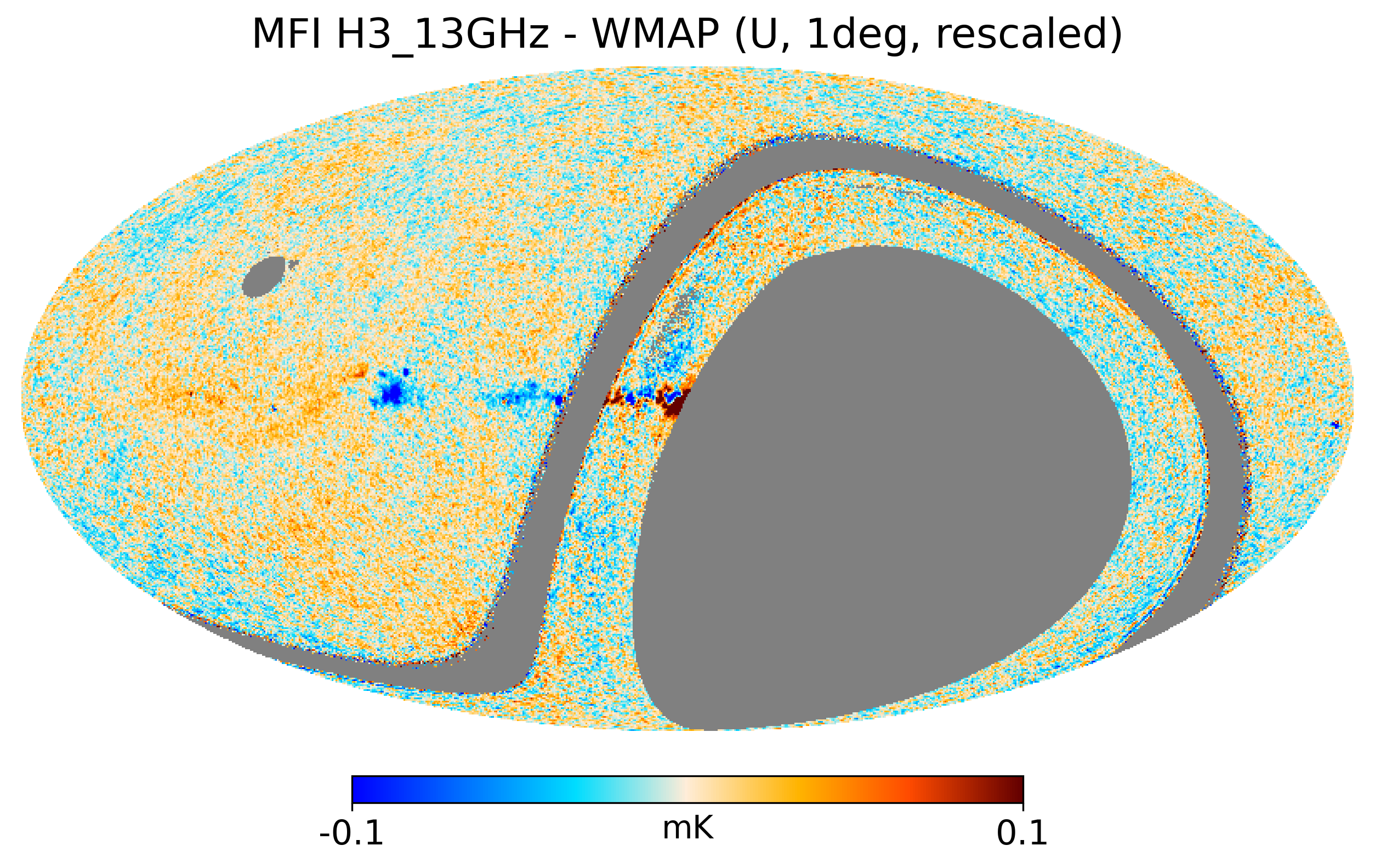}
    \caption{Inter-frequency comparison of the rescaled maps shown in Fig.~\ref{fig:interfreqMFI}. Top (bottom) row shows differences of Stokes Q (U) maps. 
    First column shows the difference between the rescaled 11 and 13\,GHz MFI maps. Second and third column show the MFI 11\,GHz minus WMAP-K, and  MFI 13\,GHz minus WMAP-K maps, respectively.  All maps use the same colour scale as in Fig.~\ref{fig:interfreqMFI}, saturated at $\pm 0.1$\,mK.  }
    \label{fig:diff_interfreqMFI}
    \end{figure*}
    
    \begin{figure}
    \centering
       \includegraphics[width=0.95\columnwidth]{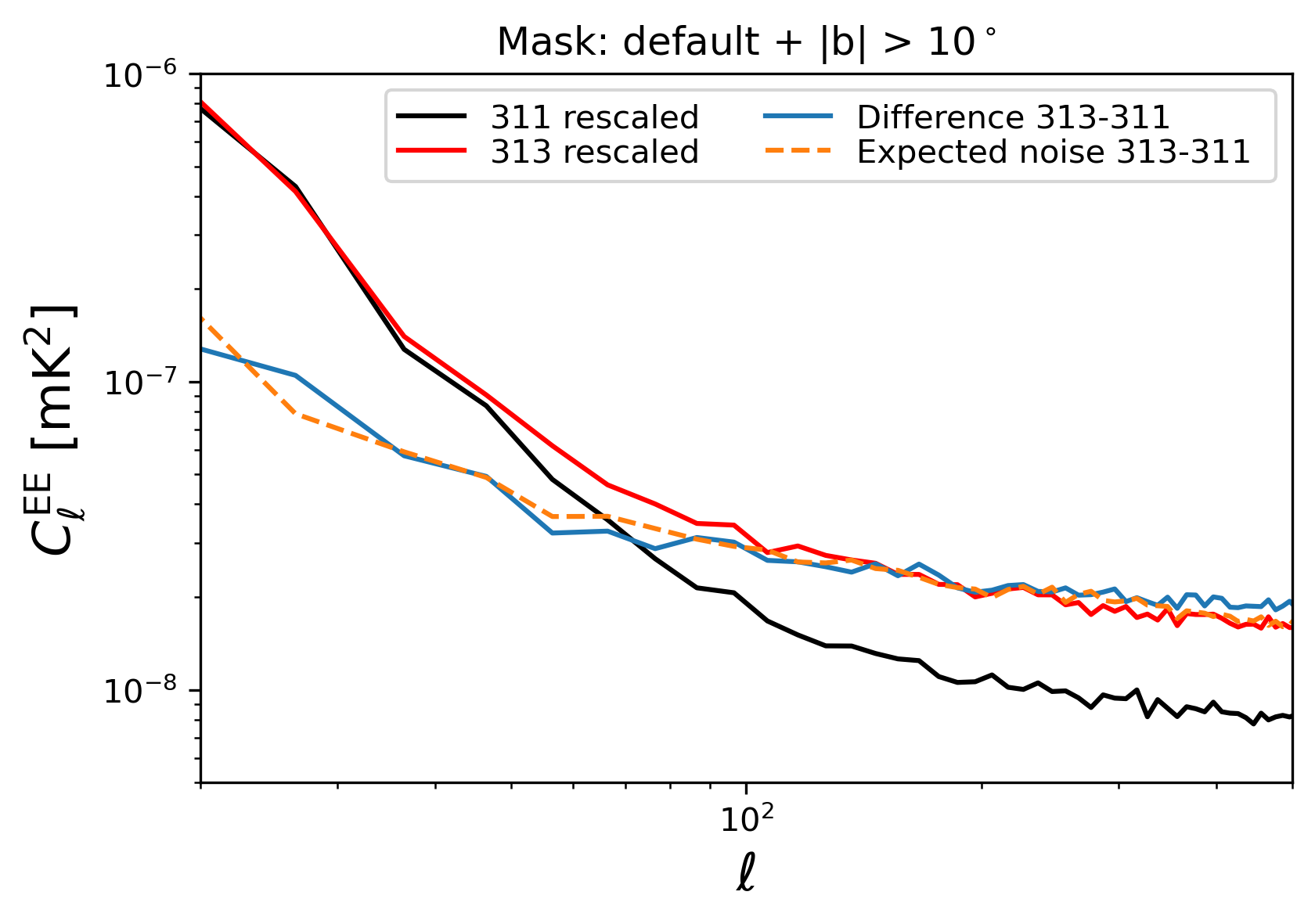}%
    \caption{EE power spectra of the inter-frequency comparison of the MFI rescaled maps  313$-$311, shown in the first column of Fig.~\ref{fig:diff_interfreqMFI}. Black and red solid lines show the EE power spectra of the rescaled  MFI 11 and 13\,GHz maps, respectively. Blue solid line is the power spectrum of the difference map 313$-$311, while the yellow dashed line shows the expected noise level for that difference map, assuming an average inter-frequency correlation of 32.8\,\% (see Sect.~\ref{sec:noisecorr_freqs}).  }
    \label{fig:aps_diff_interfreqMFI}
    \end{figure}
    
    As an additional validation test, here we present an inter-frequency comparison of the MFI wide survey maps, together with a comparison with external data.  For this test, we rely on the assumption that the average spectral index of the polarized synchrotron emission in the QUIJOTE maps is $\beta=-3.1$ (see discussion below in Sect.~\ref{sec:properties}). We then rescale the MFI wide survey maps at 1 degree resolution to the central frequency of the WMAP K-band map, $\nu=22.8$\,GHz, accounting for colour correction factors both for MFI and WMAP maps. Figure~\ref{fig:interfreqMFI} shows the rescaled MFI polarization maps at 11 and 13\,GHz compared to WMAP-K, while Figure~\ref{fig:diff_interfreqMFI} shows the differences for pairs of those maps (313$-$311, 311$-$WMAP, 313$-$WMAP).  A visual inspection shows that there is obvious polarized emission in the Galactic plane which is not consistent with the $\beta=-3.1$ spectral index, mainly towards the Galactic centre or the Fan region ($l \approx 135^\circ$). In the maps we can also identify some residual intensity to polarization leakage in the Cygnus area (around $l \approx 80^\circ$). However, the large scale emission far from the Galactic plane is largely suppressed in this difference, showing a good consistency of the MFI and WMAP-K maps. The residual emission in the difference map 313-311 is basically consistent with the expected noise level for the difference of both maps, as shown in Fig.~\ref{fig:aps_diff_interfreqMFI}. In this comparison, we use the EE power spectra for the rescaled maps using the default QUIJOTE mask with the Galactic cut $|b|>10^\circ$, and restricting the comparison to multipoles $\ell \ge 30$.

    \section{Accuracy of the wide survey calibration}
    \label{sec:cal}
    In this section we assess the overall calibration uncertainty of the QUIJOTE MFI wide survey maps in intensity and polarization, 
    using the information described in the pipeline paper \citep{mfipipeline} to account for known systematics, and also presenting a set of consistency checks 
    based on the null test maps, in order to evaluate the impact of unknown systematics. 
    Table~\ref{tab:summarycal} shows the summary of all types of uncertainties considered in this work, as well as the impact of each of them 
    in the overall calibration error budget.

\input{table_calibration.tex}

    \subsection{Statistical uncertainty and known systematics}
    \label{subsec:knownsys}
    
    \subsubsection{Calibration model}
    \label{subsec:calmodel}
    An important contribution to the global systematic uncertainty budget comes from calibration uncertainties, and in particular, the calibrator model. As discussed in \cite{mfipipeline}, the two main amplitude calibrators of QUIJOTE MFI are Tau A and Cas A, which are amongst the brightest sources on the sky in this frequency range. 
    As explained in subsection~\ref{sec:recal}, the wide survey maps have been recalibrated using flux densities extracted on these maps at the position of Tau A. These flux densities are measured with sensitivities better than 0.3\,$\%$ in all frequencies (see also Table~\ref{tab:sources_vs_models} and Sect.~\ref{sec:sources}) while the internal calibration accuracy of QUIJOTE is better than $1\,\%$ as shown below in subsection~\ref{sec:internalcal}. Therefore in our case the dominant error component is associated with the calibration models that are used as reference. As will be discussed in detail in G\'enova-Santos \& Rubi\~no-Mart\'{\i}n in prep. (see also subsection~\ref{sec:recal}), using different tests we estimate that the Tau A model has an uncertainty of $\approx 4\,\%$ in our frequency range. We believe this value is dominated by calibration errors of the different data that are used to model this spectrum. In the case of Tau A there is also an important contribution due to the modelling of its secular decrease, which leads to errors when data taken at different epochs are combined to model its spectrum. We decide to set a conservative overall calibration uncertainty of 5\,\%. The reliability of this number is supported by the tests on radiosources and planets presented in Section~\ref{sec:sources}, as well as other calibration tests based on the detection of primary CMB anisotropies shown in Sect.~\ref{sec:othercal}. 
    
    \subsubsection{Colour corrections}
    \label{subsec:coco}    
    The overall 5\,\% calibration uncertainty would strictly apply to any analysis performed in our maps on sources or regions with a power-law spectrum with index $\alpha =-0.3$, as that of Tau A (our primary calibrator). For a different spectrum, uncertainties in the colour corrections must be factored in. These are mainly associated with errors in the measurement or characterisation of the instrument bandpasses. MFI bandpasses were last measured in 2020, for the instrumental configuration corresponding to period 6. The statistical uncertainties of these measurements are very low, such that they lead to errors in the global calibrated antenna temperature below $0.01\%$ for a range of spectral indices $\alpha\in [-3,+3]$ and for all horns and frequencies. On the other hand, MFI suffered various modifications over its lifetime (see Table~\ref{tab:periods}), which may have introduced modifications in the actual bandpass shapes of periods 1, 2 and 5 with respect to period 6. 
    
    For the MFI wide survey, we conservatively assign errors to the colour corrections by comparing the last bandpass measurement from period 6 with a previous one performed in 2013 during period 1. Through comparing the colour correction coefficients obtained in both cases we find that channel 219 presents the largest differences, and in this case the error scales approximately as $\epsilon \times|\alpha +0.3|\,\%$, with $\epsilon=1.03$. Note that the error increases as the spectral index of the observation, $\alpha$, departs from that of the primary calibrator. 
    For 311 and 313, we obtain $\epsilon \approx 0.01$ and $\epsilon \approx 0.53$ respectively. For 217 we have  $\epsilon \approx 0.51$, while for horn 4 we have values between $\epsilon = 0.2$--$0.4$. We must note that these uncertainties are somewhat conservative, as differences between the two measured bandpasses may not be entirely real, but could also be due to shortcomings in the 2013 measurements, which are deemed much less reliable than those of 2020 due to measurement techniques (see details in \cite{mfipipeline}). 
    Taking this into account, and the fact that errors in the other channels are smaller, as a conservative choice for this paper, in Table~\ref{tab:summarycal} we have assigned an overall  $0.5\times|\alpha +0.3|\,\%$  error to colour corrections for 11 and 13\,GHz, and $1\times|\alpha +0.3|\,\%$ for 17 and 19\,GHz.
    
    Note that these errors in the colour correction coefficients should impact the consistency checks presented in Section~\ref{sec:sources}, where we compare with models flux densities of sources with spectral indices ranging between $-0.3$ and $-1.2$, and of planets 
    with $\alpha\approx 2$, or those presented below in this section where we correlate our maps with templates tracing the CMB anisotropies or the CMB dipole that also have $\alpha\approx 2$. In the former case, we find differences of $\sim 5\%$ which we are confident are due to uncertainties in the source calibration models. For the CMB anisotropies and CMB dipole, the differences are $\sim 3\,\%$ and $\sim 10\,\%$ respectively, and are driven by statistical noise (see Sect.~\ref{sec:othercal}).

    \subsubsection{Beams}
    \label{subsec:beams}
    One of the instrumental aspects that are most carefully characterised in CMB experiments are the beams and derived window functions, as they have a direct impact on the amplitude of the derived power spectrum and thence on cosmological parameters. In QUIJOTE MFI this is even more important as its calibration is tied to unresolved point sources. Comparison between beam radial profiles derived from observations of bright point sources and the numerical optical simulation based on CST software\footnote{\url{https://www.3ds.com/products-services/simulia/products/cst-studio-suite/}} described in \cite{mfipipeline} demonstrates an accuracy in the determination of the intensity beam typically below the $2$\,\% level (with respect to the centre of the main beam). Given that the MFI maps are (re)calibrated using a beam-fitting photometry on point sources, errors in the beams will directly impact the global map temperature scale. We confidently estimate the error in this temperature scale to be below 2\,\%. Note though that in extracting flux densities of point sources using the same beam-fitting photometry that is used for the main calibration, these errors would be largely suppressed. In Table~\ref{tab:summarycal}, we adopt a conservative value of 2 per cent, which corresponds to the maximum error associated with the determination of the brightness of a beam-filling emission.
    
    In polarization, a detailed description of the MFI beams can be found in \cite{mfipipeline}, where we use the CST optical simulations and the Mueller matrix formalism. Due to the MFI optical design, the cross-polar terms are significantly smaller than the copolar terms. For example, for horn 3, the cross-polar terms are less than 0.05\,\% of the copolar beams across the band. This implies that the diagonal components of the Mueller matrix ($M_{\rm II}$ and $M_{\rm QQ}$) can be considered nearly identical (with that accuracy). Moreover, the leakage terms $M_{\rm IQ}$ and $M_{\rm QI}$ are also identical in this limit, and are given by one half of the difference of the copolar beams at $0^\circ$ and $90^\circ$. As shown in \cite{mfipipeline}, these terms have a quadrupolar structure with two positive and two negative lobes, with typical peak amplitudes (relative to the copolar peak) of $\lesssim 1$\,\%. As shown below in Sect.~\ref{sec:sources}, when studying bright compact sources in the MFI wide survey, these patterns are clearly visible around Tau A (in Stokes $U$ parameter, because most of the signal appears in $Q$) and Cas A (in this case, as the source is essentially unpolarized, they are seen both in $Q$ and $U$ maps, rotated by $45^\circ$). When integrated on scales larger than the beam, these patterns average to zero, and thus have minimum impact on the photometry analyses \citep[see also][for the case of Planck beams]{Leahy2010}. For example, for the MFI 311 map, the impact on a photometry measurement using either aperture photometry in 1 deg, or beam fitting, is well below 0.05\,\% across the full frequency band. Thus, we neglect this contribution to the overall calibration error due to beam uncertainties, and in Table~\ref{tab:summarycal} we adopt the same calibration uncertainty in polarization as for intensity beams.

    \subsubsection{Intensity-to-polarization leakage}
    \label{subsec:itop}
    Despite of the fact that the MFI is a true polarimeter, in the sense that the polarization signal is produced directly for each individual horn and frequency band, there are several known systematic effects that may lead to spurious polarization signals, particularly in bright regions in intensity. In the previous subsection we have already discussed, for bright point sources, the intensity-to-polarization leakage (hereafter IPL) terms due to beam non-idealities. Here, we discuss the IPL terms arising from the bandpass mismatch between the two pairs of channels that contribute to a given polarization timeline. 
    For the MFI instrument, the $r$-factors in equations~\ref{eq:mfi_response_u} and \ref{eq:mfi_response_c} are determined using Tau A observations \citep[see details in][]{mfipipeline}. When observing a sky region with a bright intensity emission, the effective $r$-factor might change depending on the spectral index of the sky emission, particularly if it differs from that of Tau A ($\alpha= -0.3$). Using the detailed measurements of the bandpasses, we have estimated that for spectral indices typical of Galactic emission ($\alpha \in [-1.5, 0]$), the amount of signal leaked into Stokes $Q$ or $U$ due to this effect is typically below $0.2$\,\% of the intensity signal. For a CMB spectrum ($\alpha \approx 2$), it is still below 0.5\,\%.
    
\begin{figure}
    \centering
       \includegraphics[width=\columnwidth]{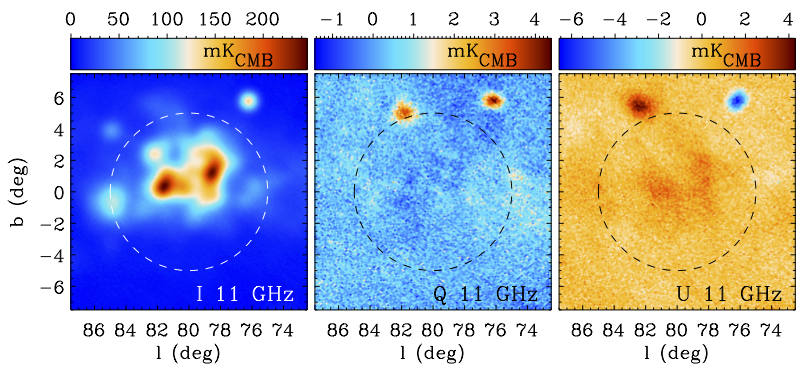}
       \includegraphics[width=\columnwidth]{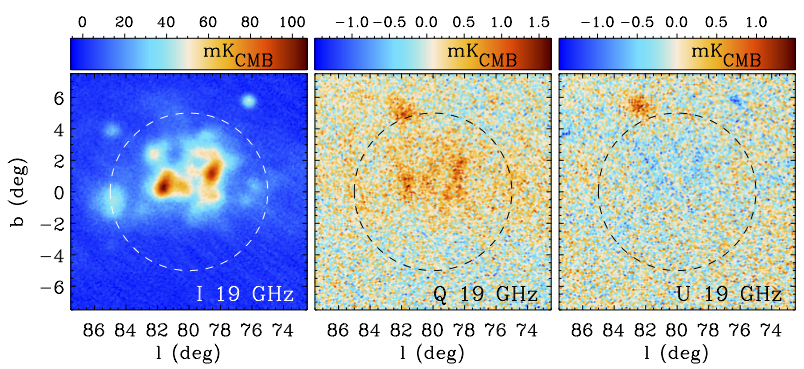}
    \caption{Minimaps of $15^\circ \times 15^\circ$ around the Cygnus region, located at Galactic coordinates $(l,b)=(80^\circ,0^\circ)$. We show the horn 3 11\,GHz (top) and horn 4 19\,GHz maps (bottom) at their original resolution. The circle indicates the region where the IPL is computed (see text for details). The two bright compact objects in the polarization maps located outside the circle, W63 and Cygnus A, are discussed in Sect.~\ref{sec:sources}. }
    \label{fig:cygnusregion}
    \end{figure}

Here, we provide an independent confirmation of the order of magnitude of the IPL in the MFI wide survey maps using the sky emission in the Cygnus region, located at Galactic coordinates $(l,b)=(80^\circ,0^\circ)$. Figure~\ref{fig:cygnusregion} shows this area in more detail. As the intensity emission in this region is dominated by free-free, it is expected to be almost unpolarized. We use a cross-correlation analysis (similar to the one used in Sect.~\ref{sec:sysmaps}) to obtain the correlation coefficient $\alpha$ that minimizes $Q-\alpha I$ within a region centred at $(l,b)=(80^\circ,0^\circ)$ with a radius of $5^\circ$. The values are always below 1 per cent for all cases, as expected. For 311, we find 0.10\,\% and 0.65\,\% for Stokes Q and U, respectively. The largest values are found for 419, where we obtain 0.91\% and $-0.41$\% for Stokes Q and U. This effect in the Cygnus area is clearly seen in the maps of Fig.~\ref{fig:diff_interfreqMFI} and \ref{fig:cygnusregion}. The values reported in Table~\ref{tab:summarycal} correspond to the most conservative case (Stokes Q or U) at each frequency, in absolute value.
    
    \subsubsection{Polarization efficiency}
    \label{subsec:poleff}
    As discussed in Sect.~\ref{sec:recal}, the calibration of the polarization efficiency of the MFI wide survey data is done in two steps. First, we use laboratory measurements taken at the end of period 6 to calibrate the polar efficiency of each individual MFI channel. In addition, we use the wide survey data in period 6 to add also the correction factors to these polar efficiencies associated with a possible error in the determination of the $r$-factors. These procedures provide a determination of the polarization efficiency in period 6 with a relative accuracy of 2\,\%. Then, in a second step these values from period 6 are transferred to the other two periods that are used in the construction of the MFI wide survey polarization maps (i.e. 2 and 5), using beam fitting photometry (BF1d) measurements on Tau A.  The error budget for these factors is given by the accuracy of the flux density extraction, which is found to be of the order of 1\,\% for horn 3, and 2\,\% for horns 2 and 4. 
    As they correspond to systematic errors, we adopt the conservative approach of adding them linearly, and we quote an overall 3\,\% error in the polar efficiency for horn 3, and 4\,\% for horns 2 and 4. In the following subsections we evaluate unknown systematic effects in the polarization maps, noting that in those cases, the global errors include the polar efficiency error. In addition, in Sect.~\ref{sec:sources} we also discuss the polarization fraction of Tau A and Cyg A, and the polarized flux in W63, as further consistency tests for this polar efficiency calibration.

    \subsection{Internal calibration of the wide survey and consistency checks: evaluating unknown systematics}
    \label{sec:internalcal}
    Following the methodologies outlined in \citet{Planck2013-iii} and \citet{Planck2013-v}, we use internal consistency checks based on null test maps and other data splits of the wide survey in order to estimate the impact of systematic effects in the overall calibration. This is particularly useful for assessing the impact of "unknown systematics", i.e. those for which we do not have specific measurements or numerical simulations. For the MFI wide survey, and given that we want to focus on the relative calibration of the instrument, we use as a reference the set of null test maps and data splits labelled as "with common baselines" in Sect.~\ref{sec:nulltests}.

    \subsubsection{Unknown systematics in real space}
    \label{sec:sysreal}

\input{table_systematics.tex}

    Uncertainties due to (unknown) calibration or systematics effects at the pixel scale have been calculated using the HMDM for common baselines, degraded to $\nside=64$. At this resolution, each pixel roughly corresponds to the beam size. The reference mask for the analysis is the default one (sat+NCP+lowdec) as defined in Sect.~\ref{sec:masks}. 
    
    Table~\ref{tab:systematics} lists the rms values and peak-to-peak (p-p) variation for the HMDM. Following \cite{Planck2013-iii}, the p-p values are computed as the difference between the 99\,\% and the 1\,\% quantiles in the pixel value distribution, in order to neglect possible outliers\footnote{Note that for a Gaussian distribution, we should have p-p=$4.65\sigma$. }. 
    A comparison between these numbers for the half-mission null tests and those for the ring null tests is useful for 
    checking residual calibration and/or systematic effects on large angular scales. Given that the ring null test maps cancel out possible variations in scales longer than 30\,s (i.e. the duration of one azimuth scan), they can be used as our best estimate of the noise level, which includes white noise and $1/f$ on degree scales. 
    Any variation on scales longer than one minute, due either to calibration uncertainties in the gain model or systematic effects, will appear as a signal excess in the HMDM. As illustration, the top panel in Fig.~\ref{fig:noise311b} shows the ring null-test difference maps for the 311 (horn 3 at 11\,GHz) case. 
    The results of this comparison are shown in Table~\ref{tab:systematics}. Column 5 presents the rms value for the ring difference maps, and column 6 shows the signal excess in the half-mission difference maps. 
    Comparing these values with those in Tables~\ref{tab:noiseps} and \ref{tab:noise3} for the noise levels for the wide survey, 
    we find that in polarization, the rms excess due to unknown systematics is well below the white noise levels, with typical values in the range 5--20$\mu$K. 
    In intensity, we find a similar situation for horn 3 and the 17\,GHz  frequency maps of horns 2 and 4. For the two maps at 19\,GHz (horns 2 and 4), the residuals are slightly larger than the white noise levels, but still well below the total noise contribution in those channels (column 5). As a reference, for horn 3, the residuals at beam scales are of the order of $\sim 50 \mu$K. 
    These numbers are used to complete the main table~\ref{tab:summarycal}, appearing as "unknown systematics" in real space. As a conservative choice, the values for horns 2 and 4 are combined linearly instead of using a quadratic combination.

    \subsubsection{Unknown systematics in harmonic space}
    \label{sec:sysharm}
    We use the ratio of cross-power spectra of the null test maps with some external maps, as the reference tool to validate the calibration in harmonic space. The use of cross-spectra to external maps minimises the effects of noise bias on the power spectrum estimation. 
    In practice, given two maps 1 and 2 that we want to compare, we compute
    \begin{equation}
    \label{eq:A12}
        A_{1,2}= \Bigg<  \Bigg< \frac{C_\ell^{1,{\rm X}} }{C_\ell^{2,{\rm X}}} \Bigg>_{\ell} \Bigg>_{\rm X},
    \end{equation}
    where $C_\ell^{i,{\rm X}}$ is the cross-spectrum of map $i$ (=$1,2$) with some other external map X, with X running over all possible uncorrelated external maps, and the brackets represent the (unweighted) average in a given multipole range ($<...>_{\ell}$) or over all external maps ($<...>_{\rm X}$), respectively.  
    For completeness, we also evaluate the uncertainty on this parameter ($\sigma_{A_{1,2}}$) as the standard deviation of those ratios over the external maps, 
    \begin{equation}
    \label{eq:sigA12}
        \sigma_{A_{1,2}}= \frac{1}{\sqrt{n_{\rm X}}} std_X \Bigg(  \Bigg< \frac{C_\ell^{1,{\rm X}} }{C_\ell^{2,{\rm X}}} \Bigg>_{\ell} \Bigg), 
    \end{equation}
    where $n_{\rm X}$ is the number of external maps involved in the analysis.

    In this section, all cross-spectra are obtained using \xpol. The reference mask adopted for this computation is the default one (sat+NCP+lowdec), which preserves the declination range $6^\circ \le \delta \le 70^\circ$. This mask is apodized using a $5^\circ$ cosine function, as implemented in the \namaster\ library \citep{namaster}. All maps have been smoothed to a common resolution of one degree. 
    For MFI, the ratios are evaluated and averaged within the multipole range $\ell=30$ to $\ell=200$. The lower value of $\ell=30$ guarantees that the pseudo-$C_\ell$ estimation is not affected by mode coupling due to incomplete sky coverage, and constitutes a conservative choice regarding possible large scale residuals due to RFI and atmosphere, as discussed in the previous section. 
    As external maps, we decided to use low frequency maps ($\le 70$\,GHz) from satellites, in order to have similar foreground components to the signal in the QUIJOTE maps. In particular, we use the 9-year WMAP maps \citep{Bennett2013} for bands K, Ka, Q and V, and the PR2 Planck-LFI maps at 30, 44 and 70\,GHz corrected from bandpass leakage \citep{Planck2015-i}.

    \paragraph{Intra-nulltest calibration.} 
    We first evaluate the relative calibration of the wide survey, using the six null test maps described in Sect.~\ref{sec:nulltests}, namely half (mission), rings, halfring, daynight, pwv and tbem. For each case, we compare the relative calibration of the two maps in each pair $h_1$ and $h_2$, as in equation~\ref{eq:A12}, and we evaluate the error bar using equation~\ref{eq:sigA12}.
    
    \begin{figure}
    \centering
    \includegraphics[width=0.9\columnwidth]{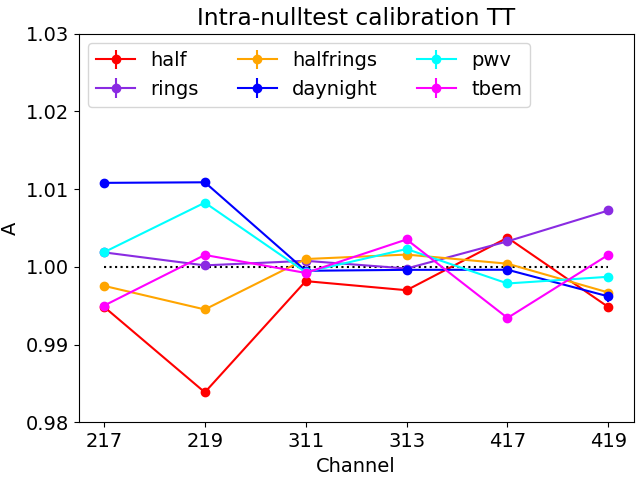}
    \includegraphics[width=0.9\columnwidth]{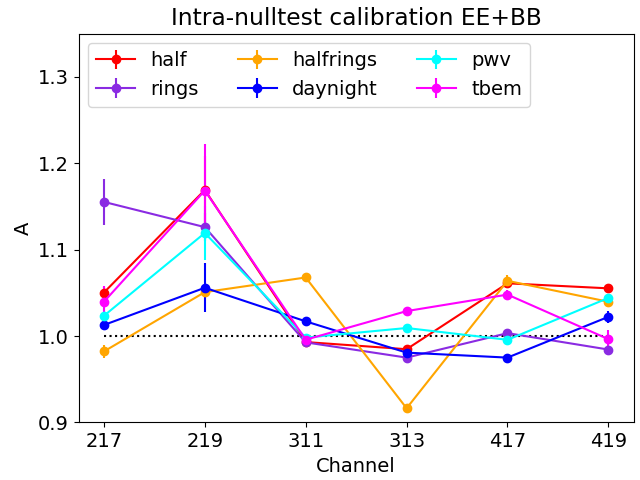}
    \caption{Intra-nulltest calibration of the MFI widey survey. We show the consistency of the null test maps, for intensity (TT, top) and polarization (average of EE and BB, bottom). }
    \label{fig:intranulltest}
    \end{figure}
    
    Fig.~\ref{fig:intranulltest} shows the result both for intensity (TT) and polarization (average of EE and BB) data. 
    In intensity, we find a good consistency of all the different data splits well within one per cent. At 11 and 13\,GHz, the maximum discrepancy is found to be 0.3\,\%. The average of the six null test cases is consistent with one (perfect relative calibration) within 0.2\,\%. At 17 and 19\,GHz, the maps from horn 4 present a maximum discrepancy of 0.7\,\%, and the scatter of the six measurements stays within 0.5\,\%. Horn 2, which is known to be the noisiest one, presents the larger discrepancy 
    of $-1.6$\,\% for the half-mission null test, and the average of the six values is consistent with one within 1\,\%. 
    
    In polarization, we find larger values of the scatter, as expected due to the lower signal-to-noise ratios of these maps, although we remind that in this case our analysis also probes possible time variations of the polarization efficiency values on top of the global calibration. 
    For horn 3, the maximum discrepancy is associated with the halfring null test, which presents deviations of +7\,\% for 311, and -8\,\% for 313. However, we note that this null test is expected to be noisier than the others, due to the lower number of independent crossings in each half. The average of the six measurements is fully consistent with one, and has a scatter of 2.9\,\% and 3.8\,\% for 11 and 13\,GHz, respectively. For horn 4, we find a maximum discrepancy of 6.4\,\%. The average of the six measurements is again consistent with one, and the scatter is 3.8\,\% and 2.8\,\% for 17 and 19\,GHz. Finally, for horn 2, as in intensity, we find the largest scatter of the measurements. The largest discrepancy is found to be 17\,\% but with a large error bar. The average of the six measurements is slightly biased towards positive values of $A$ for 219, but not significantly (two sigmas). The scatter of the measurements is 5.9\,\% and 5.2\,\% for 217 and 219, respectively. 
    
    In summary, the internal calibration scale of the MFI wide survey seems to be consistent within 0.7 per cent in intensity for all horns, reaching 0.2\,\% for horn 3. In polarization, we find consistency within 3--4 per cent in for horns 3 and 4, and within 10\,\% for horn 2. To put in context these values, it is useful to compare them with the expected scatter in the $A$ values in the case of a perfectly calibrated instrument with the realistic noise levels of the MFI wide survey. For this purpose, we have repeated this analysis using simulations including realistic $1/f$ noise levels as in Sect.~5 of \citet{destriper}. According to these simulations, the expected scatter of the six null tests in intensity is within 0.1--0.2\,\%, while in polarization we expect 2\,\% for horn 3 and horn 4 at 17\,GHz, and we could have up to 5--6\,\% for horn 2 and horn 4 at 19\,GHz. We stress that these numbers are driven by the $1/f$ noise in the maps, and therefore they represent the actual sensitivity of this method to detect calibration errors. Any calibration uncertainty due to systematic effects in the real data will add to these values. 
    
    When comparing these values from simulations with those found for real sky measurements, we find that they are consistent in intensity, but the real data produce slightly larger scatter in polarization. This small excess of uncertainty in the polarization values from the real maps can be ascribed to polarization efficiency systematic errors. As a conservative approach, we decided to quote as calibration uncertainty in Table~\ref{tab:summarycal} the final numbers obtained from this test, thus including also the $1/f$ noise contribution.

\begin{figure}
    \centering
    \includegraphics[width=0.9\columnwidth]{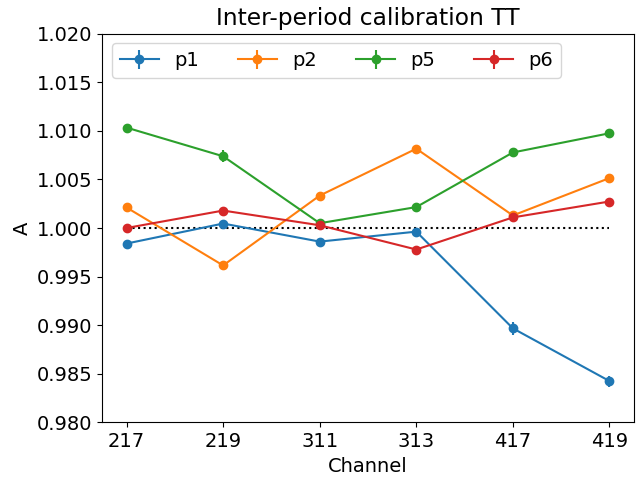}
    \includegraphics[width=0.9\columnwidth]{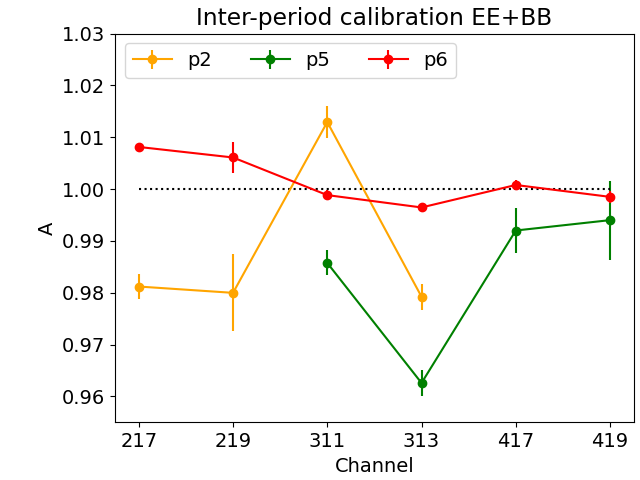}
    \caption{Inter-period consistency checks, in intensity (TT, top) and polarization (average of EE and BB, bottom). We show the $A$ factor computed as in equation~\ref{eq:A12}, when comparing the map per period (i.e. using the data of that given period only) to the total final map, for each horn and frequency.  }
    \label{fig:interperiod}
    \end{figure}

    \paragraph{Inter-period calibration.}
    We now evaluate the time stability of the wide survey calibration, using the four maps per period described in Sect.~\ref{subsec:perperiods}, again for the case of "common baselines".  We also note that period 1 only has observations at high elevations, so in order to have a common sky coverage for this comparison in the four maps, we restrict the analysis in this particular case to a sky mask covering the declination range $8^\circ \le \delta \le 50^\circ$. As usual, this extended mask is apodized using a $5^\circ$ cosine function, as implemented in the \namaster\ library \citep{namaster}. 
    Fig.~\ref{fig:interperiod} shows the comparison of the $A$ factors for the four maps by period used for the wide survey (periods 1, 2, 5 and 6), when compared to the total final map for each horn and frequency. 
    
In intensity, the internal consistency is found to be again better than 1\,\%. The largest discrepancy in absolute value is found for the map 419 in period 1, at the level of -1.5\,\%. The standard deviation of the four $A$ values for each horn and frequency is found to be $\sim 0.5$\,\% for channels in horns 2 and 3, and $0.7$--$1$\,\% for horn 4. 
    In polarization, we recall that some periods are not used for the final maps. In particular, period 1 is not used in polarization, period 2 is not used for horn 4, and period 5 is not used for horn 2. The maximum discrepancy with respect to the final map is found in 313 for period 5, at the level of $-3.7$\,\%. Taken as a whole, these values suggest that the calibration scale is stable within 1 per percent in intensity, and within 2 per cent in polarization, during the six years of observations covered by the wide survey. 
    
    \paragraph{Inter-horn calibration for horns 2 and 4. }
    
\begin{figure}
    \centering
    \includegraphics[width=0.9\columnwidth]{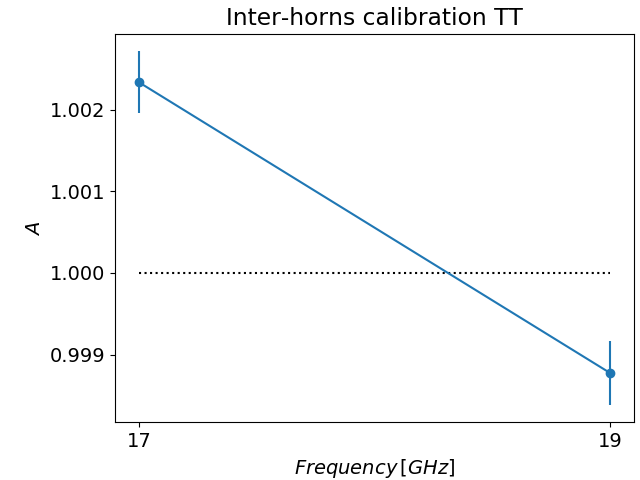}
    \includegraphics[width=0.9\columnwidth]{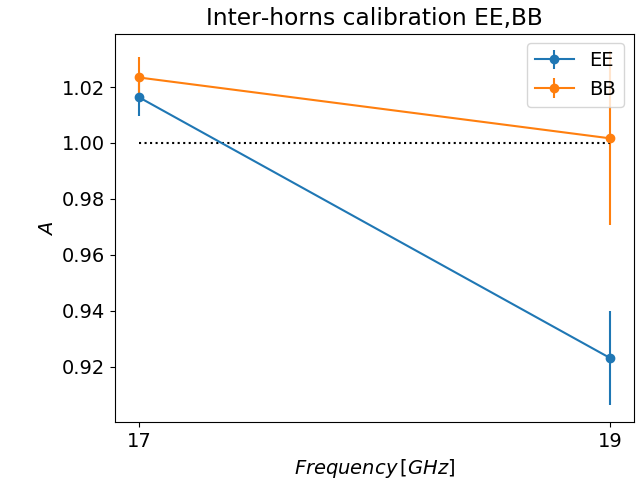}
    \caption{Inter-horn consistency check between horns 2 and 4, in intensity (top) and polarization (bottom).  }
    \label{fig:interhorn}
    \end{figure}

    Given that the frequencies of 17 and 19\,GHz are observed with horns 2 and 4, we also carry out an inter-horn comparison of the final wide survey maps at these frequencies using the same methodology as above, and where the $A$ factor in equation~\ref{eq:A12} now compares  the ratio of the two maps of a given frequency from the two horns. In this case, we obtain two values, $A_{217,417}$ and $A_{219,419}$. The results are displayed in Fig.~\ref{fig:interhorn} both for intensity (TT) and polarization (EE and BB, here plotted separately). 
    We find that the relative calibration of the wide survey between horns 2 and 4 is consistent within 0.2 per cent in intensity. 
    In polarization, this test is not providing very restrictive results due to the high noise levels of horn 2 in comparison to horn 4. Nevertheless, we can conclude that the relative calibration of the two 17\,GHz maps is found to be consistent within 2 per cent, while for 19\,GHz we find consistency within 4 per cent if we average the values for EE and BB. In this later case, our simulations show that the separated values for EE or BB alone might differ by more than 4 per cent in the ideal case of a perfect calibration, due to the (white plus $1/f$) noise levels.
    
    \subsubsection{Summary of the internal calibration tests}
    The overall calibration uncertainty quoted for the QUIJOTE MFI wide survey maps is 5\,\% in intensity for all frequency maps, 5\,\% in polarization for 11 and 13\,GHz, and 6\,\% in polarization for the combined 17 and 19\,GHz maps (see last two rows in Table~\ref{tab:summarycal}). These values are mainly limited by the physical modelling of the point-sources  (Tau A, Cas A) used to calibrate the experiment. In intensity, all the tests in this section show that the internal consistency of the calibration and gain model, which spans 6 years of measurements, is within the one per cent level. In polarization, the internal consistency tests show that the calibration is controlled at the 2--3 per cent level for frequencies 11, 13 and 17\,GHz, while for 19\,GHz, and particularly for horn 2, this uncertainty could be up to 6\,\%. However, we note that in this later case, the quoted uncertainty includes calibration errors, polarization efficiency uncertainties and $1/f$ noise contributions.

    \subsection{Other calibration tests}
    \label{sec:othercal}
    
    \subsubsection{CMB anisotropies}
    CMB anisotropies in intensity can be measured in the QUIJOTE MFI wide survey maps using a cross-correlation with an external CMB template. We follow the methodology described and validated in Section~6.5 of \cite{destriper}, and use a template fitting method with two templates: a reference CMB map ($\mathbf{m}_{\rm CMB}$), and a "foreground" map to account for chance alignments between the CMB and the Galactic foregrounds ($\mathbf{f}$). The basic assumption is that the QUIJOTE map ($\mathbf{m}_{\rm MFI}$) can be written as a linear combination of these two maps as
    \begin{equation}
    \mathbf{m}_{\rm MFI} = A \mathbf{m}_{\rm CMB} + B \mathbf{f} + \mathbf{n} ,
    \end{equation}
    where $A$ and $B$ are the parameters of the linear combination, and $\mathbf{n}$ represents a noise component. Using the cross spectra of the QUIJOTE maps with both external templates, $C_\ell^{\rm MFI, CMB}$ and $C_\ell^{\rm MFI, f}$, we can extract both $A$ and $B$ parameters. 
    As shown in \cite{destriper}, this method produces unbiased results for the CMB reconstruction ($A=1$), provided that there is a perfect consistency with the calibration of the CMB map. Thus, the method can be used as an additional calibration test. 
    
    \begin{figure}
        \centering
        \includegraphics[width=0.9\columnwidth]{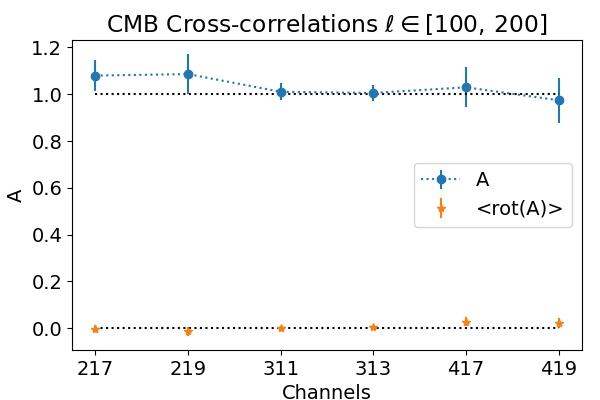}
        \caption{Relative amplitude of the CMB signal in the QUIJOTE MFI maps, using cross-correlations with the Planck SMICA map. Error bars are obtained using rotations of the CMB map. For consistency, we show that the average signal of the cross-correlation with rotated CMB maps is consistent with zero, as expected.}
        \label{fig:cmb}
    \end{figure}
    
    Here, we use as a reference the SMICA 2018 map \citep{Planck2018-iv}, but we have checked that consistent values are obtained using other versions of the Planck CMB map (NILC, COMMANDER, SEVEM). As foreground template, we use the WMAP 9-year K-band map \citep{Bennett2013}, after subtracting the CMB component. The analysis mask is the same as in \cite{destriper}, which combines the default QUIJOTE analysis mask (NCP+sat+lowdec) with the Planck common confidence mask for temperature analyses \citep{Planck2018-iv}, apodized with a simple 2-degree smoothing. All cross-spectra in this section are computed using \xpol. 
    Error bars are obtained using rotations of the CMB map in steps of $\Delta l = 18^\circ$, as in \cite{destriper}. The analysis is carried out in the multipole range $[100,200]$, but consistent results are obtained in other ranges (e.g. we also tested $[30,200]$, although the overall significance is lower in this case due to the larger $1/f$ contribution of lower multipoles). 
    The final results are shown in Figure~\ref{fig:cmb} and Table~\ref{tab:cmb}. The CMB signal is detected in all channels, with a significance larger than 10-sigma in all cases. These error bars are consistent with the level of $1/f$ noise in the QUIJOTE maps (see Table~4 in \cite{destriper}). We note that, due to the strongly correlated noise in the MFI intensity maps, estimates from the same horn tend to deviate in the same direction. 
    All values are consistent with $A=1$, providing an independent confirmation of the calibration scale of the maps. 
    Finally, we also provide a combined measurement of the CMB signal present in the QUIJOTE MFI maps, using a weighted average combination of all channels
    and accounting for the noise correlation between frequencies of the same horn. The overall result ($1.02\pm 0.03$) provides a 35-sigma detection of the CMB anisotropies in the QUIJOTE MFI intensity maps, and shows a consistent calibration with Planck within three per cent. 
    
    \begin{table}
    \caption{Relative amplitude ($A$) of the CMB component in the QUIJOTE-MFI wide survey maps with respect to the SMICA Planck map, obtained with cross-correlations in the multipole range $100$--$200$. Error bars are obtained using rotations of the CMB map. }
    \label{tab:cmb}
    \centering
    \begin{tabular}{ccc}
    \hline
    Channel & A &  Uncertainty \\
    \hline
    217 & 1.080 &  0.068 \\
    219 & 1.086 &  0.086 \\
    311 & 1.010 &  0.037 \\
    313 & 1.005 &0.033 \\
    417 & 1.030 &  0.086 \\
    419 & 0.974 &  0.097 \\
    \hline
    Combined & 1.019 & 0.029  \\
    \hline
    \end{tabular}
    \end{table}

    \subsubsection{CMB dipole}
    \label{sec:dipole}
    
    \begin{figure}
        \centering
        \includegraphics[width=0.9\columnwidth]{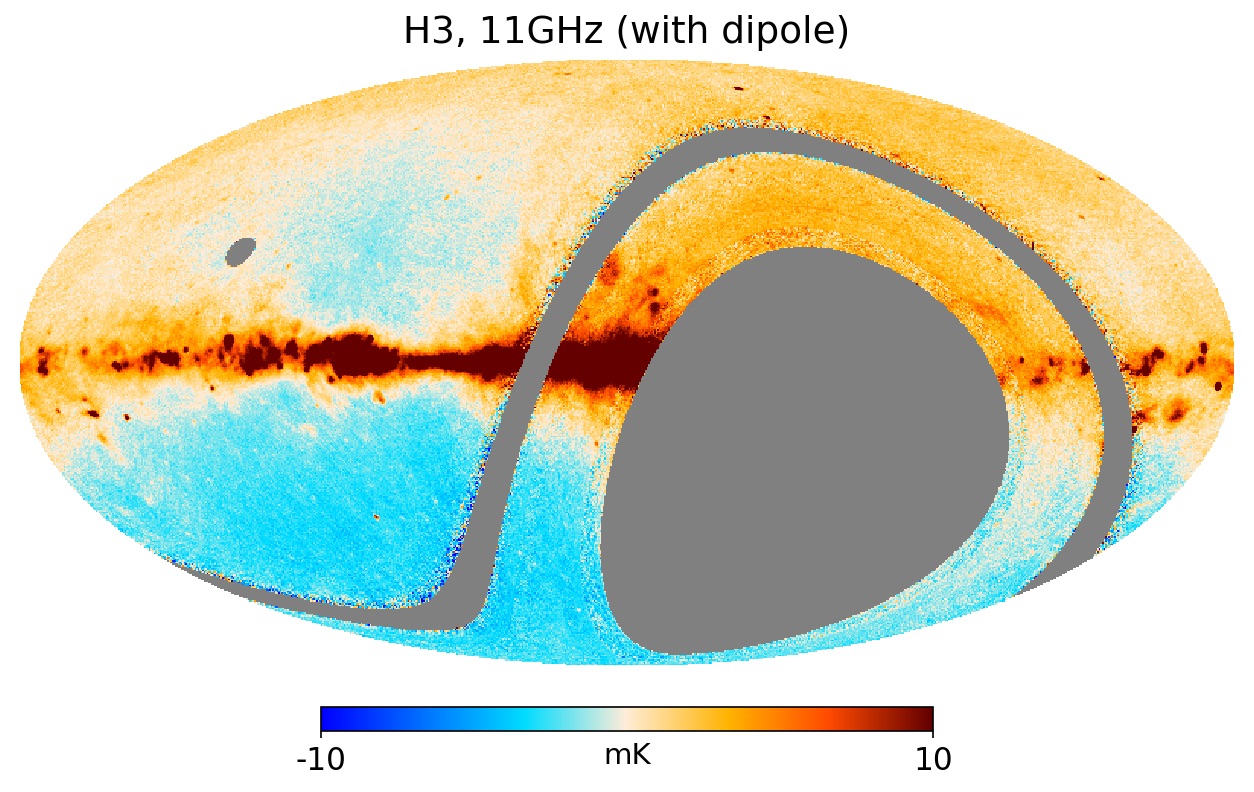}
        \caption{MFI wide survey 311 (horn 3 at 11\,GHz) map, with the dipole component not removed from the map. For display purposes, the map has been downgraded to resolution $\nside=256$. }
        \label{fig:mfidipole}
    \end{figure}
    
    As an additional calibration test, we present here the detection of the CMB dipole in the MFI wide survey maps, using a cross-correlation technique similar to the one used in the previous subsection for the CMB anisotropies. For this analysis, specific MFI wide survey maps are generated excluding the dipole removal and the atmospheric correction steps in the post-processing stage of the pipeline. Figure~\ref{fig:mfidipole} shows one example of these maps, for the case of horn 3 at 11\,GHz. 
    
    We use a template fitting method in real space with three templates: a reference CMB dipole template map ($\mathbf{m}_{\rm dip}$), a "foreground" map to account for the Galactic component ($\mathbf{f}$), and a constant map accounting for a residual monopole term ($C$). As in the previous section, we assume that the MFI wide survey maps ($\mathbf{m}_{\rm MFI}$) can be written as a linear combination of those three templates as
    \begin{equation}
    \mathbf{m}_{\rm MFI} = A \mathbf{m}_{\rm dip} + B \mathbf{f} + C +  \mathbf{n} , 
    \end{equation}
    where $A$, $B$ and $C$ are the three coefficients to be obtained and $\mathbf{n}$ represents the noise component.
    The dipole template map $\mathbf{m}_{\rm dip}$ is prepared following the methodology outlined in Sect.~4.4.2 of \citet{destriper}, including both the solar and orbital CMB dipole terms with the measured amplitudes by the Planck collaboration. 
    The dipole prediction is generated at the TOD level, and then this is projected into a sky map using the \picasso\ map-making algorithm. 
    For the Galactic template, we use again the WMAP 9-year K-band map after subtracting the CMB component. For this analysis, all maps are degraded to a common resolution of one degree. The analysis mask combines the default QUIJOTE analysis mask (NCP+sat+lowdec), the Planck confidence CMB mask for temperature analyses \citep{Planck2018-iv}, and a Galactic mask $|b|<30^\circ$, in order to avoid a possible bias in the dipole determination due to the Galactic emission. 
    
    \begin{table}
        \caption{Fitting for the CMB dipole in the MFI wide survey maps. We present the relative amplitude with respect to the expected CMB dipole, and the associated uncertainty. See text for details. }
        \label{tab:dipole}
        \centering
        \begin{tabular}{c|c|c}
        \hline
        Channel & Relative amplitude & Uncertainty \\
        \hline
        217 & 1.04  &  0.22\\
        219 &  0.97 &  0.47\\
        311 &  0.88 &  0.09\\
        313 & 0.92  &  0.12\\
        417 &  0.99 &  0.30\\
        419 &  1.23 &  0.67\\
        \hline
        Combined & 0.92 & 0.09 \\
        \hline
        \end{tabular}
    \end{table}
    
    We first validate the methodology using end-to-end simulations of the MFI wide survey including the dipole component and realistic $1/f$ noise levels as in \cite{destriper}. We find that our approach provides unbiased estimates of the dipole amplitude (i.e. $A=1$) for all MFI frequency maps, with typical errors of few percent. 
    We have also tested the impact of the three different corrections that are applied to the maps (RFI, FDEC and ATMOS) on the reconstructed dipole amplitude $A$. In summary, we find that including or not the RFI and FDEC corrections does not bias the recovered $A$ value. However, the ATMOS correction significantly affects the recovered amplitude, especially in the high frequency MFI bands. This is expected because the atmospheric templates are built on approximately one hour timescales, and on those scales the CMB dipole component is a very stable signal in the azimuth scans (rings). Because of this, the ATMOS correction is not applied for this analysis. 

    The measured values in real data are presented in Table~\ref{tab:dipole}, for each one of the MFI wide survey maps separately. Error bars have been estimated using the following methodology. We rely on the null test maps for independent baselines as the most representative method to capture large angular scale noise in the maps. Thus, we repeat the analysis and detect the CMB dipole in the half1/2, pwv1/2, tbem1/2 and daynight1/2 maps. The reported values correspond to the average dipole of the 8 cases, and the error bar is the scatter of the 8 measurements, taken to be a representative error of the method. 
    We have tested that we obtain almost identical results if we carry out the analysis on maps with no FDEC and/or RFI corrections. 
    
    Finally, we also present the weighted average combination of all channels, accounting for the correlation between frequencies of the same horn. The value is $A=0.92 \pm 0.09$, which corresponds to a 10-sigma detection of the CMB dipole, and it is consistent with the Planck calibration within nine per cent.

    \subsubsection{Bright point sources and planets}
    Bright radio sources and planets have been used extensively as a basic calibration test for MFI wide survey maps in several stages of the pipeline. Indeed, the maps in each period are recalibrated in order to match the Tau A model in intensity (Sect.~\ref{sec:recal}). Below in Sect.~\ref{sec:sources} we present a detailed study of few bright objects (Tau A, Cas A, Cyg A, 3C274, W63, Jupiter and Venus), which could be seen as a further validation test of the overall calibration scale of the experiment.

    \subsection{Setting the zero levels}
    \label{sec:zerolevels}
    The QUIJOTE MFI wide survey intensity maps produced by our default pipeline are insensitive to the true absolute zero level (monopole) of the sky emission. A monopole signal is essentially unconstrained for QUIJOTE MFI, as a global constant added to the full TOD database is not changing the map-making solution after the basic TOD processing. Indeed, in the post-processing stage maps are corrected of any residual monopole and dipole signals. 
    
    In order to estimate the zero levels of these maps in intensity, we follow a methodology similar to the one adopted by WMAP \citep{Bennett2003}, 
    and we assume a plane-parallel model for the Galactic emission. In that case, the zero level of the maps can be estimated by fitting 
    a cosecant model of the form:
    \begin{equation}
        \Delta T = A \csc(|b|) + B.
    \end{equation}
    For this analysis, we use the smoothed maps at $1^\circ$ angular resolution, and degrade them to $\nside=64$ in order to have 
    approximately independent pixels.  We carry out the fit independently in both hemispheres, using the Galactic latitude 
    ranges $15^\circ < b < 90^\circ$ and $-90^\circ < b < -15^\circ$ for the northern and southern hemispheres, respectively. 
    We mask the satellite band, and in the case of the northern sky, our analysis also excludes the region in Galactic longitude corresponding to the North Polar Spur ($0^\circ \le l \le 35^\circ$). Error bars are computed using the scatter of the results around the mean value, when adding realistic noise simulations. For this analysis, we use 100 of the simulations described in Sect.~\ref{sec:noisesims}. 
    The reference results adopted here correspond to the northern hemisphere, due to the larger sky fraction covered by the QUIJOTE MFI footprint. 
    For QUIJOTE MFI 11\,GHz (horn 3), we have $B=-0.74\pm 0.20$\,mK, where the error bar includes both the effect of varying sky emission and the noise variance contained in the simulations. Similarly, for QUIJOTE MFI 13\,GHz (horn3) we have $B=-0.59\pm 0.22$\,mK. The results for the southern hemisphere are consistent with those ($-0.59\pm 0.27$\,mK and $-0.42\pm 0.26$\,mK for 11 and 13\,GHz, respectively), although they have larger error bars. 
    For the other two frequency bands (17 and 19\,GHz), and both for horns 2 and 4, the zero levels are statistically consistent with zero in both hemispheres (with typical error bars of $1.2$--$1.3$\,mK). These values are inserted in Table~\ref{tab:summarycal}. 
    Finally, we note that there are other methods in the literature for deriving the zero levels of radio maps \citep[see e.g.][]{Wehus2017}, which could be applied here. However, we emphasize that those analyses should be done carefully, due to the special filtering of large angular scales (FDEC) applied to the MFI wide survey maps.

    \subsection{Polarization angle}
    \label{sec:polangle}
    As described in \citet{mfipipeline}, the reference angle for each MFI observation is calibrated using daily Tau A observations. 
    Our calibration scheme provides a reference angle for each period and channel, as this value changes across the spectral band, from horn to horn, and also with the instrument configuration. As this daily calibration might suffer from $1/f$ noise uncertainties, the final QUIJOTE MFI wide survey maps are recalibrated again using Tau A in each period (see Sect.~\ref{sec:recal}). 
    Here, we can evaluate the error budget associated with the polarization angle in the wide survey maps using Tau A. As a reference method, we use aperture photometry in the polarization maps smoothed to 1 degree. We adopt an integration radius of $r_1=1.5^\circ$ for the primary aperture, and an outer annulus between $r_1$ and $r_2=\sqrt{2}r_1$ to correct for the local background contribution. The photometry results are described in Table~\ref{tab:sources_vs_models} and Sect.~\ref{sec:sources}. 
    Table~\ref{tab:crabangle} presents the error budget in the polarization angle obtained using two methodologies. First, column 2 presents the scatter (standard deviation) of the Tau A angle measurements obtained from the null test maps with independent baselines (half1/2, pwv1/2, ring1/2, daynight1/2 and halfring1/2). On the other hand, column 3 presents the statistical error obtained from the propagation of the errors from the photometry measurement in the final maps.  
    As a conservative approach, we keep the highest value of each pair as representative of the error budget in the angle determination from Tau A. We see that the uncertainty changes from $0.5^\circ$ for 311, to $1.7^\circ$ for 419. 
    
    \begin{table}
        \caption{Error budget for the polarization angle in the wide survey, based on Tau A photometry. We include the error budget from the scatter of the measurements in the different null tests (column 2) and the statistical error obtained from the photometry method (column 3).  }
        \label{tab:crabangle}
        \centering
        \begin{tabular}{ccc}
        \hline
        Channel & Error (null tests) & Error (stat.)\\
          & (deg) & (deg)\\
        \hline
           217&   0.71&     0.91\\    
           219&     0.98&     0.96\\      
           311&     0.44&     0.50\\    
           313&     0.34&     0.67\\     
           417&      1.18&     0.64\\   
           419&      1.70&     0.59\\
           \hline
          Comb. 17\,GHz  &  0.96 &  0.53 \\
          Comb. 19\,GHz  &  1.43 &  0.51 \\     
        \hline
        \end{tabular}
    \end{table}

    \input{table_polangle.tex}

    As a further consistency check for the polarization angle calibration, we compare the measured MFI wide survey polarization angle maps with those 
    from WMAP 9-year K-band map \citep{Bennett2013} and Planck PR4 LFI30 data \citep{NPIPE}.  
    Table~\ref{tab:polangle} presents the results of this comparison, including also an internal comparison to the MFI 311 map. The analysis is carried out smoothing all maps to 1 degree resolution, and degrading them to $\nside=64$, in order to match approximately the beam scale in one pixel. We use the standard analysis mask (NCP+sat+lowdec), but in addition, we keep only those high signal-to-noise pixels with a nominal uncertainty in the MFI 311 angle $\sigma_{\phi_{311}} \le 2^\circ$. In order to avoid bright regions that might bias the comparison, pixels that have an absolute value in $Q$ or $U$ that is greater than 2\,mK in the WMAP K-band after being rescaled to 11.1\,GHz using a spectral index of $-3.0$ are also flagged. Finally, we also exclude the bright Cygnus area removing all pixels within 5 degrees around the location $(l,b)=(80^\circ, 0^\circ)$. The resulting analysis area has $\fsky = 0.124$. 

    In order to correct for residual zero level differences between the MFI and the WMAP/Planck maps (e.g. due to unresolved point sources), we use a TT plot technique between WMAP-K and each MFI Stokes Q and U map within the analysis mask, and we remove the fitted zero levels from the MFI maps. We note that the resulting values are basically consistent with zero (within the error), but of the order of 20\,$\mu$K for 311 and 313. Although small, they might introduce measurable differences (at the level of a degree) in our analysis. 
    For each MFI map, we compute the weighted mean of the difference between the two angles (e.g. $\phi_{\rm MFI} - \phi_{\rm WMAP}$ for the first case), using as weights the inverse variance of the angle, which in turn is derived from the $Q$ and $U$ weight maps. 
    Error bars in Table~\ref{tab:polangle} are generated with a Monte Carlo method using 100 of the noise simulations described in Sect.~\ref{sec:noisesims}. We add each noise simulation to the corresponding MFI map, and repeat the same procedure. The error bar corresponds to the standard deviation of the 100 values. 
    In general, all the measured differences are statistically consistent with zero given the noise uncertainty. MFI 311 (horn 3 at 11\,GHz) is consistent with both WMAP-K and LFI30 within the quoted uncertainty of $0.6^\circ$. The situation is similar for the 19\,GHz maps (both horns 2 and 4). However, we note that there is a moderate tension with the MFI 313, which deviates in the case of LFI30 up to $3.3$ sigmas, and the 17\,GHz cases, which deviates $2.4$ sigmas for the combined map of horn 2 and horn 4.  
    In order to investigate this possible discrepancy, we have repeated the analysis but using all the different null test maps with independent baselines (half1/2, pwv1/2, ring1/2, daynight1/2, halfring1/2). The error is now computed as the standard deviation of all those values. The result for the MFI 313 comparison with LFI30 now gives $-2.0\pm 0.9$, showing that maybe the error in this case is slightly underestimated. 
    While we are still finding a discrepancy, the significance is now reduced to $2.2$ sigmas. Another point that we have studied is the possible impact of Faraday rotation in this comparison. Using the Galactic Faraday depth maps from \citet{HE20}, we estimate that in our analysis region the mean rotation measure is $-11.9$\,rad\,m$^{-2}$. This would introduce differences of the order of approximately $-0.4^\circ$ between MFI311/MFI313 and LFI30. Although this value is not enough to explain the discrepancy, it helps to further decrease the tension below the 2 sigma level.

    The final results in Table~\ref{tab:summarycal} contain the worst case value based on the three values reported in this section (two values for Tau A in Table~\ref{tab:crabangle}, and the standard deviation of the comparison with WMAP/Planck in Table~\ref{tab:polangle}).

    \section{Simulations}
    \label{sec:sims}
    
    \subsection{Sky signal}
    \label{sec:skysims}
    
    Some of the analyses in this paper make use of sky simulations.  Our reference sky simulations were developed within the context of the RADIOFOREGROUNDS project\footnote{\url{www.radioforegrounds.eu}}, and are described in detail in Sect.~5.2 of \cite{destriper}. They contain different foreground components from the Planck FFP10 sky model \citep{Planck2018-ii, Planck2018-iii}, a CMB realization, and the CMB dipole contribution. For some applications, these sky simulations are projected into the MFI wide survey TODs, and the \picasso\ map-making code is used to generate synthetic maps with the same flagging and number of hits as in the real wide survey data. These simulated data can also include a noise contribution, injected at the TOD level. As explained in this paper, this approach has been extensively used to validate some aspects of the pipeline (map-making, transfer function, null tests, determination of the CMB dipole, etc.). These sky simulations are also used below to evaluate the statistical errors associated with the power spectra (see Sect.~\ref{sec:spectra}).

    \subsection{Simulated noise maps}
    \label{sec:noisesims}
    
    In addition to the end-to-end noise simulations that have been produced as explained in the previous subsection, we also construct noise simulations for the different channels (i.e., pair frequency-horn) maps, starting from the HMDM of totally independent splits. The simulations aim to account for the measured anisotropic behaviour, spatial correlations and the correlations between the two frequency channels of the same horn in the wide survey maps ($\sim 60$--$80$\% in intensity, and $\sim20$\% in polarization, as seen in Sect.~\ref{sec:noisecorr_freqs}). 
    
    The anisotropic behaviour follows the properties of the corresponding Local Variance (LV) maps per channel and per Stokes parameter. These LV maps are estimated from the HMDM maps, by assigning, at each pixel at resolution $\nside=512$, the variance computed from the surrounding pixels at a given distance (39 arcmin; this value has been chosen as a compromise to have enough pixels to provide an accurate estimation and, at the same time, to preserve as much information at small scale as possible). The estimation of the variance takes into account only those pixels which are within the observed sky at each channel. Each one of the HMDM are normalized by dividing them by the square root of the corresponding LV maps. 
    The non-observed pixels of the normalized HMDM are filled with a Gaussian random realization with unit dispersion, building in this manner extended-normalized HMDM maps.
    
    We now compute the noise spatial correlation by computing the TT, TE and BB angular power spectra (APS) of the extended-normalized HMDM maps, and from there, we derive a model of these APS. This is done by estimating a smoothed version of the observed APS of the extended-normalized HMDM maps for each channel, using a polynomial fit (of order 4), and by defining the maximum multipole that provides a variance (at the map level) as close as possible to the one of the corresponding extended-normalized HMDM maps. Following this process, we end-up with a model for the noise correlations that provides the right power level.
    
\begin{figure}
    \centering
    \includegraphics[width=0.95\columnwidth]{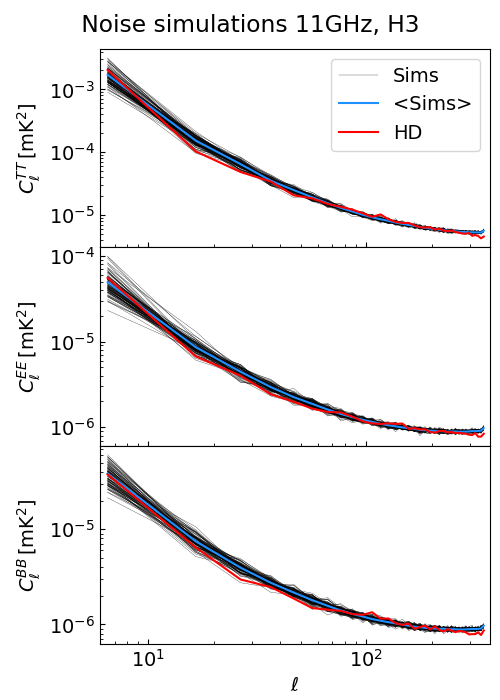}
    \caption{Power spectra (TT, EE, and BB) for 100 noise simulations of the 311 map (horn 3 at 11\,GHz).  The red line shows the reference noise power spectrum for the half mission difference maps (labelled as HD) which was used to generate the simulations. The individual power spectrum for each simulation is shown in light grey, and the average of those 100 simulations in light blue. }
    \label{fig:noisesims}
    \end{figure}

    A noise simulation is now generated by drawing a Gaussian random map in harmonic space, following the corresponding models of the noise APS for each frequency map. The maps (T, Q and U) are further multiplied by the square root of the corresponding LV map. 
    We use the correlation coefficient between frequency maps of the same horn to further modify the simulated map of the second member of the pair (e.g., the 13\,GHz frequency channel in the case of horn 3). In particular, we construct the final version of the simulated map of the second member of the pair as a linear combination of the first member of the pair and the initial version of the second map, taking into account the correlation coefficient. In this way, all the pixel-based statistics are maintained for the two members of the pair, as well as the correlation. 
    The APS of the first member are also maintained, but, eventually, we modify the APS properties of the second member and the cross-correlation. The correlation at pixel level is imposed for T, Q and U. Notice that for the Stokes parameters this is done as if they were scalars. Nevertheless, the properties of the polarization intensity are preserved, although we are not able to reproduce the observed cross-correlation in $P$. We find that this approximation is the most adequate for our further analyses, since most of them are addressed in the pixel domain. 
    As illustration, Figure~\ref{fig:noisesims} shows the power spectra for a subset of 100 noise simulations for horn 3 at 11\,GHz.

    \section{Power spectra of the wide survey maps}
    \label{sec:spectra}
    
    \begin{figure}
    \centering
    \includegraphics[width=\columnwidth]{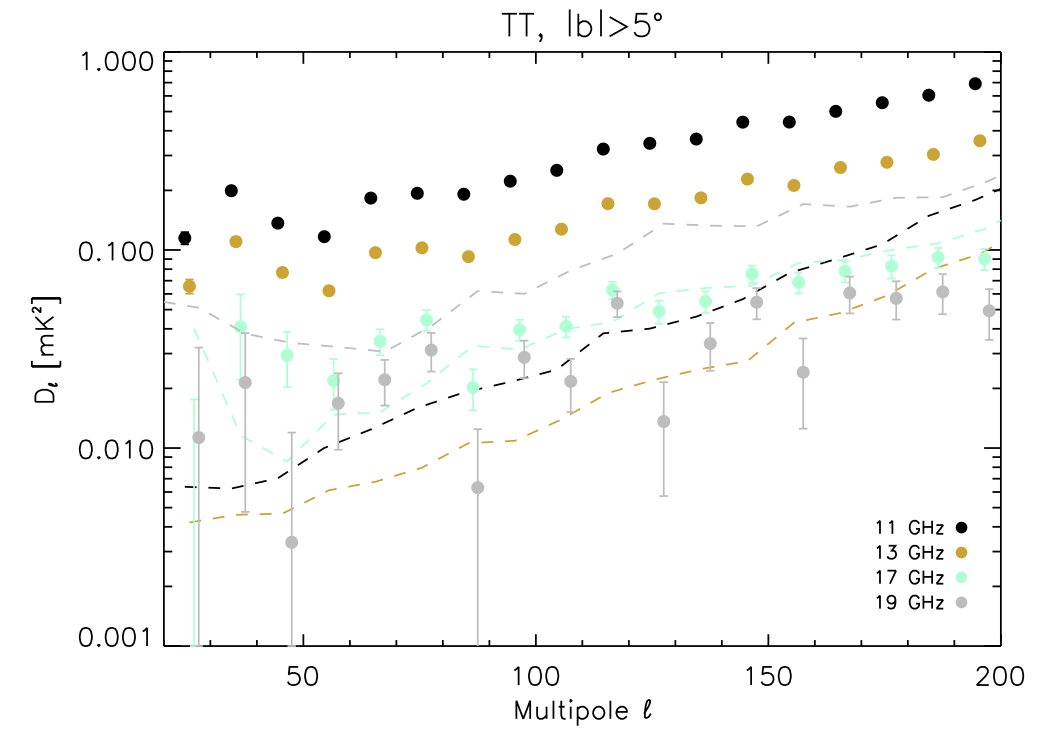}
    \includegraphics[width=\columnwidth]{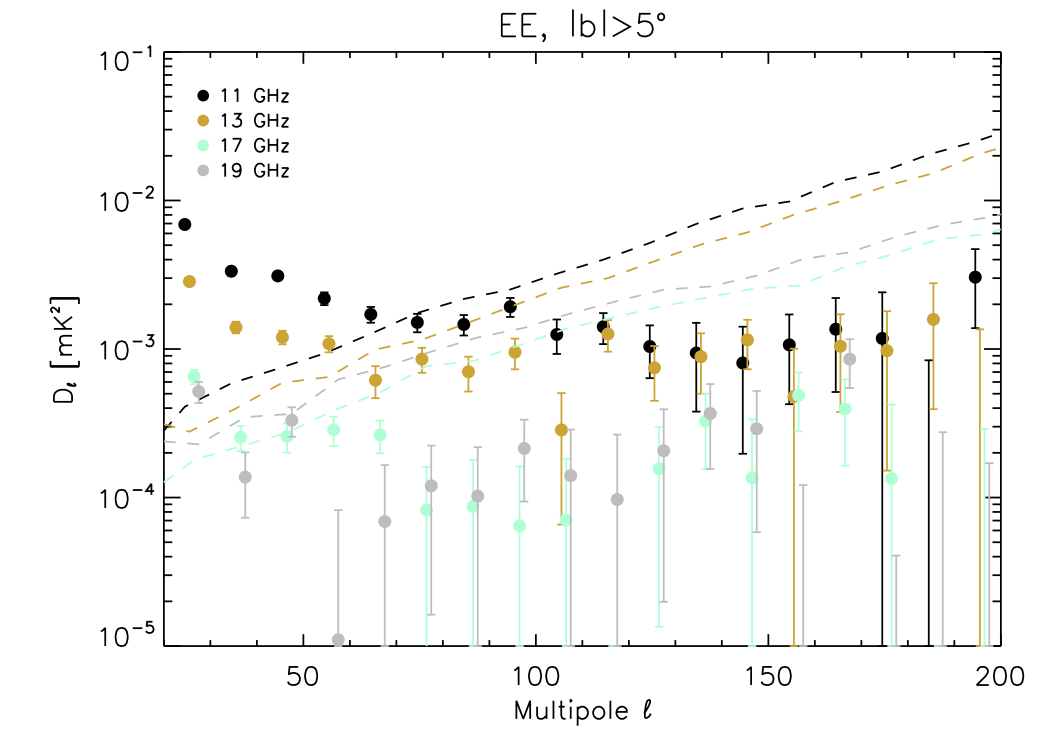}
    \includegraphics[width=\columnwidth]{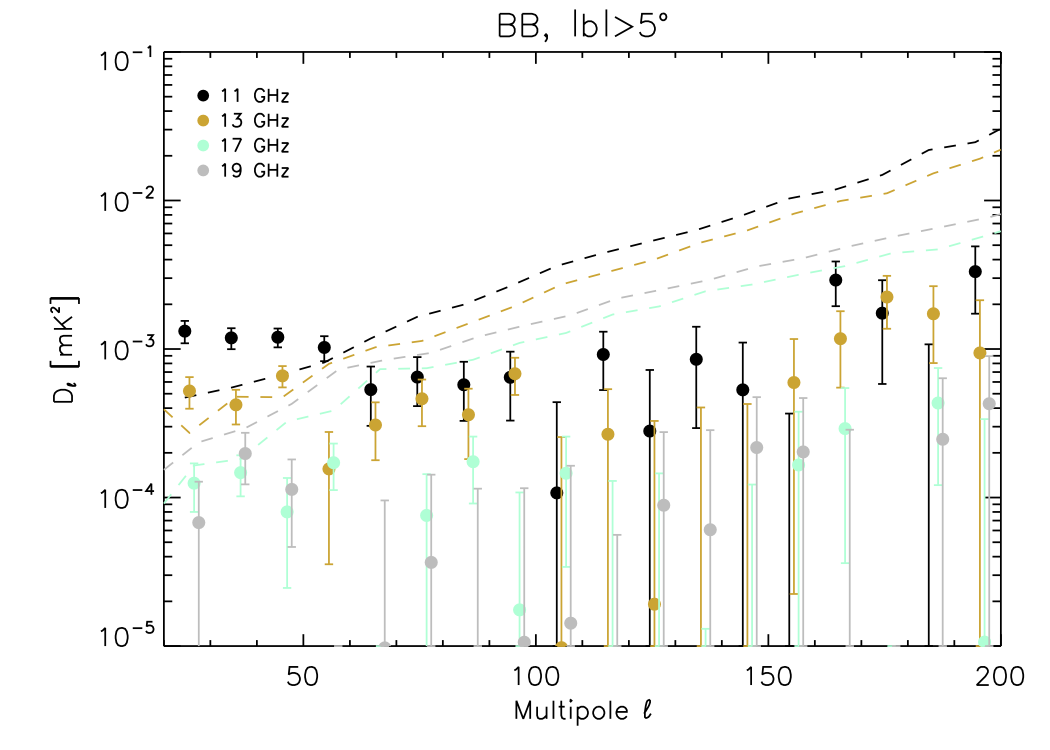}
    \caption{TT, EE and BB spectra for $|b|> 5^\circ$, and for all frequencies (11, 13, 17, 19\,GHz), represented as solid circles with their corresponding uncertainties. As a reference, dashed lines depict the noise spectra $N_\ell$ for each case, using the same colour scheme. }
    \label{fig:spectraTEB_allfreqs}
    \end{figure}
    
    In this section we study the main properties of the auto- and cross-spectra of the MFI wide survey maps. We consider three masks, corresponding to 
    different Galactic latitude cuts ($|b|>5^\circ$, $10^\circ$ and $20^\circ$), which are always combined with the default QUIJOTE analysis mask (NCP+sat+lowdec).
    As usual, each of these three masks is apodized with a five degree apodization kernel and the cosine function implemented in \citet{namaster}. 
    All spectra have been computed with the \namaster\ code, enabling for the option of "purification" of E and B modes, which allows a better reconstruction of 
    the E and B mixing matrix for cut-sky spectra. In Appendix~\ref{app:aps} we discuss the validity of the use of this pseudo-C$_\ell$ approach for the wide survey maps. 
    
    Throughout this section, all power spectra have been corrected by the MFI beam window functions, as well as the pixel window function (which in this case corresponds to a \healpix\ map with $\nside = 512$). Noise levels ($N_\ell$) are estimated from the half-mission difference maps (with independent baselines), and then subtracted from the corresponding power spectra of the maps, in order to obtain the spectrum of the sky signal, $C_\ell^{\rm sky} = C_\ell^{\rm map} - N_\ell$. We have tested that using another estimate of the noise power spectra (e.g. the average of several null test difference maps) produces consistent results to those presented in this section. 
    All spectra are binned using $\Delta \ell = 10$. In all figures in this section, we represent band power values 
    $D_\ell = \ell (\ell+1)C_\ell^{XY}/2\pi$, where $X, Y \in \{T,E,B\}$. 
    
    Uncertainties in the power spectra of the maps $\sigma(C_\ell^{\rm map})$  are estimated using 100 simulations including sky signal (Sect.~\ref{sec:skysims}) and realistic noise simulations (Sect.~\ref{sec:noisesims}). The same noise simulations are also used to estimate the uncertainties in the noise level, $\sigma(N_\ell)$. The quoted uncertainties in $\sigma(C_\ell^{\rm sky})$ are obtained as the quadratic sum of both $\sigma(C_\ell^{\rm map})$ and $\sigma(N_\ell)$. 
    
    Figure~\ref{fig:spectraTEB_allfreqs} shows the (auto) power spectra (TT, EE, BB) of the wide survey maps, for the 
    particular case of the Galactic mask with $|b|>5^\circ$, combined with the default QUIJOTE analysis mask (NCP+sat+lowdec). 
    We have a high significance detection of TT, particularly for the two lowest frequencies. At these frequencies, the polarized emission is dominated by Galactic synchrotron. The EE synchrotron signal is clearly detected at large angular scales ($\ell \lesssim 100$) for 11 and 13\,GHz, and the BB signal is also significantly detected in that range for 11\,GHz. In the next three subsections we discuss the angular and frequency dependence of these spectra. A multi-frequency analysis of the power spectra of the MFI wide survey maps, in combination with WMAP and Planck data, will be presented in a separate paper \citep{Synchwidesurvey}.

    \subsection{Fitting the EE and BB auto-spectra at 11\,GHz}
    
    \begin{figure}
    \centering
    \includegraphics[width=\columnwidth]{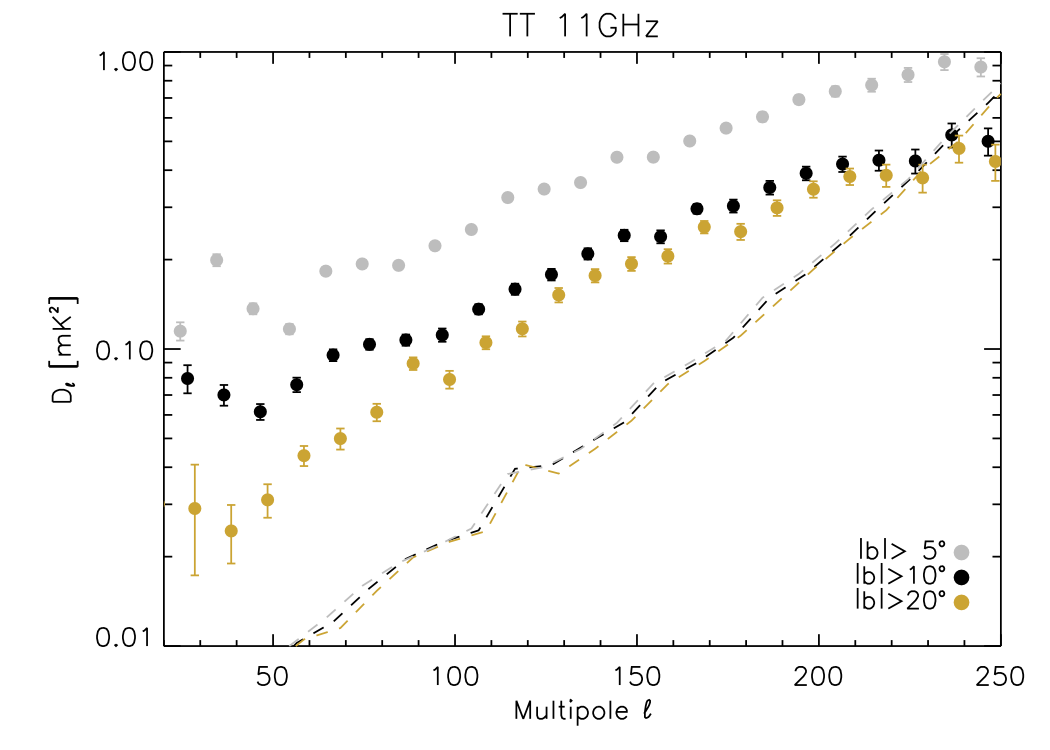}
    \includegraphics[width=\columnwidth]{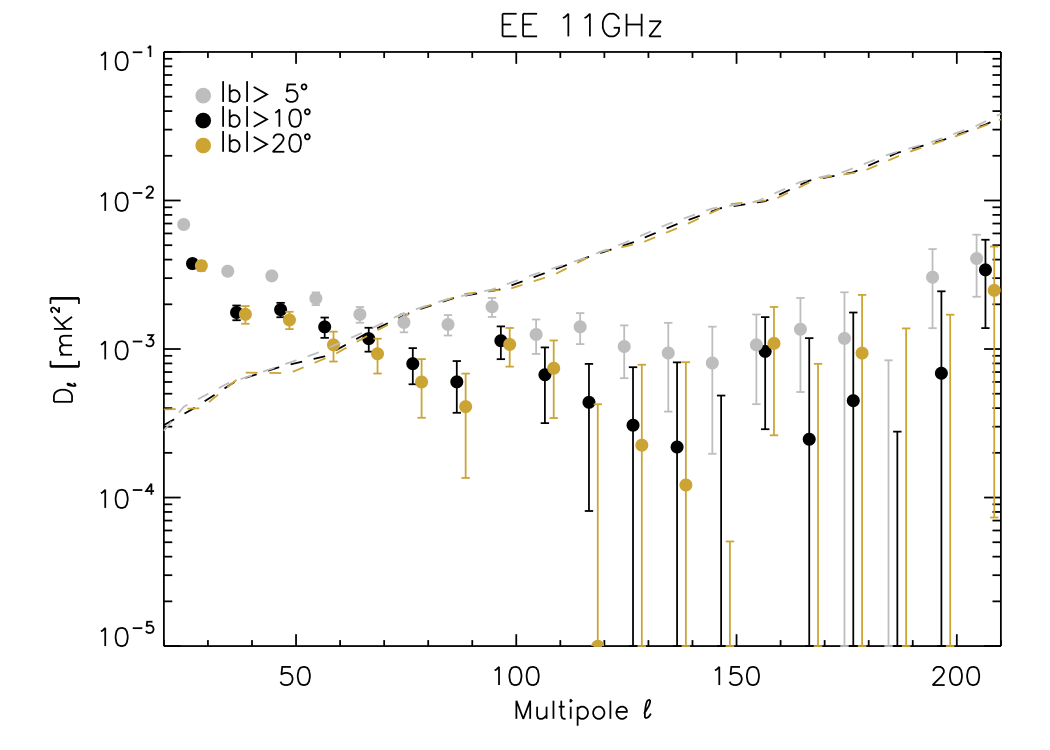}
    \includegraphics[width=\columnwidth]{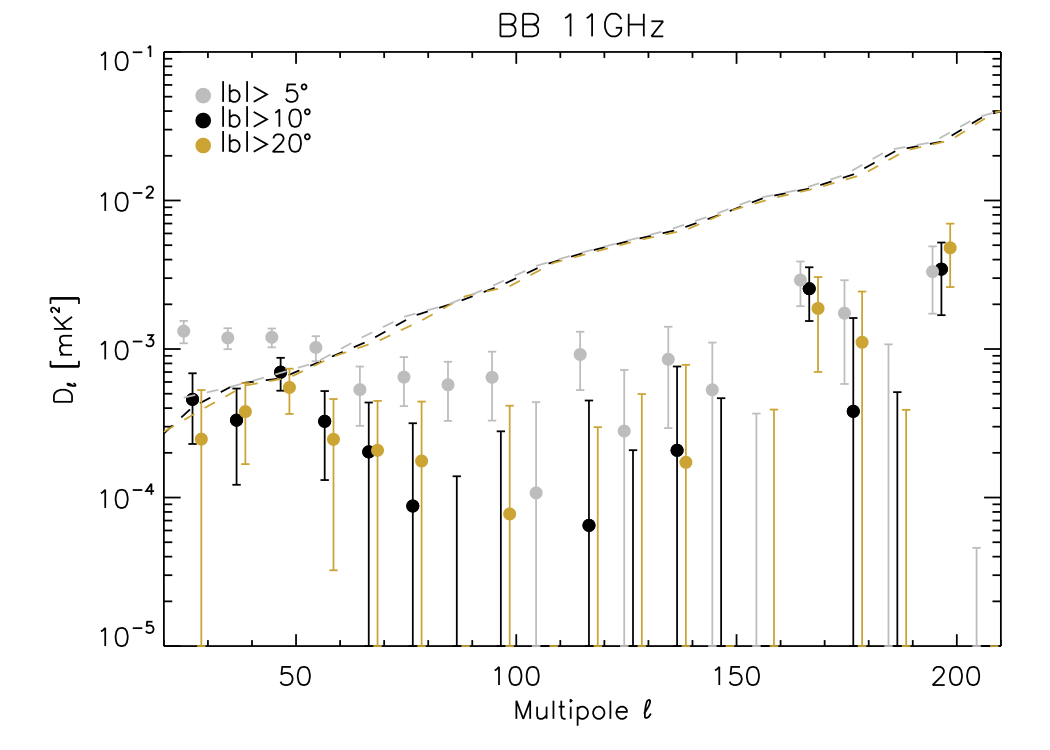}
    \caption{TT, EE and BB spectra for QUIJOTE MFI 11GHz, as a function of the Galactic cut. Dashed lines represent the corresponding noise spectra $N_\ell$ for each case, using the same colour scheme. }
    \label{fig:spectra311_TTEEBB_bcut}
    \end{figure}
    
    Figure~\ref{fig:spectra311_TTEEBB_bcut} shows the TT, EE and BB auto-spectra at 11\,GHz, for three masks with Galactic latitude cuts $|b|>5^\circ$, $10^\circ$ and $20^\circ$. We focus here on the polarization spectra, EE and BB. Following \cite{Krachmalnicoff2018}, we fit for these spectra in the multipole range $30 < \ell <300$, using the following parameterization 
    \begin{equation}
    \label{eq:cl_model}
    C_\ell^{\rm XX} = A_{\rm XX} \Bigg( \frac{\ell}{80}\Bigg)^{\alpha_{\rm XX}} + c_{\rm XX},
    \end{equation}
    where $X \in \{E,B\}$, $A_{\rm XX}$ is the amplitude of the spectrum at the pivot multipole $\ell=80$, $\alpha_{\rm XX}$ is the slope of the multipole dependence, and $c_{\rm XX}$ is a global constant which represents the contribution of unresolved (Poisson distributed) radio sources.

\input{table_powerspectra.tex}

    The power spectra are fitted using the {\sc EMCEE} ensemble sampler \citep{emcee}, and using a standard Gaussian likelihood function.
    Our best-fit results, obtained from the marginalised posterior distributions for each parameter, are given in Table~\ref{tab:ps}. First, we fit for the EE and BB power spectra separately. In all three cases, the global constants $c_{\rm EE}$ and $c_{\rm BB}$ are statistically consistent with zero, as expected given the noise levels of the wide survey maps, and the expected contribution from radio sources at these frequencies, estimated to be $\lesssim 30$\,$\mu$K.deg at 11\,GHz \citep{Puglisi2018, sourceswidesurvey}. 
    Both the EE and BB spectra present similar values of the slope, and no dependence on the Galactic latitude cut is observed. When combining the ratios of the EE and BB signals, we find that $A_{\rm BB}/A_{\rm EE}$ is of the order of 0.2 for the two higher Galactic cuts ($|b|>10^\circ$ and $|b|>20^\circ$), and we obtain $0.34\pm0.10$ for the lowest cut ($|b|>5^\circ$). 
    In order to increase the significance of this measurement, and based on these results, we repeat the analysis now assuming 
    that both EE and BB spectra have the same slope ($\alpha_{\rm EE}=\alpha_{\rm BB}$) and Poissonian terms contributions ($c_{EE}=c_{BB}$). In this case, we can fit simultaneously for the EE and BB spectra using four parameters ($A_{\rm EE}$, $\alpha_{\rm EE}$, $c_{\rm EE}$ and $A_{\rm BB}/A_{\rm EE}$). 
    The results for the amplitudes and slopes are consistent with the values obtained in the previous case. Regarding the ratio of the amplitudes, we have now a higher significance, with $A_{\rm BB}/A_{\rm EE}=0.26\pm0.08$ for the $|b|>20^\circ$ case. In summary, the MFI wide survey data at 11\,GHz show more power in the EE spectra than BB, with a typical BB/EE ratio of a factor of 0.26. This value is approximately half of the equivalent BB/EE ratio for thermal dust emission, as derived from Planck observations at 353\,GHz \citep{PlanckInt2016-xxx, Planck2018-xi}. 
    
    Our numbers for the synchrotron emission at 11\,GHz can be compared with others in the literature. \cite{Planck2018-iv} found $\alpha_{\rm EE} = -2.84 \pm 0.05$, $\alpha_{\rm BB} = -2.76 \pm 0.09$ and $A_{\rm BB}/A_{\rm EE}=0.34$ for the synchrotron map at 30\,GHz obtained with {\sc Commander} \citep{Commander}, and analysing a sky area of $f_{\rm sky}=0.78$ and a multipole range $\ell=4$-$140$. Following a similar methodology to the one used here, \cite{Martire2021} carried out a combined analysis of WMAP-K band and Planck LFI30 data, finding very stable values for the slopes and BB/EE ratios as a function of the sky mask. For the case of a mask preserving $50$\,\% of the sky, they obtain $\alpha_{\rm EE} = -2.79 \pm 0.05$, $\alpha_{\rm BB} = -2.77 \pm 0.15$, and  $A_{\rm BB}/A_{\rm EE}=0.22\pm0.02$. In both cases, the values are consistent with our results at 11\,GHz. 
    On the other hand, using S-PASS data at $2.3$\,GHz, \cite{Krachmalnicoff2018} find significantly larger values of the BB/EE ratio for similar Galactic cuts in the southern sky, with values of $0.87\pm0.02$ for $|b|>20^\circ$, and $0.64\pm0.03$ for $|b|>30^\circ$.

    \subsection{TE, TB and EB spectra at 11\,GHz}
    
    \begin{figure}
    \centering
    \includegraphics[width=0.95\columnwidth]{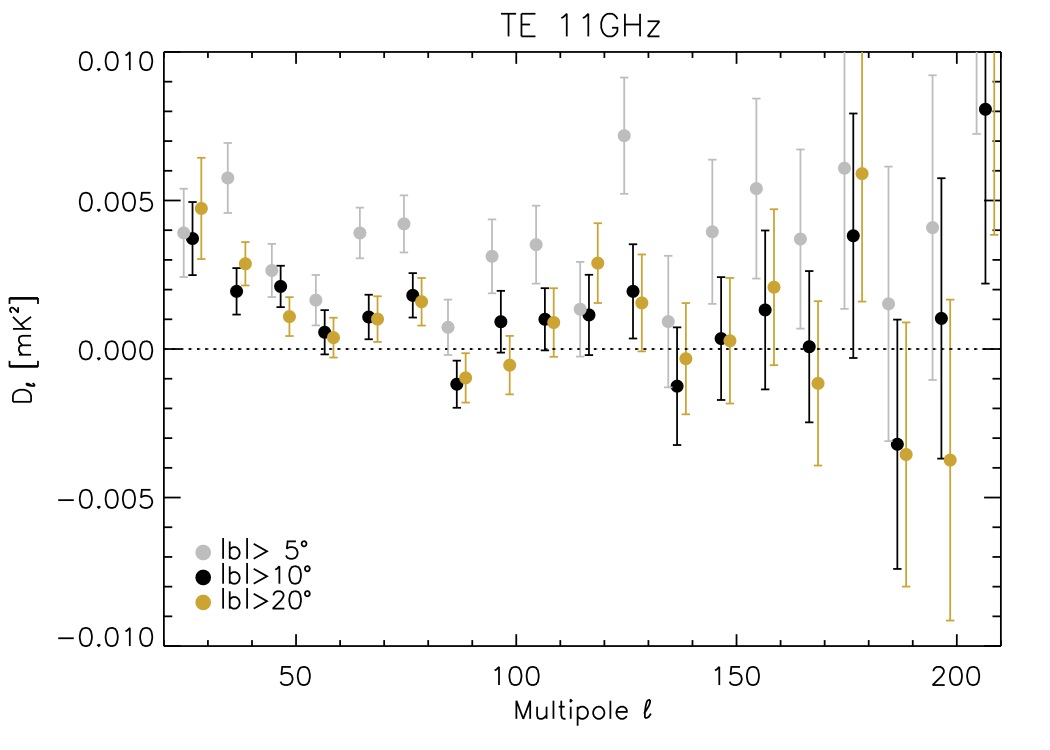}
    \includegraphics[width=0.95\columnwidth]{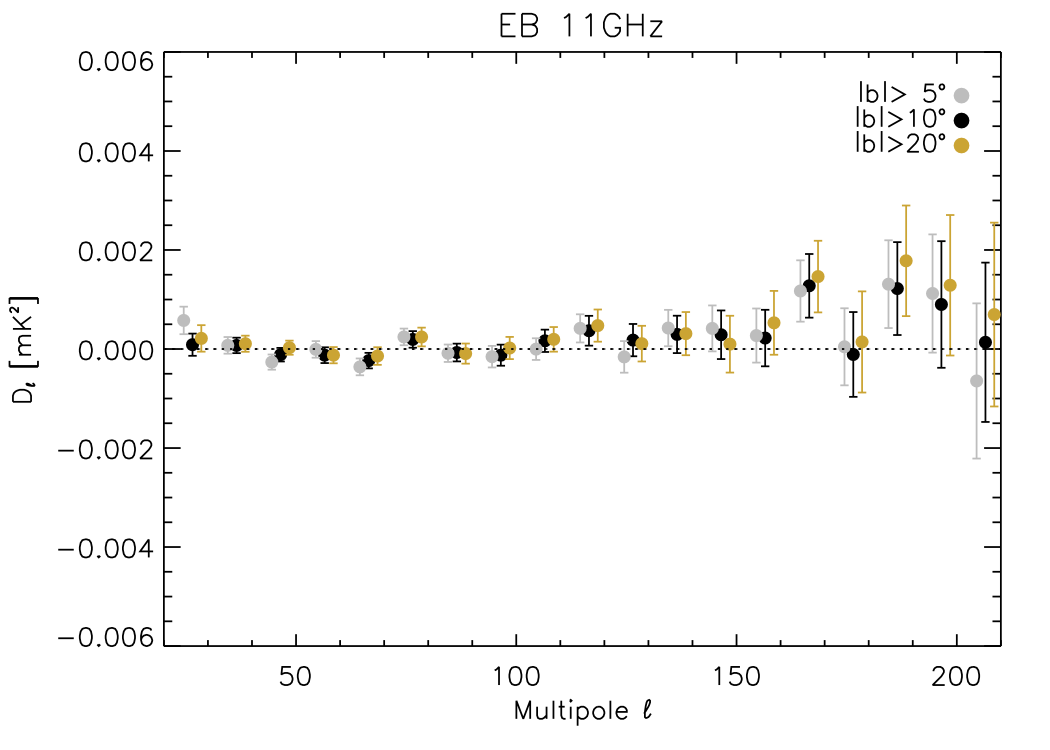}
    \includegraphics[width=0.95\columnwidth]{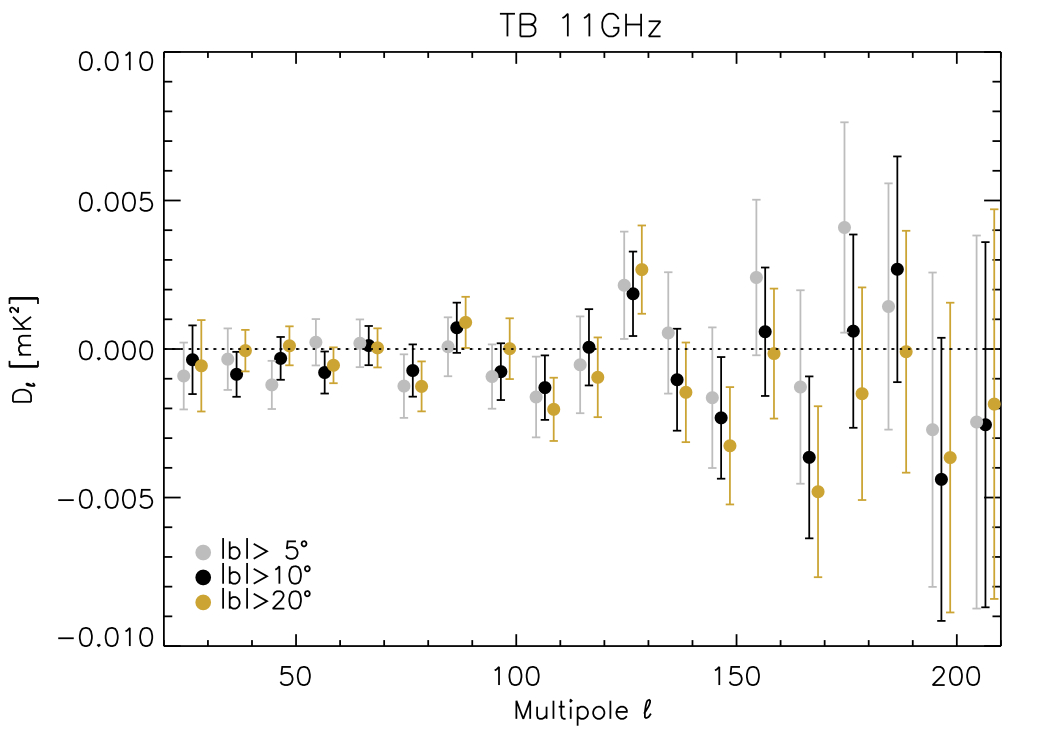}
    \caption{TE, EB and TB spectra for QUIJOTE MFI 11GHz, as a function of the Galactic cut.  }
    \label{fig:spectra311_TEEBTB_bcut}
    \end{figure}
    
    Figure~\ref{fig:spectra311_TEEBTB_bcut} shows the TE, EB and TB power spectra for the 11\,GHz map, evaluated in the same sky masks as in the previous subsection (see also Fig.~\ref{fig:spectra311_TTEEBB_bcut}). Given that the power spectra of the HMDM is statistically consistent with zero in all three cases (TE, EB and TB), we do not apply the $N_\ell$ correction in this subsection. Error bars are computed using the same methodology described above. However, for the EB spectra, we also add in quadrature the uncertainty on the power spectrum due to the polarization angle (Table~\ref{tab:summarycal}), using equation~5 in \cite{Minami2019}, and assuming that the underlying EB power spectrum is zero.

\input{table_eb_tb.tex}

    We detect a positive cross-correlation between the total intensity T and the E-mode polarization (TE$>0$) at large angular scales for the three considered Galactic cuts (up to $\ell \lesssim 80$ for $|b|>5^\circ$, and $\ell \lesssim 50$ for $|b|>10^\circ$ and $|b|>20^\circ$).
    Beyond $\ell \ga 150$, this TE cross spectrum becomes very noisy.  We also find a null correlation in TB and EB in the range $30 \lesssim \ell \lesssim 150$, as expected for a parity-invariant emission process and an accurate calibration of the polarization angle. Beyond this multipole range, the error bars increase significantly, in particular for the TB case. 

    We provide a quantitative measurement of the TB/EE and EB/EE ratios by fitting these spectra to a constant value (i.e. $C_\ell^{\rm TB} = A_{\rm TB}$, and $C_\ell^{\rm EB} = A_{\rm EB}$), in the range $ 30 \lesssim \ell \lesssim 150$. The results are presented in Table~\ref{tab:eb_tb}, where we have used the EE fits from Table~\ref{tab:ps}.  
    For the synchrotron emission, the MFI 311 maps provide upper limits on the EB signal at the level of $4$ per cent of the EE component at $\ell=80$ for the $|b|>10^\circ$ cut. These results are consistent with those found in \cite{Martire2021} for WMAP/Planck. Similarly, for the TB component we provide upper limits at the level of  20\,\% of the EE component.  We recall that for the thermal dust emission, the Planck satellite found a positive TE signal at large scales, a weakly positive TB, and a EB statistically consistent with zero \citep{PlanckInt2016-xxx,Planck2018-xi}.

    \subsection{Frequency dependence of the EE and BB signal}
    
    We carry out a simultaneous fit of all the power spectra shown in Figure~\ref{fig:spectraTEB_allfreqs}, using the parameterization from eq.~\ref{eq:cl_model}, but assuming that the amplitudes are related via a power law dependence in frequency with a temperature spectral index $\beta_{\rm s, EE}$. In practice, the amplitude at a given frequency channel $\nu$ is computed as:
    \begin{equation}
    A_{\rm EE}(\nu) =  A_{\rm EE} \Bigg( \frac{\nu}{11.1\,{\rm GHz}}\Bigg)^{2\beta_{\rm s, EE}} 
    \end{equation}
    where $A_{\rm EE} $ represents the EE amplitude in the MFI 311 map. Therefore, for this fit, we have seven parameters, namely
     $A_{\rm EE} $ and $\alpha_{\rm EE} $ for the amplitude and angular dependence of the synchrotron signal at 11\,GHz; the spectral index $\beta_{\rm s, EE}$ describing the frequency dependence, and four constant coefficients $c_{\rm EE}^{11}$, $c_{\rm EE}^{13}$, $c_{\rm EE}^{17}$ and $c_{\rm EE}^{19}$, accounting for the unresolved source contributions at each frequency.  For this analysis, we also introduce the colour correction term based on the fitted spectral index, using values reported in Table~\ref{tab:mfi_cc}. 
    For the $|b|>5^\circ$ mask, we obtain $A_{\rm EE} = 1.48 \pm 0.13$\,$\mu$K$^2$, $\alpha_{\rm EE} = -2.97 \pm 0.13$ and $\beta_{\rm s, EE} = -2.99 \pm 0.14$. 
    Similarly, we repeat the analysis for the BB power spectra, finding $A_{\rm BB} = 0.47 \pm 0.12$\,$\mu$K$^2$, $\alpha_{\rm BB} = -3.14 \pm 0.33$, and  $\beta_{\rm s, BB} = -2.79 \pm 0.35$. 
    
    In both cases, the first two parameters are in agreement with the values reported in Table~\ref{tab:ps}, taking into account that the colour correction term for the 11\,GHz map and for a spectral index of $\beta\approx-3$ is $0.967$. 
    The power spectrum of the synchrotron emission detected in the MFI wide survey maps scales with an average index of $-2.99$ for EE. The BB analysis is consistent with this value. The weighted average of the two values is $-2.96 \pm 0.13$, consistent with the result of $-2.96\pm0.09$ for the 50 per cent mask obtained in \cite{Martire2021} for the combination of WMAP-K and LFI30 data. 
    Our value also agrees with the study carried out in the next section for a real space analysis. A more detailed analysis on the reconstruction of the synchrotron spectral index with QUIJOTE MFI wide survey data is presented in two accompanying papers \citep{MFIcompsep_pol, Synchwidesurvey}.

    \section{Basic properties of the wide survey maps}
    \label{sec:properties}
    
    \subsection{Spectral index of the MFI sky emission}
    \label{sec:betas11}
    \subsubsection{Intensity}
    We first investigate the spectral dependence of the intensity emission in the MFI wide survey maps. We use as a reference the MFI 11\,GHz map, which presents the largest signal-to-noise, and we evaluate the spectral index of the sky emission when comparing it to the Haslam 408\,MHz \citep{Haslam1982} and WMAP-K 9-year maps \citep{Bennett2013}. The version of the Haslam map used here corresponds to the destriped map from \cite{Remazeilles2015}. 
    For this spectral analysis in real space, all external maps are filtered using the FDEC procedure, degraded to $2^\circ$ angular resolution, and then downgraded to $\nside=64$ resolution. Zero levels of all maps are corrected as in Sect.~\ref{sec:zerolevels}. 
    Colour corrections for MFI-311 and WMAP-K are taken into account. 
    The analysis region is restricted to the sky area covered by MFI 11\,GHz, but excluding the satellite band (satband) as described in Sect.~\ref{sec:masks}.
    For each $\nside=64$ pixel $p$ within the allowed mask, we solve for the spectral index $\beta(p)$ using a standard gaussian likelihood 
    function $\mathcal{L}$, which for the case of Haslam and MFI 11\,GHz reads
    \begin{equation}
    \label{eq:chi2_spec}
    -2\ln\mathcal{L}(p)= \frac{ \Bigg[ I_{408}(p) \Big(\frac{11.1}{0.408} \Big)^{\beta(p)} - cc_{11}(\beta(p)) I_{11}(p) \Bigg]^2 }{\sigma(p)^2},
    \end{equation}
    where $cc_{11}$ is the colour correction for MFI 11\,GHz, and the noise term $\sigma$ is evaluated using 1000 noise simulations for 
    MFI (see Sect.~\ref{sec:noisesims}), and accounting for a 10 per cent calibration error in the Haslam map.
    Similarly, for the spectral index in intensity between MFI 11\,GHz and WMAP-K, we use the same approach, accounting for the WMAP noise levels in the evaluation of the noise term.
    
    \begin{figure}
        \centering
        \includegraphics[width=0.95\columnwidth]{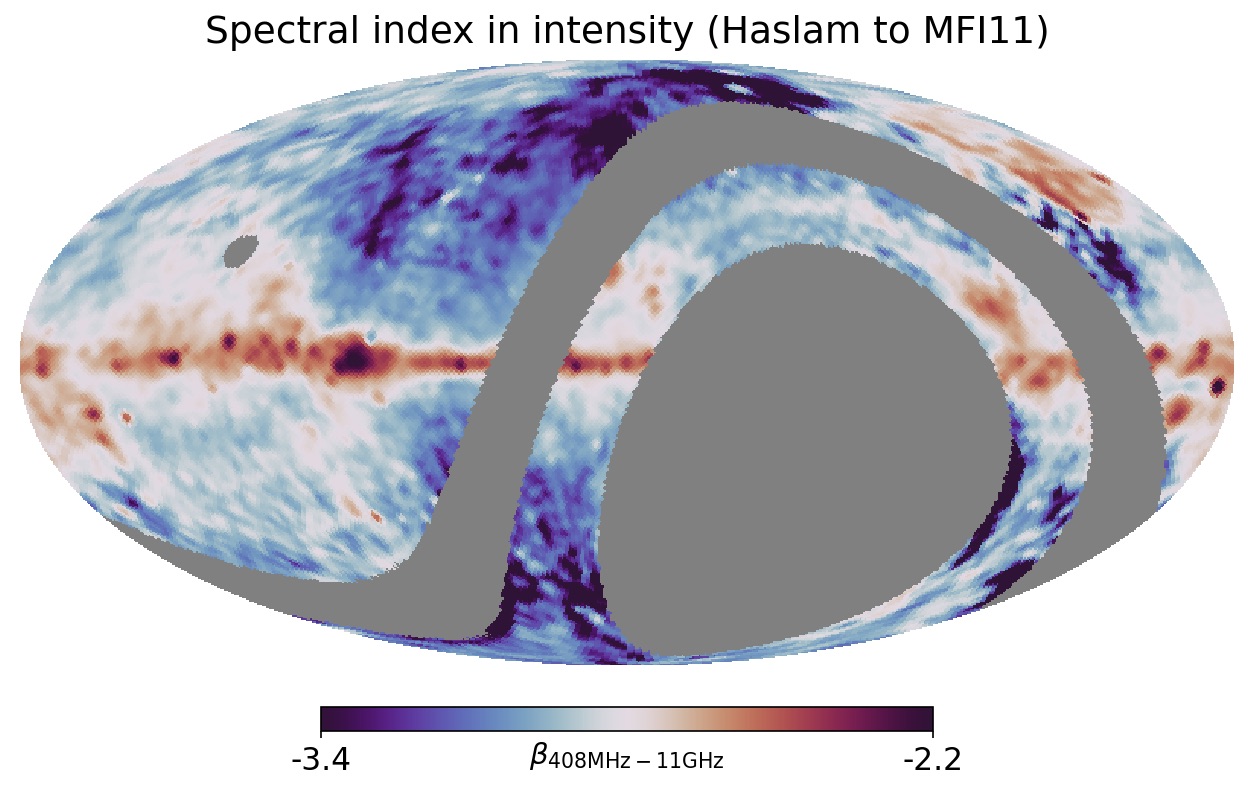}
        \includegraphics[width=0.95\columnwidth]{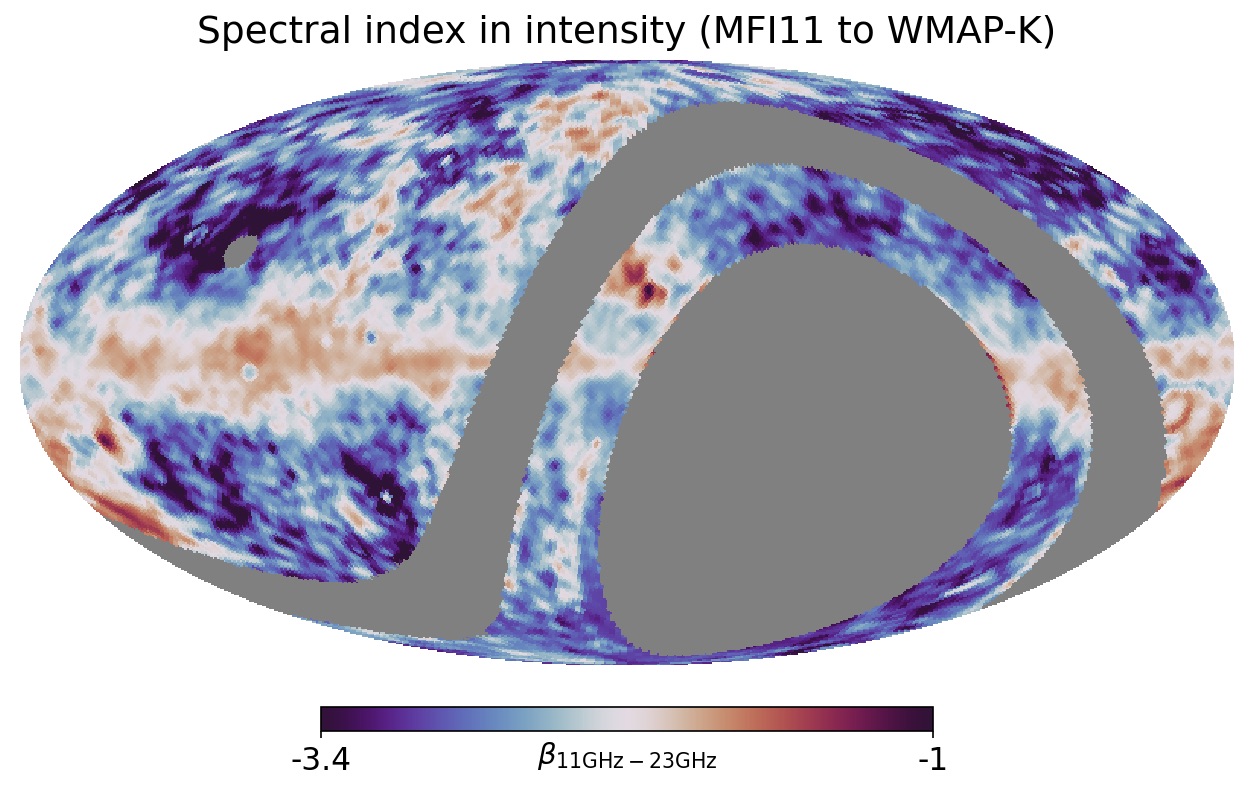}
        \caption{Spectral index of the intensity emission in the QUIJOTE 11\,GHz map. Top: Spectral index of $\beta_{408{\rm MHz}-11{\rm GHz}}$. The average index is $\beta = -2.9$. As expected, the Galactic plane regions have a flatter index, while the regions off the plane have steeper values.  Bottom: Spectral index of $\beta_{11{\rm GHz}-23{\rm GHz}}$. The average spectral index in this case is $\beta \approx -2.6$. In this colour scale, dark red corresponds to AME dominated regions.  }
        \label{fig:beta_int}
    \end{figure}
    
    Figure~\ref{fig:beta_int} shows the results for the case of $\beta_{408{\rm MHz}-11{\rm GHz}}$ (top panel) and $\beta_{11{\rm GHz}-23{\rm GHz}}$ (bottom panel), both for the intensity emission. Figure~\ref{fig:betas} shows a histogram with the distribution of spectral indices in both maps. 
    The median intensity spectral index $\beta_{408{\rm MHz}-11{\rm GHz}}$ in the full analysis mask is $-2.90$, with a standard deviation of the values across the map of 0.20. This value is consistent with the expectation for the average synchrotron emission at these frequencies \citep[see e.g.][]{Platania1998, deOliveira1999, Cosmosomas2006}. Moreover, the spatial dependence confirms the well-known steepening of the spectral index at high Galactic latitudes \citep[see e.g. the 408\,MHz--23\,GHz spectral index map in][]{Bennett2003}. 
    
    The $\beta_{11{\rm GHz}-23{\rm GHz}}$ intensity spectral index presents a much broader distribution of values, due to the presence of multiple spectral components (AME, free-free and synchrotron). In order to avoid extreme values for low signal-to-noise (high Galactic latitude) pixels, in this case we also add a broad gaussian prior $\beta=-3.1\pm0.5$ to the likelihood in equation~\ref{eq:chi2_spec}. We have checked that this has a minimal impact in the final histogram. The median spectral index in this case is $-2.59$, and the standard deviation of the values is $0.43$. Some of the bright AME dominated regions (Perseus, Lambda Orionis and rho Ophiucus) are clearly visible in dark red colour, while free-free dominated regions (e.g. Cygnus area) appear as light red. 
    A more detailed study of the spectral properties of the sky emission in intensity along the Galactic plane ($|b| \le 10^\circ$) in the MFI wide survey maps is carried out in an accompanying paper \citep{ameplanewidesurvey}. We also present a component separation analysis of the full MFI maps in \cite{MFIcompsep_int}. 
    
    \subsubsection{Polarization}
    
    \begin{figure}
        \centering
        \includegraphics[width=0.95\columnwidth]{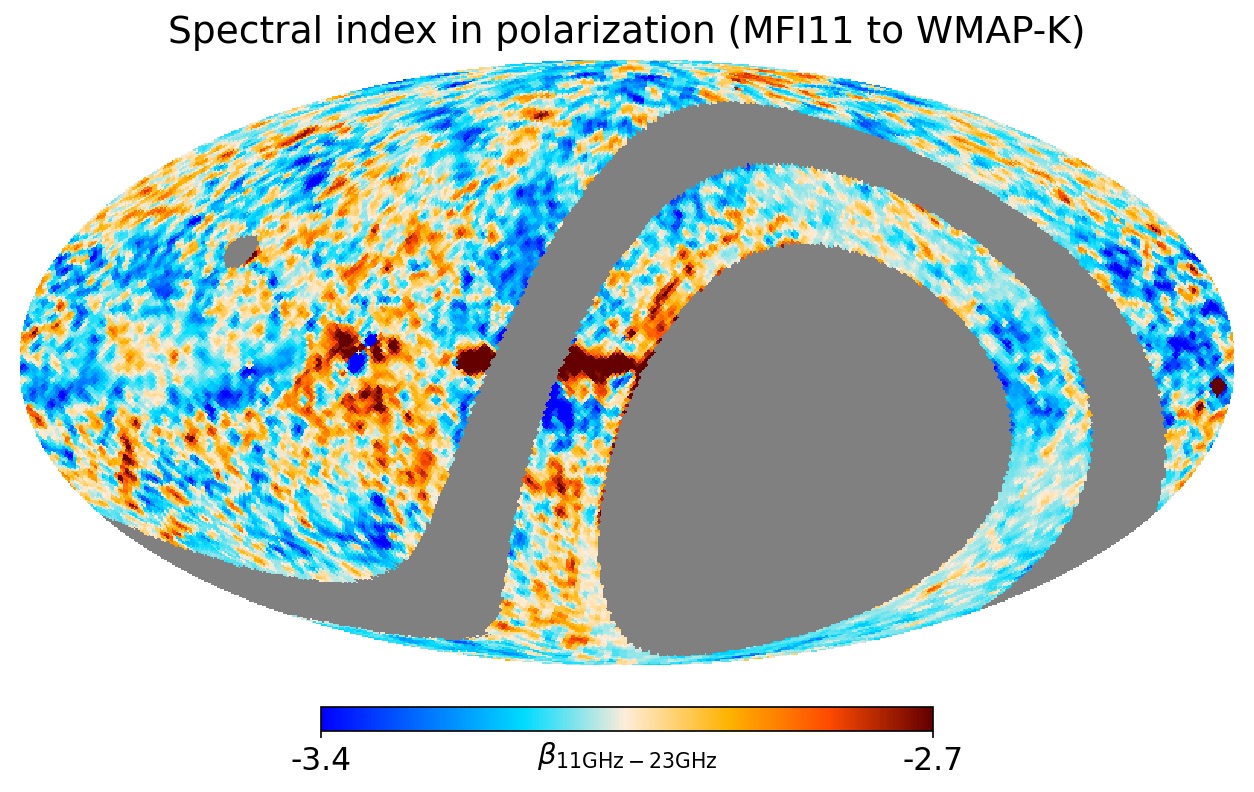}
        \includegraphics[width=0.95\columnwidth]{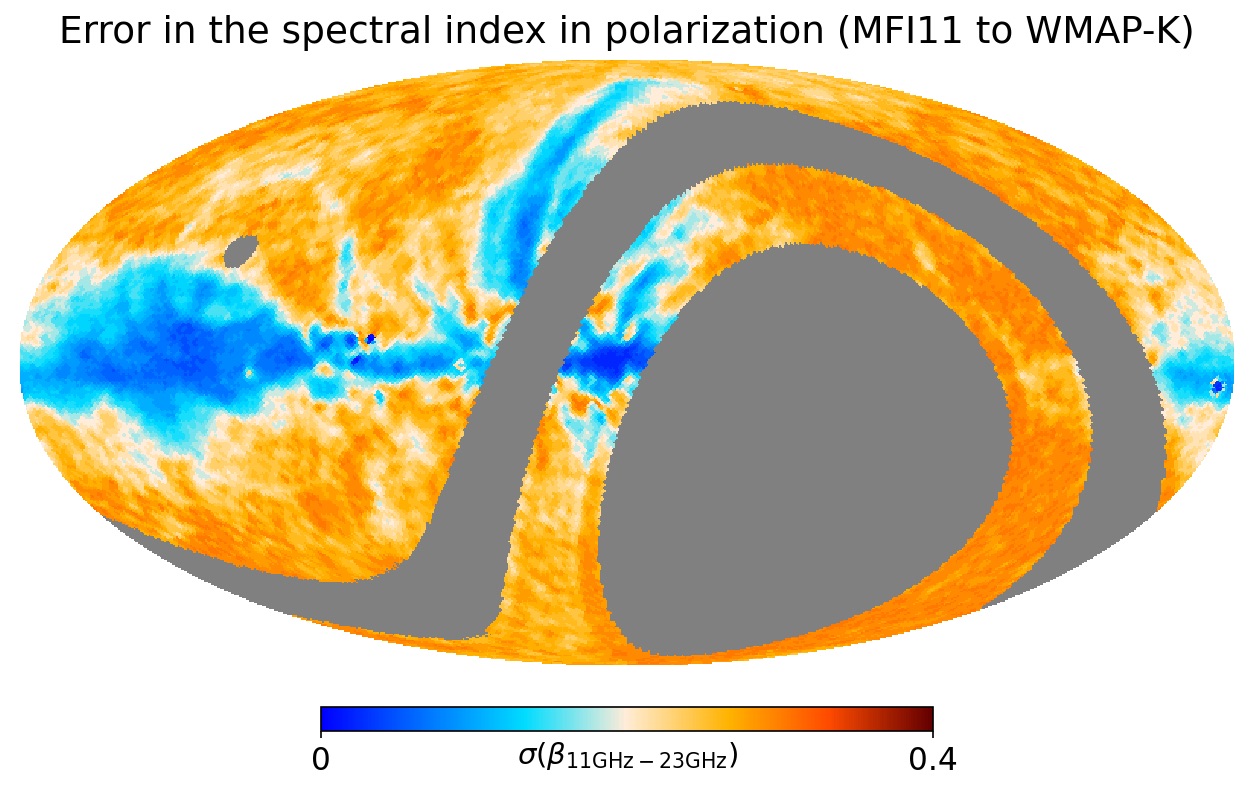}
        \caption{Top: Spectral index map of the polarized emission between QUIJOTE 11\,GHz and WMAP 23\,GHz. Bottom: the associated error map.  }
        \label{fig:beta_pol}
    \end{figure}
    
\begin{figure}
        \centering
        \includegraphics[width=8cm]{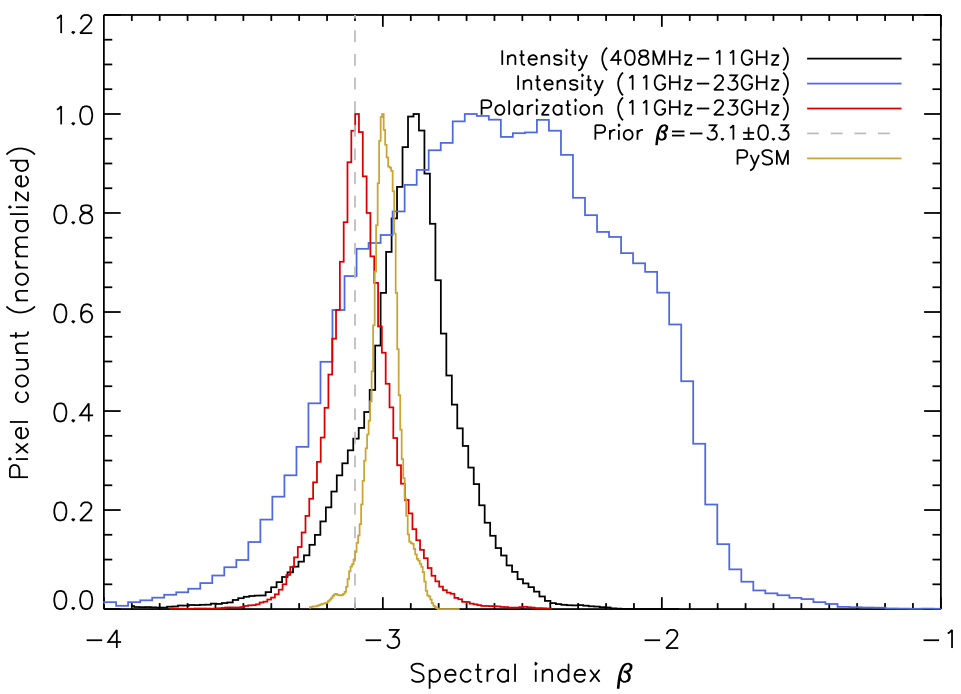}
        \caption{Histogram of spectral index values obtained from Figures~\ref{fig:beta_int} and \ref{fig:beta_pol}.  We show in dashed lines the mean of the prior adopted in the determination of the spectral index in polarization. For comparison, we also include the histogram of spectral index values from the PySM synchrotron model 1 \citep{PySM2}. We recall that in the intensity case for $\beta_{11{\rm GHz}-23{\rm GHz}}$ (blue line), the 11\,GHz map contains free-free and AME in addition to synchrotron, and thus the histogram presents a different shape with a broader distribution (see text for details). }
        \label{fig:betas}
    \end{figure}

    In polarization, the $\beta_{11{\rm GHz}-23{\rm GHz}}$ spectral index presents a cleaner interpretation in this case, as we are dominated by synchrotron emission only. Figure~\ref{fig:beta_pol} presents the recovered polarization spectral index map, following the same methodology as for the intensity. 
    The fit is carried out simultaneously in Stokes Q and U parameters, and in order to obtain a stable solution for high Galactic latitude pixels, we add a Gaussian prior $\beta=-3.1\pm0.3$ to the likelihood in equation~\ref{eq:chi2_spec}. The bottom panel in that figure shows the associated error map, derived from the posterior distribution.
    Fig.~\ref{fig:betas} includes also the histogram of these polarization $\beta_{11{\rm GHz}-23{\rm GHz}}$  values, showing that the median value is $-3.09$, and the standard deviation is $0.14$. 
    For comparison, we also include in this figure the histogram of spectral index values for the PySM synchrotron model 1 \citep{PySM2}, which in turn corresponds 
    to "Model 4" of \cite{MivilleDeschenes2008} calculated from a combination of Haslam and WMAP 23\,GHz polarization data using a model of the Galactic magnetic field. We find that in the same sky mask, the PySM spectral index map peaks at a higher value and presents a much narrower distribution ($-2.99\pm0.06$). 
    As a further consistency check, Appendix~\ref{app:betas13} presents the results for the same analysis carried out in this section, but using the MFI 13\,GHz map as reference. We can see that both the mean values and widths of the distributions discussed here are consistently reproduced in this case. 
    These values for $\beta_{11{\rm GHz}-23{\rm GHz}}$ in polarization are consistent with those measured in the range 22.8--100\,GHz ($\beta_{\rm s} \sim -3.1$) by other authors \citep{Dunkley2009,Fuskeland2014,Fuskeland2021,Harper2022}.
    A more detailed study of the spectral properties of the sky emission in polarization using the MFI wide survey maps in combination with WMAP and Planck, including a discussion on synchrotron spectral curvature, is carried out in an accompanying paper \citep{MFIcompsep_pol}.
    
    \subsection{E- and B-mode maps}
    As a complementary view of the relative power distribution in the E- and B-mode components for the synchrotron emission traced by the QUIJOTE MFI wide survey map, we have obtained in this section E- and B-mode maps. We use the full QUIJOTE observed area, but we mask the satellite band (satband) as described in Sect.~\ref{sec:masks}. In order to minimize the impact of E/B mixing \citep{Lewis2001}, we apodize this analysis mask using a Gaussian kernel of $2^\circ$. 
    E- and B-mode maps are then generated using the standard \healpix\ routines {\sc anafast} and {\sc synfast}, as
    \begin{align}
    \nonumber
    E(\hat{n}) = \sum_{\ell=2}^{\infty} \sum_{m=-\ell}^{\ell} a_{\ell,m}^{\rm E} Y_{\ell,m}(\hat{n}) \\
    B(\hat{n}) = \sum_{\ell=2}^{\infty} \sum_{m=-\ell}^{\ell} a_{\ell,m}^{\rm B} Y_{\ell,m}(\hat{n}) , 
    \end{align}
    where $a_{\ell,m}^{\rm E}$ and $a_{\ell,m}^{\rm B}$ are the corresponding harmonic coefficients. 
    
\begin{figure}
    \centering
    \includegraphics[width=8cm]{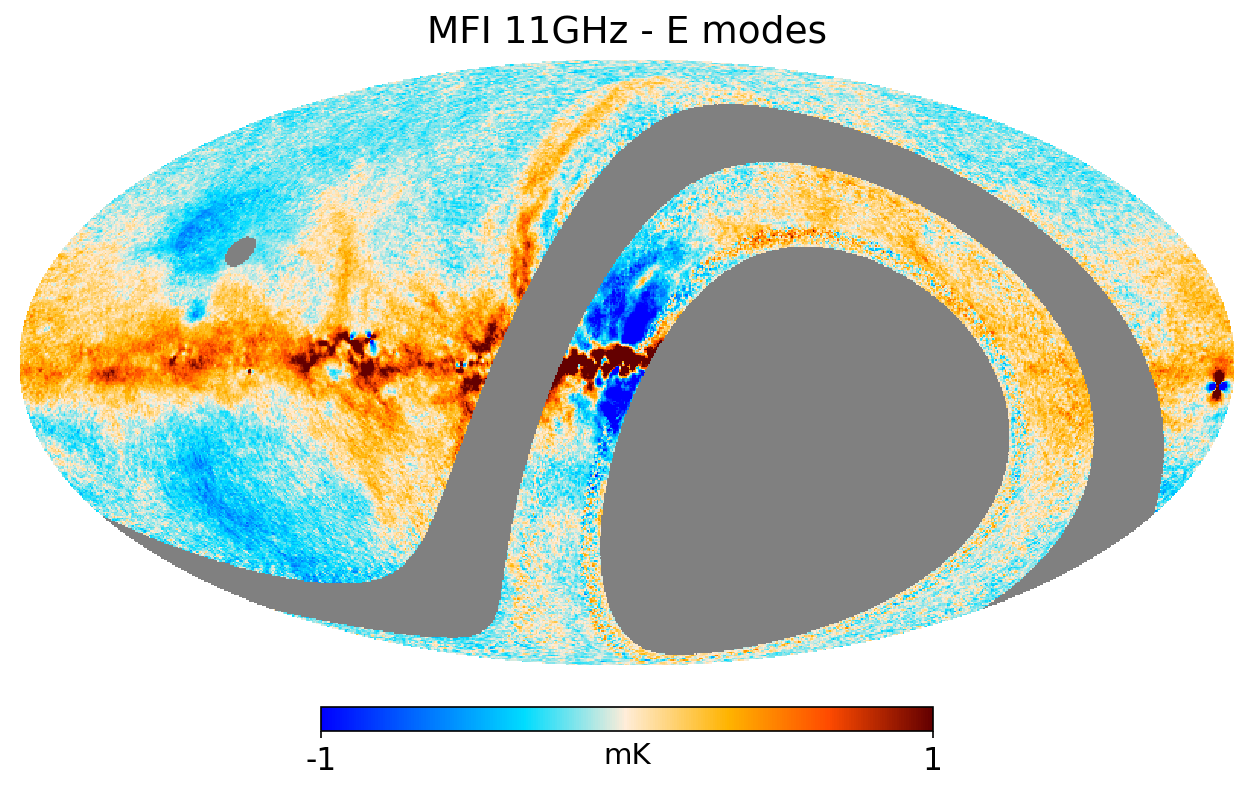}
    \includegraphics[width=8cm]{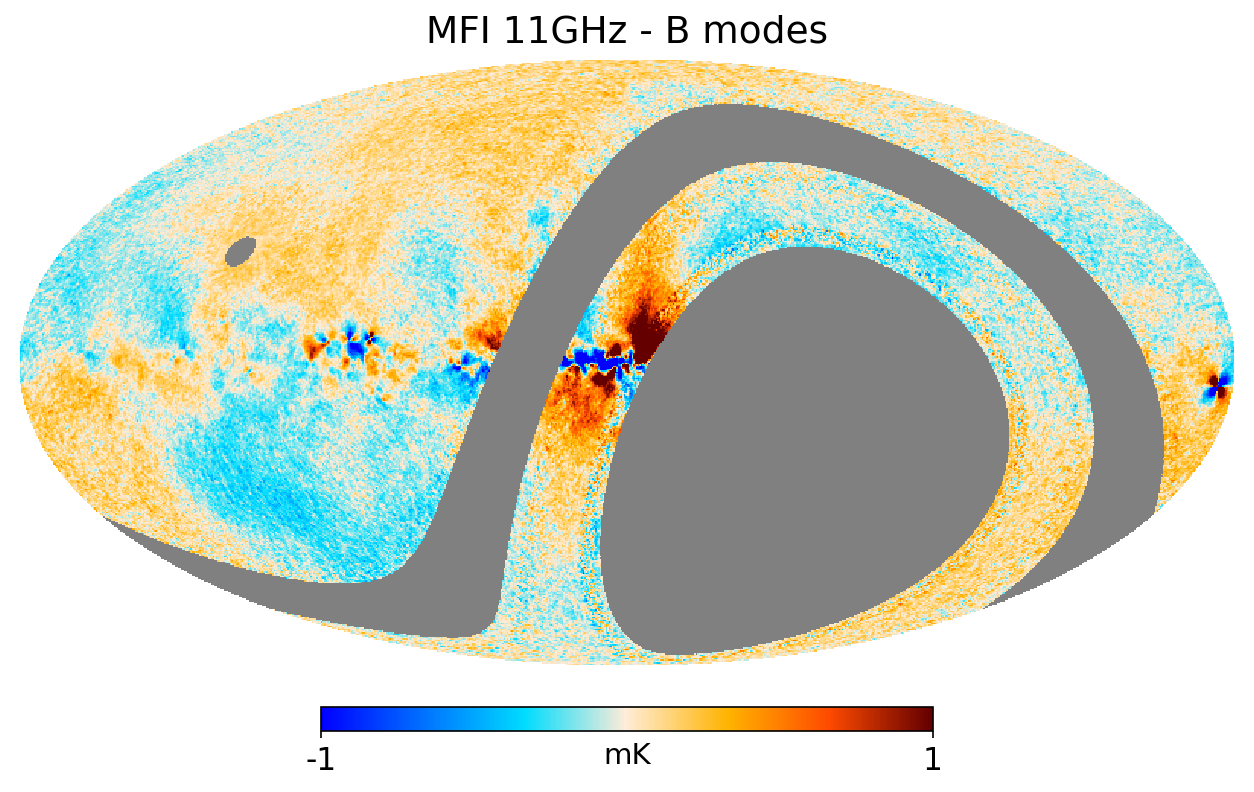}
    \caption{E and B-mode maps at 11\,GHz. Most of the brightest features in the QUIJOTE map (North Polar Spur, Fan region, Galactic plane) appear in the E-mode map.  }
    \label{fig:ebmaps}
    \end{figure}

    Figure~\ref{fig:ebmaps} shows the derived maps for MFI 11\,GHz.  As expected from the power spectrum analysis in Sect.~\ref{sec:spectra}, there is significantly more power in the E-mode than in the B-mode map. Moreover, most of the brightest synchrotron features in the polarized intensity map (North Polar Spur, Fan region, Galactic centre) appear mostly in the E-mode, as expected due to the underlying magnetic field structure. 
    Strongly polarized radio sources (Tau A, Cyg A) appear in these E- and B-mode maps with the characteristic quadrupole patterns with two positive and two negative lobes, and with the B-mode profile rotated by $45^\circ$ with respect to the E-mode map \citep[see e.g.][]{2021JCAP...03..048D}.
    
    \subsection{Bright structures in the polarized intensity maps}
    The MFI wide survey polarized intensity maps are dominated by several bright and extended structures (see Figure~\ref{fig:p11_ang11_1deg}). We discuss some of them in four accompanying papers: the Fan region \citep{FANwidesurvey}, the Haze and Galactic center \citep{hazewidesurvey}, the North Polar Spur \citep{NPSwidesurvey}, and other synchrotron loops and spurs \citep{loopswidesurvey}.
    
    \subsection{AME in the MFI wide survey maps}
    The MFI wide survey maps can be used to characterize the spectral properties of the AME, both in intensity and polarization. In particular, in \cite{ameplanewidesurvey} we present a study of the diffuse AME emission in intensity along the Galactic plane ($|b| \le 10^\circ$), while \cite{AMEwidesurvey} characterizes the SED in intensity for 52 compact sources with AME. Finally, two additional papers update the constraints in intensity and polarization of the AME in several Galactic regions \citep{W51,snrwidesurvey}.
    
    \section{Bright compact sources and planets in the wide survey}
    \label{sec:sources}
    
    Despite its coarse angular resolution a high number of point sources are detected to high significance in the QUIJOTE MFI wide survey data. In a companion paper, where we discuss radio source detectability in these maps and derived statistical properties \citep{sourceswidesurvey}, we show that we detect 235 point sources at S/N$>3$ at 11\,GHz, while 85 are detected at S/N$>5$. As a further consistency check of the global amplitude calibration, in this section we compare with models the recovered flux densities on four of the brightest sources having well characterised spectra (Tau A, Cas A, Cyg A and 3C274), and in two planets (Jupiter and Venus). We also calculate polarization flux densities in three bright polarized sources (Tau A, Cyg A and W63) to assess the accuracy of the polarization calibration.
    
    \subsection{Compact sources in intensity}
    
    Tau A (also known as the Crab nebula), Cas A and Cyg A are amongst the brightest compact sources in the microwave range, and hence they have traditionally been used to calibrate experiments operating in this frequency range, including CMB experiments \citep{Baars1977}. Using WMAP data, \citet{Weiland2011} presented updated spectrum models in the range $\sim 1$--$300$\,GHz of these three sources and of 3C274 (also known as Virgo A or M87) and 3C58. Here we will focus on Tau A, Cas A, Cyg A and 3C274, while 3C58 will be discussed in detail in \cite{FANwidesurvey}. 
    
    \begin{figure*}
    \centering
    \includegraphics[width=8.3cm]{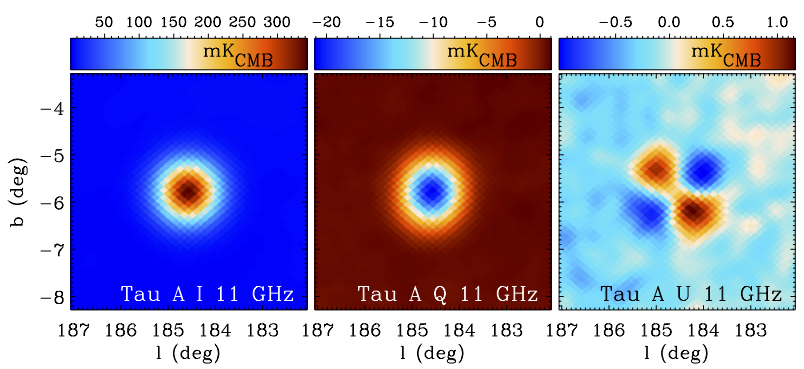}%
    \includegraphics[width=8.3cm]{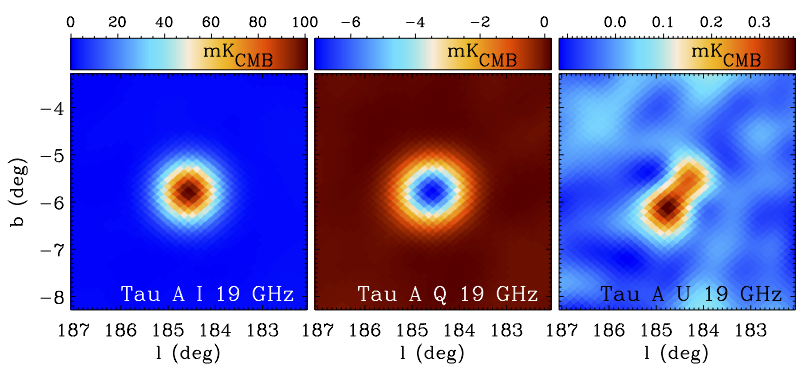}
    \includegraphics[width=8.3cm]{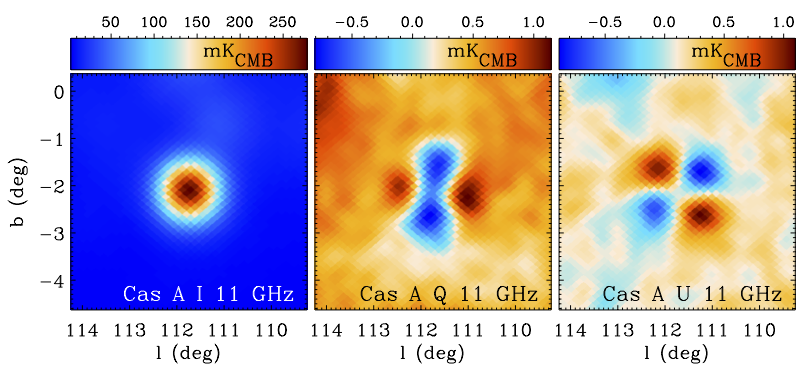}%
    \includegraphics[width=8.3cm]{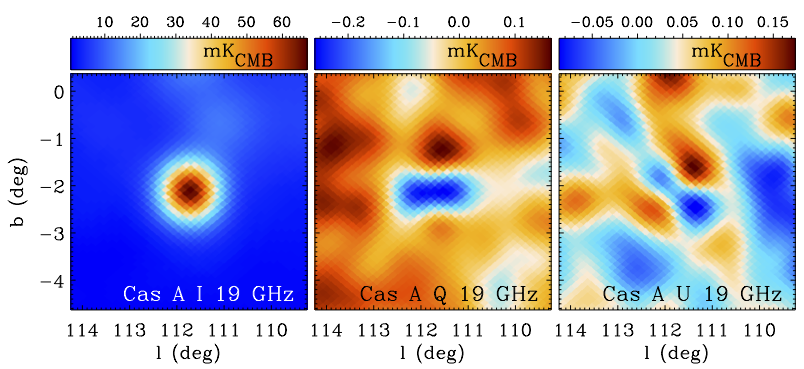}
    \includegraphics[width=8.3cm]{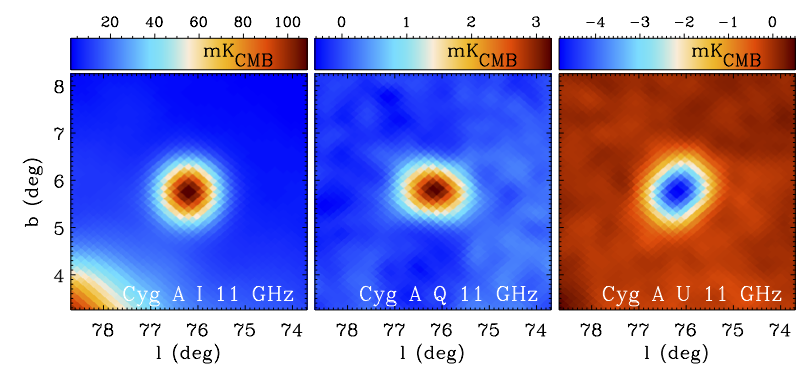}%
    \includegraphics[width=8.3cm]{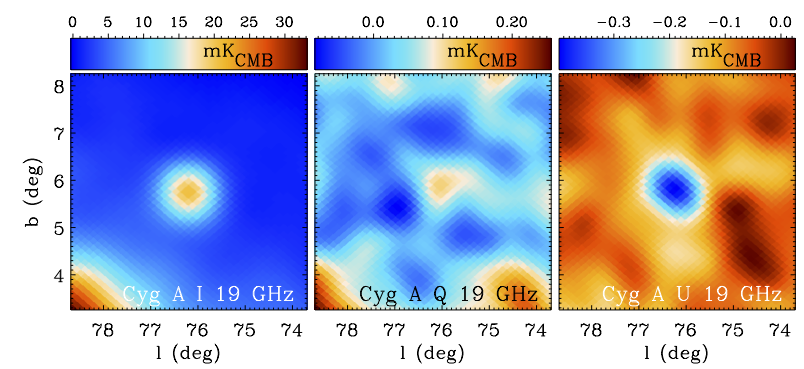}
    \includegraphics[width=8.3cm]{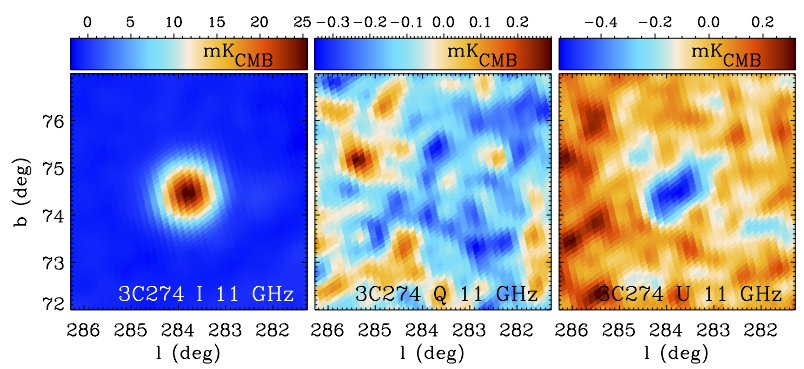}%
    \includegraphics[width=8.3cm]{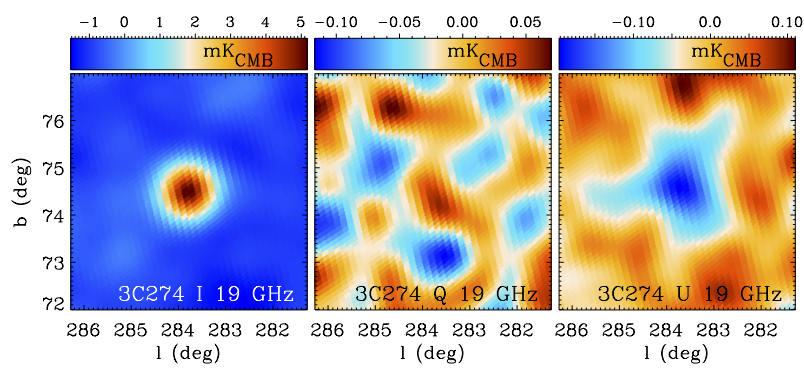}
    \caption{ Minimaps of $5^\circ \times 5^\circ$ size around four bright radiosources: Tau A (first row), Cas-A (second row), Cygnus-A (third row), and 3C274 (bottom row) at 11\,GHz (first three columns are I, Q, and U), and 19\,GHz (columns 4 to 6 are I, Q, U, respectively). For display purposes, we use the MFI maps degraded to a common angular resolution of one degree.}
    \label{fig:sources_minimaps}
    \end{figure*}
    
    Figure~\ref{fig:sources_minimaps} shows the MFI wide survey maps on the positions of these four sources (Tau A, Cas A, Cyg A and 3C274) at 11\,GHz and 19\,GHz, smoothed to a common angular resolution of $1^\circ$. We note that Tau A and Cas A are the two main calibrators of QUIJOTE MFI, and thus we have much more sensitive data on these two sources obtained in raster mode. However we focus here on the wide survey maps only, in order to provide another consistency check for the calibration scheme. 
    
    We extracted total-intensity flux densities on these maps using a beam-fitting photometry (BF1d), consisting in fitting a $1^\circ$-FWHM Gaussian beam superimposed on a flat background. We applied colour corrections following the methodology described in \citet{mfipipeline}, and using for each source a spectral index derived from the model. We compare these flux densities with spectral emission models that we have specifically derived for these sources, and  which will be presented in a separate paper (G\'enova-Santos \& Rubi\~no-Mart\'{\i}n, in preparation). While in that paper we discuss models extracted with different photometry techniques, here we compare with models derived from WMAP and Planck maps convolved to a common resolution of $1^\circ$ and using the same BF1d technique that we applied to QUIJOTE MFI. In particular, the Tau A model was used in Sect.~\ref{sec:recal} to recalibrate the wide survey maps.
    As it will be discussed in depth in G\'enova-Santos \& Rubi\~no-Mart\'{\i}n (in prep.), the uncertainties of these models are of the order of $3$--$5$\,\%, and are driven not by the statistical noise of the individual observations which is well below this value, but by systematic effects and calibration uncertainties of the fitted data, which lead to higher model-fitting residuals than would be expected in the presence of just statistical errors. In the cases of Tau A and Cas A, modelling of their secular decrease also introduces significant uncertainty.
    
    \input{table_sources.tex}

    Final QUIJOTE MFI flux densities, for each horn and frequency, and relative deviation with respect to the fitted intensity models, are quoted in Table~\ref{tab:sources_vs_models}. All values are referred to date 2016.3 (1 April, 2016), which roughly corresponds to the middle of the wide survey observations. It can be seen that in most cases the measured flux densities deviate less than $3$--$5$\,\% with respect to the models, while in the case of Tau A, which is the main amplitude calibrator, the deviations are within $1$\,\% (the difference is not exactly zero due to the way the different periods are calibrated and combined; see section~\ref{sec:recal}). The level of these deviations is expected given the typical model uncertainties, and therefore these results give full confidence to our global calibration strategy and the quoted uncertainty (see Table~\ref{tab:summarycal}).
    A detailed discussion on the variability of these four sources (and others in the wide survey maps) can be found in \cite{sourceswidesurvey}.

    \subsection{Planets in intensity}
    
    Venus and Jupiter are also detected to high significance in the QUIJOTE MFI wide survey data. Owing to its orbital motion Venus declination varies roughly between $\pm 27^\circ$. Given Tenerife's latitude ($28.3^\circ$ N), when its declination is close to $27^\circ$ it is always visible in any of the elevations considered in the wide-survey. On the contrary, when it reaches its minimum declination of $-27^\circ$ it culminates at elevation $35.5^\circ$, and therefore it is only picked up in observations at elevations 30 or $35^\circ$. The distance between this planet and the Earth changes between $0.27$ and $1.74$\,A.U., meaning that there is a factor $\approx 42$ variation between its minimum and maximum brightness. At 19\,GHz its flux density is expected to vary between $10.9$ and 445\,Jy. Then, during its inferior conjunction it is amongst the brightest sources on the sky at the QUIJOTE MFI frequencies. In the case of Jupiter, being an external planet, this variation is much smaller. Its distance to Earth varies between 4.1 and 6.4\,A.U., producing a variation of its flux density at 19\,GHz between 26.1 and 61.1\,Jy. Between 2012 and 2016 its declination was always positive, reaching $23^\circ$, meaning that it was picked up in most of the wide survey data. Between 2016 and 2018 its declination dropped below zero, reaching $-22^\circ$, and therefore during this period it was only visible on the wide survey observations performed at low elevations. 
    
    \begin{figure}
    \centering
    \includegraphics[width=8cm]{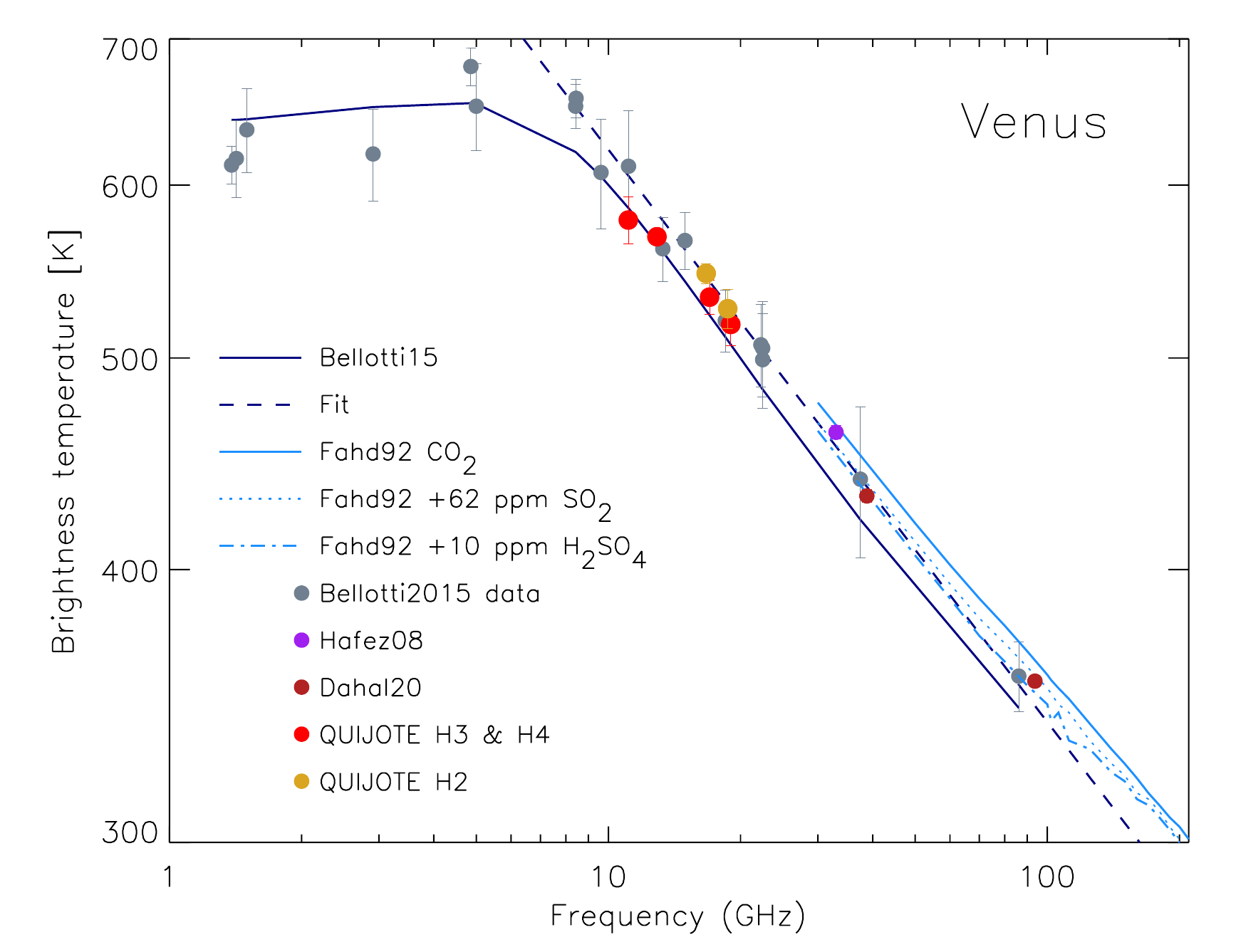}
    \includegraphics[width=8cm]{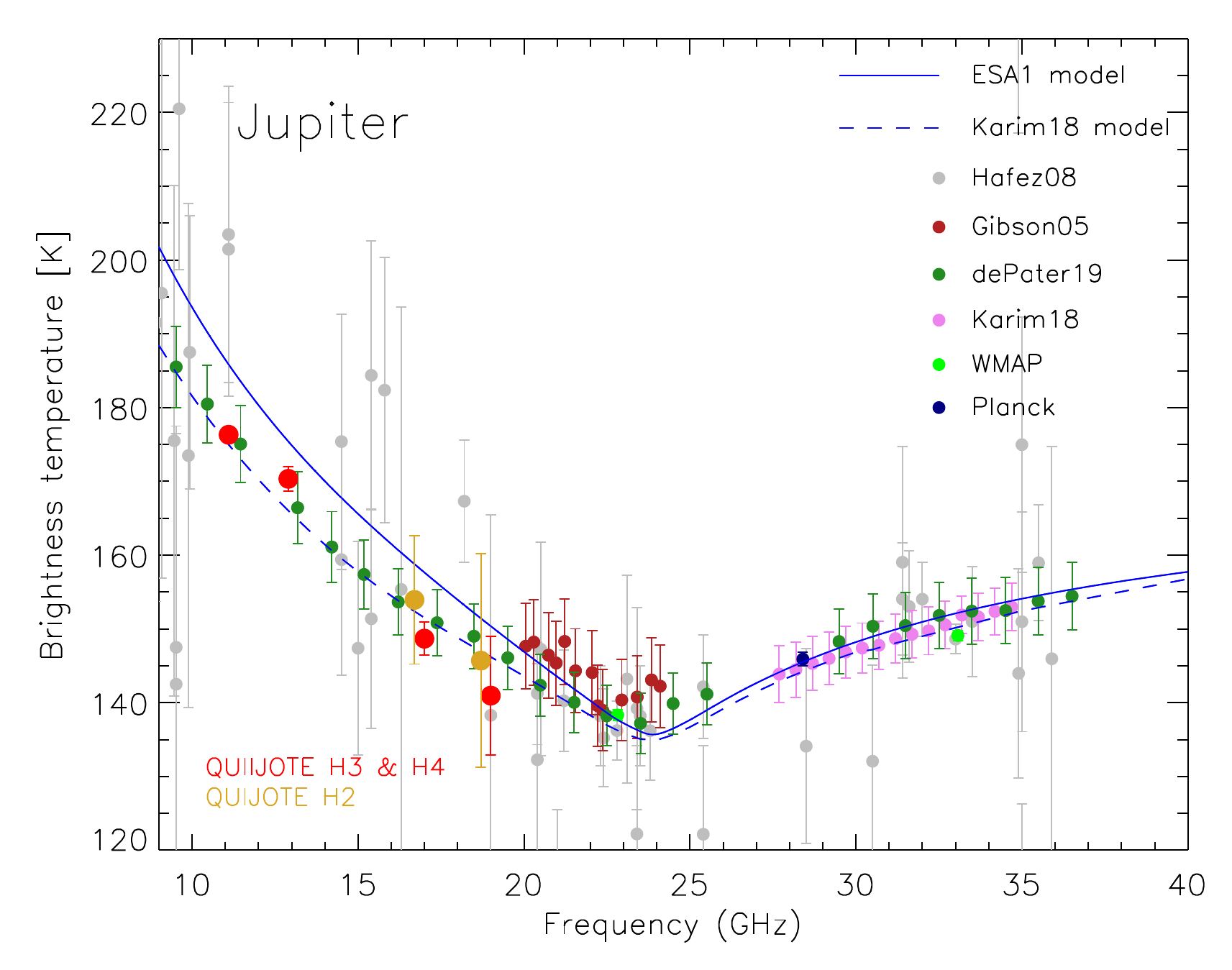}
    \caption{ Venus (top) and Jupiter (bottom) brightness temperatures derived from the QUIJOTE MFI wide survey data (red and yellow) in comparison with ancillary data, and with various models. Venus data have been obtained from \citet{Bellotti2015,Hafez2008,Dahal2021}, while the plotted models are from \citet{Bellotti2015, Fahd1992}. We also show a power-law fit to the observed data in the range 7--100\,GHz that we use to compare with the QUIJOTE MFI measurements. Jupiter data come from \citet{Hafez2008, Gibson2005, dePater2019, Karim2018, Weiland2011, Planck2015-v, PlanckInt2017-lii}. We plot the model by \citet{Karim2018} and the ESA1 model. }
    \label{fig:planets}
    \end{figure}
    
    While further details will be given in a future paper where we will discuss planets and other bright astronomical sources, here we briefly describe the procedure we have developed to estimate planets' brightness temperatures. We implemented a specific map-making in which we rotate the coordinates of QUIJOTE MFI wide survey data to planet-centred coordinates, to produce planet-centred maps. We use the same final calibrated data that were used to produce the final maps that are presented in this paper. In order to account for the $1/d^2$ effect we define distance bins (3 bins for Jupiter and 6 for Venus), and produce individual maps for each bin. We have verified in the final map that the (symmetrized) beam shape is well preserved, this being a health check both for the tailored map-making that we use here as well as for the pointing model. On these maps we apply a beam-fitting photometry to derive flux densities for each distance bin and for each horn/frequency. These flux densities are then colour-corrected using a Rayleigh-Jeans spectrum (spectral index $\alpha=2$). In addition to data maps for each redshift bin we produce maps of $1/d^2$ using the same noise weights and flags that are applied to the data. These maps are later used to calculate an effective distance at the position of the planet. Using this information we fit the flux densities measured in each bin to a $1/d^2$ law in order to derive the final brightness temperatures.
    
    Our Venus and Jupiter brightness temperatures derived from QUIJOTE MFI are listed in Table~\ref{tab:planets_vs_models} and plotted in Figure~\ref{fig:planets}, in comparison with other data at similar frequencies, as well as with various models giving the spectral dependency of the brightness temperatures of these planets. In both cases we have corrected for the planet absorption of the CMB monopole, and therefore the quoted values represent the intrinsic brightness temperature of the planets. In the case of Venus, it is seen that the ancillary measurements seem a bit high with respect to the \citet{Bellotti2015} and \citet{Fahd1992} models and therefore we performed a power-law fit to the data in the range 7--100\,GHz (dashed line in the figure) and use this fit as a reference to compare with the QUIJOTE MFI values. In the case of Jupiter we use as reference the model of \citet{Karim2018}, which seems to trace better the ancillary data, and in particular the VLA data from \cite{dePater2019} below the ammonia absorption at 23\,GHz. As can be seen in Table~\ref{tab:planets_vs_models}, both for Venus and Jupiter the QUIJOTE MFI measurements deviate always less than 5\,\% from the models (note that in some cases the statistical error bar is larger than this value), which bestows confidence to our calibration strategy.

    \begin{table*}
        \caption{Brightness temperatures (in Kelvin) of Jupiter and Venus extracted from the QUIJOTE MFI wide survey data. Inside parentheses we quote the percent deviation with respect to predictions from spectral models.}
        \label{tab:planets_vs_models}
        \centering
        \begin{tabular}{c|c|c|c|c|c|c}
        \hline
             Planet  &   311 ($11.1$\,GHz) & 313 ($12.9$\,GHz) & 217 ($16.7$\,GHz) & 417 ($17.0$\,GHz) & 219 ($18.7$\,GHz) & 419 ($19.0$\,GHz)  \\
             \hline
    Jupiter          &  $ 176.3\pm 0.4$ (+0.8)  &	$170.3\pm 1.7$ (+2.6)  & $153.9\pm 8.7$ (+1.0)    &  $148.7\pm 2.2 $ (-1.9)  &  $145.7\pm 14.5$ (-0.9)  &   $141.0\pm 8.0 $ (-3.6)   \\
    Venus	         &  $578.3\pm 14.4$ (-4.6)  &	$568.2\pm 4.7$ (-2.5)  & $546.6\pm 5.7$ (+0.4)    &  $533.1\pm 9.5 $ (-1.6)  &  $526.7\pm 10.9$ (-0.4)  &   $518.0\pm 11.5$ (-1.6)  \\
        \hline
        \hline
        \end{tabular}
    \end{table*}

    \subsection{Polarized sources}
    
    Figure~\ref{fig:sources_minimaps} also shows wide-survey polarization maps of Tau A, Cas A, Cyg A and 3C274, projected in Galactic coordinates and convolved to an angular resolution of one degree. 
    Clear polarized emission is seen in Tau A, mainly concentrated in the $Q$ map, as expected due to its polarization angle \citep[see e.g.][]{Weiland2011}. The $U$ map shows the typical cloverleaf pattern (with the expected peak-to-peak amplitude of $\sim 1$\,\% with respect to the total intensity) arising from the differences between the two co-polar beams \citep{mfipipeline}. This pattern is also visible in the $Q$ and $U$ maps of Cas A, more notably at 11\,GHz. Due to it being a very young shell-type supernova remnant (SNR), the magnetic field of Cas A is expected to be radial \citep{Anderson1995}. Being $\sim 5'$ across, this source is unresolved by the QUIJOTE MFI beam and therefore we expect zero integrated polarization. Clear polarized emission is also seen in Cyg A. A rotation of the polarization angle is apparent between 11 and 19\,GHz, which is due to the two jets of this radio galaxy having different rotation measures, the so-called Laing-Garrington effect \citep{Laing1988}. In the case of 3C274, we only have a marginal polarization detection in the $U$ maps. This is expected given our noise levels (between $0.5$--$1$\,Jy), and the fact that the measured polarization fraction at 23\,GHz is approximately 4\,\% \citep{Weiland2011}. 
    
    In order to minimize systematic effects introduced by differences between the two co-polar beams, we extract flux densities in polarization through an aperture photometry technique on maps smoothed to one-degree angular resolution (AP1d). The circular aperture radius ($r_1$) is taken to be $r_1=1.5^\circ$ for Tau A and 3C274, and $r_1=1.3^\circ$ for Cas A and Cyg A, due to the larger foreground contamination in the surroundings of the latter two sources. In the case of Cas A, we also mask the region centred at Galactic coordinates $(l,b)=(111.11^\circ, -0.53^\circ)$, using an exclusion radius of $0.7^\circ$. The background emission in all four cases is corrected using the mean of the signal in the annulus between $r_1$ and $r_2=\sqrt{2}r_1$. 
    Table~\ref{tab:sources_vs_models} shows the Stokes $Q$ and $U$ flux densities measured on Tau A, Cas A, Cyg A and 3C274. 
    
    We now discuss the first three cases in detail, as well as the bright polarized emission in W63. 
    For this discussion, we also apply the same methodology (i.e. AP1d for polarization and BF1d for intensity) to derive the photometry values for these sources using WMAP 9-year data \citep{Bennett2013} and Planck 2018 maps \citep{Planck2018-i} at the common one-degree resolution. Specially for the cases of Tau A and Cas A, and for Planck LFI, we correct for the intensity-to-polarization leakage due to bandpass mismatch following the methodology described in Appendix C of \cite{Planck2016-xxvi}, using the maps of projection factors described in \cite{Planck2015-ii}, and the spectral index of each source derived from the intensity SED. 
    
    \subsubsection{Tau A and Cyg A}
\begin{figure}
    \centering
    \includegraphics[width=0.95\columnwidth]{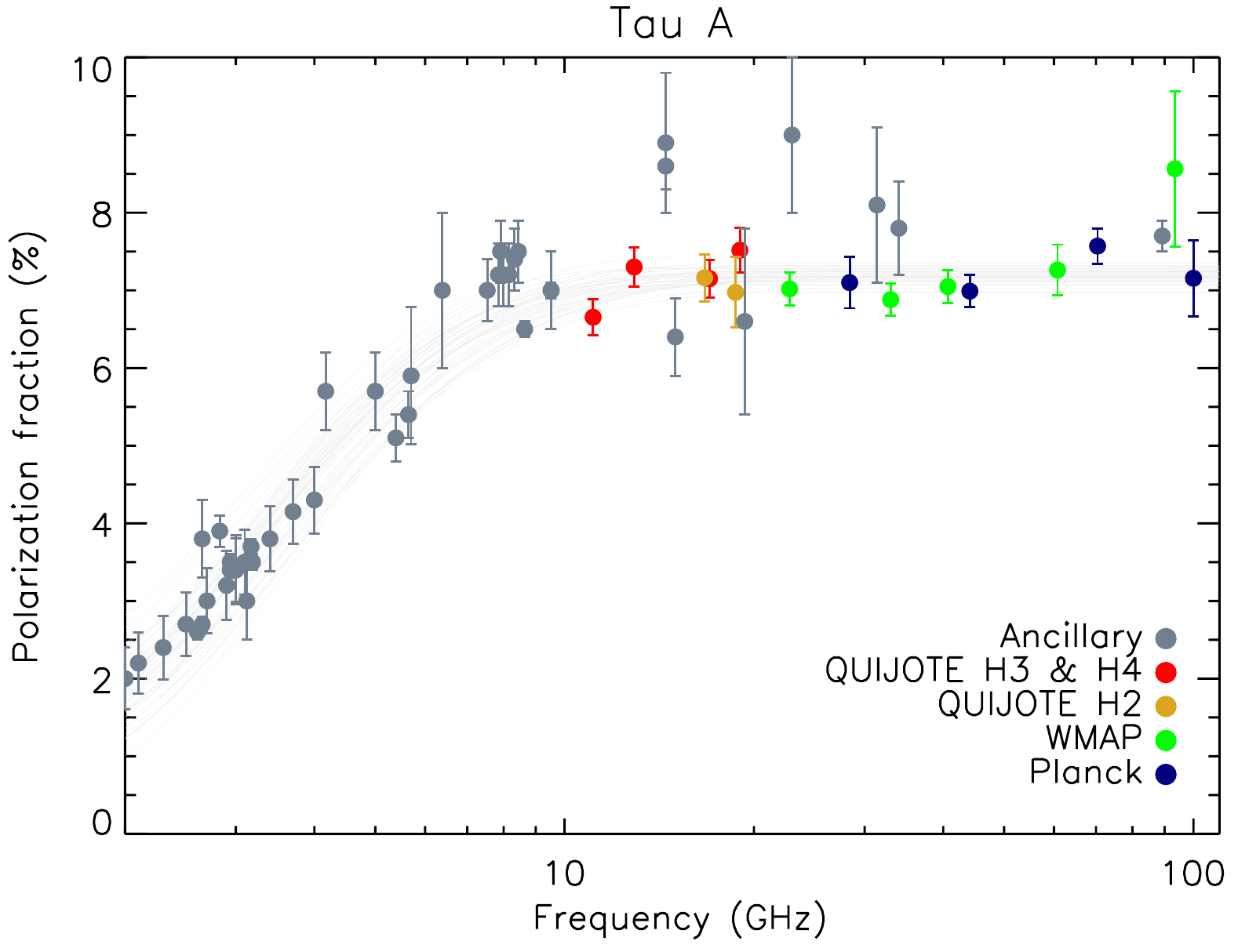}
    \includegraphics[width=0.95\columnwidth]{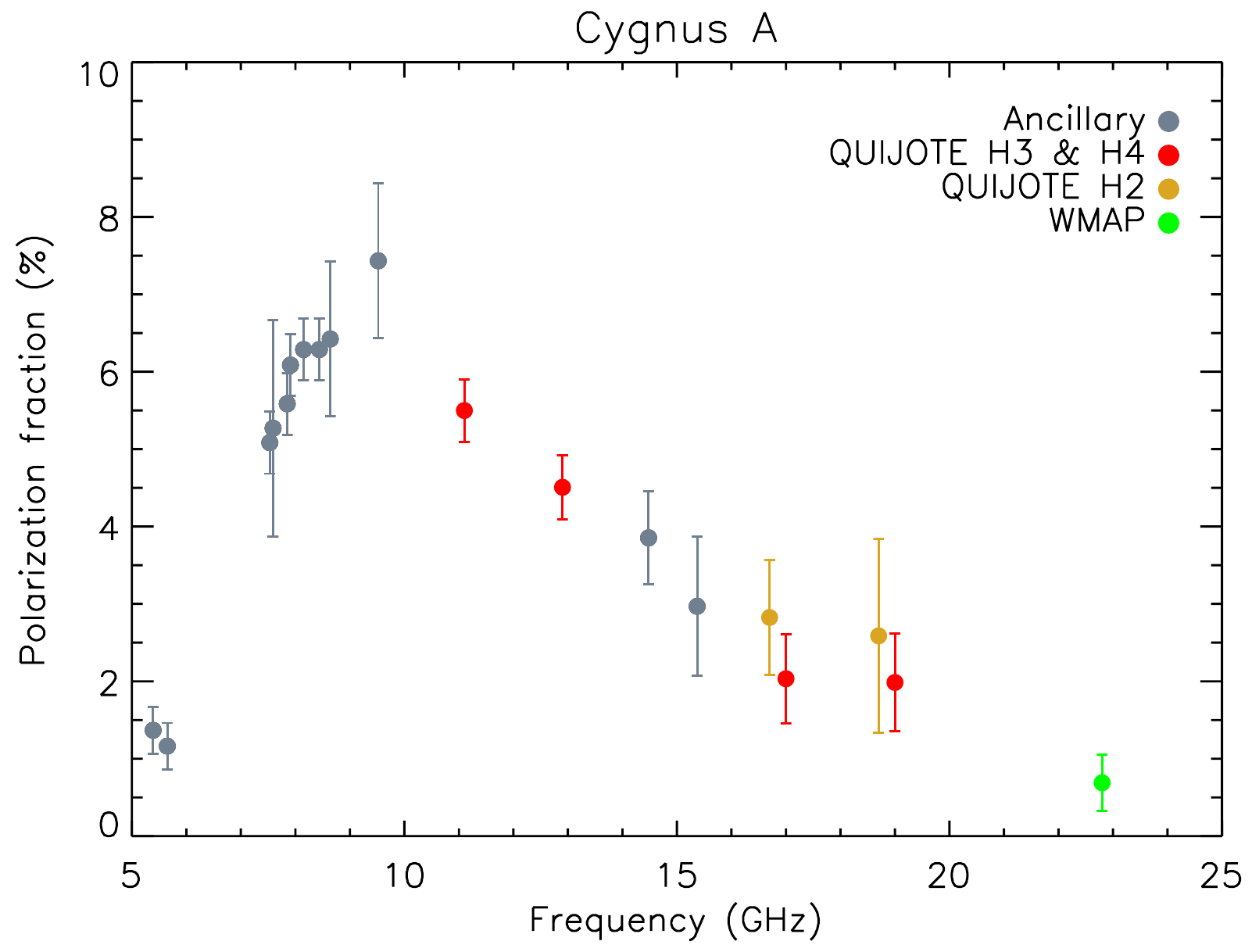}
    \caption{Consistency checks on polarized sources detected on the QUIJOTE MFI the wide survey. We show polarization fractions measured in Tau A and in Cyg A, in comparison with our WMAP and Planck results obtained using the same methodology, and with ancillary measurements (see the complete list of references in the main text). In the case of Tau A we overplot in grey models for the polarization fraction that account for Faraday rotation. The Cyg A data show the Laing-Garrington effect arising from different rotation measures in the two lobes of this galaxy.}
    \label{fig:sources_pol}
    \end{figure}

    Figure~\ref{fig:sources_pol} shows our results for the polarization fractions in Tau A and Cyg A at one degree resolution. 
    We include also our WMAP (for both sources) and Planck (only for Tau A) measurements, as well as ancillary measurements both for Tau A \citep{Kuzmin1959, MayerSlo1959, 1962AJ.....67Q.581M, Davies1963, 1964ApJ...140..656H, MayerMcSlo1964, MorrisBerge1964, 1966ApJ...144..437B, Gardner1966, 1967ApJ...147..908H, 1967AJ.....72..230S, Satoh1967, Hobbs1968, HollingerHobbs1968, MayerHollinger1968, 1968ApJ...154..817S, 1969ApJ...158..145J, Dmitrenko1970, Wright1970, 1975A&A....44..187G, Hafez2008, Aumont2010} and for Cyg A \citep{1962AJ.....67Q.581M, 1964ApJ...140..656H, 1966SvA....10..214S, 1966ApJ...144..437B, 1966AJ.....71R.864M,1967ApJ...149..707H}. 
    
    For Tau A, we show in solid grey lines Monte Carlo realizations of a simple model for the spectral dependency of its polarization fraction that accounts for the Faraday depolarization \citep{Burn1966}, and that is consistent with both the ancillary measurements at low frequencies ($\nu \lesssim 10$\,GHz) and existing measurements of the Faraday dispersion in the region \citep[e.g.][]{1991ApJ...368..231B}. This model will be described in detail in a separate paper (G\'enova-Santos \& Rubi\~no-Mart\'{\i}n, in preparation).
    We note that in our recalibration strategy of the MFI wide survey maps, we use Tau A for fixing the intensity calibration scale and the polarization angle, while the polarization amplitude is essentially given by independent polarization efficiency measurements. Thus, this analysis provides a consistency test on the MFI polarization calibration. 
    
    Cyg A data points clearly show the effect of the Faraday depolarization produced by the Laing-Garrington effect. It is evident from this plot that the maximum alignment between the polarization directions of the two jets occurs at $\approx 10$\,GHz, and then the measured polarization fractions decrease in both sides of the spectrum. Modelling this effect is complicated and beyond the scope of this paper. The QUIJOTE MFI measurements are in good agreement with the other measurements, again providing confidence in our calibration strategy.
    
    \subsubsection{Cas A} 
\begin{figure}
    \centering
    \includegraphics[width=0.95\columnwidth]{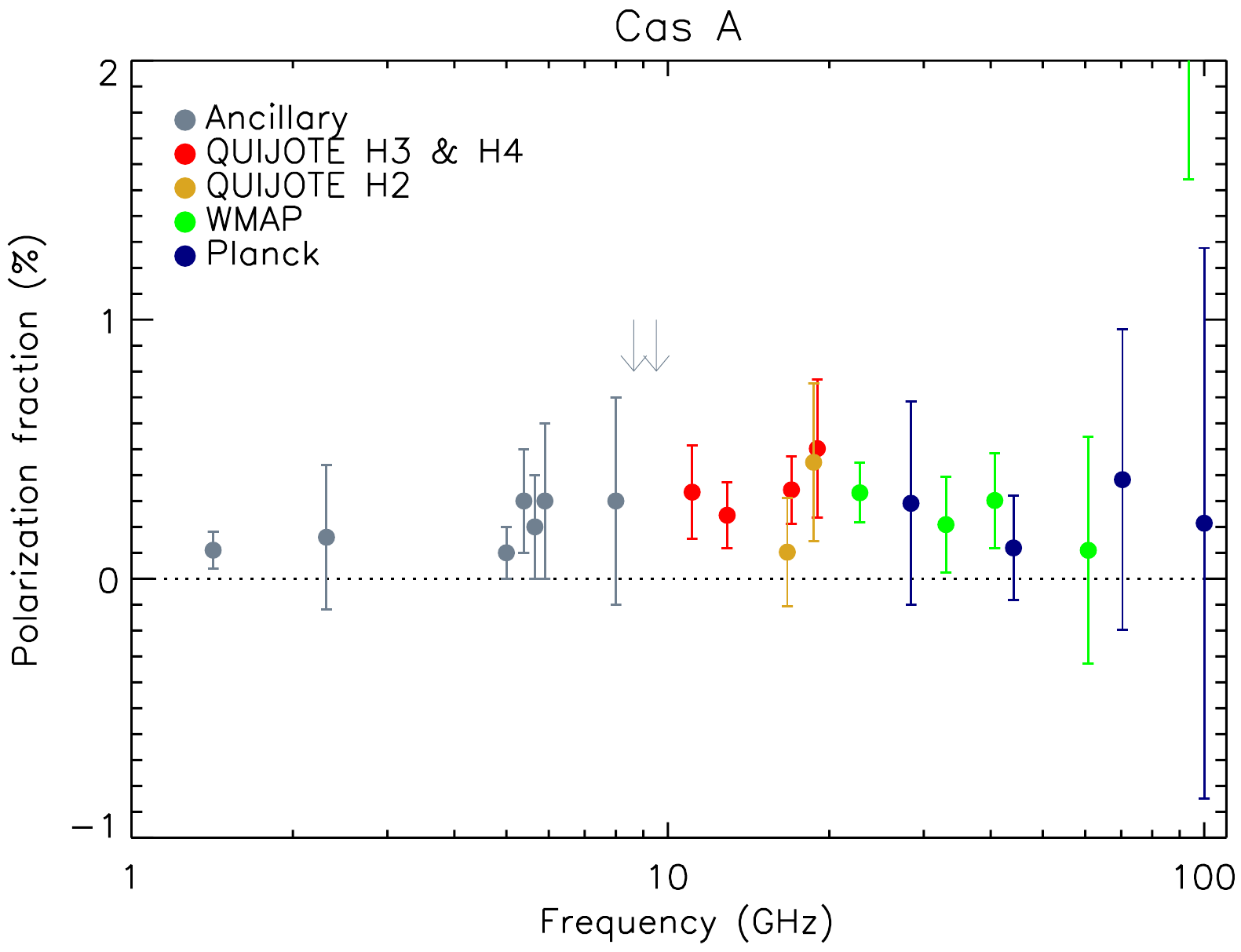}
    \caption{Polarization fractions measured on Cas A in QUIJOTE MFI wide survey data, in comparison with other measurements. WMAP and Planck results are obtained using the same methodology as for the MFI maps values. The complete list of ancillary measurements is given in the main text. Upper limits are represented with arrows. At degree-beam scales, the polarized emission of Cas A is expected to be zero, so these measurements serve as a consistency check for the overall intensity-to-polarization leakage of the MFI wide survey maps. }
    \label{fig:cass}
    \end{figure}

    Figure~\ref{fig:cass} shows the polarization fraction measured in Cas A with QUIJOTE MFI. We also include our photometry results for WMAP and Planck, and ancillary data from the literature \citep{1962AJ.....67Q.581M, 1964ApJ...140..656H, 1967ApJ...147..908H, 1967AJ.....72..230S, 1968ApJ...154..817S, 2014ARep...58..626V}. 
    All values are noise-debiased using the PMAS estimator \citep{Plaszczynski2014}. The intensity-to-polarization leakage due to the co-polar beam asymmetry is almost cancelled in the integrated flux densities thanks to the positive and negative structure of this pattern, leading to integrated polarization fractions in QUIJOTE of around $\sim 0.3\,\%$. Note that similar levels are detected in WMAP, and could also be due to beam effects as discussed in \cite{Weiland2011}. At face value, these numbers can be considered as a conservative upper limit on the overall intensity-to-polarization leakage in the MFI wide survey maps. 
    
    \subsubsection{W63 region}
\begin{figure}
    \centering
    \includegraphics[width=0.95\columnwidth]{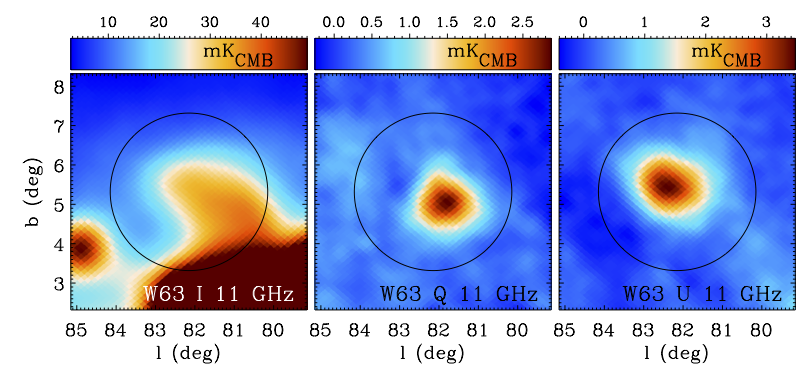}
    \includegraphics[width=0.95\columnwidth]{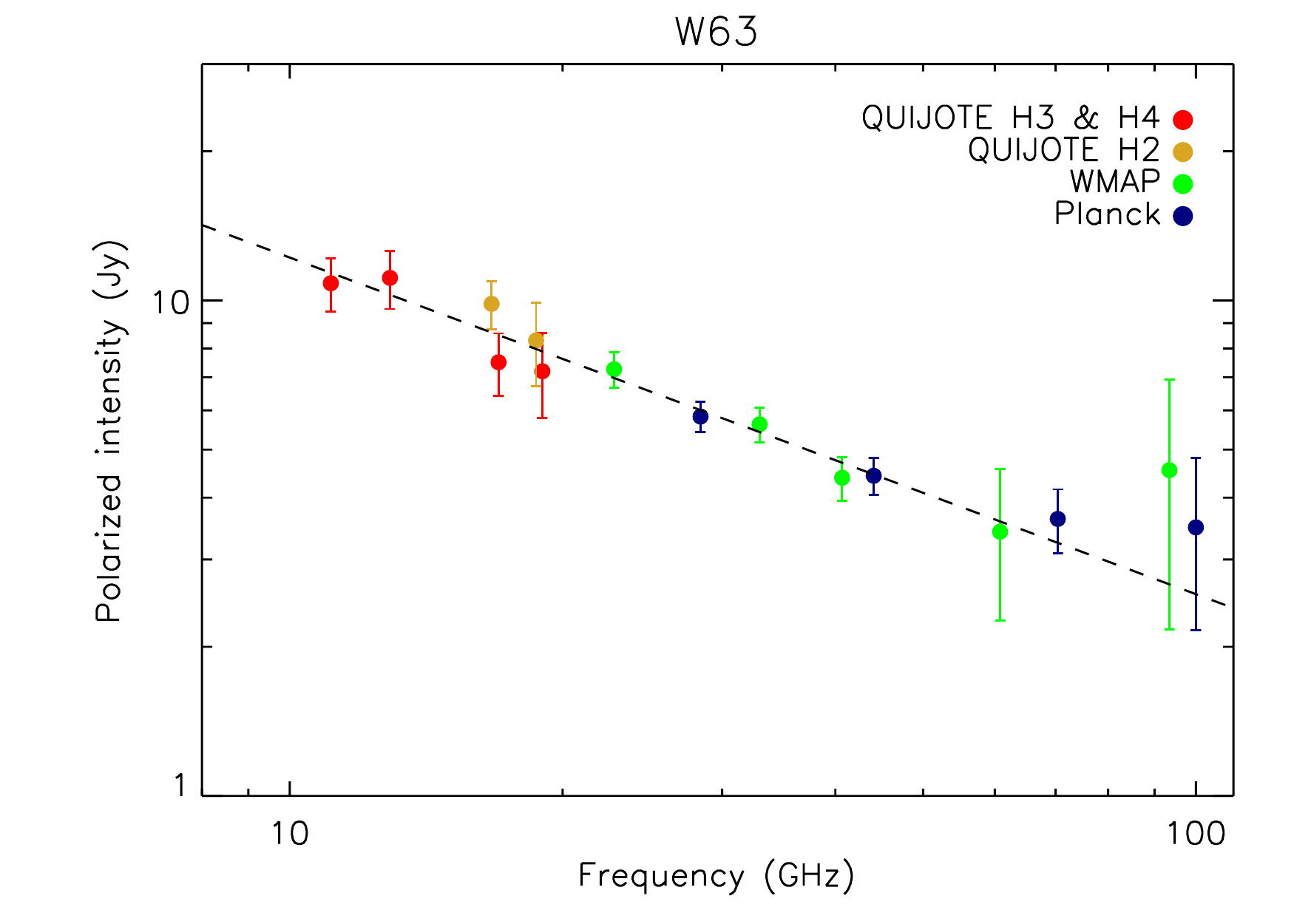}
    \caption{Top: Minimaps of $6^\circ \times 6^\circ$ size around W63 for the MFI 11\,GHz I, Q, and U maps at one degree angular resolution. A circle with radius of $2^\circ$ indicates the integration area for the photometry analysis. Bottom: Polarized intensity measurements on W63 with the QUIJOTE MFI wide survey data, in comparison with WMAP and Planck measurements. We overplot with a dashed line a power-law fit representing the spectrum of the synchrotron emission fitted to the WMAP and Planck data. }
    \label{fig:w63}
    \end{figure}

    As an additional test of the polarization calibration of the MFI, we also investigate the polarized intensities of W63, another SNR which appears as a very bright extended structure in the polarization maps at these frequencies. The top panel in Fig.~\ref{fig:w63} shows the MFI 11\,GHz Stokes I, Q and U maps for this object. 
    The total-intensity emission of W63 is practically embedded inside the emission of the Cygnus X star-forming complex, so it is difficult to extract reliable total-intensity flux density estimates in this case. However, the polarization signal is reasonably isolated. Thus, we only discuss its polarized flux density here.  
    In order to capture all the flux in the region, we use an aperture radius of $r_1=2^\circ$. As in the previous cases, we carry out this analysis in the smoothed maps at one degree resolution (i.e. AP1d photometry). The bottom panel in Fig.~\ref{fig:w63} shows the SED in polarized intensity $P=\sqrt{Q^2 + U^2}$ derived from our photometry measurements, including also our results for WMAP and Planck applying the same methodology. All values are noise-debiased using the PMAS estimator. Error bars account for the photometry error plus the corresponding calibration uncertainties added in quadrature. 
    For MFI, we use the values reported in Table~\ref{tab:summarycal}, while for WMAP and Planck data we adopt the conservative value of 3\,\%, as done for similar analyses \citep[see e.g.][]{planck2015galacticcloudsAME,Taurus, LambdaOrionis}.
    The WMAP and Planck polarized intensity flux in W63 can be fitted to a power-law $P=6.97 (\nu/22.8{\rm GHz})^{-0.68}$\,Jy, that is depicted by the dashed line. 
    The QUIJOTE MFI data are consistent within 1-sigma with the fitted model, which gives additional confidence to our calibration strategy in polarization.

    \section{Data release and description of the data products}
    \label{sec:release}
    
    Together with this paper, there is a series of further publications containing scientific results derived from the QUIJOTE-MFI wide survey maps presented here. The titles of all the papers in the series begin with "QUIJOTE scientific results", and comprise:
    \begin{enumerate}
    \item[IV.] A northern sky survey at 10--20\,GHz with the Multi-Frequency Instrument (this paper).
    \item[V.] The microwave intensity and polarization spectra of the Galactic regions W49, W51 and IC443 \citep{W51}.
    \item[VI.] The Haze as seen by QUIJOTE \citep{hazewidesurvey}.
    \item[VII.] Galactic AME sources in the QUIJOTE-MFI North Hemisphere Wide-Survey \citep{AMEwidesurvey}.
    \item[VIII.] Diffuse polarized foregrounds from component separation with QUIJOTE-MFI \citep{MFIcompsep_pol}.
    \item[IX.] Radio sources in the QUIJOTE-MFI wide survey maps \citep{sourceswidesurvey}.
    \item[X.] AME variability along the Galactic Plane in the QUIJOTE-MFI wide survey \citep[][in prep.]{ameplanewidesurvey}.
    \item[XI.] Polarized synchrotron loops and spurs in the QUIJOTE-MFI wide survey \citep[][in prep.]{loopswidesurvey}.
    \item[XII.] Analysis of the polarized synchrotron emission at the power spectrum level in the MFI wide survey \citep[][in prep.]{Synchwidesurvey}.
    \item[XIII.] SNRs in the QUIJOTE-MFI wide survey \citep[][in prep.]{snrwidesurvey}.
    \item[XIV.] The FAN region as seen by QUIJOTE-MFI \citep[][in prep.]{FANwidesurvey}.
    \item[XV.] The North Polar Spur as seen by QUIJOTE-MFI \citep[][in prep.]{NPSwidesurvey}.
    \item[XVI.]  Diffuse intensity foregrounds from component separation with QUIJOTE-MFI \citep[][in prep.]{MFIcompsep_int}.
    \end{enumerate}
    In addition, we have a dedicated paper describing the MFI data processing pipeline \citep{mfipipeline}. 
    The distribution of released data products associated with the QUIJOTE-MFI wide survey papers contain the following items:
    \begin{itemize}
        \item Four frequency maps (11, 13, 17, 19\,GHz) in intensity and polarization, both at native and one degree resolution. Maps at 11 and 13\,GHz correspond to those produced from MFI horn 3. Maps at 17 and 19\,GHz correspond to the weighted average of horns 2 and 4, as described in Sec.~\ref{sec:maps}. 
        \item The associated weight and hit maps for each frequency map at native resolution. 
        \item One set of null tests maps (half1/2 for independent baselines). 
        \item Instrument Model (IMO), containing  central frequencies, beams properties, beam profiles and window functions for each MFI horn, bandpasses and colour corrections. 
        \item The default analysis mask (sat+NCP+lowdec), as well as the satellite mask (sat). The later is applied to all the released maps.
    \end{itemize}
    %

    \section{Conclusions}
    \label{sec:conclusions}
    
    This paper presents and characterizes the properties of the QUIJOTE wide survey maps of the northern sky carried out with the MFI instrument. They result from approximately $9\,000$\,h of observations spread over six years between 2013 and 2018, and include four frequency maps at  $11.1$, $12.9$, $16.8$ and $18.8$\,GHz, with angular resolutions between 55 and 39 arcmin. The maps cover around $29\,000$\,deg$^2$ with sensitivities in linear polarization (Stokes Q and U parameters) within 35--40\,$\mu$K per 1-degree beam. Although the MFI instrument is not optimized for intensity measurements, we also present the corresponding intensity maps at those four frequencies, with sensitivities in the range 65--200\,$\mu$K per 1-degree beam. 

    Together with the description of the specific aspects of the MFI pipeline related to the production of the wide survey maps, we have presented a detailed validation of the maps, a characterization of residual systematic effects (Sect.~\ref{sec:validation}), and an extensive study of their calibration accuracy (Sect.~\ref{sec:cal} and Table~\ref{tab:summarycal}). The overall calibration uncertainty of the polarization maps is 5\,\% for the two lowest frequency channels, and 6\,\% for the highest ones. 
    These final maps and other derived data products are part of a public data release associated with this paper. 
    
    Although a full description of the science results obtained from these maps are given in the accompanying papers listed in Sect.~\ref{sec:release}, this paper presents some global properties of the Galactic foregrounds at these frequencies, and in particular, the polarized synchrotron emission.
    The average synchrotron spectral index in polarization between 11\,GHz and the WMAP 23\,GHz is found to be $\beta=-3.07 \pm 0.16$, showing a much broader distribution (by a factor $\sim 2.7$) than the one adopted in current synchrotron sky models \citep[e.g.][]{MivilleDeschenes2008}. 
    Most of the large-scale polarized synchrotron features in the MFI maps appear in the E-mode map, which shows significantly more power than the B-mode at these frequencies. Based on the analysis of the angular power spectra of the measured polarized signal, we find that the BB/EE ratio at multipole scales of $\ell=80$ is $0.26\pm0.07$ for a Galactic cut $|b|>10^\circ$. This value is consistent with that found for WMAP/Planck low frequency maps \citep{Martire2021}, but it is significantly different from the values obtained for the S-PASS polarized signal at 2.3\,GHz \citep{Krachmalnicoff2018}, suggesting that probably there is some contribution of Faraday rotation and/or depolarization at lower frequencies than those probed by QUIJOTE MFI.
    We also find a positive correlation in the TE spectrum for 11\,GHz at large angular scales ($\ell \lesssim 80$), while the EB and TB signals are consistent with zero in the multipole range $30 \lesssim \ell \lesssim 150$, as expected for the synchrotron emission, as its polarization orientation is dictated by the Galactic magnetic field lines. 
    
    The MFI instrument was decommissioned in 2018. At this moment, QT2 is operating with a combination of the TGI and FGI instruments in a single cryostat. In addition to QUIJOTE, there are two other CMB polarization experiments at the Teide Observatory and providing a similar sky coverage: GroundBird and LSPE-STRIP. GroundBird \citep{Groundbird} is a MKIDs array with two bands centered at 145 and 220\,GHz, installed back in 2019. STRIP \citep{LSPE} is part of LSPE, a combined programme of ground-based and balloon-borne polarization observations. STRIP will operate in the 42 and 90\,GHz bands, and will be installed at the Teide Observatory in 2023. The QUIJOTE collaboration is developing a new instrument at these frequencies, called MFI2, with an expected sensitivity three times better than the former MFI \citep{MFI2}. The new MFI2 is now in the final integration phase, and it is using a digital back-end based on Field Programmable Gate Arrays (FPGAs), that will allow us to identify and filter the RFI signals from geostationary satellites directly in the data processing stage. A new wide survey at these frequencies (10--20\,GHz) will be carried out with MFI2 at the first QUIJOTE telescope (QT1) starting 2023.

    \section*{Data Availability}
    All data products described in Sect.~\ref{sec:release} can be freely downloaded from the QUIJOTE web page\footnote{\url{http://research.iac.es/proyecto/quijote}}, as well as from the RADIOFOREGROUNDS platform\footnote{\url{http://www.radioforegrounds.eu/}}. They include also an Explanatory Supplement describing the data formats. Maps will be submitted also to the Planck Legacy Archive (PLA) interface and the LAMBDA site. 
    Any other derived data products described in this paper (null test maps, simulations, etc) are available upon request to the QUIJOTE collaboration.

    \section*{Acknowledgements}
    
    \input{quijote_acknow}

    This research made use of computing time available on the high-performance computing systems at the IAC.
    We thankfully acknowledge the technical expertise and assistance provided by the Spanish Supercomputing Network 
    (Red Espa\~nola de Supercomputaci\'on), as well as the computer resources used: the Deimos/Diva Supercomputer, located at the IAC.
    This research used resources of the National Energy Research Scientific Computing Center, which is supported by the Office of Science of the U.S. Department of Energy under Contract No. DE-AC02-05CH11231. 
    The PWV data used in the tests presented in Section~4 comes from the Iza\~na Atmospheric Observatory (IZO), and have been made available to us by the Iza\~na Atmospheric Research Center (AEMET). 
    SEH and CD acknowledge support from the STFC Consolidated Grant (ST/P000649/1). FP acknowledges support from the 
    Spanish State Research Agency (AEI) under grant number PID2019-105552RB-C43. DT acknowledges the support from the Chinese Academy of Sciences (CAS) President's International Fellowship Initiative (PIFI) with Grant N. 2020PM0042. 
    Some of the presented results are based on observations obtained with Planck (\url{http://www.esa.int/Planck}), an ESA science mission with instruments and contributions directly funded by ESA Member States, NASA, and Canada. We acknowledge the use of the Legacy Archive for Microwave Background Data Analysis (LAMBDA). Support for LAMBDA is provided by the NASA Office of Space Science.
    Some of the results in this paper have been derived using the \healpix\  \citep{healpix} and {\sc healpy} \citep{healpy} packages. We also use {\tt Numpy} \citep{numpy}, {\tt Matplotlib} \citep{matplotlib} and the {\sc sklearn} module \citep{scikit-learn}.

    
    
    \bibliographystyle{mnras}
    \bibliography{quijote,mfiwidesurvey} 

    
    

    \appendix
    
    
    \section{Data flagging in the MFI wide survey}
    \label{app:dataflagged}
    Tables~\ref{tab:flagged1}, \ref{tab:flagged2},  \ref{tab:flagged5} and  \ref{tab:flagged6} show the percentage of data used (and flagged) for each period, elevation and horn in the MFI wide survey.

\input{table2_flagged_period1.tex}

\input{table2_flagged_period2.tex}

\input{table2_flagged_period5.tex}

\input{table2_flagged_period6.tex}

    \section{Impact of FDEC filtering on polarization maps}
    \label{app:fdec}
    
    In this appendix, we investigate the impact of the FDEC filtering on some of the scientific analyses carried out in this paper and in other papers of the associated release (Sect.~\ref{sec:release}). In particular, we consider here a photometry method (aperture photometry) and correlation method (the so called TT plot).
    
    For this study, we use the sky signal simulations presented in Sect.~\ref{sec:skysims}. Figure~\ref{fig:appfdec} shows the simulated (noiseless) sky maps in polarization at 11\,GHz used as reference. We apply the FDEC filter to these maps, and show in the same figure the resulting filtered maps, as well as the residual maps (i.e. difference between the original and the filtered map). 
    As shown in Sect.~\ref{sec:transfer}, the FDEC filtering effectively corresponds to a high-pass filter, which removes the zero mode for any line of constant declination on a map in local (equatorial) coordinates. The image illustrates again that the effective transfer function of the FDEC filter leaves unaltered all scales with $\ell \ga 30$, because the residual maps only contain large scale features. 
    
    \subsection{Impact of FDEC on photometry methods: aperture photometry}
\begin{figure*}
    \centering
       \includegraphics[width=6cm]{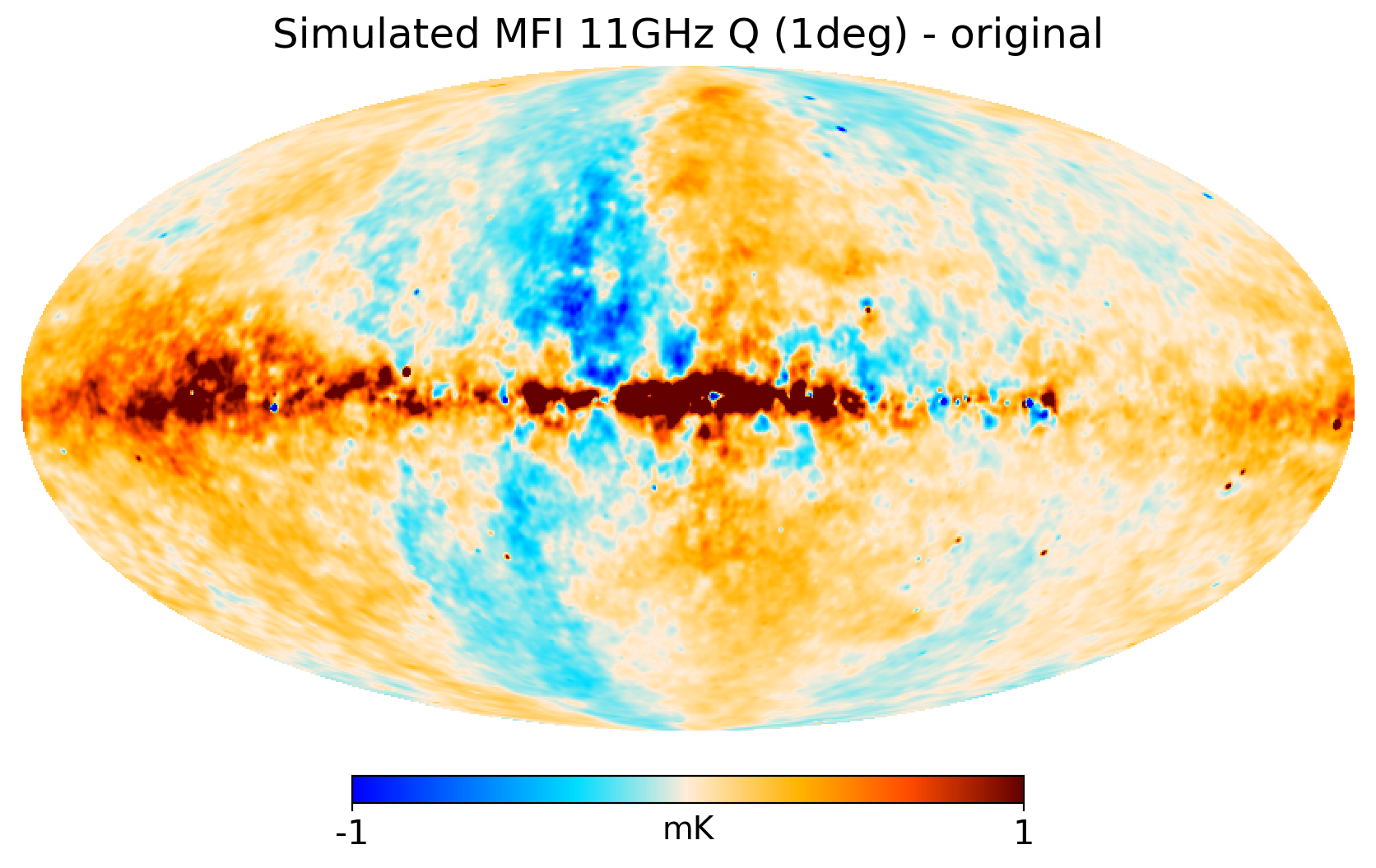}%
       \includegraphics[width=6cm]{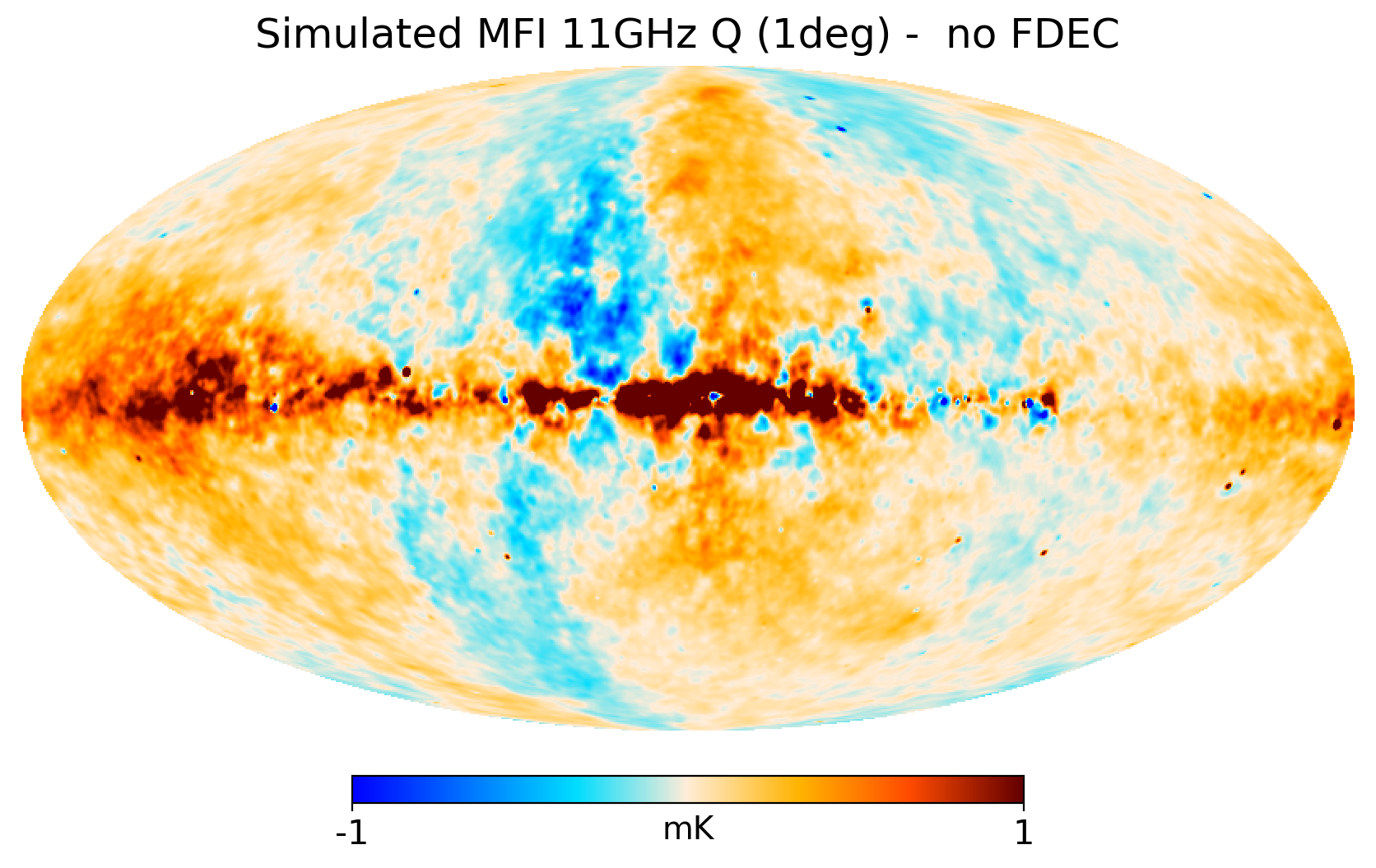}%
       \includegraphics[width=6cm]{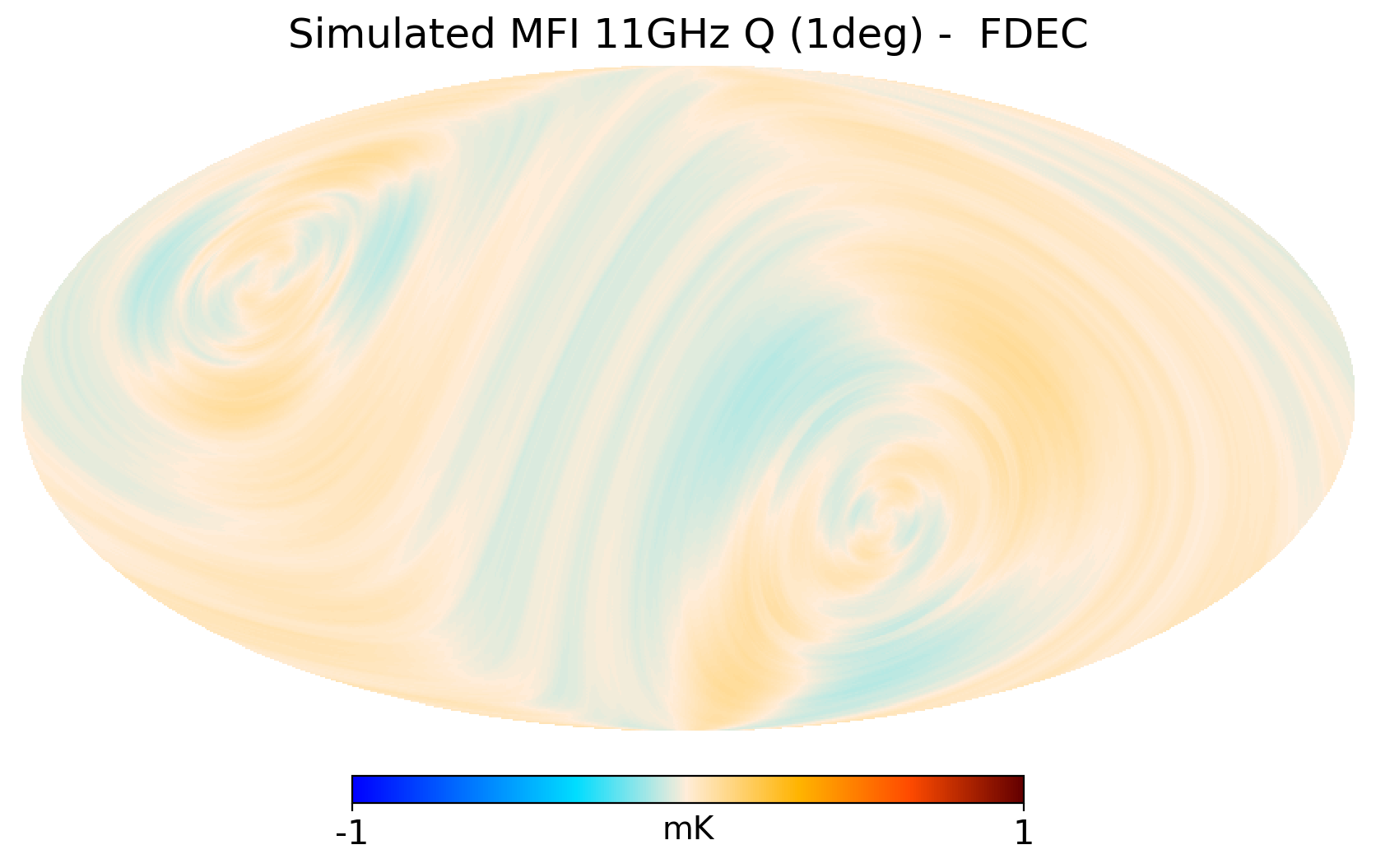}
       \includegraphics[width=6cm]{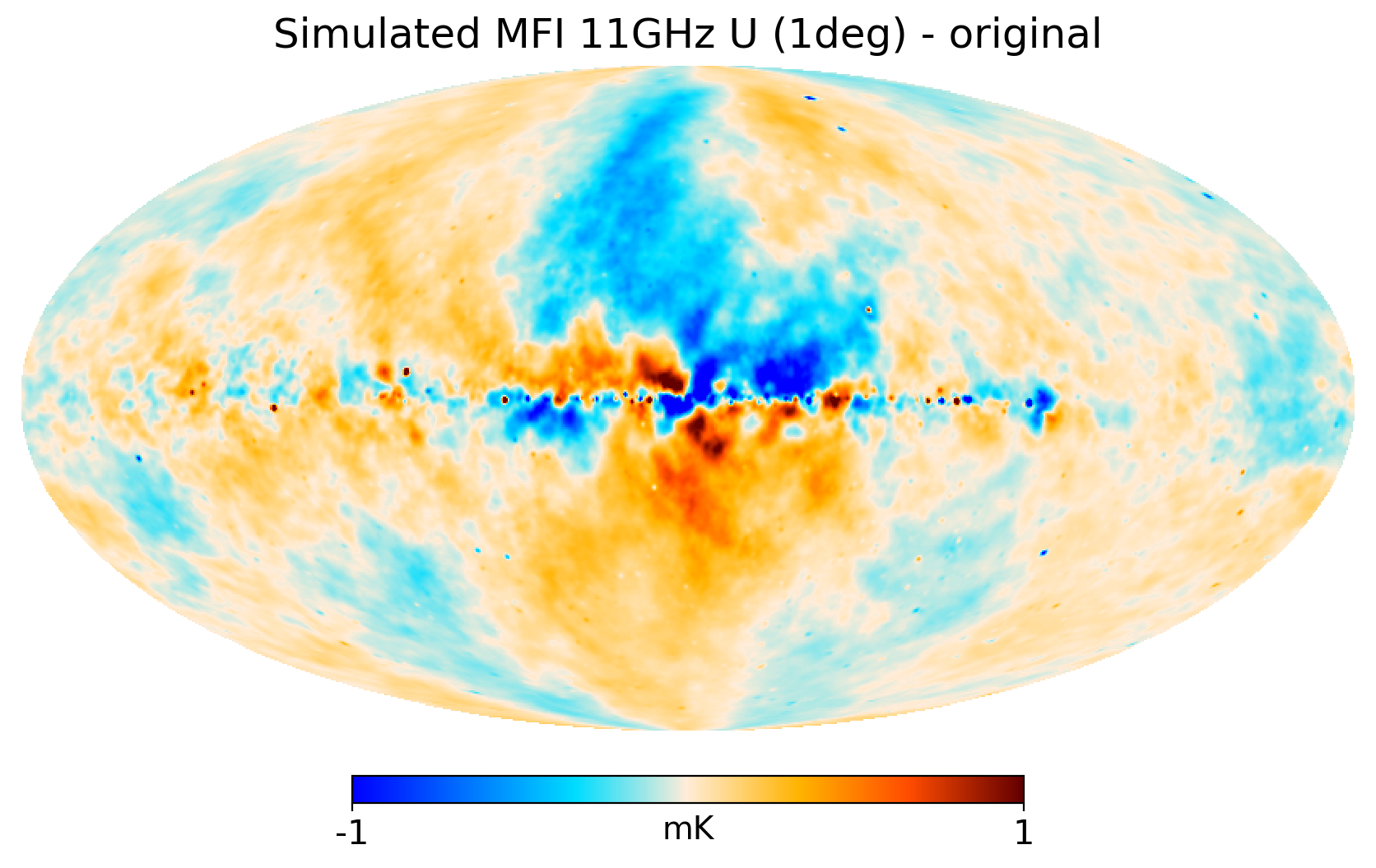}%
       \includegraphics[width=6cm]{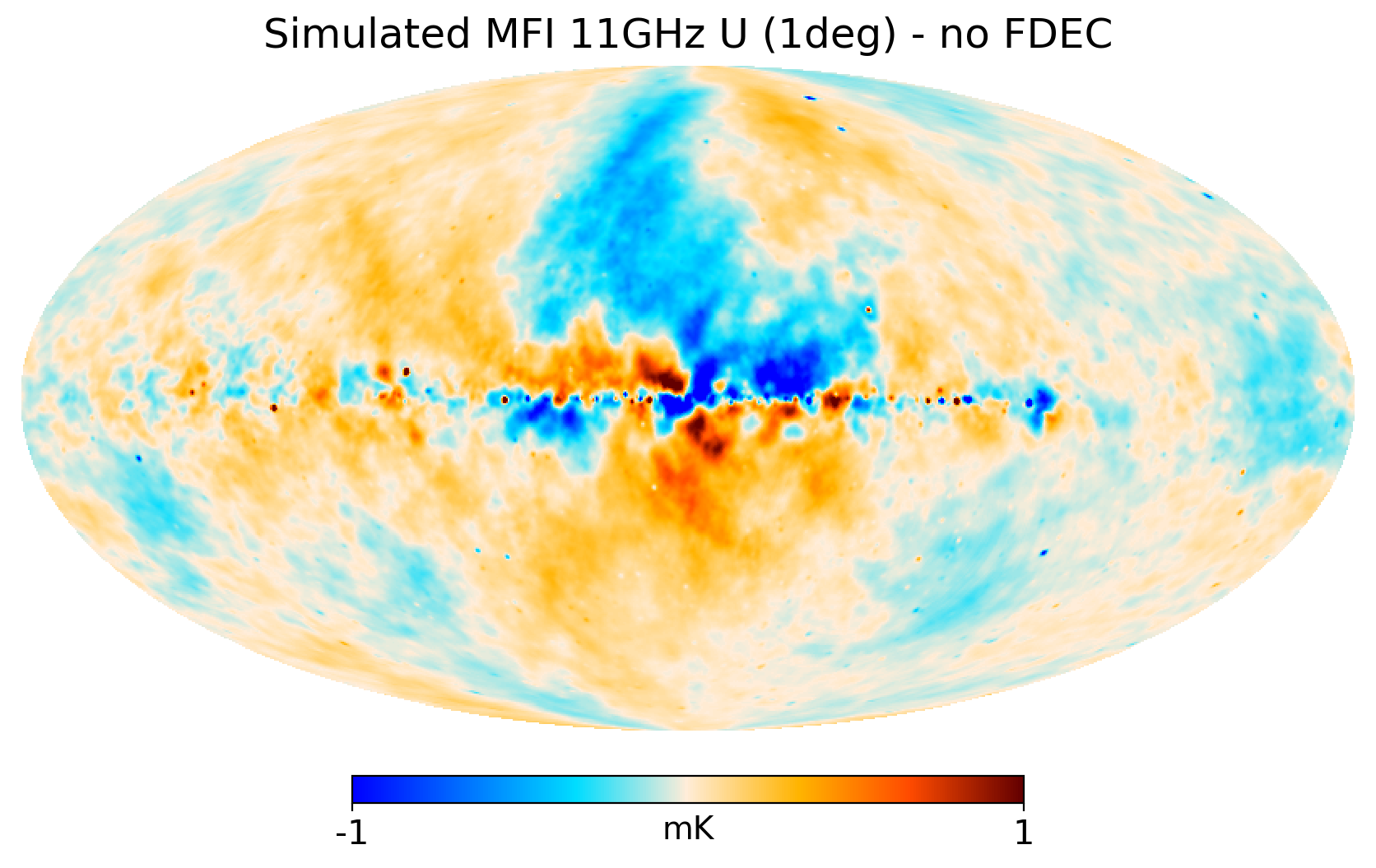}%
       \includegraphics[width=6cm]{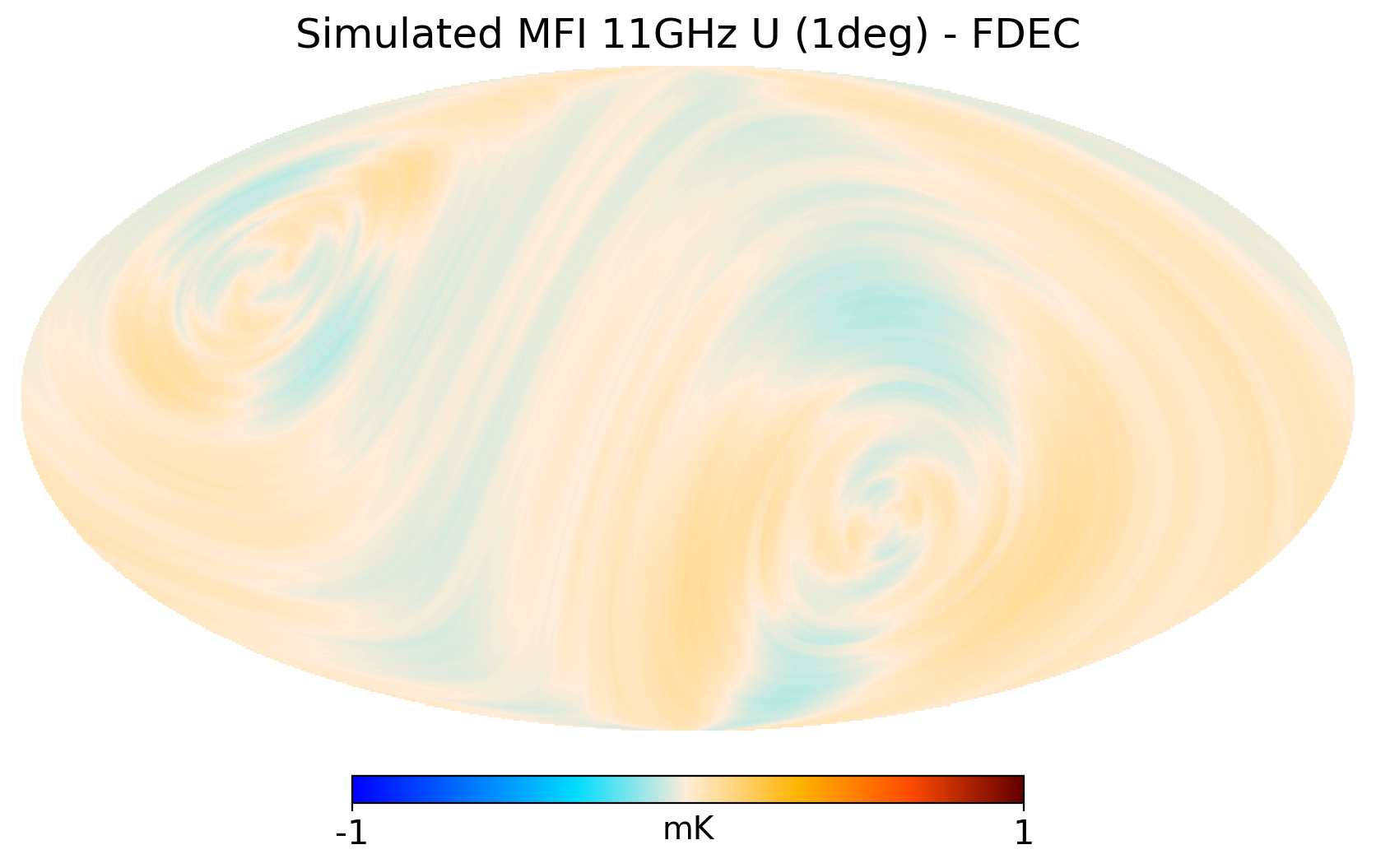}
    \caption{Example of application of FDEC filtering in simulations of the polarized MFI signal. Top (bottom) row corresponds to Stokes Q (U) parameters. Left column shows the simulated MFI 11\,GHz map at 1 deg resolution; middle column corresponds to the same map, after applying the FDEC filtering; and last column shows the difference of the previous two maps. All maps use the same colour scale, saturated at $\pm 1$\,mK.   }
    \label{fig:appfdec}
    \end{figure*}

    From Fig.~\ref{fig:appfdec}, one would expect that all analyses in real space involving "local" analyses (e.g. the photometry extraction of a compact source with a local determination of the background), should be unaffected by the FDEC filtering. 
    To test this hypothesis, we take as a reference one of the photometry methods used in this paper: the aperture photometry method (AP1d) described in Sect.~\ref{sec:sources}. Then, we apply AP1d to all possible pixels in the simulated maps within the MFI wide survey sky mask, both to the original and to the filtered maps. For this analysis, we use a reference aperture of $r_1=1^\circ$, and the background is estimated in the annulus between $r_1$ and $r_2=\sqrt{2}r_1$. 
    We find that the maximum difference between the photometry on both Stokes Q and U parameters obtained in the original map and the filtered one is 0.06\,Jy, while the standard deviation of the difference of the two photometry methods is 0.007\,Jy. Both values are significantly smaller than the typical error in the photometry (see e.g. the results presented in Table~\ref{tab:sources_vs_models}, in Sect.~\ref{sec:sources}), thus confirming that we can safely neglect any impact on the photometry due to the FDEC filtering. For completeness, we repeat the analysis for a larger aperture of $r_1=2^\circ$, and find that in this case the maximum difference is 0.4\,Jy, with a standard deviation of 0.037\,Jy. 
    
    \subsection{Impact of FDEC on correlation analyses: recovery of the spectral index }
    We now evaluate the impact of the FDEC filtering on the recovery of spectral index of the sky emission. To this end, we use the same simulation set described above, taking as a reference the simulated maps at 11 and 23\,GHz. We now apply two different methods to reconstruct the spectral index $\beta$ of the sky emission between 11 and 23\,GHz.
    
\begin{figure}
    \centering
       \includegraphics[width=0.95\columnwidth]{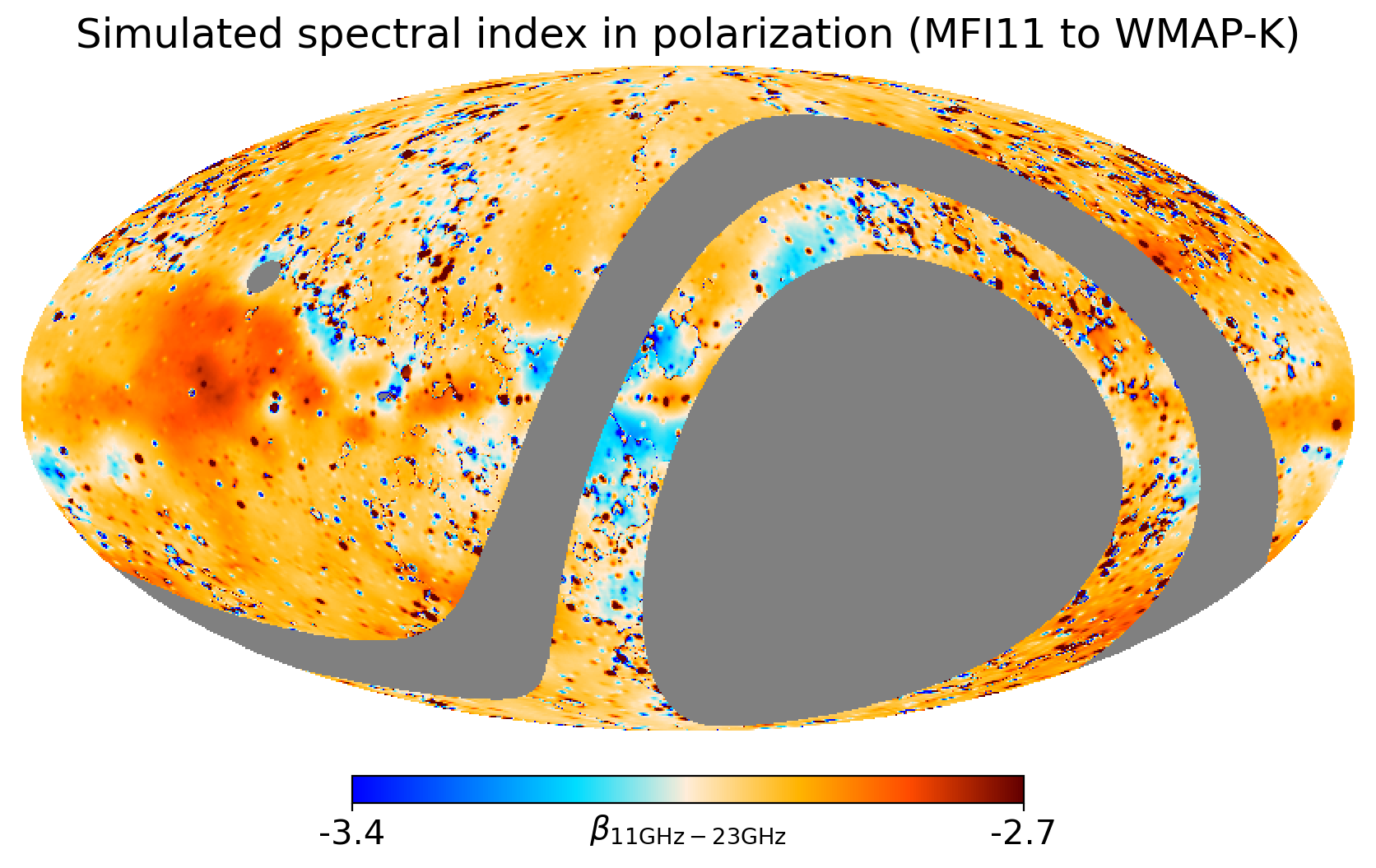}
       \includegraphics[width=0.95\columnwidth]{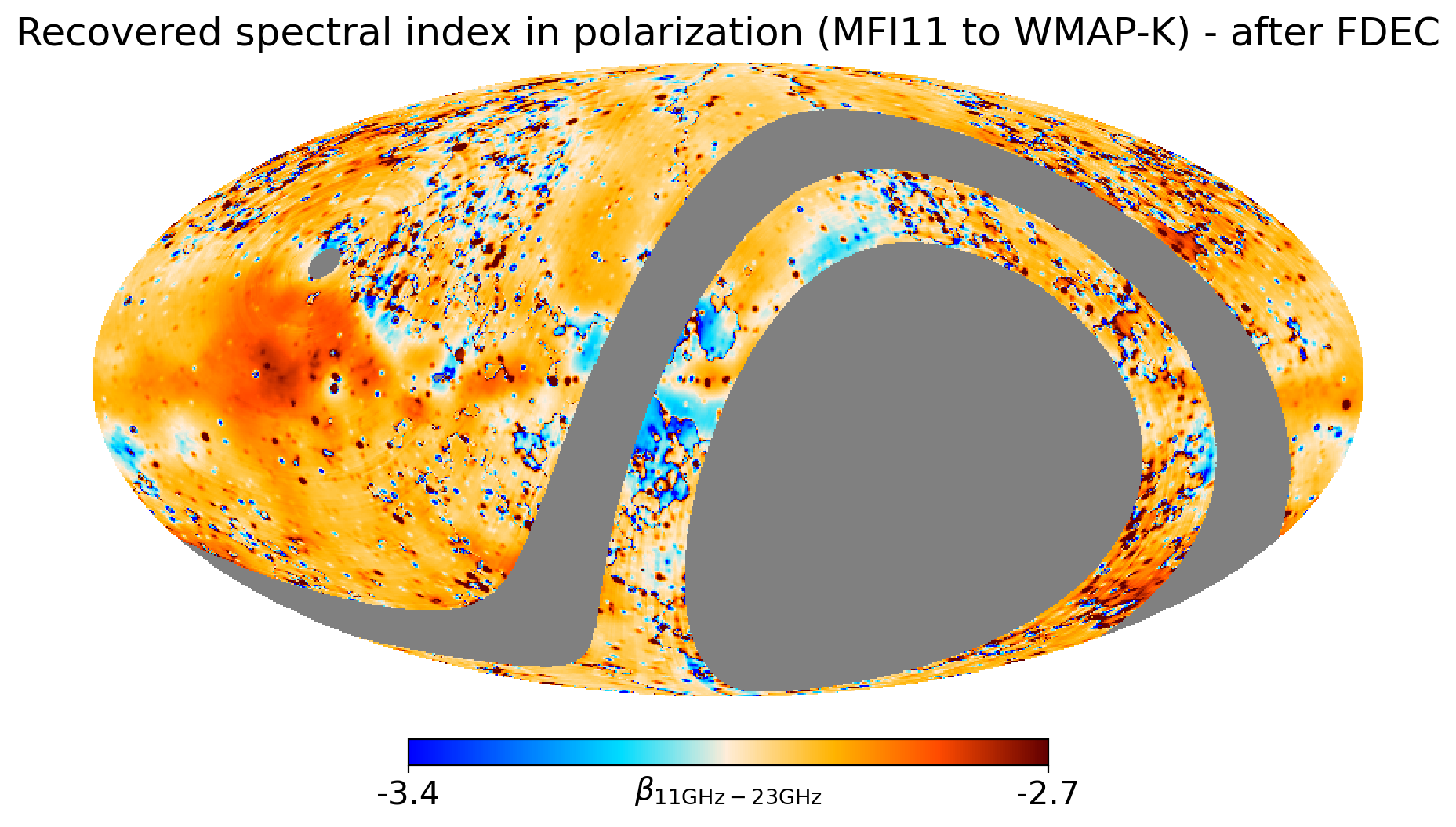}
    \caption{Impact of the application of FDEC filtering in the reconstruction of the spectral index in real space. We use simulations of the polarized sky signal in MFI 11\,GHz and WMAP 23\,GHz. Top panel shows the (true) underlying spectral index of the simulated signal between 11 and 23\,GHz, within the MFI observing mask. Bottom panel shows the reconstructed spectral index after applying the FDEC filtering to both simulated maps (11 and 23\,GHz).  }
    \label{fig:appfdec2}
    \end{figure}

    First, we carry out a direct evaluation of the spectral index at the pixel level ($\nside=512$) in the original (unfiltered) maps, and also in the filtered ones.  
    This methodology is similar to the one used in Sect~\ref{sec:properties}. It is important to emphasize that both maps (the simulated MFI 11\,GHz and the simulated WMAP 23\,GHz) have to be filtered with the FDEC, in order to have consistent scales between the two. The results are shown in Fig.~\ref{fig:appfdec2}. The reconstructed spectral index is fully consistent with the input one. If we restrict the comparison to pixels with high emission (polarized intensity at 11\,GHz greater than 0.1\,mK), and we compare the reconstructed maps after degrading to $2^\circ$ (to be consistent with Sect~\ref{sec:properties}), we find that the median difference $\Delta \beta$ between the reconstructed and original spectral index is 0.0005, while the standard deviation of the difference is 0.02. 
    
\begin{figure}
    \centering
       \includegraphics[width=0.95\columnwidth]{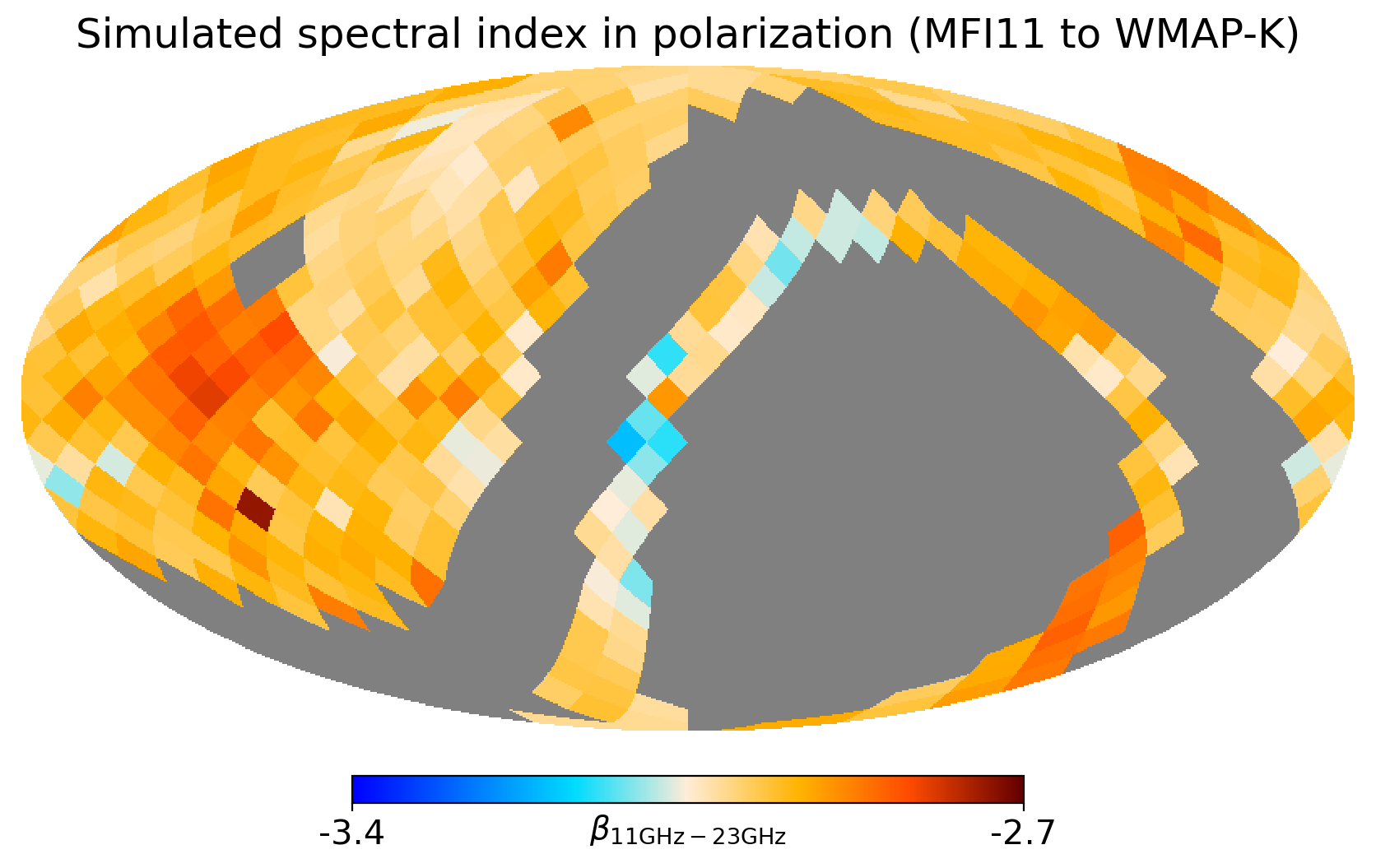}
       \includegraphics[width=0.95\columnwidth]{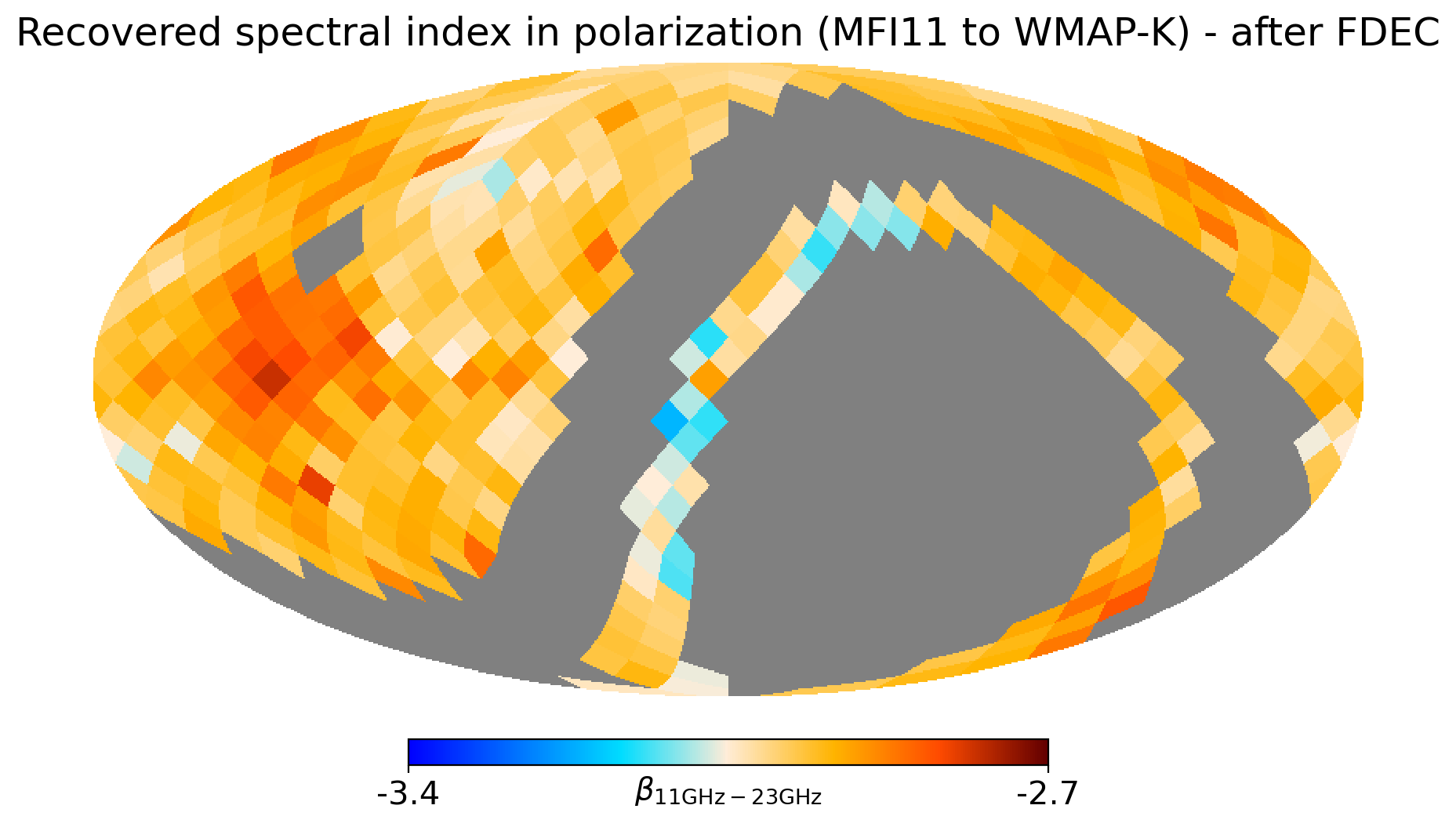}
    \caption{Impact of the application of FDEC filtering in the reconstruction of the spectral index using correlation analysis (TTplot). We carry out the correlation analysis in regions defined by \healpix\ pixels of $\nside=8$, and extract the spectral index of the polarized sky signal between MFI 11\,GHz and WMAP 23\,GHz. Top panel shows the (true) underlying spectral index of the simulated signal within the MFI observing mask. Bottom panel shows the reconstructed spectral index after applying the FDEC filtering to both simulated maps (11 and 23\,GHz).  }
    \label{fig:appfdec3}
    \end{figure}

    Second, we use a correlation analysis method (also called TT plot) to recover the spectral index of the emission between 11 and 23\,GHz. For this analysis, we degrade the simulated maps at 1 degree resolution to $\nside=64$, in order to have approximately independent pixels. Then, we divide the observed sky in patches of $\sim 7.3^\circ$, using as a reference the pixels of a $\nside=8$ \healpix\ map. Within each patch, we carry out a TT plot analysis assuming a typical error in each map corresponding to 3 per cent of the sky signal, and accounting for errors in both axes \citep[see e.g.][]{Fuskeland2014}. Fig.~\ref{fig:appfdec3} shows the obtained results from the original maps (top panel), and the FDEC filtered maps (bottom panel). As expected, the spatial distribution of the reconstructed index has a good correspondence with the maps shown in Fig.~\ref{fig:appfdec2}. A numerical comparison of both maps in Fig.~\ref{fig:appfdec3} gives that median difference $\Delta \beta$ between the two maps is -0.0007, and the standard deviation of the difference is 0.02. 
    
    Summarising, we can obtain an unbiased reconstruction of the spectral index of the sky signal, provided that both maps are filtered in the same way using FDEC. In practice, this means that when doing these type of analyses using QUIJOTE MFI wide survey maps and external ancillary data, we must filter first the external maps using the same procedure as for the MFI maps. If the FDEC filtering is not applied to the external ancillary data, we find that for these simulations the standard deviation of the reconstructed spectral index can be as large as 0.3. This issue is further discussed in other papers in the 
    series \citep[see e.g. Appendix~C in][]{MFIcompsep_pol}.

    \section{QUIJOTE MFI wide survey maps per horn at original resolution}
    \label{app:maps}
    
\begin{figure*}
    \centering
    \includegraphics[width=6cm]{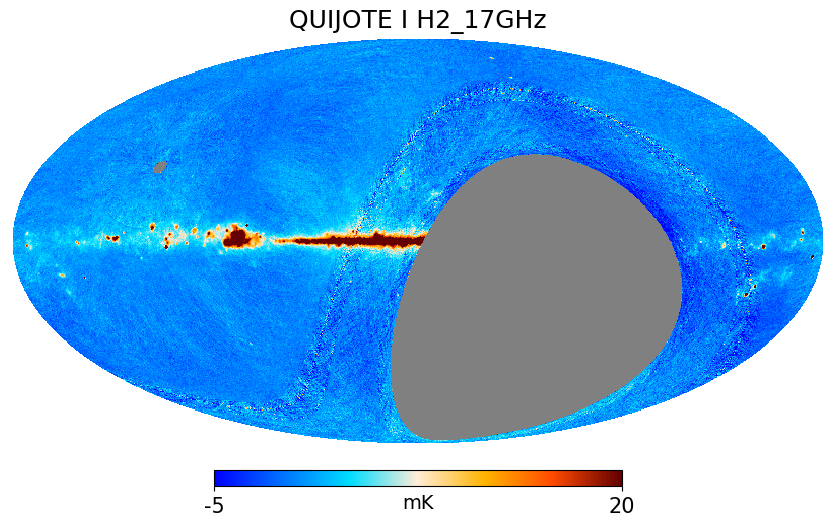}%
    \includegraphics[width=6cm]{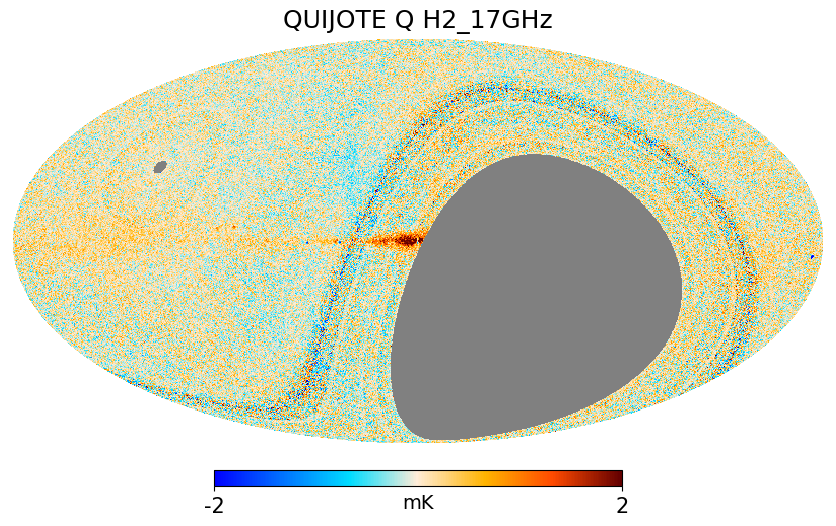}%
    \includegraphics[width=6cm]{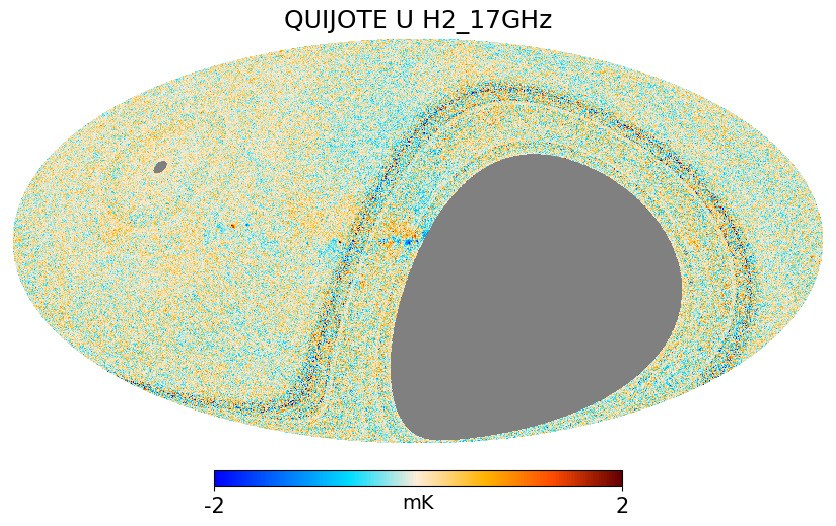}
    \includegraphics[width=6cm]{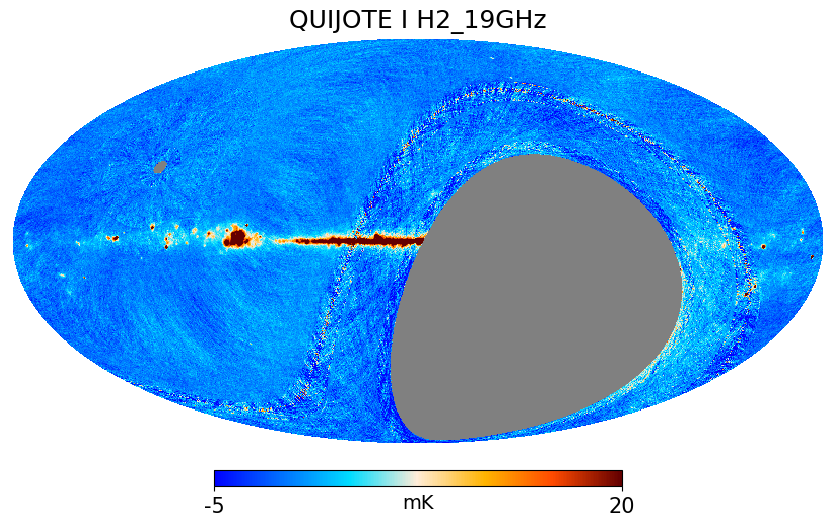}%
    \includegraphics[width=6cm]{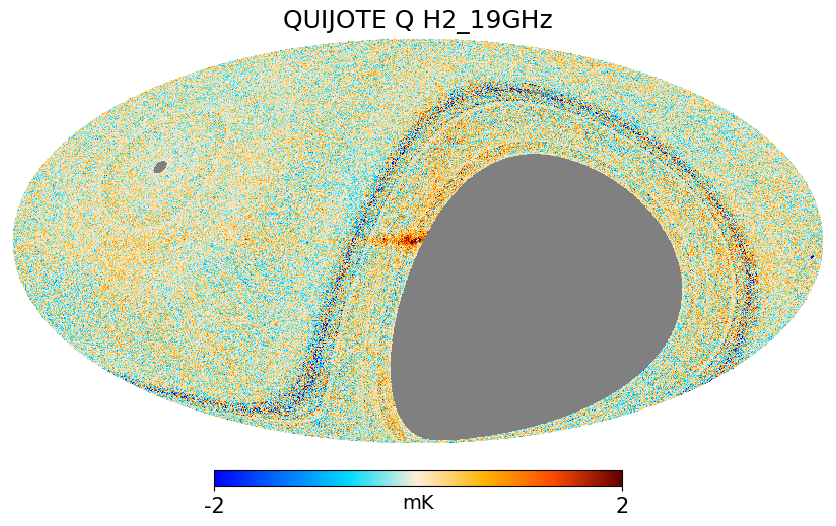}%
    \includegraphics[width=6cm]{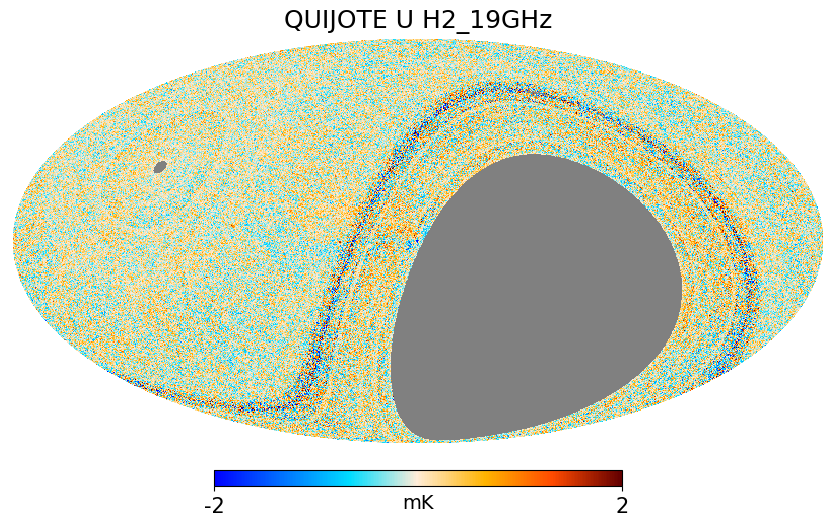}
    \caption{Original resolution QUIJOTE MFI wide survey maps for horn 2. Maps are shown in Galactic coordinates. All figures use the same linear colour scale, saturated at 20\,mK$_{\rm CMB}$ for intensity (first column) and 2\,mK$_{\rm CMB}$ in polarization for Stokes Q (second column) and Stokes U (third column) parameters. For display purposes, maps are downgraded to $\nside=256$. }
    \label{fig:h2maps}
    \end{figure*}

\begin{figure*}
    \centering
    \includegraphics[width=6cm]{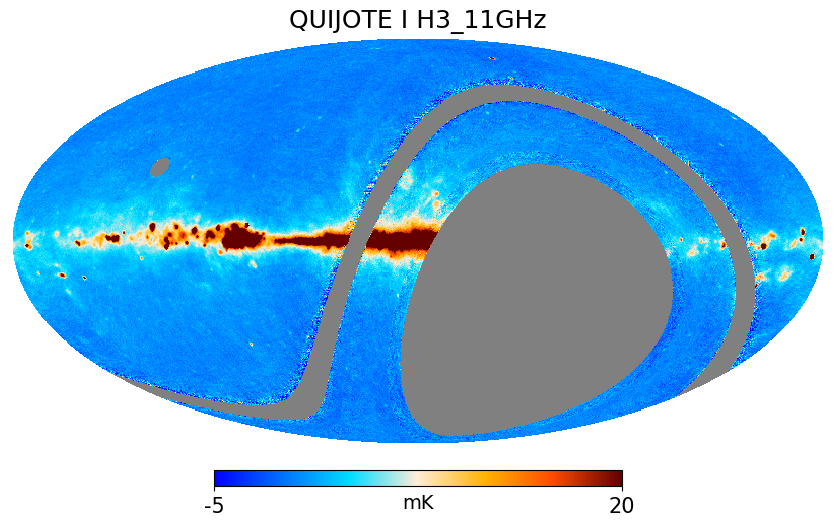}%
    \includegraphics[width=6cm]{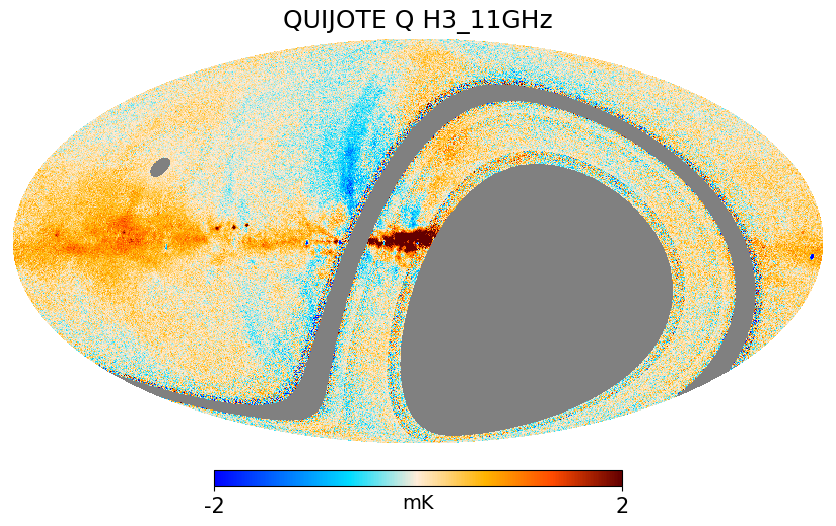}%
    \includegraphics[width=6cm]{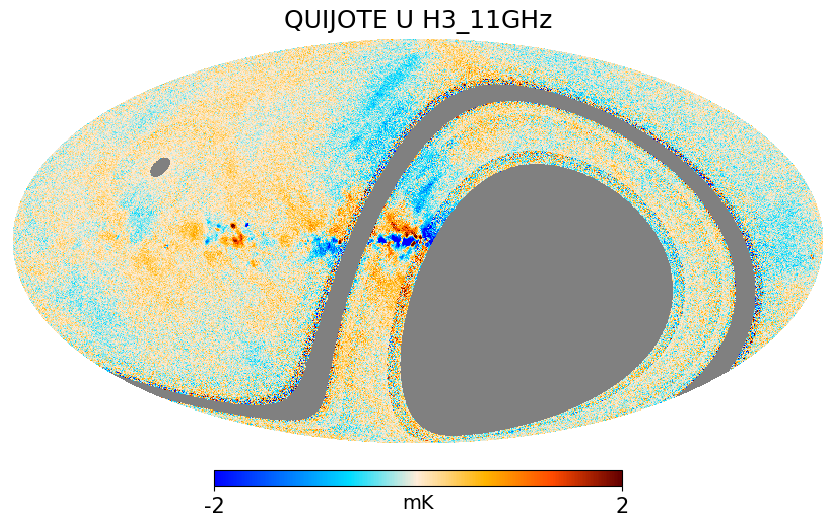}
    \includegraphics[width=6cm]{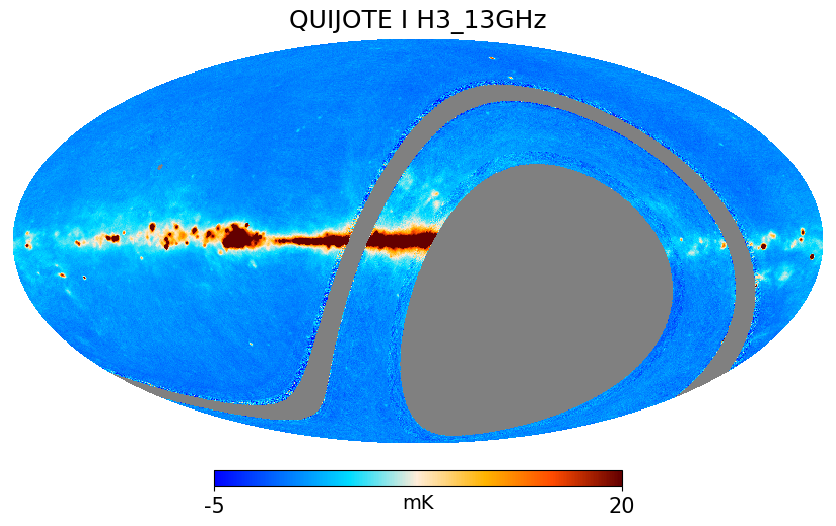}%
    \includegraphics[width=6cm]{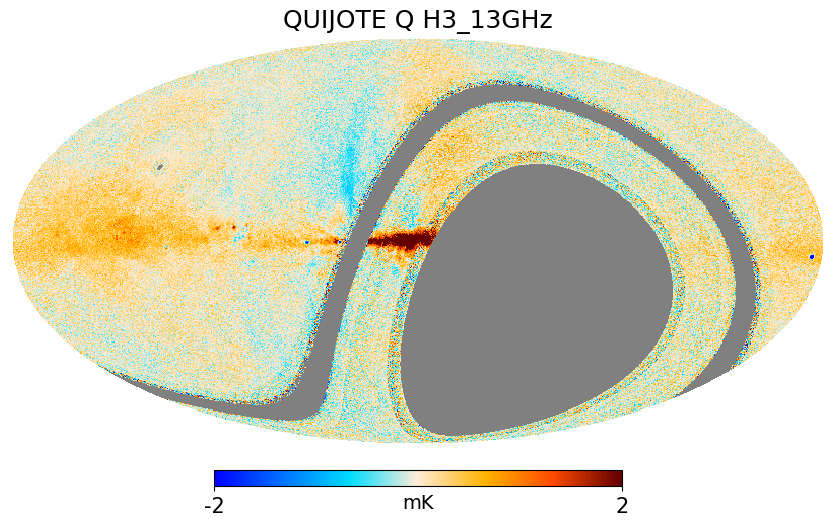}%
    \includegraphics[width=6cm]{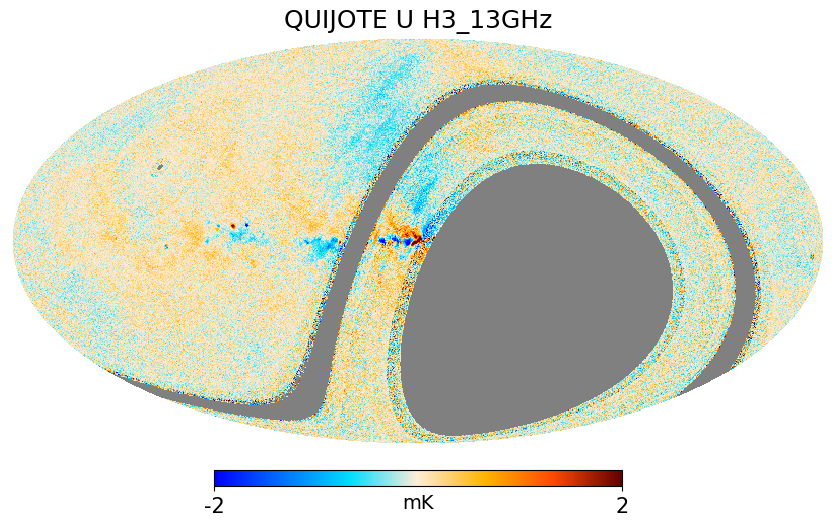}
    \caption{Same as Fig.~\ref{fig:h2maps}, but for QUIJOTE MFI wide survey maps for horn 3. }
    \label{fig:h3maps}
    \end{figure*}

\begin{figure*}
    \centering
    \includegraphics[width=6cm]{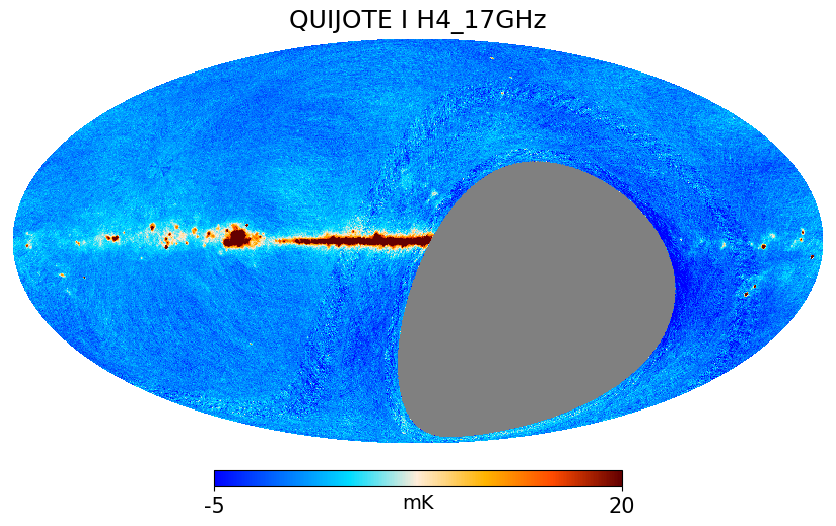}%
    \includegraphics[width=6cm]{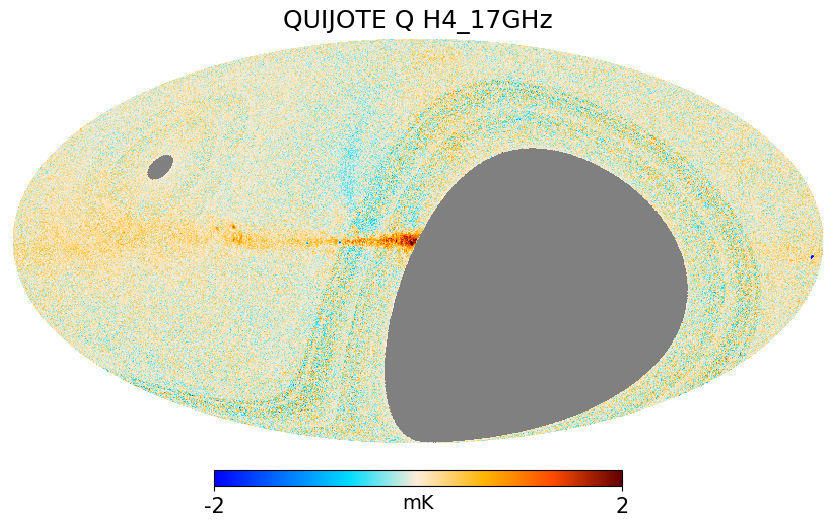}%
    \includegraphics[width=6cm]{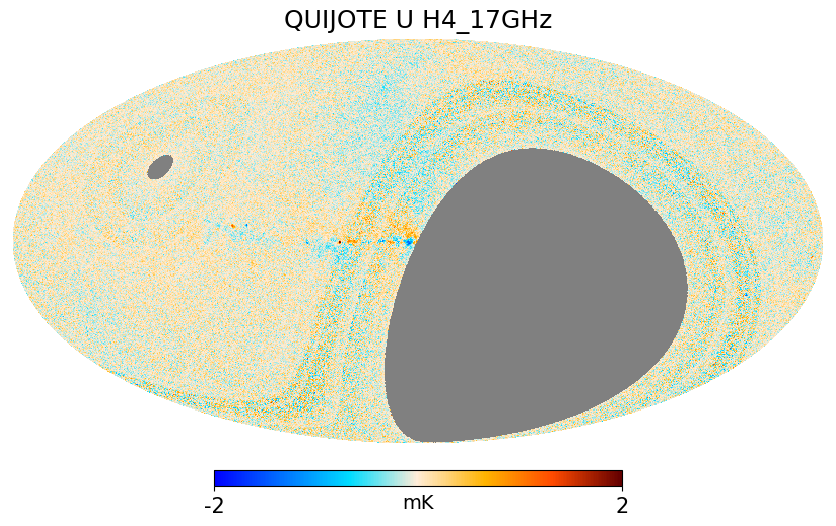}
    \includegraphics[width=6cm]{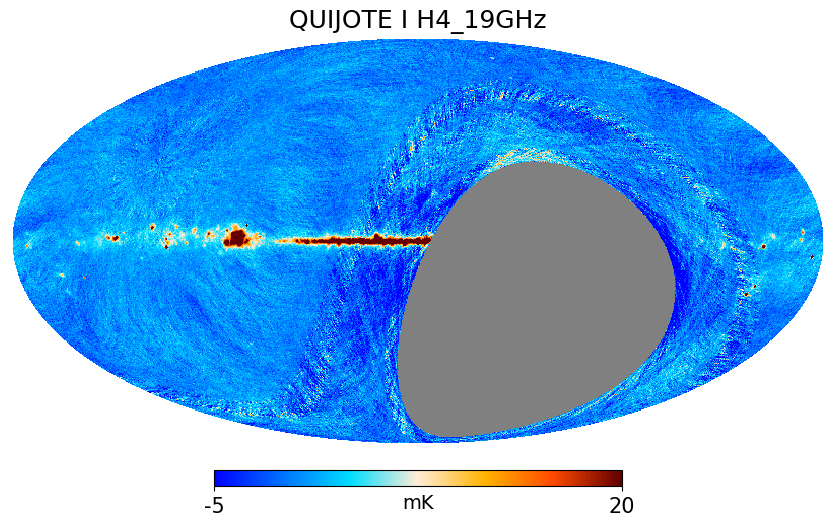}%
    \includegraphics[width=6cm]{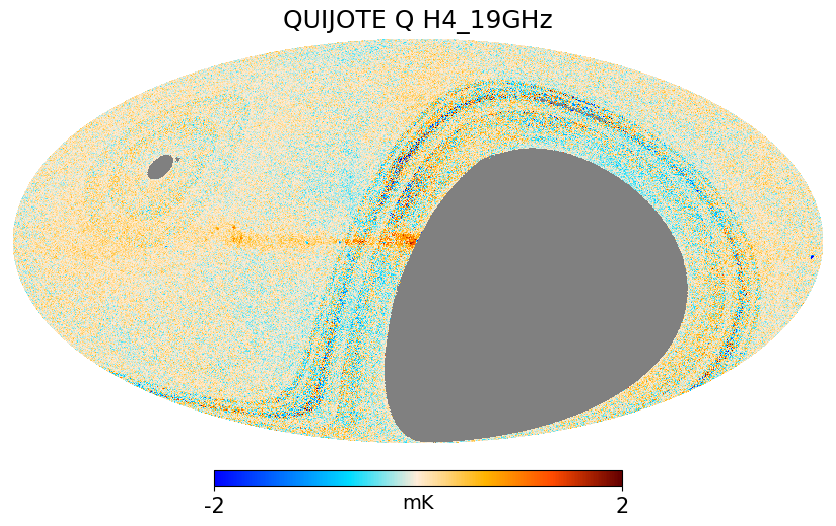}%
    \includegraphics[width=6cm]{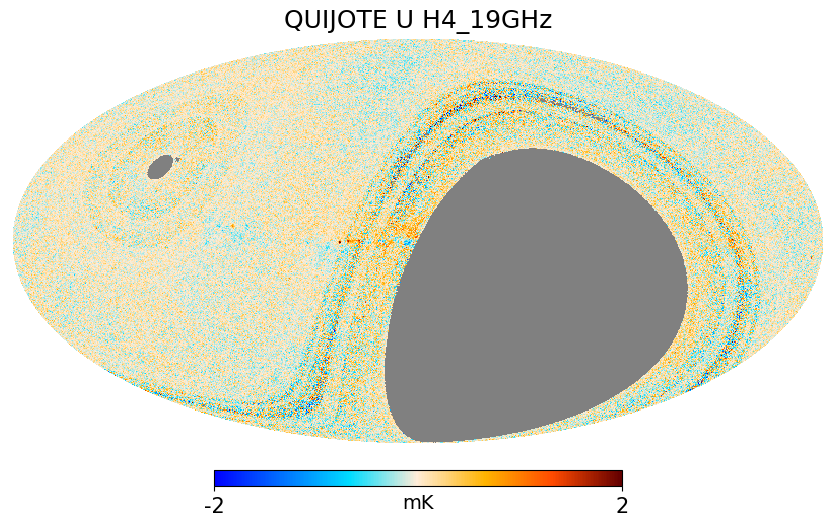}
    \caption{Same as Fig.~\ref{fig:h2maps}, but for QUIJOTE MFI wide survey maps for horn 4.  }
    \label{fig:h4maps}
    \end{figure*}

\begin{figure*}
    \centering
    \includegraphics[width=6cm]{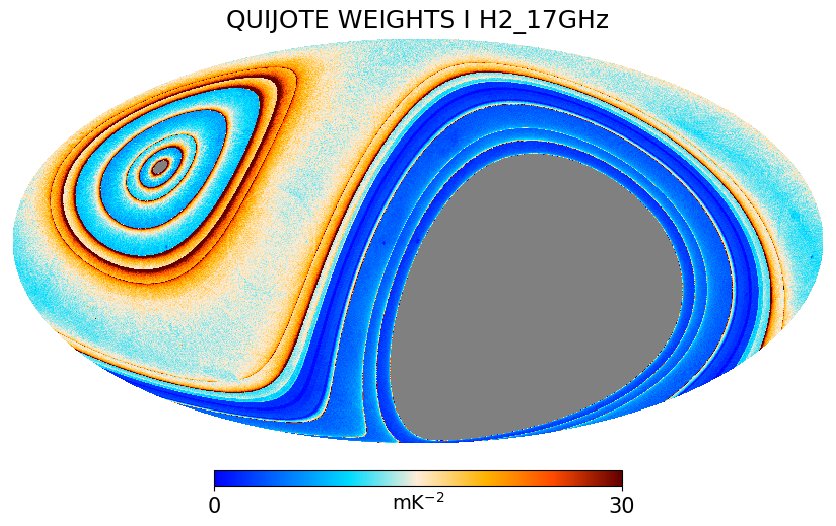}%
    \includegraphics[width=6cm]{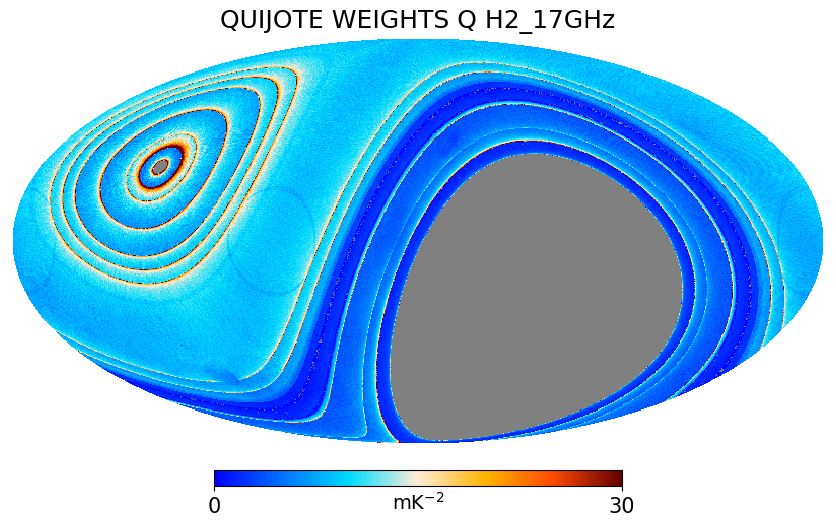}%
    \includegraphics[width=6cm]{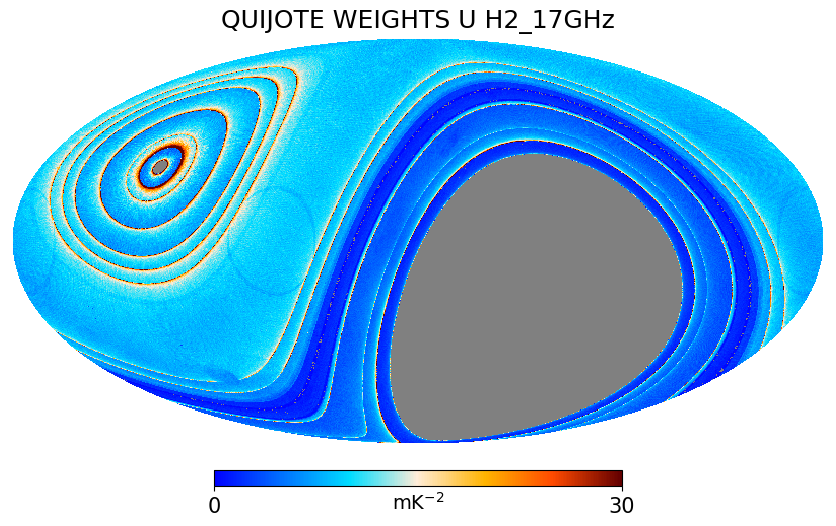}
    \includegraphics[width=6cm]{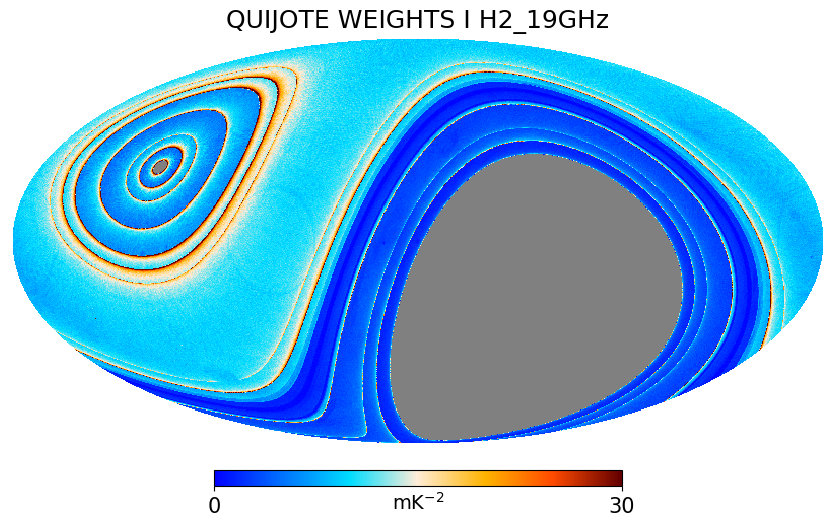}%
    \includegraphics[width=6cm]{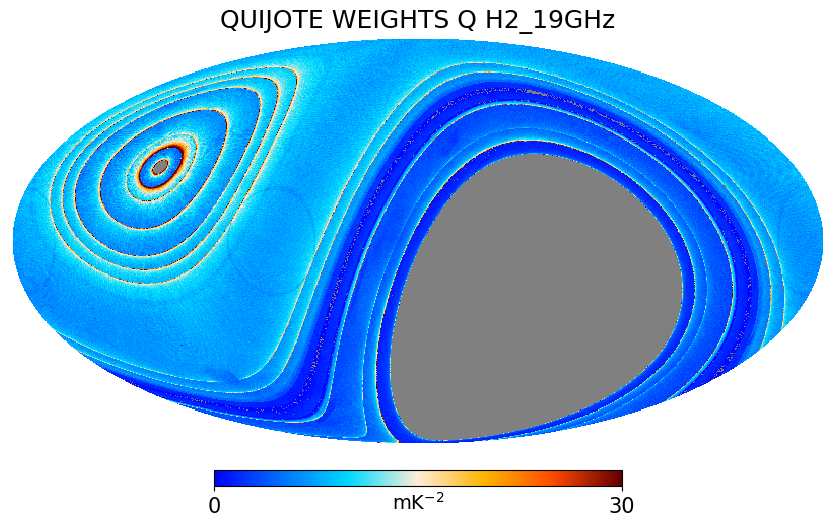}%
    \includegraphics[width=6cm]{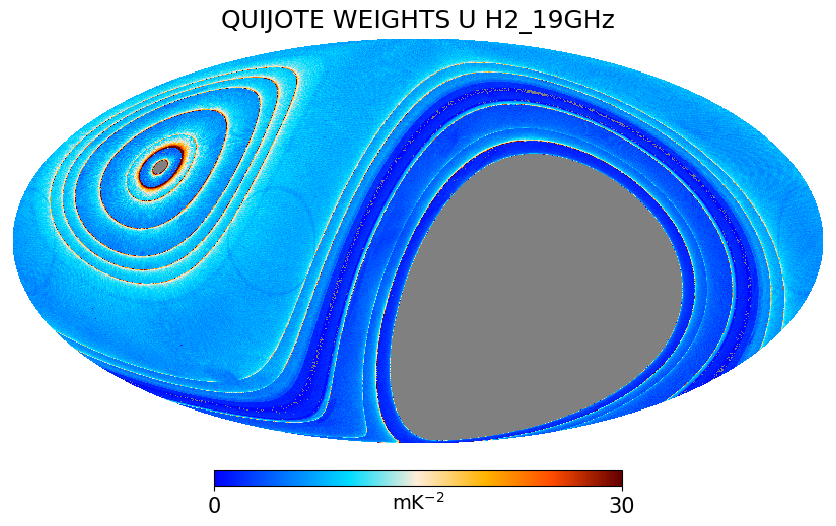}
    \caption{QUIJOTE MFI wide survey weight maps for horn 2. Top row is 17\,GHz, and bottom row is 19\,GHz. Each row shows, from left to right, the weight maps for Stokes I, Q and U.  }
    \label{fig:h2wei}
    \end{figure*}

\begin{figure*}
    \centering
    \includegraphics[width=6cm]{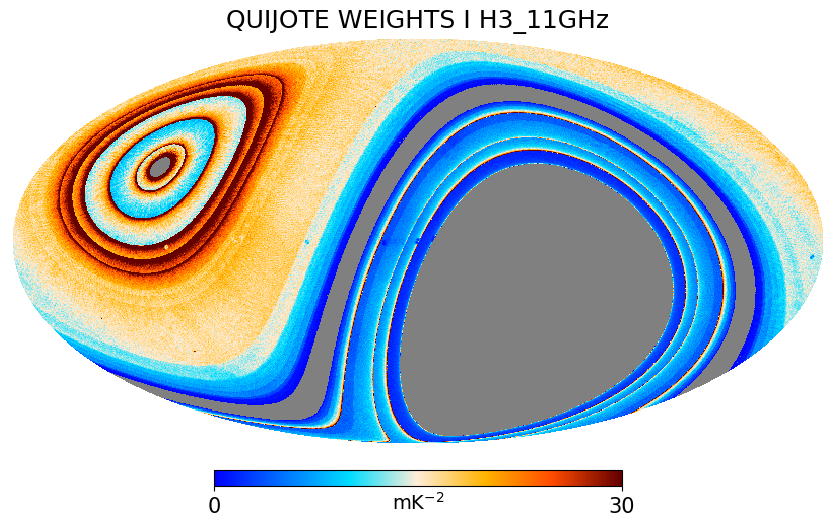}%
    \includegraphics[width=6cm]{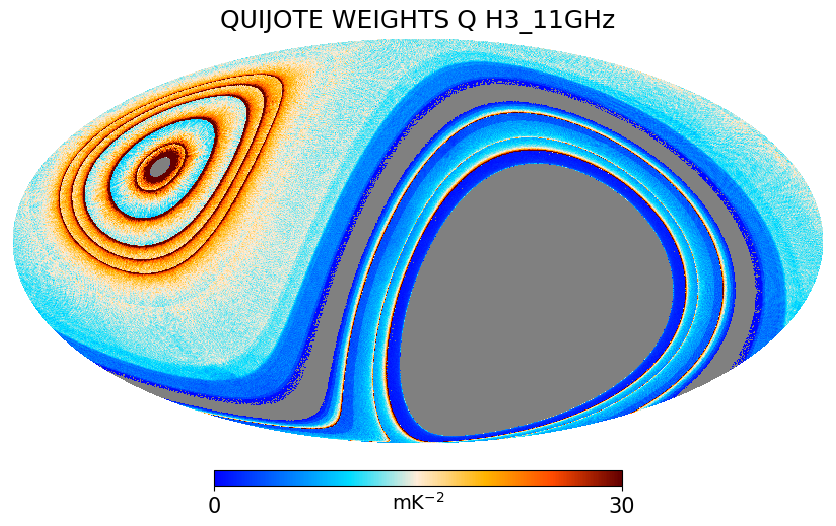}%
    \includegraphics[width=6cm]{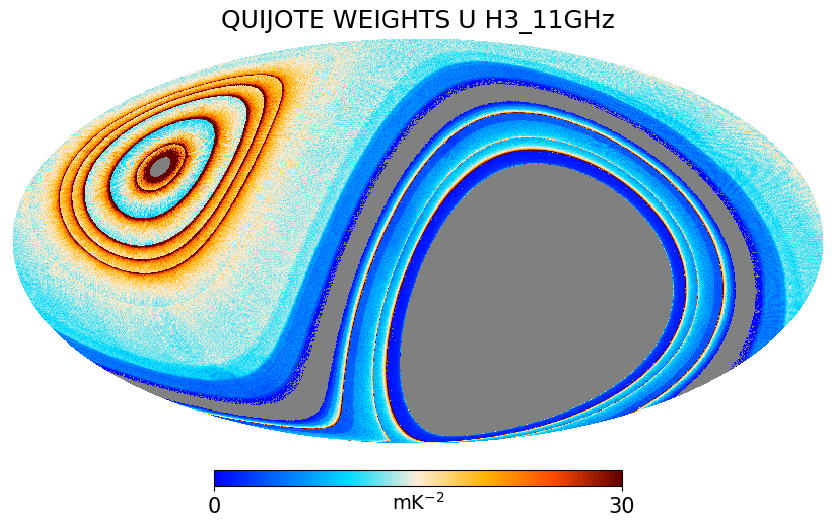}
    \includegraphics[width=6cm]{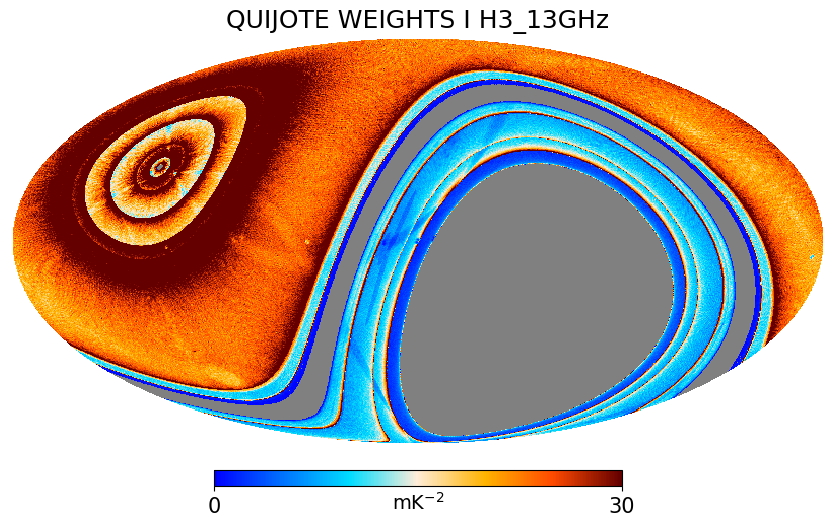}%
    \includegraphics[width=6cm]{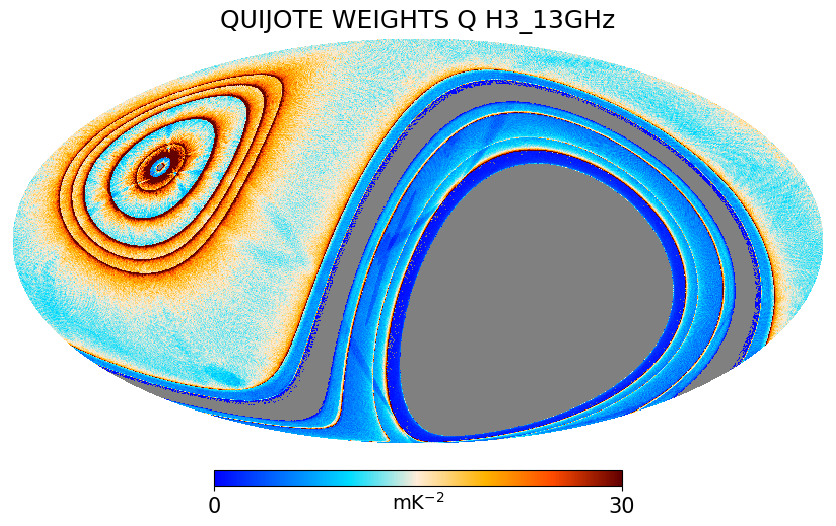}%
    \includegraphics[width=6cm]{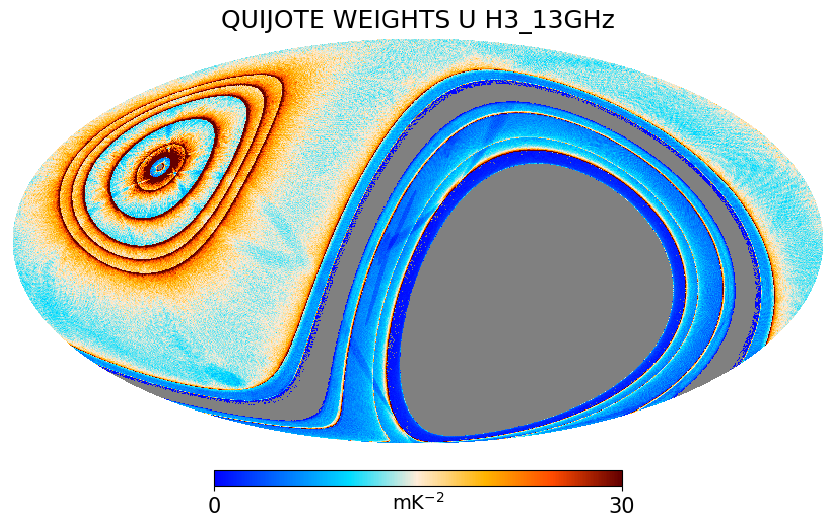}
    \caption{QUIJOTE MFI wide survey weight maps for horn 3.  }
    \label{fig:h3wei}
    \end{figure*}

\begin{figure*}
    \centering
    \includegraphics[width=6cm]{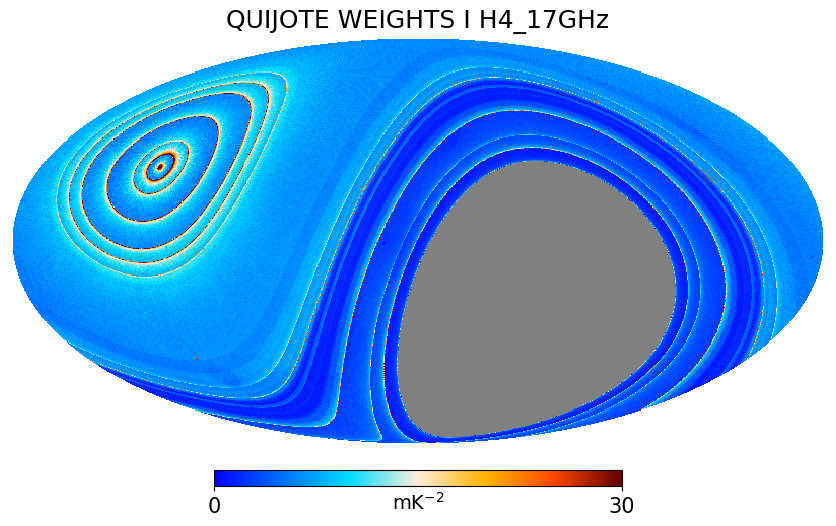}%
    \includegraphics[width=6cm]{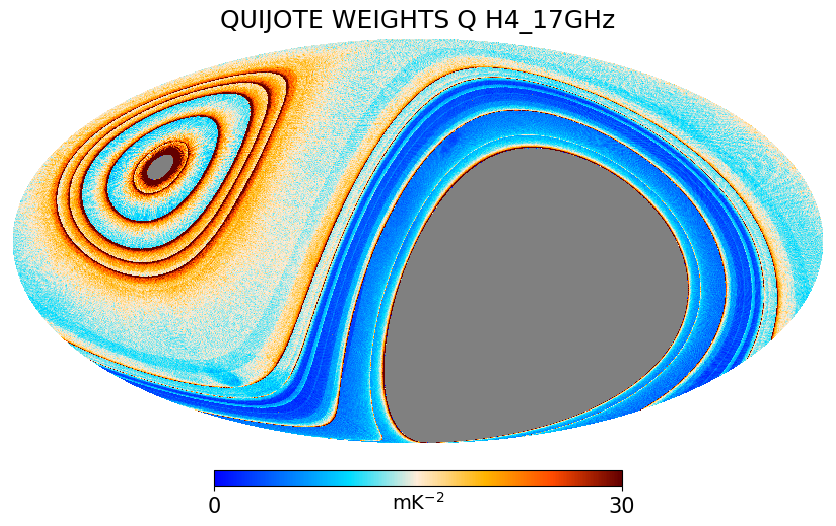}%
    \includegraphics[width=6cm]{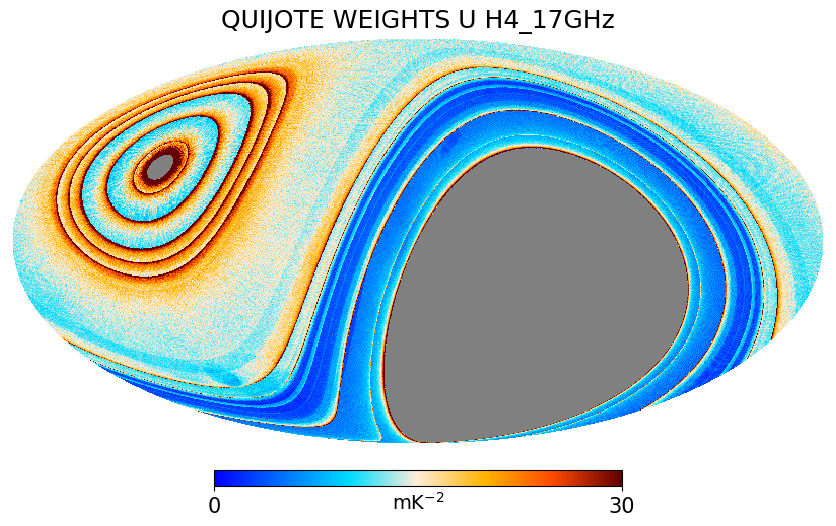}
    \includegraphics[width=6cm]{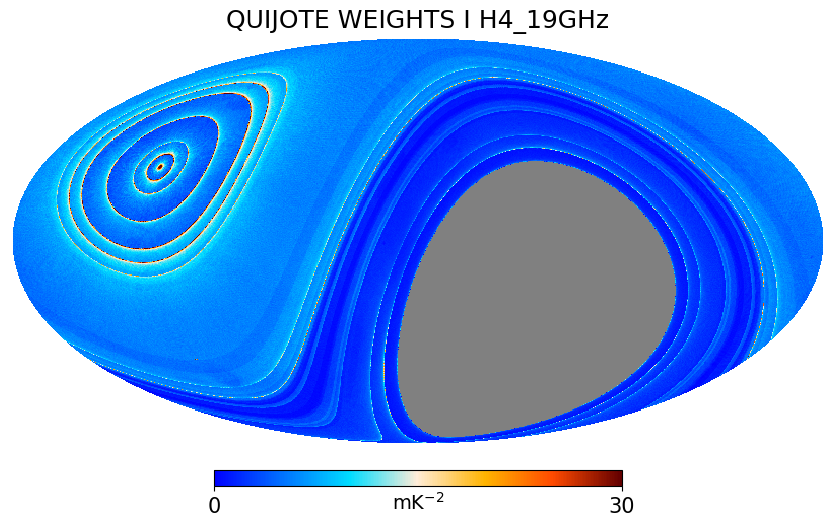}%
    \includegraphics[width=6cm]{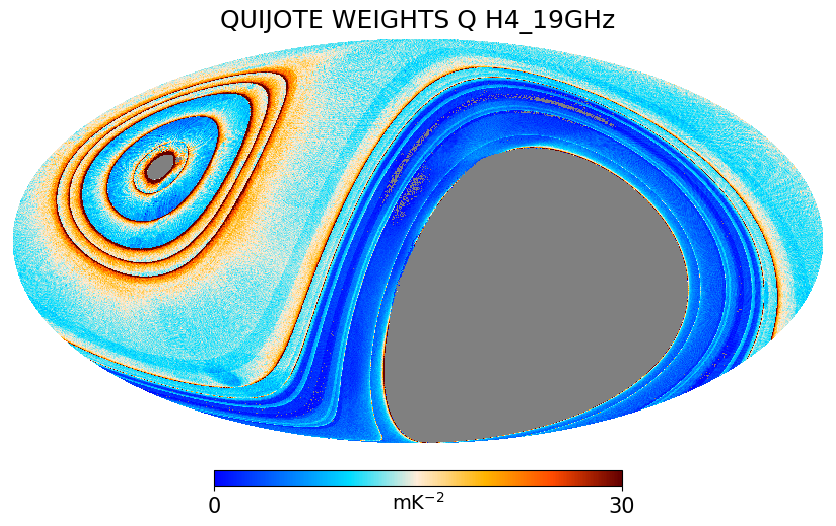}%
    \includegraphics[width=6cm]{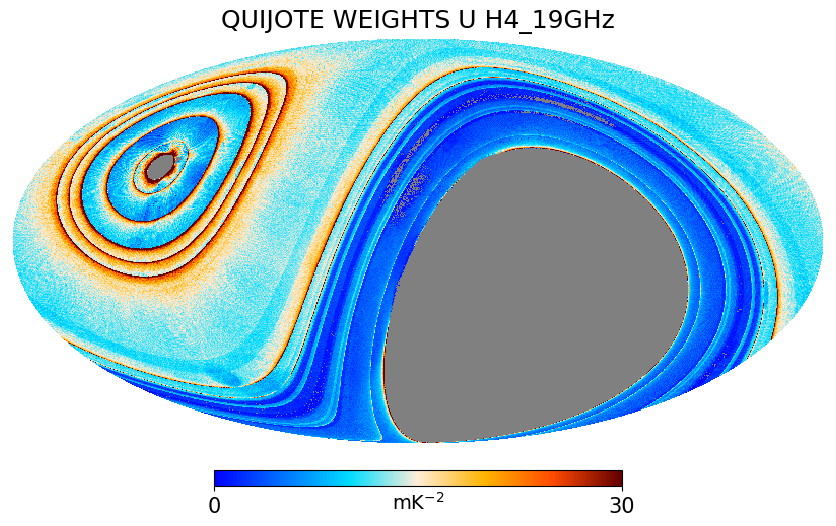}
    \caption{QUIJOTE MFI wide survey weight maps for horn 4.  }
    \label{fig:h4wei}
    \end{figure*}

\begin{figure*}
    \centering
    \includegraphics[width=6cm]{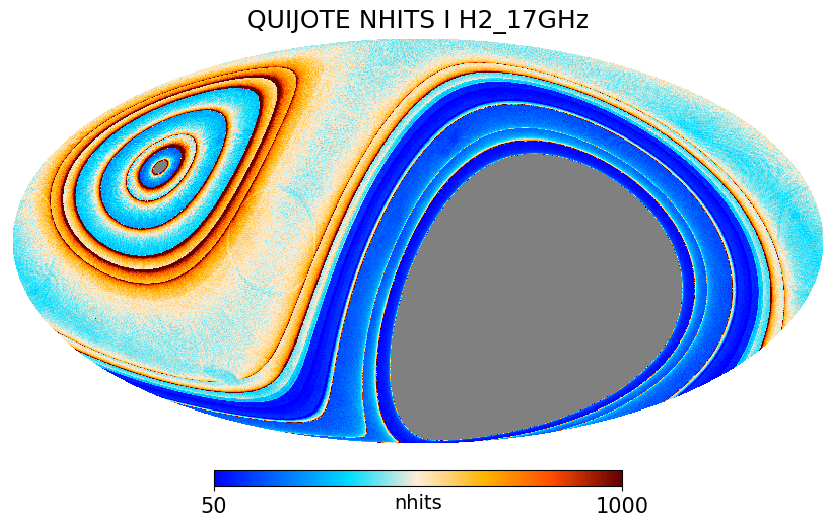}%
    \includegraphics[width=6cm]{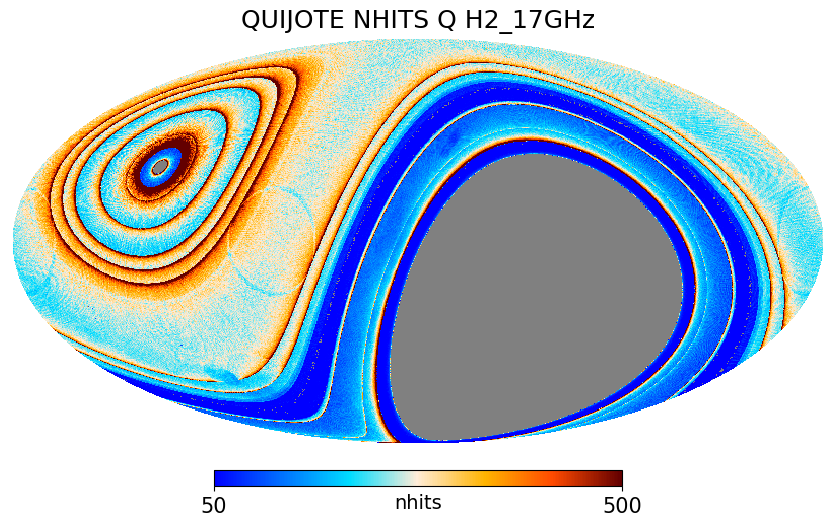}%
    \includegraphics[width=6cm]{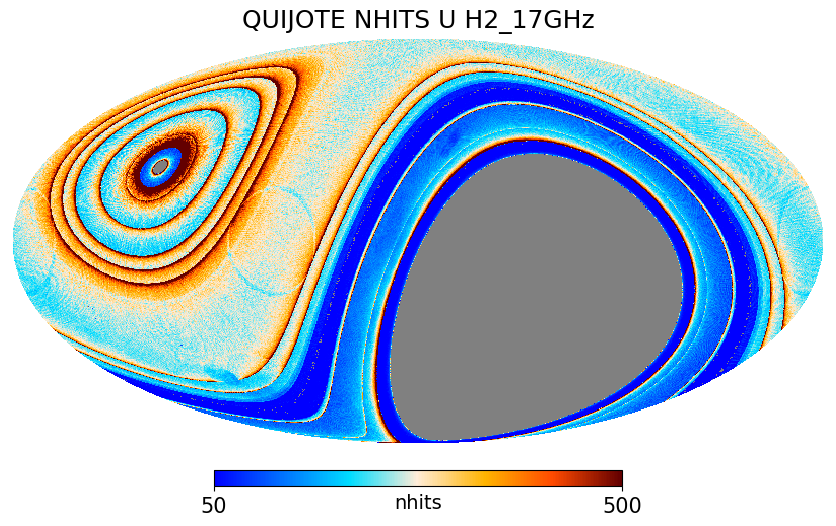}
    \includegraphics[width=6cm]{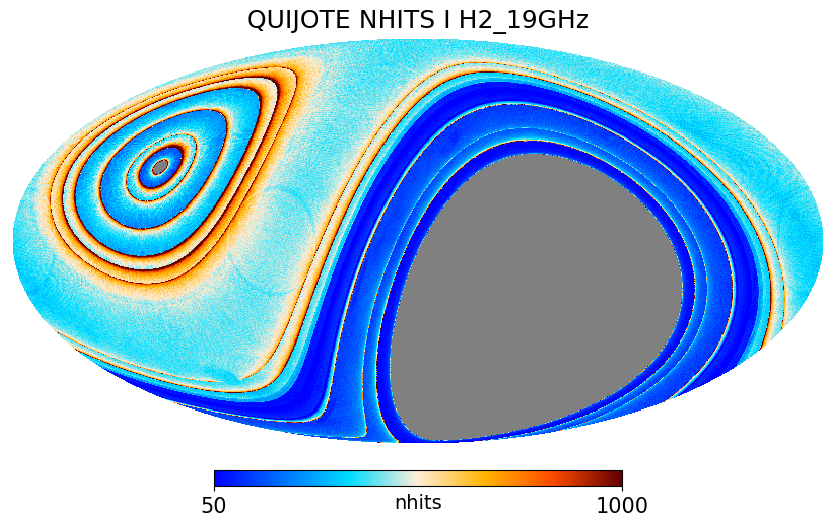}%
    \includegraphics[width=6cm]{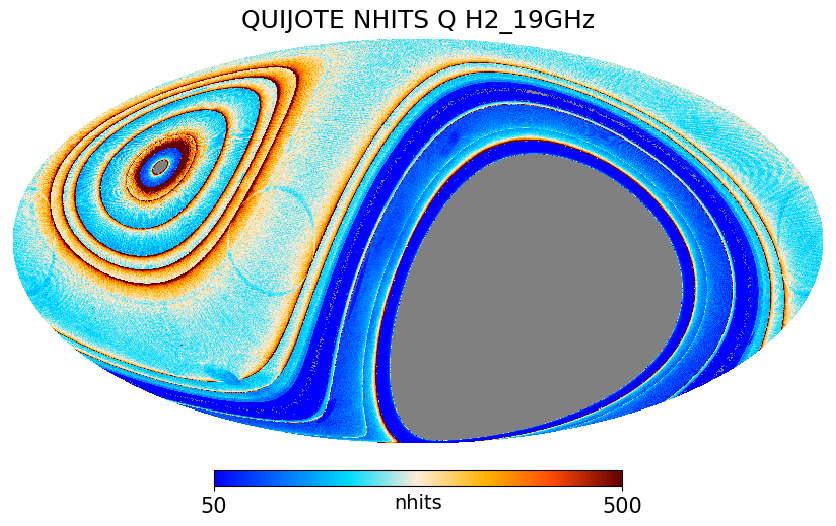}%
    \includegraphics[width=6cm]{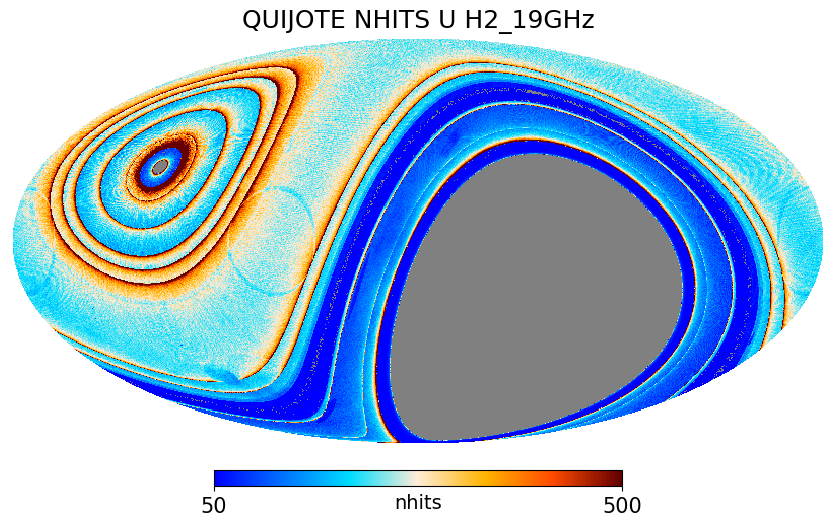}
    \caption{QUIJOTE MFI wide survey hit maps for horn 2. They show the total number of 40\,ms samples in each \healpix\ pixel of $\nside=512$ resolution. }
    \label{fig:h2nhits}
    \end{figure*}

\begin{figure*}
    \centering
    \includegraphics[width=6cm]{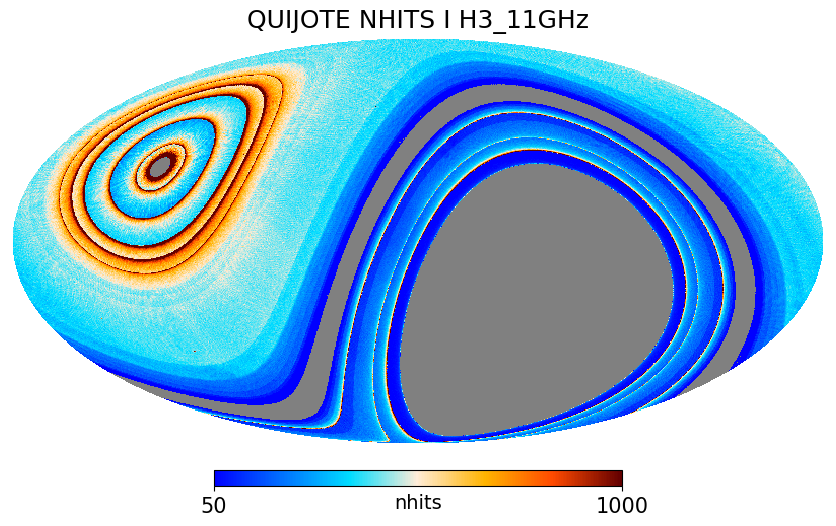}%
    \includegraphics[width=6cm]{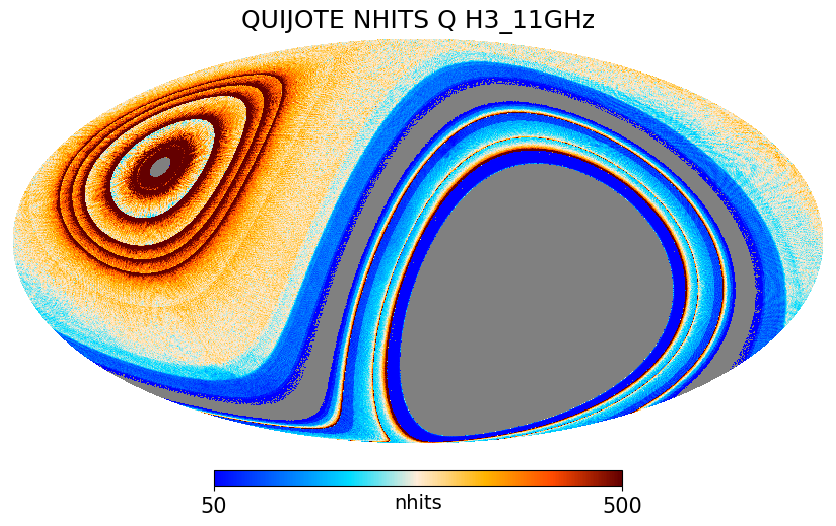}%
    \includegraphics[width=6cm]{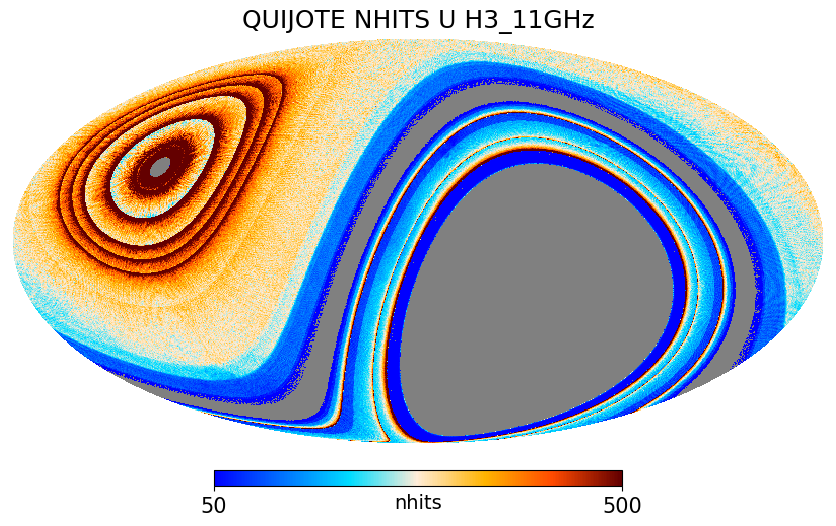}
    \includegraphics[width=6cm]{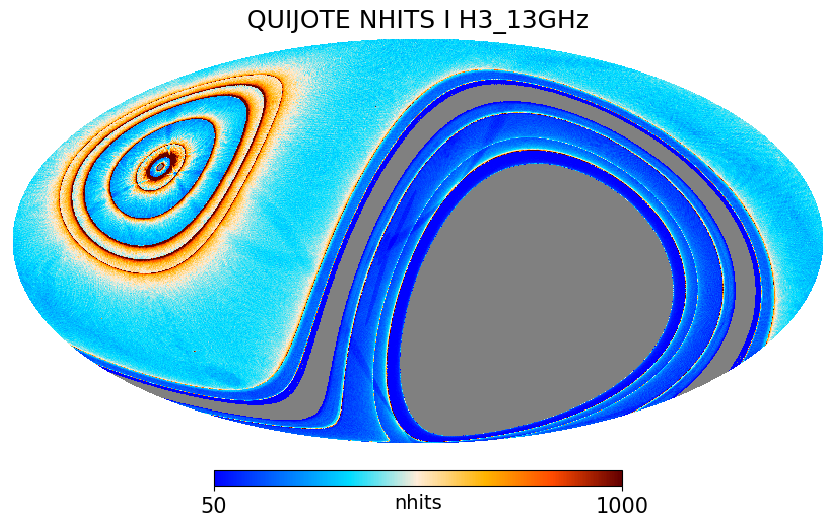}%
    \includegraphics[width=6cm]{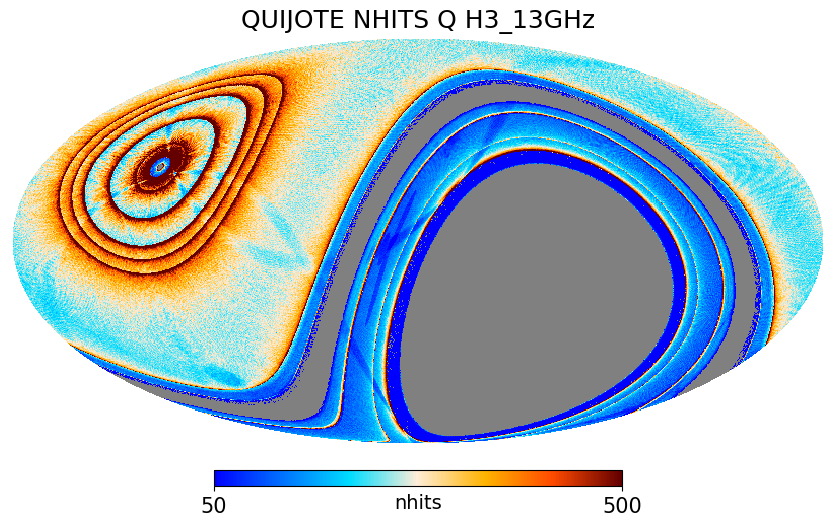}%
    \includegraphics[width=6cm]{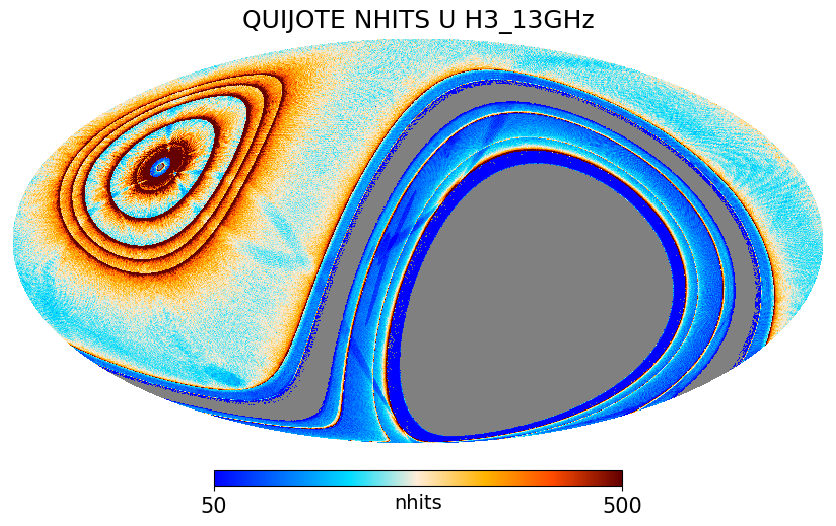}
    \caption{QUIJOTE MFI wide survey hit maps for horn 3.  }
    \label{fig:h3nhits}
    \end{figure*}

\begin{figure*}
    \centering
    \includegraphics[width=6cm]{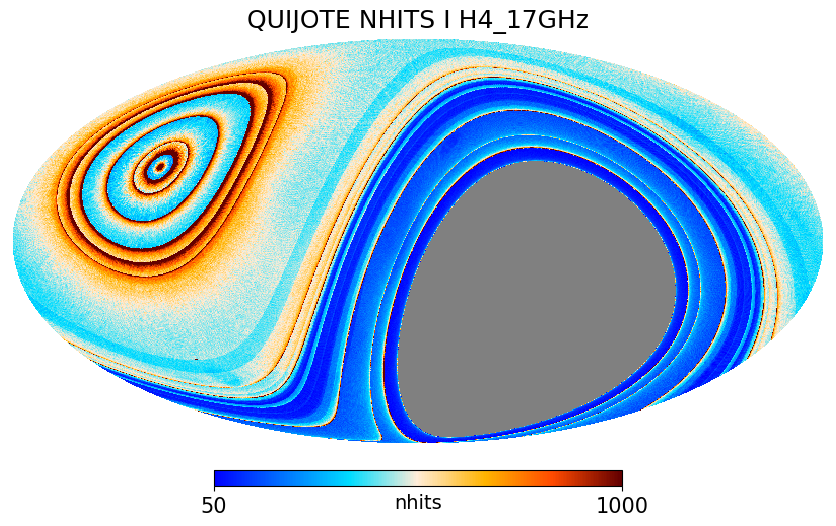}%
    \includegraphics[width=6cm]{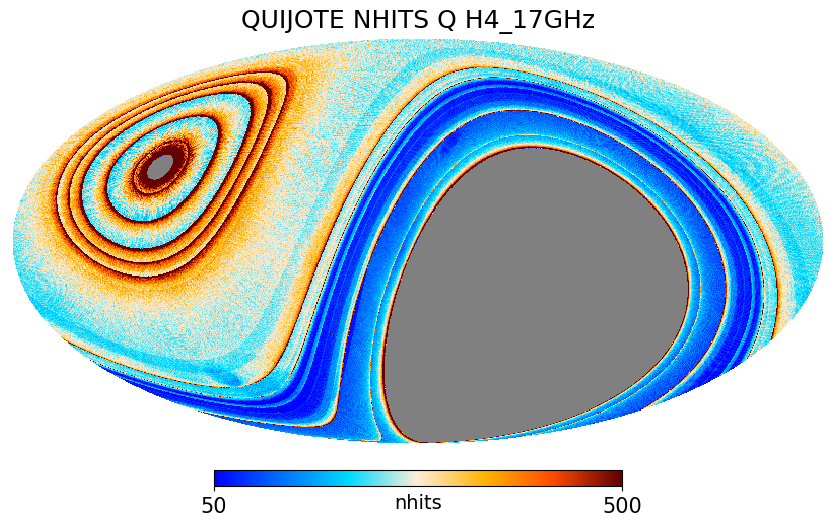}%
    \includegraphics[width=6cm]{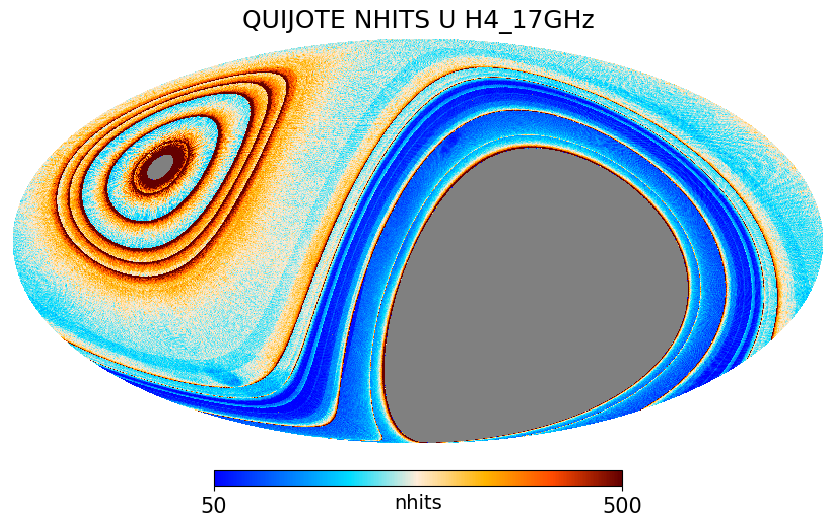}
    \includegraphics[width=6cm]{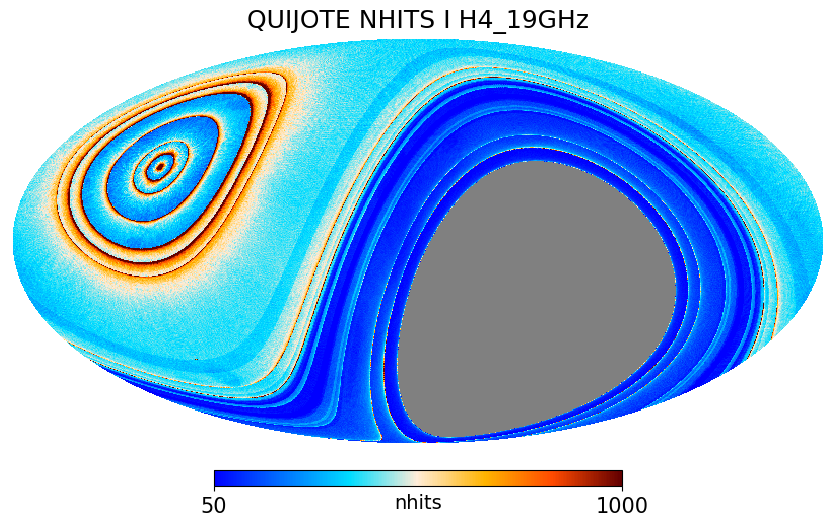}%
    \includegraphics[width=6cm]{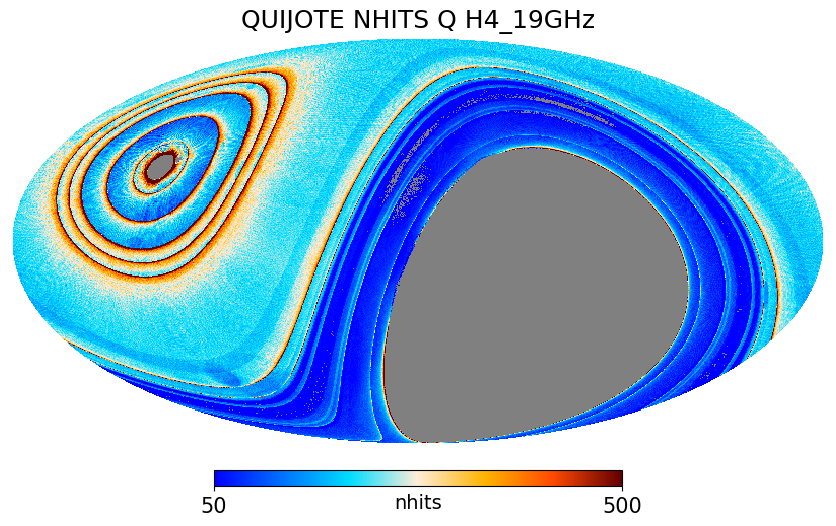}%
    \includegraphics[width=6cm]{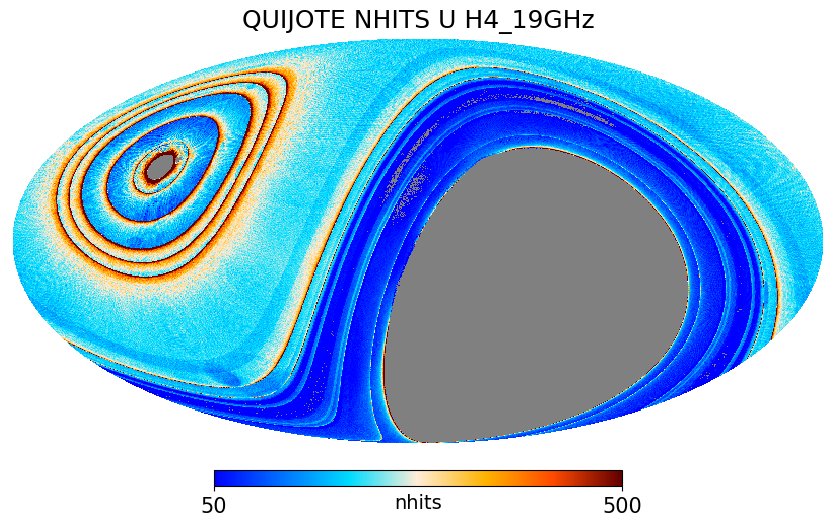}
    \caption{QUIJOTE MFI wide survey hit maps for horn 4.  }
    \label{fig:h4nhits}
    \end{figure*}

\begin{figure*}
    \centering
    \includegraphics[width=6cm]{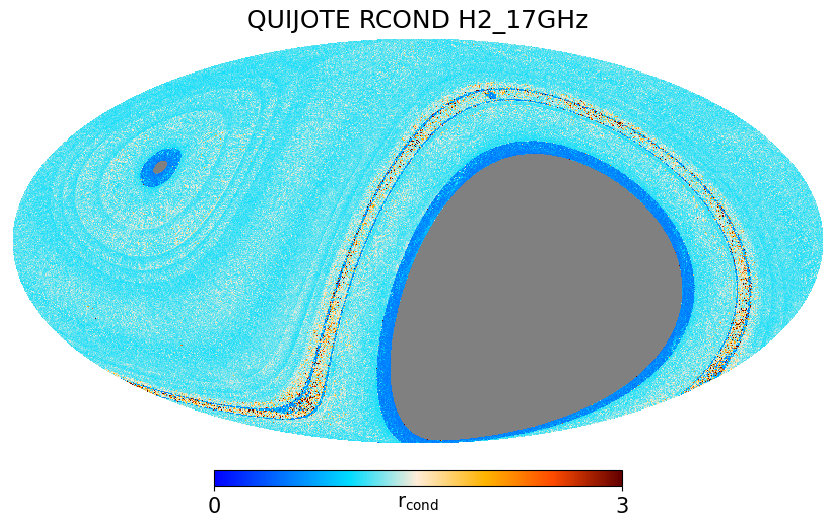}%
    \includegraphics[width=6cm]{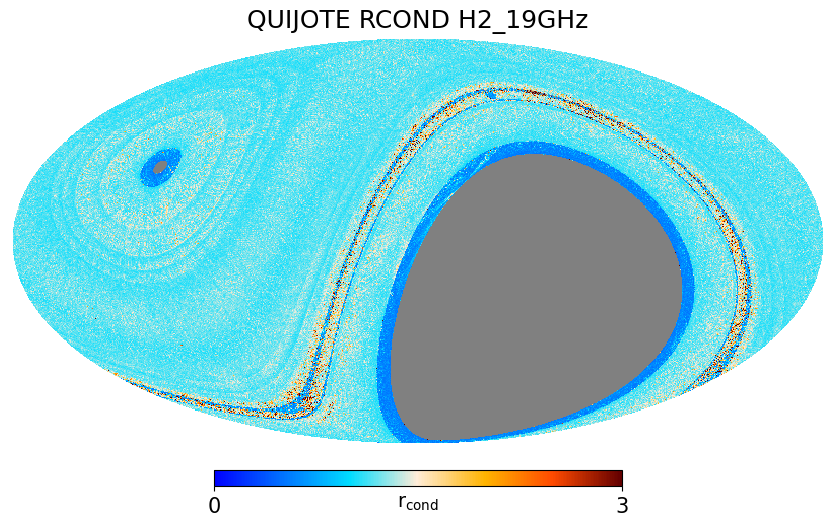}%
    \includegraphics[width=6cm]{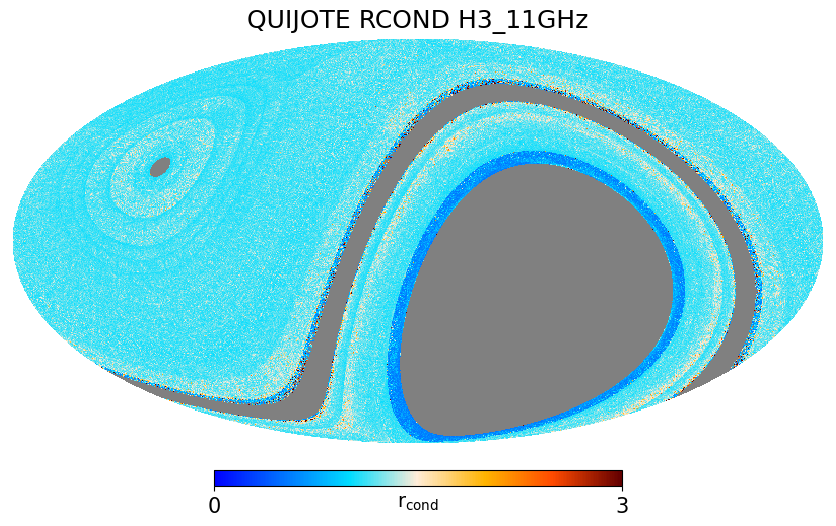}
    \includegraphics[width=6cm]{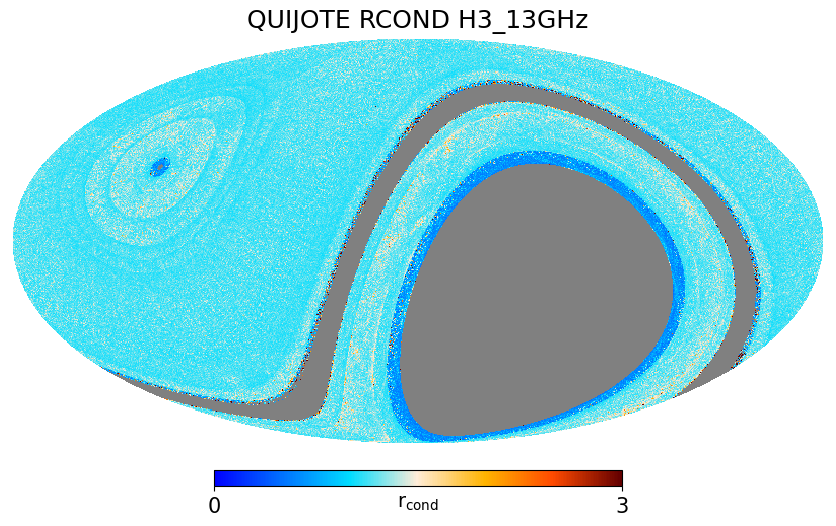}%
    \includegraphics[width=6cm]{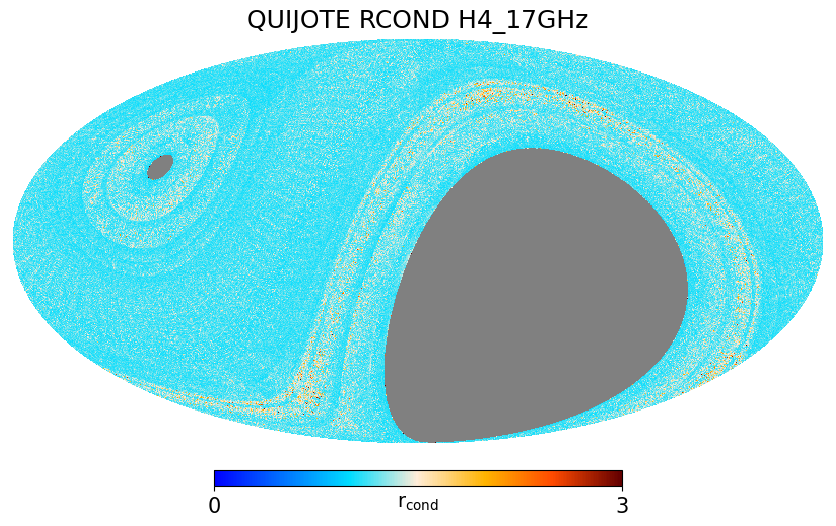}%
    \includegraphics[width=6cm]{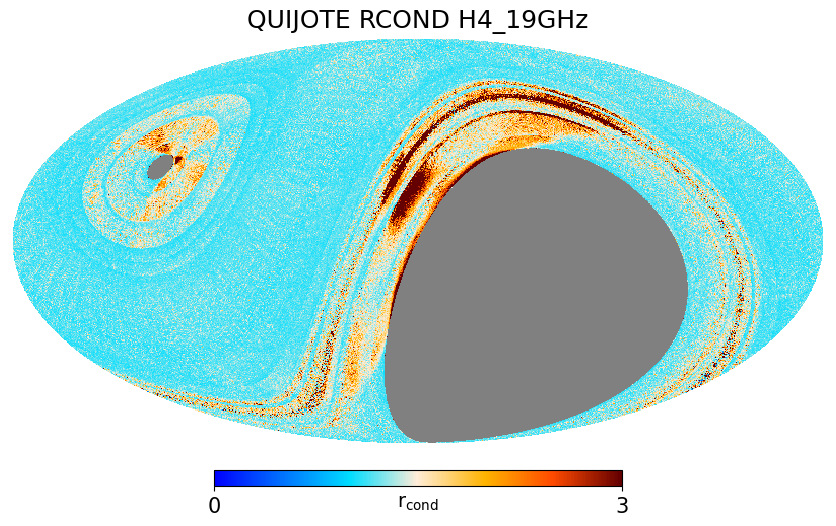}
    \caption{QUIJOTE MFI wide survey $\rcond$ maps for all four horns. During the post-processing stage, all pixels with $\rcond > 3$ are removed from the final wide survey polarization maps.  }
    \label{fig:rcond}
    \end{figure*}

\begin{figure*}
    \centering
    \includegraphics[width=6cm]{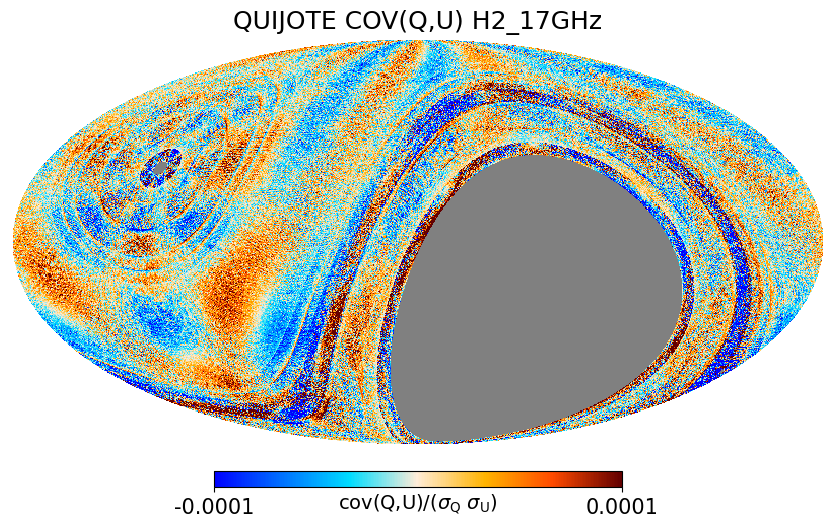}%
    \includegraphics[width=6cm]{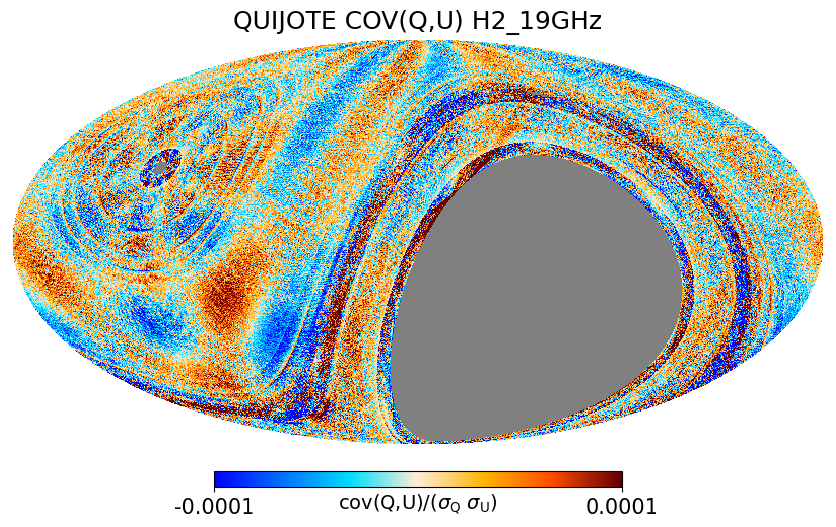}%
    \includegraphics[width=6cm]{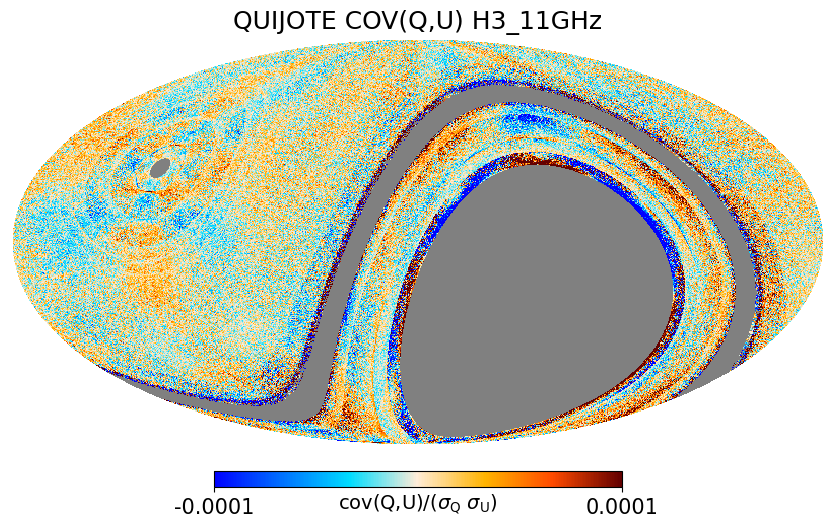}
    \includegraphics[width=6cm]{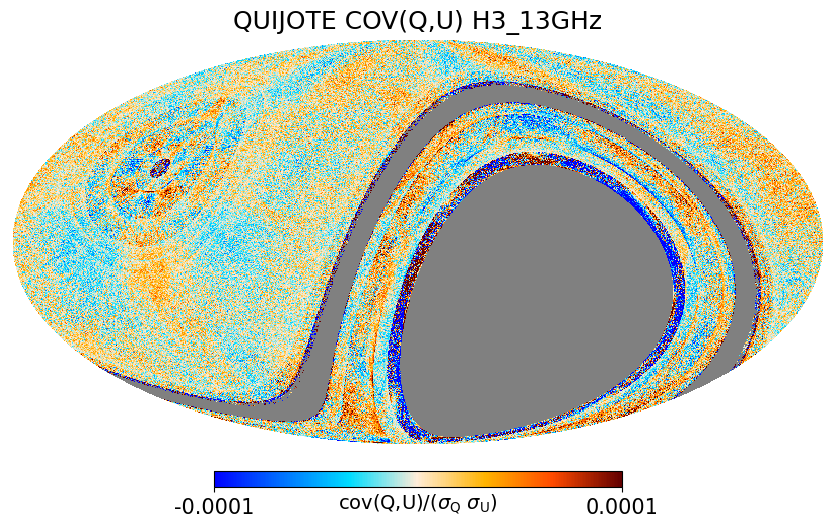}%
    \includegraphics[width=6cm]{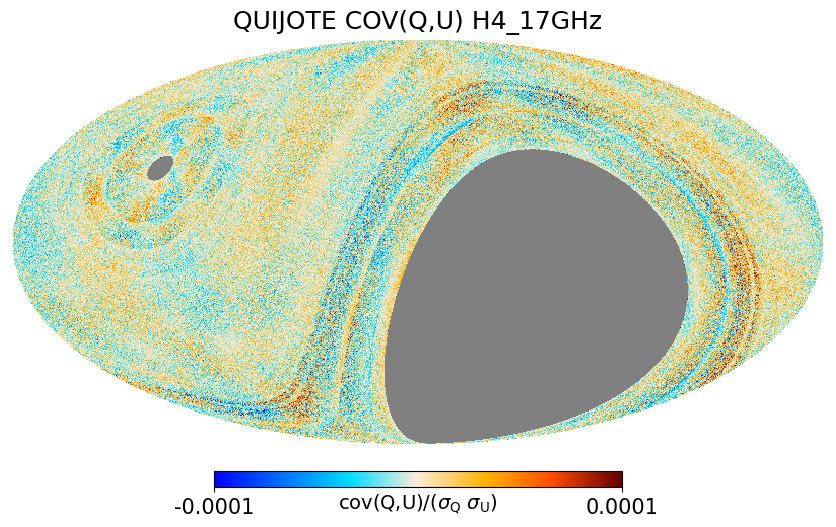}%
    \includegraphics[width=6cm]{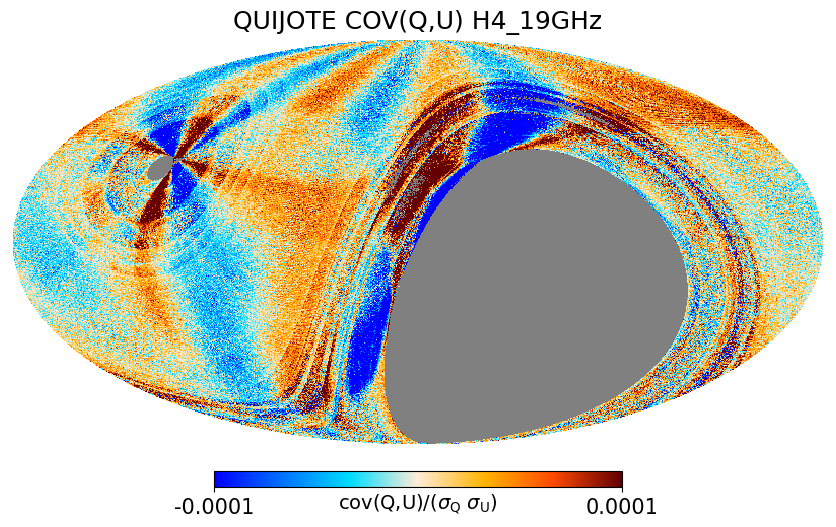}
    \caption{QUIJOTE MFI wide survey normalized covariance ($cov(Q,U)$) maps for all four horns.   }
    \label{fig:covqu}
    \end{figure*}

    Figures~\ref{fig:h2maps}, \ref{fig:h3maps} and \ref{fig:h4maps} show the final MFI wide survey maps at their original resolution (quoted as beam FWHM in  Table~\ref{tab:mfi_parameters}), obtained for horns 2, 3 and 4 respectively. The intensity maps of horns 2 and 4 show some large angular-scale residual patterns, particularly visible in the highest frequency map (19\,GHz). These are due to a combination of residual instrumental and atmospheric $1/f$ noise. 
    Figures~\ref{fig:h2wei}, \ref{fig:h3wei} and \ref{fig:h4wei} show the corresponding weight maps at the original resolution. 
    Figures~\ref{fig:h2nhits}, \ref{fig:h3nhits} and \ref{fig:h4nhits} show the maps with the number of individual TOD samples in each pixel (the so called "hit maps", $\nhit$). They correspond to the total number of 40\,ms samples in each \healpix\ pixel of $\nside=512$ resolution. 
    The ring structures correspond to lines of constant declination, and indicate the edges of the declination limits of observations performed at different elevations. Due to projection effects, the number of hits is significantly larger in those boundaries. 
    In the low declination band of the maps, particularly for negative declinations, the number of hits is significantly lower due to the combined effect of smaller number of observations at low elevations (mainly $30^\circ$, $35^\circ$ and $40^\circ$) and projection effects. We recall that the number of hits in intensity is larger than in polarization, due to the fact that some intensity data are not used in polarization, as shown in Table~\ref{tab:elevations} (period 1 is not used for any polarization maps, data from period 2 are not used in polarization for horn 4, and data from period 5 are not used in polarization for horn 2). 
    Finally, Fig.~\ref{fig:rcond} shows the $\rcond$ maps in polarization, and Fig.~\ref{fig:covqu} shows the normalized covariance $cov(Q,U)$, both at original resolution. 
    
    \subsection{Signal-to-noise of the QUIJOTE MFI maps}
    \label{app:snr}
    From the maps at original resolution shown in Figs.~\ref{fig:h2maps}--\ref{fig:h4maps}, and the noise variance maps estimated from the inverse of the weights presented in Figs.~\ref{fig:h2wei}--\ref{fig:h4wei} and rescaled by the factors reported in Table~\ref{tab:noise}, we can produce signal-to-noise maps for the MFI wide survey. To this end, we downgrade these maps to a \healpix\ resolution of $\nside =64$, which roughly corresponds to the beam size of the maps. Table~\ref{tab:snr} presents some basic statistics about the fraction of $\nside =64$ pixels in the maps observed about a certain signal-to-noise significance. As a reference, the 11\,GHz polarized intensity map has 52\,\% of its pixels with a signal-to-noise ratio larger than 3. 
    
    \begin{table}
    \caption{Fraction of $\nside = 64$ pixels with signal-to-noise ratio (SNR) above a certain threshold in the four QUIJOTE-MFI frequency maps (horns 2 and 4 have been combined).  We report the SNR both for the intensity (I) and the (noise debiased) polarized intensity ($P=\sqrt{Q^2+U^2}$) maps. }
    \label{tab:snr}
    \centering
    \begin{tabular}{@{}lcccc}
    \hline
         &   11\,GHz   &13\,GHz   &  17\,GHz & 19\,GHz \\
    \hline
     & \multicolumn{4}{c}{Intensity (I)}\\
    \cline{2-5}
     SNR$>1$    &   0.88  &    0.90 &     0.86 &     0.82\\
     SNR$>2$    &   0.78  &    0.81 &     0.72 &     0.64\\
     SNR$>3$    &   0.70  &    0.73 &     0.59 &     0.49\\
     SNR$>4$    &   0.64  &    0.66 &     0.48 &     0.36\\
     SNR$>5$    &  0.58  &    0.60 &     0.39 &     0.26\\
    \hline
     & \multicolumn{4}{c}{Polarized intensity (P)}\\
     \cline{2-5}
     SNR$>1$  &        0.87  &    0.82  &    0.69 &     0.70\\
     SNR$>2$   &       0.70  &    0.60  &    0.38 &     0.38\\
     SNR$>3$    &      0.52  &    0.42  &    0.19 &     0.16\\
     SNR$>4$     &     0.38  &    0.29  &    0.10 &     0.06\\
     SNR$>5$      &    0.29  &    0.21  &    0.06 &     0.02\\
    \hline
    \end{tabular}
    \end{table}

    \section{Impact in the polarization TOD of an error in the determination of the \lowercase{$r$}-factor }
    \label{app:rfactors}
    
    We illustrate this effect using the particular case of uncorrelated channels in the first MFI configuration, but the result is equivalent for correlated channels and for all MFI configurations.  We follow the notation introduced in \cite{mfipipeline}, and used in equation~\ref{eq:mfi_response_i_u}. Following the notation of \cite{Jarosik2003}, the MFI response for the two uncorrelated channels, $x$ and $y$, in the first MFI configuration is given by 
    \begin{align}
    V_{\rm x}  &= \frac{s_{\rm x} g_1^2}{2} \Big[ I +\rho_{\rm x} (Q \cos \theta - U \sin \theta ) \Big] \\
    V_{\rm y}  &= \frac{s_{\rm y} g_2^2}{2} \Big[ I +\rho_{\rm y} (-Q \cos \theta + U \sin \theta ) \Big] 
    \end{align}
    where $\theta$ stands for the argument of the cosine and sine in MFI receivers (i.e. $\theta = 4\theta_{\rm pm}+2\gamma_{\rm p}$, as in equations~\ref{eq:mfi_response_u} and  \ref{eq:mfi_response_c}), $s_{\rm x}$ and $s_{\rm y}$ represent the responsivities of the detectors in the two branches, $g_1$ and $g_2$ represent the voltage gains of the two amplifiers in each MFI polarimeter, and $\rho_{\rm x}$ and $\rho_{\rm y}$ are the polar efficiencies in each branch. The $r$-factor is defined as 
    \begin{equation}
    r_{\rm u}\equiv \frac{ s_{\rm x}g_1^2}{s_{\rm y}g_2^2}.
    \end{equation}
    
    If we are using an incorrect $r$-factor $r'_{\rm u} = r_{\rm u} + \epsilon$, where $r_{\rm u}$ is the correct underlying value, then we have 
    \begin{equation}
    V_{\rm x} - r'_{\rm u} V_{\rm y} =  s_{\rm x} g_1^2 \Bigg[  \Big(\frac{\rho_{\rm x} +\rho_{\rm y} }{2} + \frac{\epsilon}{2r_{\rm u}}\rho_{\rm y}\Big) (Q \cos \theta - U \sin \theta ) - \frac{\epsilon}{2r_{\rm u}}I  \Bigg].
    \end{equation}
    We find that this error on the $r$-factor translates into an effective modification of the polar efficiency, and the appearance of a constant offset factor in the polarization timeline. For the particular case of $\rho_{\rm x}=\rho_{\rm y}$, then the effective polar efficiency is rescaled by the factor
    \begin{equation}
    \rho_{\rm x} \rightarrow \rho_{\rm x} \Big(1 +  \frac{\epsilon}{2r_{\rm u}} \Big).
    \end{equation}
    Finally, we note that for the intensity timeline, the same effect generates an overall calibration shift, and a small polarization-to-intensity leakage term.
    The first term is absorbed once we carry our a recalibration of the instrument, while the second one can be safely ignored, as the polarization fraction of the sky emission is already small (typically well below 10 per cent). 
    
    \section{Power spectrum estimators for MFI wide survey maps}
    \label{app:aps}
    Throughout this paper, we have been using two power spectrum estimation codes, both based on a pseudo-C$_\ell$ approach: \xpol\ \citep{xpol} and \namaster\ \citep{namaster}. In this appendix, we show that both methods produce consistent results for the typical sky masks adopted in this paper. 
    For this comparison, we take as a reference case the MFI 11\,GHz wide survey map and the default QUIJOTE mask (NCP+sat+lowdec) combined with the Galactic cut $|b|>10^\circ$. 
    In addition, we justify the use of the pseudo-spectra approach by comparing these results with those from an optimal estimator based on a fast implementation of a quadratic maximum-likelihood (QML) estimator \citep[{\sc ECLIPSE},][]{eclipse}. Running this QML code is computationally very expensive, so the comparison is limited to this case only. 
    
\begin{figure*}
        \centering
        \includegraphics[width=1.9\columnwidth]{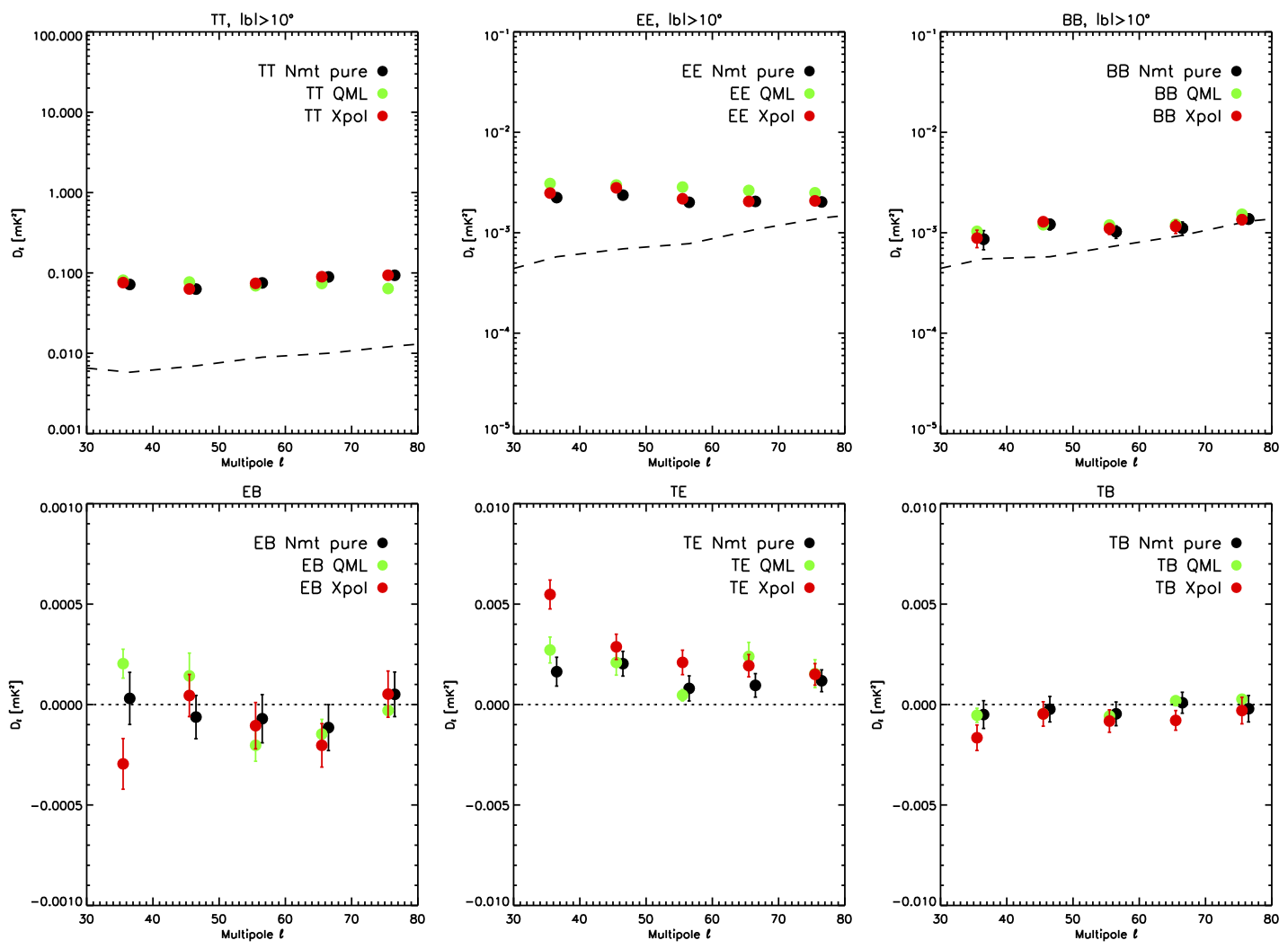}
        \caption{Comparison of the angular power spectrum estimators used in this work. Using the default QUIJOTE analysis mask together with the Galactic cut $|b|>10^\circ$, we evaluate the TT, EE, BB auto-spectra and the TE, EB and TB cross-spectra of the MFI 11\,GHz map, using \xpol\ and \namaster\ (both based on pseudo-C$_\ell$ formalism), and {\sc ECLIPSE} (based on a QML approach). As a reference, in the first three panels we also show the corresponding noise power spectrum. For display purposes, the different data points have been shifted by $\Delta \ell=1$. See text for details. }
        \label{fig:aps_comp}
    \end{figure*}

    Figure~\ref{fig:aps_comp} shows the (binned) low multipoles points of the angular power spectra and cross-spectra ($30 \le \ell \le 80$) computed with those three codes using the same mask. For the case of \namaster\, we use the "purification" option. The conclusion is that, within the multipole range used in this paper ($\ell \ge 30$), all methods provide consistent results, so it is justified to use the pseudo-C$_\ell$ approach for our computations. 
    In this work, we use equally \xpol\ or \namaster\ for TT, EE and BB. For the cross-spectrum analysis in Sect.~\ref{sec:spectra}, we use the \namaster\ code, as it provides slightly closer results to the (optimum) QML solution.

    \section{Spectral index of the MFI 13\,GHz sky emission}
    \label{app:betas13}
\begin{figure}
        \centering
        \includegraphics[width=0.95\columnwidth]{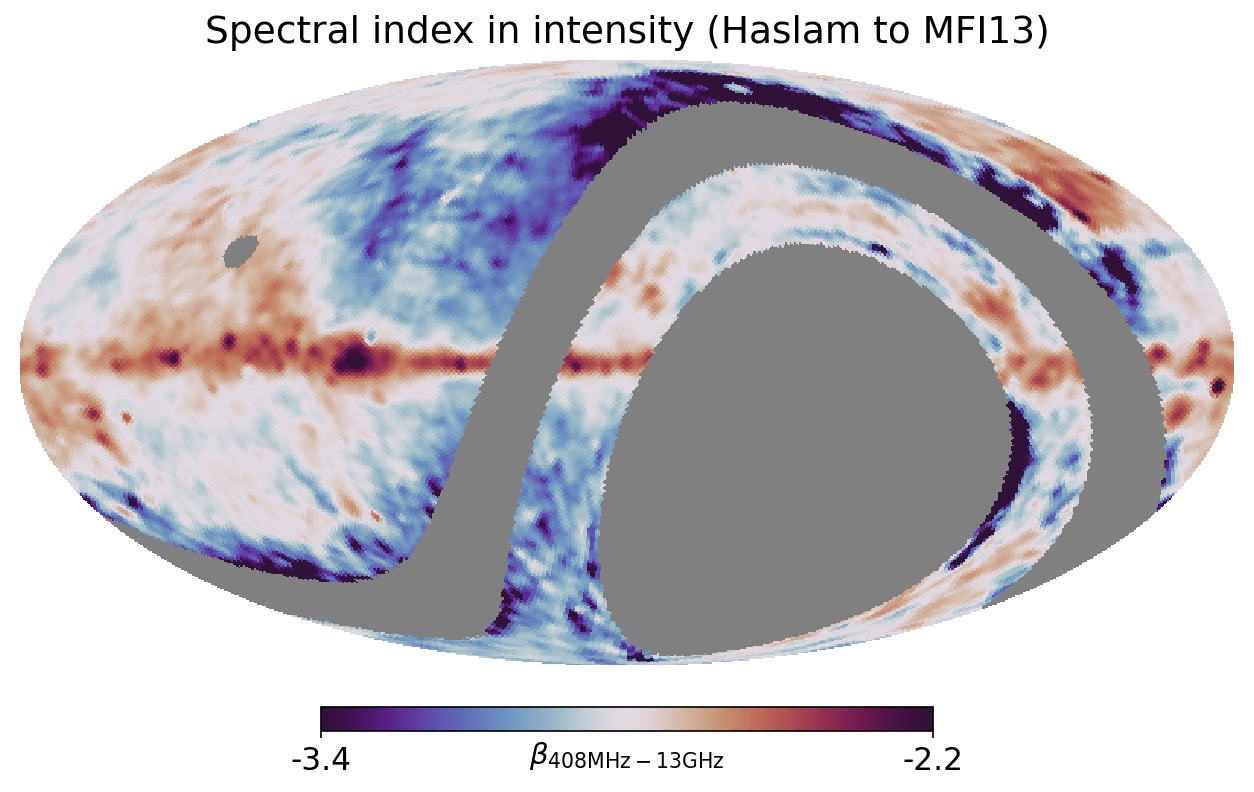}
        \includegraphics[width=0.95\columnwidth]{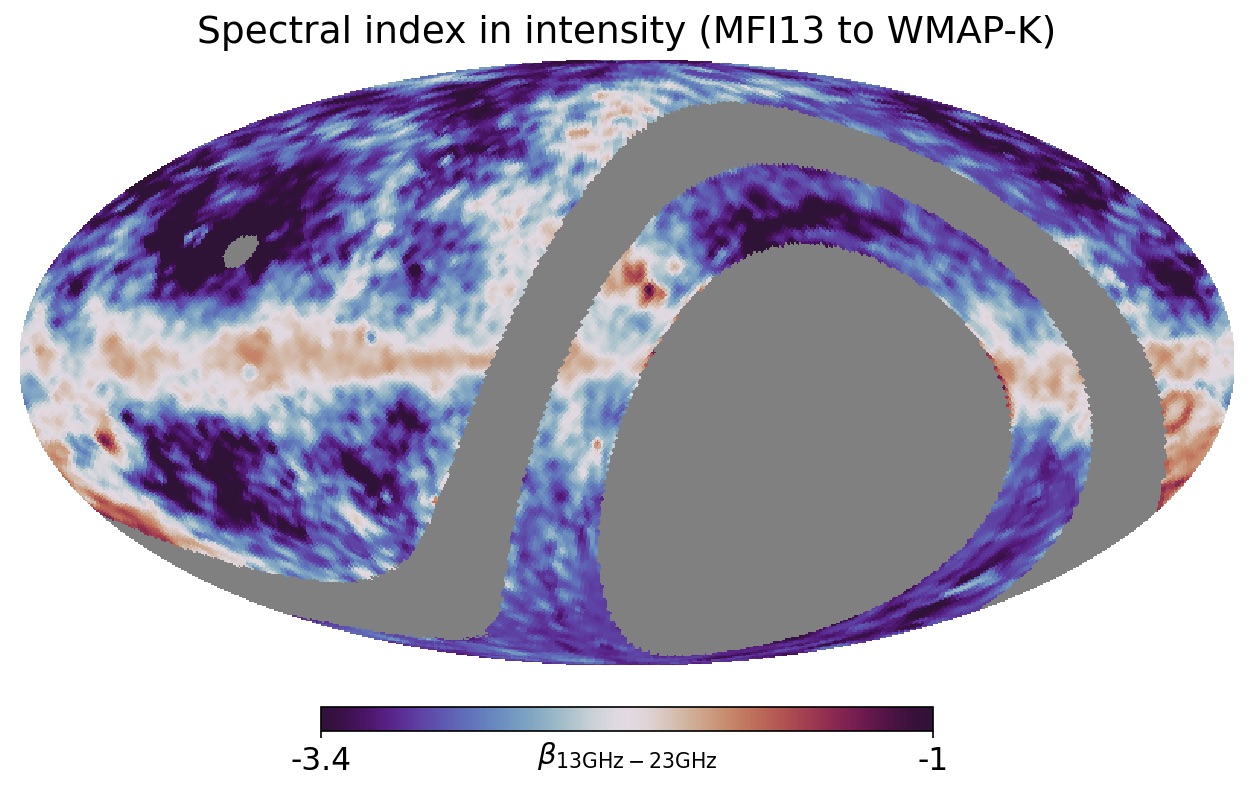}
        \caption{Spectral index of the intensity emission in the QUIJOTE 13\,GHz map. Top: Spectral index of $\beta_{408{\rm MHz}-13{\rm GHz}}$. The average index is approximately $-2.8$. Bottom: Spectral index of $\beta_{13{\rm GHz}-23{\rm GHz}}$. The average spectral index is also $\beta=-2.8$. In this colour scale, dark red corresponds to AME dominated regions.  }
        \label{fig:beta13_int}
    \end{figure}

\begin{figure}
        \centering
        \includegraphics[width=0.95\columnwidth]{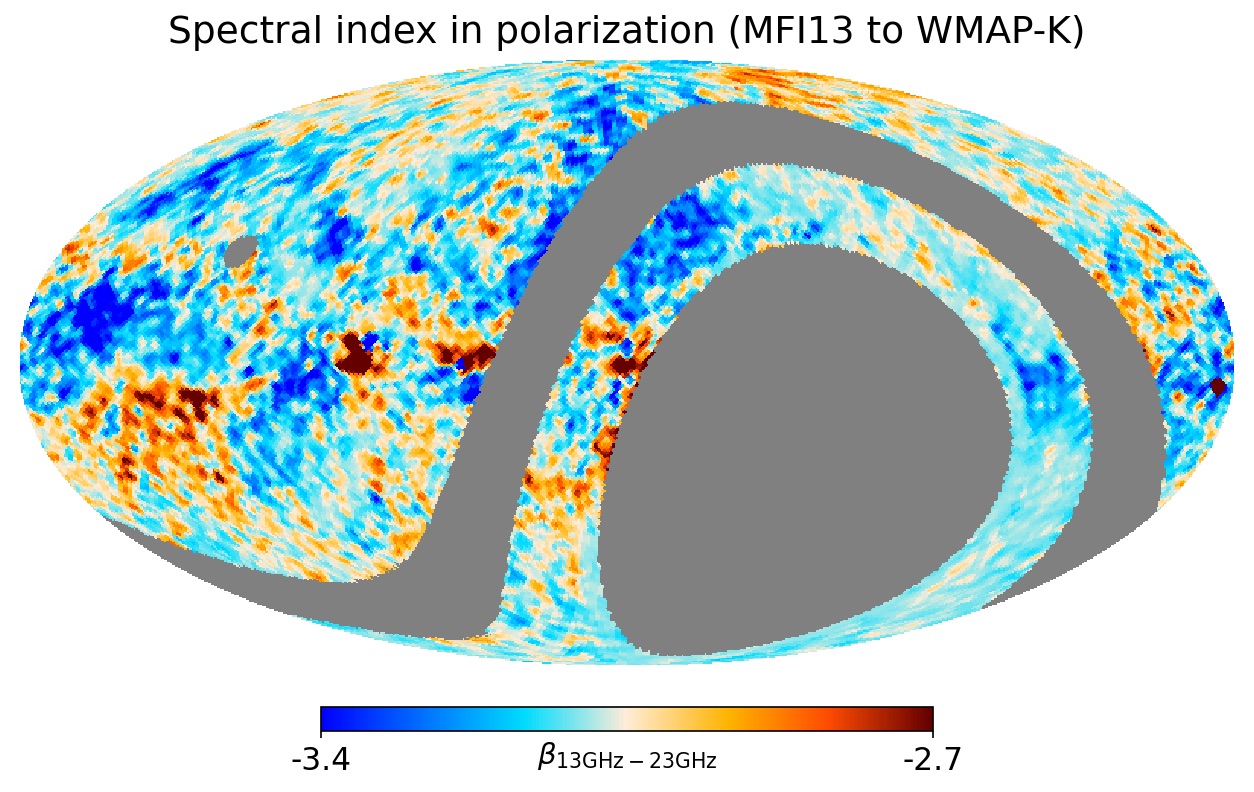}
        \includegraphics[width=0.95\columnwidth]{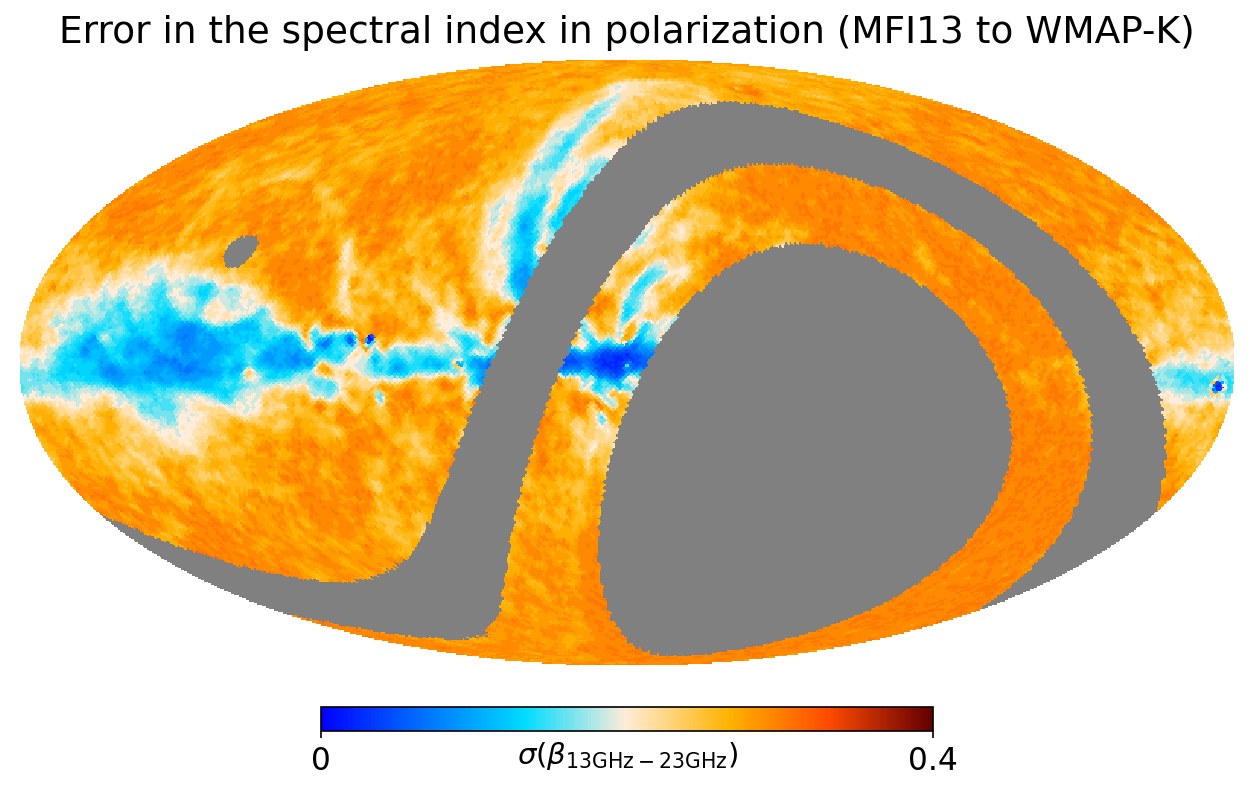}
        \caption{Top: Spectral index map of the polarized emission between QUIJOTE 13\,GHz and WMAP 23\,GHz. Bottom: error map. }
        \label{fig:beta13_pol}
    \end{figure}

\begin{figure}
        \centering
        \includegraphics[width=0.95\columnwidth]{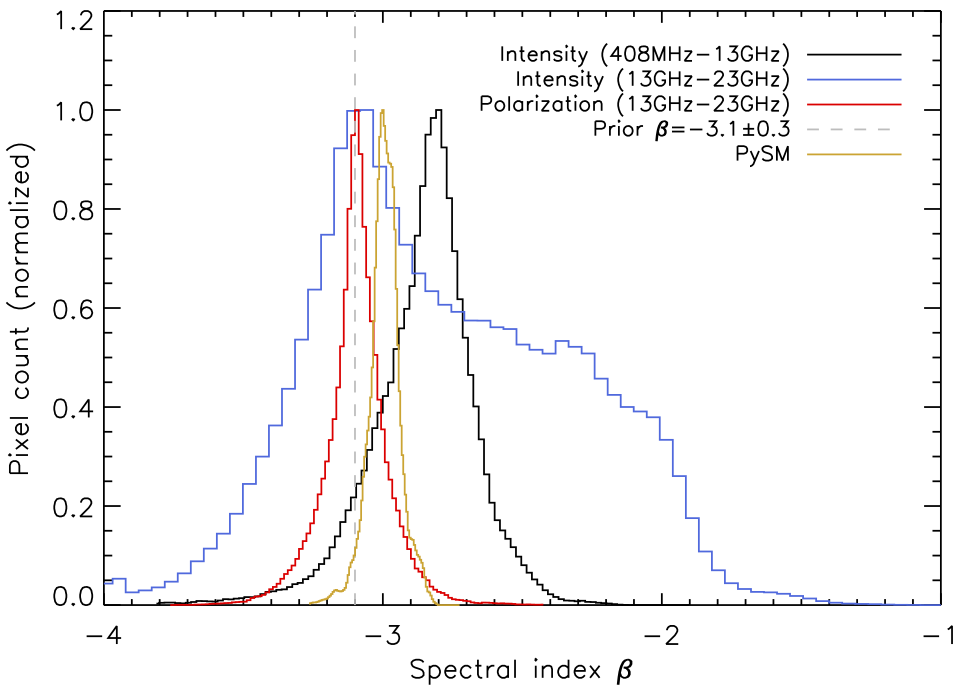}
        \caption{Histogram of spectral index values obtained from Figures~\ref{fig:beta13_int} and \ref{fig:beta13_pol}. We show in dashed lines the mean of the prior adopted in the determination of the spectral index in polarization. For comparison, we also include the histogram of spectral index values from the PySM synchrotron model 1 \citep{PySM2}. }
        \label{fig:betas13}
    \end{figure}

    In this appendix we repeat the same analysis carried out in Sect.~\ref{sec:betas11}, but using now as a reference the MFI 13\,GHz map. 
    Figure~\ref{fig:beta13_int} shows the result for the intensity spectral index in $\beta_{408{\rm MHz}-13{\rm GHz}}$ (top panel) and $\beta_{13{\rm GHz}-23{\rm GHz}}$ (bottom panel), while Fig.~\ref{fig:beta13_pol} presents the polarization spectral index map $\beta_{13{\rm GHz}-23{\rm GHz}}$.
    Again, in Fig.~\ref{fig:betas13} we show the histogram with the distribution of spectral indices in both maps. 
    
    In general, all results are consistent with those obtained using MFI 11\,GHz as the reference map. In intensity, the median spectral index $\beta_{408{\rm MHz}-13{\rm GHz}}$ in the full analysis mask is $-2.83$, with a standard deviation of the values across the map of $0.19$; and the $\beta_{13{\rm GHz}-23{\rm GHz}}$ spectral index has a median of $-2.83$ and a standard deviation of $0.46$. We note that in this latter case, there is a peak around $-3.1$, which is due to the adopted prior. In polarization, the $\beta_{13{\rm GHz}-23{\rm GHz}}$ spectral index presents a median value $-3.09$, and the standard deviation is $0.13$. 
    
    \bsp	
    \label{lastpage}
    \end{document}

%% file: authors_and_institutes_mfiwidesurvey.tex
\author[Rubi\~{n}o-Mart\'{\i}n et al.]{
J.~A. Rubi\~{n}o-Mart\'{\i}n,$^{1,2}$\thanks{E-mail: jalberto@iac.es}
F. Guidi,$^{1,2,3}$
R.~T. G\'enova-Santos,$^{1,2}$
S.~E. Harper,$^{4}$
D. Herranz,$^{5}$
\newauthor
R.~J. Hoyland,$^{1,2}$
A.~N. Lasenby,$^{6,7}$
F. Poidevin,$^{1,2}$
R. Rebolo,$^{1,2,8}$
B. Ruiz-Granados,$^{1,2,9}$
\newauthor
F. Vansyngel,$^{1,2}$
P. Vielva,$^{5}$
R.~A. Watson,$^{4}$
E. Artal,$^{10}$
M. Ashdown,$^{6,7}$
R.~B. Barreiro,$^{5}$
\newauthor
J.~D. Bilbao-Ahedo,$^{5,11}$
F.~J. Casas,$^{5}$
B. Casaponsa,$^{5}$
R. Cepeda-Arroita,$^{4}$
E. de la Hoz,$^{5,11}$
\newauthor
C. Dickinson,$^{4}$
R. Fern\'andez-Cobos,$^{5,12}$ 
M. Fern\'andez-Torreiro,$^{1,2}$
R. Gonz\'alez-Gonz\'alez,$^{1,2}$
\newauthor
C. Hern\'andez-Monteagudo,$^{1,2}$
M. L\'opez-Caniego,$^{13,14}$
C. L\'opez-Caraballo,$^{1,2}$
\newauthor
E. Mart\'inez-Gonz\'alez,$^{5}$
M.~W. Peel,$^{1,2}$
A.~E. Pel\'aez-Santos,$^{1,2}$ 
Y. Perrott,$^{6,15}$ 
L. Piccirillo,$^{4}$
\newauthor
N. Razavi-Ghods,$^{6}$ 
P. Scott,$^{6}$
D. Titterington,$^{6}$ 
D. Tramonte,$^{1,2,16,17}$
R. Vignaga.$^{1,2}$
\\
$^{1}$Instituto de Astrof\'{\i}sica de Canarias, E-38205 La Laguna, Tenerife, Spain\\
$^{2}$Departamento de Astrof\'{\i}sica, Universidad de La Laguna,
E-38206 La Laguna, Tenerife, Spain\\
$^{3}$Institut d'Astrophysique de Paris, UMR 7095, CNRS \& Sorbonne
Universit\'e, 98 bis boulevard Arago, 75014 Paris, France.\\
$^{4}$Jodrell Bank Centre for Astrophysics, Alan Turing Building, 
Department of Physics and Astronomy, School of Natural Sciences, The 
University of Manchester,\\ Oxford Road, Manchester M13 9PL, Manchester, UK\\
$^{5}$Instituto de F\'{\i}sica de Cantabria (IFCA), CSIC-Univ. de Cantabria, Avda. los
Castros, s/n, E-39005 Santander, Spain\\
$^{6}$Astrophysics Group, Cavendish Laboratory, University of Cambridge, 
J J Thomson Avenue, Cambridge CB3 0HE, UK\\
$^{7}$Kavli Institute for Cosmology, University of Cambridge, Madingley Road, Cambridge CB3 0HA, UK\\
$^{8}$Consejo Superior de Investigaciones Cient\'{\i}ficas, E-28006
Madrid, Spain\\
$^{9}$Departamento de F\'{\i}sica. Facultad de Ciencias. Universidad de C\'ordoba. Campus de Rabanales, Edif. C2. Planta Baja.  E-14071 C\'ordoba, Spain.\\
$^{10}$Departamento de Ingenieria de COMunicaciones (DICOM),
Laboratorios de I+D de Telecomunicaciones, Plaza de la Ciencia s/n,
E-39005 Santander, Spain\\
$^{11}$Departamento de F\'{\i}sica Moderna, Universidad de Cantabria,
Avda. de los Castros s/n, 39005 Santander, Spain\\
$^{12}$Departamento de Matem\'aticas, Estad\'{\i}stica y Computaci\'on,
Universidad de Cantabria, Avda. los Castros, s/n, E-39005 Santander,
Spain\\
$^{13}$Aurora Technology for the European Space Agency (ESA), European
Space Astronomy Centre (ESAC),  Camino Bajo del Castillo s/n, 28692
\\Villanueva de la Ca\~nada, Madrid, Spain\\
$^{14}$Universidad Europea de Madrid, 28670, Madrid, Spain\\
$^{15}$School of Chemical and Physical Sciences, Victoria University
of Wellington, PO Box 600, Wellington 6140, New Zealand\\
$^{16}$Purple Mountain Observatory, CAS, No.10 Yuanhua Road, Qixia District, Nanjing 210034, China\\
$^{17}$NAOC-UKZN Computational Astrophysics Center (NUCAC), University
of Kwazulu-Natal, Durban 4000, South Africa.
}

%% file: table1_elevations.tex
\begin{table*}
\caption{List of telescope elevations used for the wide survey observations with
  the QUIJOTE MFI instrument. The second column indicates the total observing time ($T_{\rm observed}$) in hours dedicated to each elevation. Columns 3 to 6 show the
total observing time for the actual subset of observations used for the final intensity ($T_{\rm used, I}$) and polarization
  ($T_{\rm used, P}$) maps. In the later case, different subsets of data are used for each particular horn. 
  Observations are separated in periods (column 7), which correspond to specific epochs (column 8) and instrumental configurations (see text for details). }
\label{tab:elevations}
\centering
\begin{tabular}{@{}crrrrrcc}
\hline
\hline
Elevation ($^\circ$) & $T_{\rm observed}$ (h) & $T_{\rm used,I}$ (h) & $T_{\rm used,P,H2}$ (h) & $T_{\rm used,P,H3}$ (h) & $T_{\rm used,P,H4}$ (h) & Period & Range of Dates \\
\hline
\hline
  30&   121.9&     0.0&     0.0&     0.0&     0.0& 1&    06/2013--07/2013\\
  60&   986.5&   986.5&     0.0&     0.0&     0.0& 1&    05/2013--03/2014\\
  65&   665.2&   665.2&     0.0&     0.0&     0.0& 1&    05/2013--03/2014\\
  70&   394.9&     0.0&     0.0&     0.0&     0.0& 1&    06/2013--03/2014\\
\hline
  30&   829.4&   829.4&   829.4&   829.4&     0.0& 2&    08/2014--03/2015\\
  40&   489.3&   489.3&   489.3&   489.3&     0.0& 2&    08/2014--01/2015\\
  50&   564.7&   564.7&   564.7&   564.7&     0.0& 2&    08/2014--10/2015\\
  60&    91.7&    91.7&    91.7&    91.7&     0.0& 2&    06/2014--09/2014\\
  65&   128.9&   128.9&   128.9&   128.9&     0.0& 2&    08/2014--10/2014\\
\hline
  30&   200.1&     0.0&     0.0&     0.0&     0.0& 5&    08/2016--10/2016\\
  40&   324.6&   324.6&     0.0&   324.6&   324.6& 5&    08/2016--10/2016\\
  50&   488.6&   488.6&     0.0&   488.6&   488.6& 5&    08/2016--10/2016\\
  60&   198.4&   198.4&     0.0&   198.4&   198.4& 5&    08/2016--09/2016\\
\hline
  35&  1998.6&  1998.6&  1998.6&  1998.6&  1998.6& 6&   12/2017--06/2018\\
  50&   326.7&   326.7&   326.7&   326.7&   326.7& 6&   03/2017--04/2017\\
  60&   552.5&   552.5&   552.5&   552.5&   552.5& 6&   12/2016--02/2017\\
  65&   430.7&   430.7&   430.7&   430.7&   430.7& 6&   03/2017--04/2017\\
  70&   400.8&   400.8&   400.8&   400.8&   400.8& 6&   02/2017--04/2017\\
\hline
	TOTAL: &  9193.6&  8476.6& 5813.3& 6824.9 & 4720.9&  &  \\
\hline
\hline
\end{tabular}
\end{table*}

%% file: table_periods.tex
\begin{table*}
\caption{Definition of the four observing periods during which we carried out wide survey observations 
  with the QUIJOTE MFI instrument. Last column indicates the instrument configuration and main changes.
  Configuration 1 corresponds to the original MFI design \citep{MFIstatus12}, while configuration 2 corresponds 
 to the installation of $90^\circ$-hybrids \citep{status2016SPIE}. See text for details. }
\label{tab:periods}
\centering
\begin{tabular}{@{}ccccc}
\hline
\hline
Period & From       & To         & Effective year & Comments \\
       & (dd/mm/yyyy) & (dd/mm/yyyy) &  & \\
\hline
1  & 12/11/2012 & 10/04/2014 & 2013.7 & Configuration 1 for all horns. No extended shielding. \\
2  & 11/04/2014 & 30/11/2015 & 2014.9 & Horn 1 in configuration 2. Extended shielding installed. \\
5  & 01/05/2016 & 14/10/2016 & 2016.7 & All horns in configuration 2. Horn 1 not operative. \\
6  & 15/10/2016 & 01/11/2018 & 2017.8 & All horns in configuration 2. Horn 1 not operative. \\ 
\hline
\hline
\end{tabular}
\end{table*}

%% file: table_general_mfi_parameters.tex
\begin{table*}
\caption{QUIJOTE-MFI basic peformance parameters. Values for 11 and 13\,GHz correspond to horn 3 of MFI. Values for 17 and 19\,GHz have been obtained as the weighted average of horns 2 and 4, using the relative weights described in Table~\ref{tab:weih2h4}. }
\label{tab:mfi_parameters}
\centering
\begin{tabular}{@{}lcccc}
\hline
\hline
                                                   Parameter  &  11 GHz  &  13 GHz  &  17 GHz  &  19 GHz\\

\hline
\hline
MFI horns contributing to these bands      &     3    &      3    &   2,4    &    2,4 \\
Centre frequency (nominal), $\nu_0$ (GHz)  &    11.1  &    12.9   &   16.8	&   18.8  \\
Effective frequency for $\alpha=-1$, $\nu_e(\alpha=-1)$ (GHz)  &   10.98  &   12.89   &  16.85	&  18.85  \\
                                             Bandwidth (GHz)  &    2.17  &    2.20   &	 2.24	&   2.34  \\
                                          Beam FWHM (arcmin)  &   55.38  &   55.84   &  38.95	&  40.32  \\
      Main beam solid angle, $\Omega_{\rm mb}$ ($10^{-4}$sr)  &   2.748  &   2.781   &  1.362	&  1.428  \\
                                       Beam ellipticity$^{a}$, $e$  &   0.013  &   0.040   &  0.034	&  0.035  \\
       Antenna sensitivity, $\Gamma$ ($\mu$K$_{\rm CMB}$/Jy)  &   961.9  &   703.8   &  847.0	&  645.2  \\
\hline
White-noise level in timelines ($\mu$K$_{\rm CMB}$s$^{1/2}$)  &     858  &     697   &	  773	&    866  \\
Knee frequency $f_{\rm k}$ in polarization (mHz)  &     254  &     198   &	  223	&    556  \\
$1/f$ slope in polarization  &    1.95  &    1.86   &	 1.73	&   1.34  \\
Overall calibration uncertainty I (\%)  &       5  &	 5   &	    5	&      5  \\
Overall calibration uncertainty Q,U (\%)  &       5  &	 5   &	    6	&      6  \\
\hline
\hline
\multicolumn{5}{l}{$^a$ The ellipticity is defined here as $e=1 - {\rm FWHM}_{\rm min}/{\rm FWHM}_{\rm max}$.}\\
\end{tabular}
\end{table*}

%% file: table_cc_coeff.tex
\begin{table}
\caption{Colour correction coefficients, $C(\alpha,\nu_0)=c_0+c_1\alpha+c_2\alpha^2$. The colour corrected temperature is obtained as $C(\alpha,\nu_0) T$, being $T$ the uncorrected one. }
\label{tab:mfi_cc}
\centering
\begin{tabular}{@{}lcccc}
\hline
\hline
Band & $\nu_0$ & $c_0$  & $c_1$  & $c_2$ \\
\hline
\hline
11 &  11.1    &   0.981 &  0.0125 & -0.0015  \\
13 &  12.9    &   1.001 &  0.0018 & -0.0012  \\
17 &  16.8    &   1.007 & -0.0022 & -0.0007  \\
19 &  18.8    &   1.007 & -0.0020 & -0.0008  \\
\hline
\hline
\end{tabular}
\end{table}

%% file: table_poleff.tex
\begin{table}
\caption{Polarization efficiency for horns 2, 3 and 4 in period 6. Error bars for all measurements are 2 per cent. See text for details. }
\label{tab:poleff}
\centering
\begin{tabular}{@{}lcc}
\hline
 Channel & $\rho_{\rm corr}$ & $\rho_{\rm uncorr}$ \\
\hline
217 & 0.84 & 0.98 \\
219 & 0.86 & 0.96 \\
311 & 0.89 & 0.98 \\
313 & 0.83 & 0.97 \\
417 & 1.00 & 0.93 \\
419 & 0.99 & 0.91 \\
\hline
\end{tabular}
\end{table}

%% file: table_notation.tex
\begin{table}
\caption{List of periods contributing to each final MFI map per horn. Column 1 indicates the map per horn with the usual notation: the first number indicates the horn/pixel (column 2), and second and third numbers indicate the nominal frequency (column 3). Column 4 shows the list of periods contributing to the map based on the correlated channels $V_{\rm x+y}$ and $V_{\rm x-y}$. Column 5 shows the list of periods used for the map based on the uncorrelated channels $V_{\rm x}$ and $V_{\rm y}$. The final map is the combination of both correlated and uncorrelated maps. }
\label{tab:mfi_notation}
\centering
\begin{tabular}{@{}lcccc}
\hline
Map & Horn/Pixel & Nominal Freq. (GHz) & Corr & Uncorr \\ 
\hline
\multicolumn{5}{c}{Intensity}\\
\hline
311 & 3 & 11 & 1,2,5,6 & 1,2,5,6 \\
313 & 3 & 13 & 1,2,5,6 & 1,2,5,6 \\
217 & 2 & 17 & 1,2,5,6 & 1,2,5,6 \\
219 & 2 & 19 & 1,2,5,6 & 1,2,5,6 \\
417 & 4 & 17 & 1,2,5,6 & 1,2,5,6 \\
419 & 4 & 19 & 1,2,5,6 & 1,2,5,6 \\
\hline
\multicolumn{5}{c}{Polarization}\\
\hline
311 & 3 & 11 & 2,5,6 & 5,6 \\
313 & 3 & 13 & 2,5,6 & 5,6 \\
217 & 2 & 17 & 2,6 & 6 \\
219 & 2 & 19 & 2,6 & 6 \\
417 & 4 & 17 & 5,6 & 5,6 \\
419 & 4 & 19 & 5,6 & 5,6 \\
\hline
\end{tabular}
\end{table}

%% file: table_system_corr.tex
\begin{table}
\caption{Cross-correlation in real space between the half mission difference maps and the final signal maps. Columns 2--4 correspond to the case of half mission maps with common baselines, while columns 5--7 show the results for the case of independent baselines. Error bars are of the order of $0.1$ in all cases. }
\label{tab:systemcorr}
\centering
\begin{tabular}{@{}crrrrrr}
\hline
Channel &  $\alpha_{\rm T}$ & $\alpha_{\rm Q}$ & $\alpha_{\rm U}$ & $\alpha_{\rm T}$ & $\alpha_{\rm Q}$ & $\alpha_{\rm U}$\\  
        &  [$\%$] & [$\%$] &[$\%$] & [$\%$] & [$\%$] & [$\%$]\\ 
\hline
 & \multicolumn{3}{c}{Common baselines} & \multicolumn{3}{c}{Indep. baselines}\\
\hline
  217&    $0.2$ &   $0.5$  &  $0.9$  &     $-4.8$&    $0.4$ &    $0.8$\\    
  219&    $0.3$ &   $1.4$ &   $1.3$  &     $-1.8$&    $1.4$ &    $1.3$ \\   
  311&   $-0.1$ &  $0.0$ &   $0.7$  &      $0.1$ &  $-0.6$ &   $0.8$\\   
  313&   $-0.2$ &  $ 0.3$ &   $0.9$ &      $-0.1$&   $0.5$ &   $1.0$\\    
  417&    $0.3$ &  $-0.2$ &  $-0.6$ &    $-3.0$&   $-0.5$ &  $-0.6$\\    
  419&    $1.2$ &  $-0.1$ &  $0.0$   &    $-0.4$&   $-0.1$ &   $0.2$\\    
\hline
\end{tabular}
\end{table}

%% file: table6_fit2noisespectrum_nochi2.tex
\begin{table}
\caption{Noise levels from the fit to the noise power spectra based on
  the parametric equation~\ref{eq:fit_cl}, computed from the
  half-mission null tests with independent baselines. In polarization, we show the results of the
fit to the EE spectra. Results for BB are fully consistent. }
\label{tab:noiseps}
\centering
\begin{tabular}{@{}ccccc}
\hline
Channel & $C_{\rm w}$ & $\sigmadeg$ & $\alpha$ & $\ell_{\rm k}$ \\
 & [mK$^2$\,sr] & [$\mu$K] & &  \\
\hline
\multicolumn{5}{c}{Intensity (TT)}\\
\hline
217 & $6.13\times 10^{-6}$ & 133.5 & 1.50 & 228.8  \\
219 & $1.05\times 10^{-5}$ & 174.5 & 1.82 & 229.3  \\
311 & $2.56\times 10^{-6}$ &  86.3 & 1.27 & 221.4  \\
313 & $1.29\times 10^{-6}$ &  61.3 & 1.60 & 192.5  \\
417 & $1.07\times 10^{-5}$ & 176.4 & 1.45 & 230.4  \\
419 & $1.40\times 10^{-5}$ & 201.7 & 1.82 & 243.6  \\
\hline
\multicolumn{5}{c}{Polarization (EE)}\\
\hline
217 & $1.21\times 10^{-6}$ & 59.4 & 1.20 & 145.0  \\
219 & $1.87\times 10^{-6}$ & 73.7 & 1.30 & 173.7  \\
311 & $6.13\times 10^{-7}$ & 42.2 & 1.24 & 86.0  \\
313 & $4.95\times 10^{-7}$ & 37.9 & 1.35 & 75.3  \\
417 & $4.42\times 10^{-7}$ & 35.8 & 1.06 & 53.5  \\
419 & $5.02\times 10^{-7}$ & 38.2 & 1.24 & 73.2  \\
\hline
\end{tabular}
\end{table}

%% file: table3_noise.tex
\begin{table}
\caption{Recalibration factor of the noise standard deviation included in the weight maps, based on null test maps. }
\label{tab:noise}
\centering
\begin{tabular}{@{}ccccccc}
\hline
Map & H2,17 & H2,19 & H3,11 & H3,13 & H4,17 & H4,19 \\
\hline
 & \multicolumn{6}{c}{Half mission null test}\\
\hline
I &     4.974 &     5.596 &     3.424 &     3.016 &     4.695 &     5.108 \\ 
Q &     1.723 &     2.001 &     1.471 &     1.372 &     1.285 &     1.292 \\ 
U &     1.723 &     1.999 &     1.473 &     1.373 &     1.285 &     1.292 \\ 
\hline
 & \multicolumn{6}{c}{Ring null test}\\
\hline
I &     4.896 &     5.449 &     3.410 &     2.993 &     4.641 &     4.978 \\ 
Q &     1.717 &     1.994 &     1.471 &     1.370 &     1.286 &     1.289 \\ 
U &     1.716 &     1.991 &     1.473 &     1.370 &     1.285 &     1.291 \\ 
\hline
\end{tabular}
\end{table}

%% file: table4_sigma0.tex
\begin{table}
\caption{Characteristic value of the sensitivity for each channel, $\sigma_0$, in units of mK\,s$^{1/2}$. Based on the half-mission null test maps. }
\label{tab:noise2}
\centering
\begin{tabular}{@{}ccccccc}
\hline
Map & H2,17 & H2,19 & H3,11 & H3,13 & H4,17 & H4,19 \\
\hline
 & \multicolumn{6}{c}{Half mission null test}\\
\hline
I & 5.896 & 7.445 & 3.481 & 2.422 & 7.939 & 8.427 \\ 
Q & 1.878 & 2.280 & 1.371 & 1.188 & 1.101 & 1.059 \\ 
U & 1.875 & 2.273 & 1.372 & 1.188 & 1.100 & 1.064 \\ 
\hline
\end{tabular}
\end{table}

%% file: table5_rms.tex
\begin{table}
\caption{Mean noise figures in the final MFI maps, in units of $\sigmadeg$ ($\mu$K per 1-degree beam), using real-space statistics. A variance map is estimated based on the half-mission nulltest maps, computing the variance within a circle of 1 degree radius. Those values are then converted into $\sigmadeg$. }
\label{tab:noise3}
\centering
\begin{tabular}{@{}ccccccc}
\hline
Map & H2,17 & H2,19 & H3,11 & H3,13 & H4,17 & H4,19 \\
\hline
 & \multicolumn{6}{c}{Half mission null test}\\
\hline
I &   136.6 &   184.8 &    88.3 &    65.0 &   184.2 &   214.8 \\ 
Q &    59.4 &    76.4 &    40.5 &    35.9 &    34.2 &    32.7 \\ 
U &    59.4 &    76.1 &    40.6 &    35.9 &    34.1 &    32.9 \\ 
\hline
\end{tabular}
\end{table}

%% file: table_calibration.tex
\begin{table*}
\caption{Accuracy of the calibration in the QUIJOTE MFI wide survey data. Second column indicates if the type of uncertainty is applicable to intensity (I) and/or to polarization (P) maps. }
\label{tab:summarycal}
\centering
\begin{tabular}{@{}lcccccll}
\hline
\hline
Type of uncertainty & Applies to  & 11\,GHz & 13\,GHz & 17\,GHz & 19\,GHz & Method & Reference \\
\hline
Calibration model  & I,P & 5\,\%          & 5\,\% & 5\,\% & 5\,\% & Model for calibrators & Sect.~\ref{subsec:calmodel}\\
Colour corrections$^a$ & I,P & $0.5\,\%$ & $0.5\,\%$  &$1\,\%$ & $1\,\%$& Bandpass measurements & Sect.~\ref{subsec:coco} \\
Beam uncertainty   & I,P & $2\,\%$  &  $2\,\%$  &  $2\,\%$  &   $2\,\%$ & CST beam model, Tau A & Sect.~\ref{subsec:beams}\\
Zero level [mK]    & I   & $-0.74\pm 0.20$ & $-0.59\pm 0.22$ & 0 & 0 & Plane-parallel model & Sect.~\ref{sec:zerolevels}\\
\hline
I$\rightarrow$P leakage  & P & $0.65\,\%$  & $0.4\,\%$  & $0.8\,\%$ & $0.9\,\%$ & Cygnus area  & Sect.~\ref{subsec:itop} \\
Polarization efficiency  & P & $3\,\%$  & $3\,\%$  & $4\,\%$ & $4\,\%$ & Lab measurements, Tau A & Sect.~\ref{subsec:poleff} \\
Polarization angle (deg) & P & $0.6$ & $0.9$ & $1.0$ & $3.2$ & Tau A, WMAP/Planck & Sect.~\ref{sec:polangle}\\
\hline
Unknown systematics: & \\
$\qquad$ Real space ($\mu$K/beam) & I & $<53$ & $<49$  & $<118$ & $<224$ & Null tests at $\nside=64$ & Sect.~\ref{sec:sysreal}  \\
$\qquad$ Real space ($\mu$K/beam) & P & $<12$ & $<15$  & $<10$  & $<13$ & Null tests at $\nside=64$ & Sect.~\ref{sec:sysreal} \\
$\qquad$ Harmonic space ($30<\ell<200$)  & I & $0.2$\,\% & $0.3$\,\% & $0.5$\,\% & $0.7$\,\% & Null tests& Sect.~\ref{sec:sysharm} \\ 
$\qquad$ Harmonic ($30<\ell<200$)  & P & $3$\,\% & $4$\,\% & $6$\,\% & $6$\,\% & Null tests& Sect.~\ref{sec:sysharm} \\ 
\hline
Overall calibration error$^b$ & I & 5\,\%  & 5\,\%  & 5\,\%  & 5\,\%  &  & \\
Overall calibration error$^b$ & P & 5\,\%  & 5\,\%  & 6\,\%  & 6\,\%  &  & \\
\hline
\multicolumn{5}{l}{$^a$ These numbers should be multiplied by
  $|\alpha+0.3|$, being $\alpha$ the spectral index of the source. }\\
\multicolumn{8}{l}{$^b$ Obtained as the maximum value of the following errors: for intensity, calibration, beam uncertainty and unknown systematics in harmonic space; }\\
\multicolumn{8}{l}{and for polarization, we add also I$\rightarrow$P leakage and polar efficiency.}\\
\end{tabular}

\end{table*}

%% file: table_systematics.tex
\begin{table}
\caption{Systematic effects in the MFI wide survey maps, evaluated in the maps degraded to $\nside=64$. The excess signal (last column) is computed as the quadratic difference between the values for half and ring null test difference maps. See text for details. }
\label{tab:systematics}
\centering
\begin{tabular}{@{}ccrrrr}
\hline
Channel &  T,Q,U & p-p (half)  & rms (half) & rms (ring) & Excess rms \\
        &        & [$\mu$K] & [$\mu$K] & [$\mu$K] & [$\mu$K] \\
\hline
  217 &   T &     1177.8 &      249.8 &      224.6 &      109.3 \\
  217 &   Q &      410.8 &       88.8 &       86.9 &       18.5 \\
  217 &   U &      417.0 &       87.7 &       86.7 &       12.8 \\
  219 &   T &     1736.3 &      363.0 &      297.8 &      207.6 \\
  219 &   Q &      552.6 &      116.0 &      113.2 &       25.5 \\
  219 &   U &      539.7 &      115.1 &      113.4 &       19.2 \\
\hline
  311 &   T &      736.7 &      153.8 &      144.3 &       53.3 \\
  311 &   Q &      283.4 &       59.3 &       58.5 &       10.1 \\
  311 &   U &      282.5 &       59.5 &       58.3 &       12.0 \\
  313 &   T &      538.5 &      113.0 &      101.7 &       49.3 \\
  313 &   Q &      241.9 &       51.2 &       49.2 &       14.3 \\
  313 &   U &      239.1 &       51.0 &       48.7 &       15.1 \\
\hline
  417 &   T &     1586.5 &      332.8 &      304.5 &      134.3 \\
  417 &   Q &      210.4 &       45.0 &       44.5 &        6.8 \\
  417 &   U &      209.8 &       44.8 &       44.6 &        4.1 \\
  419 &   T &     2053.8 &      429.9 &      352.1 &      246.6 \\
  419 &   Q &      232.9 &       48.8 &       48.2 &        7.3 \\
  419 &   U &      233.2 &       49.5 &       48.2 &       11.3 \\
\hline
\end{tabular}
\end{table}

%% file: table_polangle.tex
\begin{table}
\caption{Comparison of the reconstructed angles in the QUIJOTE MFI wide survey data to WMAP-K (column 2), LFI30 (column 3) and MFI 311 (column 4). See text for details. }
\label{tab:polangle}
\centering
\begin{tabular}{@{}lrrr}
\hline
 Channel &  \multicolumn{1}{c}{WMAP-K}  &  \multicolumn{1}{c}{LFI30}    &  \multicolumn{1}{c}{MFI-311}  \\
         & \multicolumn{1}{c}{(deg)} & \multicolumn{1}{c}{(deg)} & \multicolumn{1}{c}{(deg)}\\
\hline
 217  &  $-2.8 \pm 1.5$  &  $-3.3 \pm 1.5$  &  $-4.2 \pm 1.5$  \\
 219  &  $ 0.8 \pm 3.0$  &  $ 0.4 \pm 3.0$  &  $-0.4 \pm 2.9$  \\
 311  &  $ 0.6 \pm 0.6$  &  $-0.5 \pm 0.6$  &  \multicolumn{1}{c}{--}  \\
 313  &  $-1.2 \pm 0.6$  &  $-2.0 \pm 0.6$  &  $-2.2 \pm 0.6$  \\
 417  &  $-1.2 \pm 1.0$  &  $-1.6 \pm 1.0$  &  $-2.3 \pm 1.0$  \\
 419  &  $ 0.5 \pm 3.6$  &  $ 0.0 \pm 3.6$  &  $-0.9 \pm 3.5$  \\
\hline
 Comb. 17\,GHz  &  $-1.6 \pm 0.9$  &  $-2.2 \pm 0.9$  &  $-2.8 \pm 0.9$  \\
 Comb. 19\,GHz  &  $ 0.9 \pm 3.2$  &  $ 0.3 \pm 3.2$  &  $-0.5 \pm 3.1$  \\
\hline
\end{tabular}
\end{table}

%% file: table_powerspectra.tex
\begin{table}
\caption{Best fit results obtained after fitting the model in equation~\ref{eq:cl_model} to the wide survey EE and BB power spectra at 11\,GHz, in the multipole range $30 < \ell <200$. No colour corrections were applied when fitting the spectra. }
\label{tab:ps}
\centering
\begin{tabular}{@{}lrrr}
\hline
Mask  &  $|b|>5^\circ$  & $|b|>10^\circ$ & $|b|>20^\circ$ \\
\hline
$f_{\rm sky}$ & 0.38 & 0.34 & 0.27\\
\hline
& \multicolumn{3}{c}{EE and BB fitted separately}\\
\hline
$A_{\rm EE}$ [$\mu$K$^2$] & $1.52\pm 0.15$ & $1.05\pm 0.18$ & $0.81\pm 0.19$ \\ 
$A_{\rm BB}$ [$\mu$K$^2$] & $0.52 \pm 0.15$ & $0.20\pm 0.12$ & $0.18\pm 0.13$ \\ 
$\alpha_{\rm EE}$ & $-3.00\pm 0.16$ & $-2.72\pm 0.26$ & $-2.96\pm 0.36$ \\
$\alpha_{\rm BB}$ & $-3.08\pm 0.42$ & $-3.13\pm 0.87$ & $-3.12\pm 1.03$ \\
$c_{\rm EE}$ [$\mu$K$^2$] & $0.07\pm 0.09$ & $-0.13 \pm 0.11$ & $-0.09\pm 0.12$ \\
$c_{\rm BB}$ [$\mu$K$^2$] & $0.10\pm 0.09$ & $-0.06\pm 0.09$ & $-0.09\pm 0.09$ \\
$A_{\rm BB}/A_{\rm EE}$ &  $0.34 \pm 0.10$ & $0.19\pm 0.12$ & $0.22\pm 0.18$\\
\hline
& \multicolumn{3}{c}{Joint EE and BB analysis}\\
\hline
$A_{\rm EE}$ [$\mu$K$^2$] & $1.49\pm 0.12$ & $0.97\pm 0.13$ & $0.78\pm 0.14$ \\
$\alpha_{\rm EE}$ ($=\alpha_{\rm BB}$) & $-3.04\pm 0.13$ & $-2.83\pm 0.21$ & $-3.03\pm 0.29$ \\
$c_{\rm EE}$ ($=c_{\rm BB}$) [$\mu$K$^2$]  & $0.09\pm 0.06$ & $-0.08 \pm 0.06$ & $-0.08\pm 0.07$ \\
$A_{\rm BB}/A_{\rm EE}$ &  $0.36 \pm 0.04$ & $0.26\pm 0.07$ & $0.26\pm 0.08$ \\
\hline
\hline
\end{tabular}
\end{table}

%% file: table_eb_tb.tex
\begin{table}
\caption{Best fit results obtained after fitting a constant model to the wide survey EB and TB power spectra at 11\,GHz, in the multipole range $30 < \ell <150$. No colour corrections are applied. }
\label{tab:eb_tb}
\centering
\begin{tabular}{@{}lrrr}
\hline
Mask  &  $|b|>5^\circ$  & $|b|>10^\circ$ & $|b|>20^\circ$ \\
\hline
$A_{\rm EB}$ [$\mu$K$^2$] & $-0.014 \pm 0.037$ & $0.002 \pm 0.038$ & $0.043 \pm 0.041$ \\
$A_{\rm EB}/A_{\rm EE}$ ($\ell = 80$) &  $-0.010 \pm 0.025$ & $0.002 \pm 0.038$ & $0.057 \pm 0.059$ \\
\hline
$A_{\rm TB}$ [$\mu$K$^2$] & $-0.17 \pm 0.24$ & $-0.15 \pm 0.20$ & $-0.21 \pm 0.19$ \\
$A_{\rm TB}/A_{\rm EE}$ ($\ell=80$) & $-0.11\pm 0.16$ & $-0.15\pm 0.20$ & $-0.28\pm 0.28$ \\
\hline
\end{tabular}
\end{table}

%% file: table_sources.tex
\begin{table*}
    \caption{Flux densities (Jy), in intensity and in polarization, extracted from the QUIJOTE MFI wide survey maps at one degree resolution on Tau A, Cas A, Cyg A and 3C274. Intensity measurements are based on BF1d photometry, while the polarization measurements used AP1d. For the intensity measurements, inside parentheses we quote the percent deviation of flux densities with respect to predictions from spectral models. Tau A and Cas A values are referred to an effective date corresponding to 1 April 2016. All flux densities include colour corrections. }
    \label{tab:sources_vs_models}
    \centering
    \begin{tabular}{c|c|c|c|c|c|c|c}
    \hline
Source & Stokes & 311 ($11.1$\,GHz) & 313 ($12.9$\,GHz) & 217 ($16.7$\,GHz) & 417 ($17.0$\,GHz) & 219 ($18.7$\,GHz) & 419 ($19.0$\,GHz) \\
\hline
\multirow{3}{*}{Tau A} &I& $ 440.0\pm  0.9$ (-0.8) & $ 427.4\pm  0.8$ (+0.7) & $ 391.2\pm  0.8$ (-0.5) &  $ 393.4\pm  0.8$ (+0.6) &  $ 377.9\pm  0.7$ (-0.6) &  $ 378.8\pm  0.8$ (+0.2) \\
                       &Q& $-29.27\pm 0.51$	   & $-31.20\pm 0.51$	     & $-28.00\pm 0.83$        &  $-28.12\pm 0.43$	  &  $-26.36\pm 1.52$	     &  $-28.42\pm 0.66$	\\
                       &U& $  0.63\pm 0.51$	   & $  0.90\pm 0.73$	     & $  1.43\pm 0.89$        &  $  1.05\pm 0.63$	  &  $  0.34\pm 0.89$	     &  $  1.87\pm 0.59$	\\
\hline
\multirow{3}{*}{Cas A} &I& $ 340.9\pm  1.8$ (-1.1) & $ 309.7\pm  1.8$ (-0.5) & $ 255.8\pm  1.9$ (-2.4) &  $ 256.3\pm  1.9$ (-1.0) &  $ 236.2\pm  2.1$ (-2.8) &  $ 235.7\pm  1.9$ (-2.0) \\
                       &Q& $ -1.18\pm 0.62$	   & $ -0.01\pm 0.53$	     & $  0.32\pm 0.57$        &  $ -0.93\pm 0.32$	  &  $ -0.34\pm 0.65$	     &  $ -1.25\pm 0.64$	\\
                       &U& $  0.15\pm 0.34$	   & $ -0.90\pm 0.39$	     & $  0.26\pm 0.47$        &  $ -0.28\pm 0.45$	  &  $  1.18\pm 0.72$	     &  $  0.29\pm 0.51$	\\
\hline                    	
\multirow{3}{*}{Cyg A} &I& $ 129.3\pm  1.0$ (-4.1) & $ 108.7\pm  1.0$ (-3.5) & $  79.2\pm  1.0$ (-4.3) &  $  78.1\pm  1.0$ (-3.5) &  $  69.5\pm  0.9$ (-3.8) &  $  67.5\pm  1.0$ (-4.8) \\
                       &Q& $  3.93\pm 0.61$	   & $  1.69\pm 0.64$	     & $ -0.55\pm 0.54$        &  $  0.41\pm 0.45$	  &  $ -1.24\pm 0.66$	     &  $  0.59\pm 0.38$	\\
                       &U& $ -5.95\pm 0.44$	   & $ -4.64\pm 0.39$	     & $ -2.23\pm 0.59$        &  $ -1.60\pm 0.45$	  &  $ -1.52\pm 0.98$	     &  $ -1.26\pm 0.44$	\\
\hline                    	 
\multirow{3}{*}{3C274} &I& $  34.2\pm  0.1$ (-5.3) & $  30.9\pm  0.1$ (-3.8) & $  25.6\pm  0.2$ (-3.1) &  $  25.9\pm  0.2$ (-0.5) &  $  22.3\pm  0.3$ (-8.1) &  $  24.0\pm  0.3$ (+0.2) \\
                       &Q& $ -0.26\pm 0.48$        & $  0.39\pm 0.52$	     & $ -0.54\pm 0.77$        &  $ -0.19\pm 0.38$        &  $  0.45\pm 1.10$	     &  $ -0.24\pm 0.48$ \\
                       &U& $ -0.74\pm 0.44$        & $ -0.81\pm 0.56$	     & $ -2.14\pm 0.72$        &  $ -0.97\pm 0.45$        &  $ -2.63\pm 1.19$	     &  $ -1.35\pm 0.47$ \\
    \hline
    \hline
    \end{tabular}
\end{table*}

%% file: quijote_acknow.tex
We thank the staff of the Teide Observatory for invaluable assistance in the commissioning and operation of QUIJOTE.
The {\it QUIJOTE} experiment is being developed by the Instituto de Astrofisica de Canarias (IAC),
the Instituto de Fisica de Cantabria (IFCA), and the Universities of Cantabria, Manchester and Cambridge.
Partial financial support was provided by the Spanish Ministry of Science and Innovation 
under the projects AYA2007-68058-C03-01, AYA2007-68058-C03-02,
AYA2010-21766-C03-01, AYA2010-21766-C03-02, AYA2014-60438-P,
ESP2015-70646-C2-1-R, AYA2017-84185-P, ESP2017-83921-C2-1-R,
AYA2017-90675-REDC (co-funded with EU FEDER funds),
PGC2018-101814-B-I00, 
PID2019-110610RB-C21, PID2020-120514GB-I00, IACA13-3E-2336, IACA15-BE-3707, EQC2018-004918-P, the Severo Ochoa Programs SEV-2015-0548 and CEX2019-000920-S, the
Maria de Maeztu Program MDM-2017-0765, and by the Consolider-Ingenio project CSD2010-00064 (EPI: Exploring
the Physics of Inflation). We acknowledge support from the ACIISI, Consejeria de Economia, Conocimiento y 
Empleo del Gobierno de Canarias and the European Regional Development Fund (ERDF) under grant with reference ProID2020010108.
This project has received funding from the European Union's Horizon 2020 research and innovation program under
grant agreement number 687312 (RADIOFOREGROUNDS).

%% file: table2_flagged_period1.tex
\begin{table*}
\caption{Fraction of data used in period 1 after applying the flags for the wide survey observations
  with the QUIJOTE MFI instrument. Column 1 indicates the elevation, columns 2 and 3 show the horn and frequency 
(0 for low and 1 for high). Columns 4 and 5 show the percentage of used data in correlated and uncorrelated channels, respectively. 
Columns 6 and 7 show the percentage of flagged data during the post-processing stage, and columns 8 and 9 show the percentage of flagged data due to Sun, Moon and planets (Mars, Venus, Jupiter). Last column indicates the range of dates when each elevation was observed. }
\label{tab:flagged1}
\centering
\begin{tabular}{@{}cccccccccc}
\hline
\hline
Elevation & Horn & Freq & Used c & Used u & Flag1 c & Flag1 u & Flag2 c & Flag2 u & Range of Dates \\
(deg)     &      &      &  (\%) &  (\%) &  (\%) & (\%) & (\%) & (\%) & \\ 
\hline
\hline
  60&  2&  0&   71.7&   73.9&   16.8&   14.1&    1.4&    1.4&     5/2013--3/2014\\
  60&  2&  1&   60.7&   49.3&   29.5&   42.6&    1.4&    1.4&     5/2013--3/2014\\
  60&  3&  0&   33.1&   58.7&   52.8&   16.7&    6.6&    6.6&     5/2013--3/2014\\
  60&  3&  1&   32.8&   62.7&   54.6&   13.8&    6.6&    6.6&     5/2013--3/2014\\
  60&  4&  0&   73.4&   74.2&    6.7&    5.7&    1.5&    1.5&     5/2013--3/2014\\
  60&  4&  1&   56.9&   63.1&   27.8&   19.9&    1.5&    1.5&     5/2013--3/2014\\
  65&  2&  0&   81.6&   84.5&   13.8&   10.7&    2.0&    2.0&     5/2013--3/2014\\
  65&  2&  1&   73.8&   55.8&   22.1&   40.9&    2.0&    2.0&     5/2013--3/2014\\
  65&  3&  0&   39.3&   70.8&   54.7&   18.5&    1.8&    1.8&     5/2013--3/2014\\
  65&  3&  1&   30.9&   72.8&   62.2&   11.5&    1.8&    1.8&     5/2013--3/2014\\
  65&  4&  0&   86.3&   87.1&    4.5&    3.6&    1.9&    1.9&     5/2013--3/2014\\
  65&  4&  1&   70.6&   77.7&   22.0&   14.1&    1.9&    1.9&     5/2013--3/2014\\
\hline
\hline
\end{tabular}
\end{table*}

%% file: table2_flagged_period2.tex
\begin{table*}
\caption{Fraction of data used in period 2 after applying the flags for the wide survey observations
  with the QUIJOTE MFI instrument. Same format as in Table~\ref{tab:flagged1}. }
\label{tab:flagged2}
\centering
\begin{tabular}{@{}cccccccccc}
\hline
\hline
Elevation & Horn & Freq & Used c & Used u & Flag1 c & Flag1 u & Flag2 c & Flag2 u & Range of Dates \\
(deg)     &      &      &  (\%) &  (\%) &  (\%) & (\%) & (\%) & (\%) & \\
\hline
\hline
  30&  2&  0&   57.1&   56.9&   37.3&   37.5&    2.0&    2.0&    8/2014--3/2015\\
  30&  2&  1&   53.1&   46.1&   41.7&   49.4&    2.0&    2.0&    8/2014--3/2015\\
  30&  3&  0&   31.0&   37.9&   64.5&   56.4&    1.7&    1.7&    8/2014--3/2015\\
  30&  3&  1&   29.4&   41.6&   66.8&   53.0&    1.7&    1.7&    8/2014--3/2015\\
  30&  4&  0&   57.4&   58.2&   37.4&   36.5&    1.8&    1.8&    8/2014--3/2015\\
  30&  4&  1&   51.2&   53.0&   44.1&   42.1&    1.8&    1.8&    8/2014--3/2015\\
  40&  2&  0&   49.2&   51.8&   43.2&   40.1&    1.9&    1.9&    8/2014--1/2015\\
  40&  2&  1&   46.7&   38.5&   46.1&   55.5&    1.9&    1.9&    8/2014--1/2015\\
  40&  3&  0&   28.6&   51.5&   65.4&   38.0&    5.2&    5.2&    8/2014--1/2015\\
  40&  3&  1&   22.0&   41.1&   73.5&   50.6&    5.2&    5.2&    8/2014--1/2015\\
  40&  4&  0&   55.1&   56.1&   36.9&   35.8&    2.0&    2.0&    8/2014--1/2015\\
  40&  4&  1&   52.4&   51.7&   40.0&   40.8&    2.0&    2.0&    8/2014--1/2015\\
  50&  2&  0&   64.7&   65.7&   22.7&   21.4&    2.0&    2.0&   8/2014--10/2015\\
  50&  2&  1&   61.2&   60.9&   27.0&   27.4&    2.0&    2.0&   8/2014--10/2015\\
  50&  3&  0&   41.9&   55.4&   47.3&   30.2&    7.0&    7.0&   8/2014--10/2015\\
  50&  3&  1&   35.8&   55.5&   54.3&   29.2&    7.0&    7.0&   8/2014--10/2015\\
  50&  4&  0&   62.4&   62.8&   20.2&   19.7&    2.1&    2.1&   8/2014--10/2015\\
  50&  4&  1&   53.1&   53.1&   31.4&   31.3&    2.1&    2.1&   8/2014--10/2015\\
  60&  2&  0&   58.3&   59.4&   31.7&   30.3&    2.0&    2.0&    6/2014--9/2014\\
  60&  2&  1&   58.6&   30.5&   31.4&   64.3&    2.0&    2.0&    6/2014--9/2014\\
  60&  3&  0&    8.8&   50.4&   87.6&   29.0&    7.0&    7.0&    6/2014--9/2014\\
  60&  3&  1&    0.0&   50.8&  100.0&   29.8&    7.0&    7.0&    6/2014--9/2014\\
  60&  4&  0&   55.5&   54.5&   28.9&   30.1&    2.0&    2.0&    6/2014--9/2014\\
  60&  4&  1&   52.8&   53.0&   32.4&   32.0&    2.0&    2.0&    6/2014--9/2014\\
  65&  2&  0&   73.8&   79.3&   21.4&   15.5&    3.3&    3.3&   8/2014--10/2014\\
  65&  2&  1&   76.0&   44.8&   19.0&   52.5&    3.3&    3.3&   8/2014--10/2014\\
  65&  3&  0&   36.9&   66.4&   56.9&   21.6&    3.2&    3.2&   8/2014--10/2014\\
  65&  3&  1&   36.1&   67.2&   56.3&   17.6&    3.2&    3.2&   8/2014--10/2014\\
  65&  4&  0&   71.6&   76.9&   20.0&   14.2&    3.2&    3.2&   8/2014--10/2014\\
  65&  4&  1&   65.3&   68.4&   27.2&   23.7&    3.2&    3.2&   8/2014--10/2014\\
\hline
\hline
\end{tabular}
\end{table*}

%% file: table2_flagged_period5.tex
\begin{table*}
\caption{Fraction of data used in period 5 after applying the flags for the wide survey observations
  with the QUIJOTE MFI instrument. Same format as in Table~\ref{tab:flagged1}.  }
\label{tab:flagged5}
\centering
\begin{tabular}{@{}cccccccccc}
\hline
\hline
Elevation & Horn & Freq & Used c & Used u & Flag1 c & Flag1 u & Flag2 c & Flag2 u & Range of Dates \\
(deg)     &      &      &  (\%) &  (\%) &  (\%) & (\%) & (\%) & (\%) & \\
\hline
\hline
  40&  2&  0&   57.6&   57.7&   33.9&   33.7&    2.0&    2.0&   8/2016--10/2016\\
  40&  2&  1&   50.9&   49.6&   41.6&   43.1&    2.0&    2.0&   8/2016--10/2016\\
  40&  3&  0&   60.8&   61.4&   27.6&   26.9&    5.1&    5.1&   8/2016--10/2016\\
  40&  3&  1&   46.1&   45.0&   45.2&   46.5&    5.1&    5.1&   8/2016--10/2016\\
  40&  4&  0&   59.5&   59.2&   32.4&   32.7&    2.0&    2.0&   8/2016--10/2016\\
  40&  4&  1&   43.4&   49.5&   50.7&   43.8&    2.0&    2.0&   8/2016--10/2016\\
  50&  2&  0&   65.6&   66.1&   21.6&   21.1&    2.3&    2.3&   8/2016--10/2016\\
  50&  2&  1&   62.0&   61.7&   26.1&   26.5&    2.3&    2.3&   8/2016--10/2016\\
  50&  3&  0&   57.3&   56.6&   28.7&   29.5&    6.9&    6.9&   8/2016--10/2016\\
  50&  3&  1&   57.8&   56.3&   26.8&   28.6&    6.9&    6.9&   8/2016--10/2016\\
  50&  4&  0&   62.5&   62.6&   20.6&   20.5&    2.1&    2.1&   8/2016--10/2016\\
  50&  4&  1&   45.5&   54.0&   41.4&   30.6&    2.1&    2.1&   8/2016--10/2016\\
  60&  2&  0&   79.4&   79.5&    7.1&    6.9&    2.0&    2.0&    8/2016--9/2016\\
  60&  2&  1&   77.0&   76.6&    9.8&   10.3&    2.0&    2.0&    8/2016--9/2016\\
  60&  3&  0&   61.9&   62.3&   13.2&   12.6&    7.0&    7.0&    8/2016--9/2016\\
  60&  3&  1&   67.2&   64.5&    7.3&   11.1&    7.0&    7.0&    8/2016--9/2016\\
  60&  4&  0&   72.7&   73.4&    7.2&    6.3&    1.9&    1.9&    8/2016--9/2016\\
  60&  4&  1&   60.3&   67.9&   23.0&   13.2&    1.9&    1.9&    8/2016--9/2016\\
\hline
\hline
\end{tabular}
\end{table*}

%% file: table2_flagged_period6.tex
\begin{table*}
\caption{Fraction of data used in period 6 after applying the flags for the wide survey observations
  with the QUIJOTE MFI instrument. Same format as in Table~\ref{tab:flagged1}. }
\label{tab:flagged6}
\centering
\begin{tabular}{@{}cccccccccc}
\hline
\hline
Elevation & Horn & Freq & Used c & Used u & Flag1 c & Flag1 u & Flag2 c & Flag2 u & Range of Dates \\
(deg)     &      &      &  (\%) &  (\%) &  (\%) & (\%) & (\%) & (\%) & \\
\hline
\hline
  35&  2&  0&   58.0&   56.6&   33.7&   35.4&    1.9&    1.9& 12/2017--6/2018\\
  35&  2&  1&   47.7&   47.2&   45.5&   46.1&    1.9&    1.9& 12/2017--6/2018\\
  35&  3&  0&   63.5&   62.0&   26.1&   27.8&    4.2&    4.2& 12/2017--6/2018\\
  35&  3&  1&   51.6&   51.4&   39.9&   40.1&    4.2&    4.2& 12/2017--6/2018\\
  35&  4&  0&   61.9&   62.6&   33.0&   32.3&    1.8&    1.8& 12/2017--6/2018\\
  35&  4&  1&   42.2&   24.5&   54.5&   73.5&    1.8&    1.8& 12/2017--6/2018\\
  50&  2&  0&   68.2&   68.3&   18.3&   18.1&    2.8&    2.8&  3/2017--4/2017\\
  50&  2&  1&   60.3&   59.9&   28.0&   28.4&    2.8&    2.8&  3/2017--4/2017\\
  50&  3&  0&   61.9&   61.6&   22.5&   23.0&    7.3&    7.3&  3/2017--4/2017\\
  50&  3&  1&   59.6&   56.6&   24.1&   27.9&    7.3&    7.3&  3/2017--4/2017\\
  50&  4&  0&   67.6&   67.6&   13.7&   13.8&    2.6&    2.6&  3/2017--4/2017\\
  50&  4&  1&   55.4&   52.7&   28.6&   32.1&    2.6&    2.6&  3/2017--4/2017\\
  60&  2&  0&   73.9&   73.9&   14.3&   14.3&    0.7&    0.7& 12/2016--2/2017\\
  60&  2&  1&   72.1&   72.0&   16.4&   16.5&    0.7&    0.7& 12/2016--2/2017\\
  60&  3&  0&   55.6&   55.6&   22.3&   22.3&    5.8&    5.8& 12/2016--2/2017\\
  60&  3&  1&   61.8&   61.6&   15.4&   15.6&    5.8&    5.8& 12/2016--2/2017\\
  60&  4&  0&   68.4&   68.5&   13.4&   13.2&    0.7&    0.7& 12/2016--2/2017\\
  60&  4&  1&   66.6&   65.8&   15.6&   16.6&    0.7&    0.7& 12/2016--2/2017\\
  65&  2&  0&   85.3&   85.2&    8.8&    8.9&    3.0&    3.0&  3/2017--4/2017\\
  65&  2&  1&   83.5&   83.2&   10.8&   11.1&    3.0&    3.0&  3/2017--4/2017\\
  65&  3&  0&   59.5&   59.5&   29.1&   29.0&    3.2&    3.2&  3/2017--4/2017\\
  65&  3&  1&   71.3&   69.3&   12.4&   14.9&    3.2&    3.2&  3/2017--4/2017\\
  65&  4&  0&   81.9&   81.9&    8.2&    8.2&    3.1&    3.1&  3/2017--4/2017\\
  65&  4&  1&   80.2&   76.9&   10.1&   13.8&    3.1&    3.1&  3/2017--4/2017\\
  70&  2&  0&   85.1&   85.1&    9.4&    9.4&    2.2&    2.2&  2/2017--4/2017\\
  70&  2&  1&   84.3&   84.3&   10.4&   10.4&    2.2&    2.2&  2/2017--4/2017\\
  70&  3&  0&   67.4&   67.5&   28.8&   28.7&    2.5&    2.5&  2/2017--4/2017\\
  70&  3&  1&   83.8&   82.6&   10.7&   12.0&    2.5&    2.5&  2/2017--4/2017\\
  70&  4&  0&   86.5&   86.5&    7.8&    7.8&    2.4&    2.4&  2/2017--4/2017\\
  70&  4&  1&   85.6&   84.8&    8.9&    9.7&    2.4&    2.4&  2/2017--4/2017\\
\hline
\hline
\end{tabular}
\end{table*}